 \documentclass [12pt,a4paper,     ]{report} 
\usepackage{graphics}
\usepackage{times}

\DeclareFontFamily{OT1}{times}{}
\DeclareFontShape {OT1}{times}{m }{n }{ <-> ptmr }{}
\DeclareFontShape {OT1}{times}{bx}{n }{ <-> ptmb }{}
\DeclareFontShape {OT1}{times}{m }{it}{ <-> ptmri}{}
\DeclareFontShape {OT1}{times}{bx}{it}{ <-> ptmbi}{}
\usepackage{amsmath}
\usepackage{amsfonts}
\usepackage{amssymb}
\usepackage{latexsym}
\newcommand{\tr}{\pmb{\perp}}         
\newcommand{\lo}{\pmb{\parallel}}     
\newcommand{\DEF}{:=}                 
\newcommand{\CON}{\overline}          
\newcommand{\REA}{\operatorname{Re}}  
\newcommand{\IMA}{\operatorname{Im}}  
\newcommand{\e  }{\operatorname{e}}   
\newcommand{\Oh}{\tfrac{1}{2}}        

\begin{document}

\title{\bf The Physics of high-intensity high-energy\\
                   Particle Beam Propagation\\
              in open Air and outer-space Plasmas}

\author{
         {\bf Andre Gsponer}\\
         {\it Independent Scientific Research Institute}\\ 
         {\it Oxford, OX4 4YS, England}
       }

\date{ISRI-82-04.56 ~~  January 11, 2009} 

\maketitle

\newpage

%

~\\ \vspace{4\baselineskip}

\begin{center}{\Large {\it 1979 --- 2009}} \end{center}

\vspace{1\baselineskip}

\begin{quote} 
\emph{``Nuclear physics has put into the hands of mankind formidable power. We are still struggling with the problem of how to use nuclear energy efficiently and safely, we are rightly alarmed at the accumulation of nuclear weapons of annihilation.  Until mankind has shown that it can deal wisely with nuclear power, it is not prepared for something entirely new.  Until the last nuclear warhead has either been dispatched to outer space or quitely burnt up as fuel in an energy-producing reactor, I would not welcome an entirely new development.  I have often said that I am in favor of supporting high energy physics, provided that the high energy physicists can promise not to produce applicable results within the next twenty-five years.  I am usually not taken seriously when I make such remarks.  I do, however, mean them very seriously.''}
\end{quote}

~\\

\hspace{6.5cm} H.B.G. Casimir,

\hspace{6.5cm} \emph{The 25th Anniversary Ceremony},

\hspace{6.5cm} CERN Courier,

\hspace{6.5cm} September 1979,

\hspace{6.5cm} page 237.

\begin{abstract}

\normalsize

This report is a self-contained and comprehensive review of the physics of propagating pulses of high-intensity high-energy particle beams in pre-existing or self-generated plasmas. Consideration is given to beams of electrons, protons, muons, and their antiparticles, as well as to neutral-hydrogen, positronium, and electron-positron-plasmoid beams.  The first part is a systematic overview of the theory pertaining to propagation, plasma self-generation, energy/current-losses, and stability of such pulses.  The second part reviews the major full-scale propagation experiments which have been carried out, in atmospheric and outer-space plasmas, to assess the validity of theoretical models.  It is found that the data available on these experiments demonstrate that range and stability are in agreement with theory.  In particular, stable self-pinched propagation of high-current charged-particle beams in the atmosphere is possible over distances equal to several Nordsieck lengths.  In order not to be deflected by Earth's magnetic field, electron-beam pulses need to be guided by a pre-formed channel, while proton-beam pulses may under suitable conditions propagate undeflected through both the low- and high-atmosphere.  In ionospheric or outer-space plasmas, very-long-range propagation across Earth's magnetic field requires GeV to TeV electrons or positron beams in order for the transverse deflection to be acceptable, while undeflected propagation is possible for plasmoid beams consisting of co-moving high-energy particle pairs such as electrons and positrons.

\end{abstract}


\tableofcontents

\medskip
\medskip

\listoffigures

\listoftables

\chapter{Introduction}
\label{int:0}

This review deals in a comprehensive manner with theoretical and experimental plasma-physics and accelerator-physics research which have been actively followed or done by the author over the past thirty years, that is starting from approximately the time of the beginning of the construction of the `Advanced Test Accelerator' (ATA) at the Lawrence Livermore National Laboratory, and of the creation of the `Accelerator Technology' (AT) division at the Los Alamos National Laboratory.

In order to understand the relevance of the key experiments, which started to give significant data in the mid-1980s, it is important to understand the theory underlying the numerous plasma-physical effects at work when a high-intensity beam-pulse of particles propagates in a background gas or through the atmosphere.  Since there is no published text book or monograph covering this subject in a systematic manner, Chapters~2 to 7 attempt to synthesize numerous published articles and many informal reports which deal with one or another aspect of this theory.  As shown by the bibliography, this meant studying many papers published over the past fifty years, often dealing only indirectly with the subject, in order to extract the pertinent information necessary to produce a consistent theory.

Chapters 2 and 3 review the envelope equations for neutral and charged particle beams.  Homogeneous and constant background conditions are assumed, and transient effects and instabilities are neglected.  Chapter~5 discusses the transient effects at the head of a charged beam when it is fired into a initially neutral gas such as the atmosphere which is turned into a plasma by the beam.  Chapter~6 examines the main possible instabilities affecting the propagation of such a beam.  Finally, Chapter~7, which is somewhat more tutorial than the previous ones, concludes the first part of the report by an exposition of the theory of plasmoid beam propagation.

In the second part of the report, Chapters 8 to 10, after a discussion of the scientific and technical prospect, the focus is on accelerator facilities and beam propagation experiments which are significant for the purpose of establishing the feasibility of generating  suitable high-intensity high-energy particle beams, and of propagating them in outer-space plasmas or through the atmosphere.  The difficulty, here, is that the openly available data is more of a qualitative than quantitative nature, which is precisely why a thorough understanding of the plasma-physics pertinent to these experiments is so important.  In these chapters, which deal with technologies at the frontier of the state-of-the-art, an effort is made to refer to the implications of the most advanced theoretical ideas and technologies, in order to show how much the possible future engineering-development of high-power particle beam generation/propagation technology still depends on ongoing research.

While we said that there appears to be no published text book or monograph covering the subject of this report in a systematic manner,\footnote{One exception appears to be the lecture notes prepared by Prof.\ K.E.\ Woehler, Department of Physics, Naval Postgraduate School, Monterey, CA, for his course PH 4959 --- Physics of directed energy weapons: Part I, Particle beam weapons (March 1981) 120\,pp; Part II, Particle accelerators (March 1981) 60\,pp. However, the level and scope of these lectures are more elementary and less comprehensive than those of the present report.  Another possible exception is a small review entitled \emph{Propagation of charged particle beams in the atmosphere} presented at the 1987 Particle Accelerators Conference \cite{LAMPE1987-}.\smallskip \\  A recent example illustrating the absence of any published comprehensive coverage of the subject is D.H.\ Whittum's report LBL-27965, \emph{A continuous plasma final focus}, first published in V. Stefan, ed., Nonlinear and Relativistic Effects in Plasmas (AIP, New York, 1992) 387--401, and reissued in 1997 by D.H.\ Whittum as ARDB Technical Note 120 (Accelerator Research Department B, SLAC, October 1997) in order to ``provide a hard-to-find summary of the zeroth-order phenomena that arise when an intense relativistic electron beam is injected into a plasma.''} there is a growing number of excellent books available on the physics of charged particle beams and their applications.  The books by R.C. Davidson \cite{DAVID1974-}, J.D. Lawson \cite{LAWSO1977-}, R.B. Miller \cite{MILLE1982-},  S. Humphries Jr.\ \cite{HUMPH1986-, HUMPH1990-}, M.V. Nezlin \cite{NEZLI1993-}, and M. Reiser \cite{REISE1994-}, are possibly the most useful in the context of the present report.

As the information and research summarized in this report extend over so many years, there are many people to thank for their direct and indirect contributions to it.  While I cannot mention all of them, I wish in particular to thank my former colleagues at CERN (where this work started): Claude Bovey, Steve Geer, Peter Jenni, Pierre Lef\`evre, Claude Metzger, Dieter M\"ohl, Emilio Picasso, Monique et Raymond S\'en\'e, Peter Sonderegger, Charling Tao, Daniel Treille, and Horst Wachsmuth; as well as Frank Barnaby and Bhupendra Jasani at SIPRI (where most of the first part of this report was written); Erik Witalis at FOA; Kosta Tsipis and late Victor Weisskopf at MIT; and last but not least, Jean-Pierre Hurni at ISRI.

%
%

\chapter{Some preliminary definitions and concepts}
\label{def:0}

	A particle beam pulse may be thought of as an ensemble of moving particles whose trajectories constitute a `bundle.'  The diameter of this bundle is small compared with its length, and the trajectories generally make a small angle with the `axis.'  The complete description of the evolution of such a system of interacting particles, especially if they propagate through a gas or plasma, is in general very complicated.  However, in many cases, the beam pulse can be characterized statistically by the RMS (i.e., `root mean squared') values of its radius, length, angular spread, energy spread, etc.  A good description is then provided by the so-called \emph{envelope equations} giving the RMS radius (or length) of the pulse as a function of time or propagation distance.

	In order to simplify the calculations, the usual treatment generally assumes that $\tilde{v}_{\tr}$ and $\tilde{v}_{\lo}$, the RMS values of the random components of the transverse (or perpendicular) and longitudinal (or parallel) velocities, are small compared with the mean longitudinal drift velocity $v=\beta c$ :
\begin{equation}\label{def:1} 
	\tilde{v}_{\tr}^2 \ll \beta^2c^2,     ~~~  ~~~ 
        \tilde{v}_{\lo}^2 \ll \beta^2c^2.
\end{equation}
This is the \emph{paraxial approximation} in which the particle's trajectories deviate only slightly from parallel straight lines.  In such a model, the beam particles's momentum $p$, total energy $W$, and kinetic energy $K$, are slowly changing parameters with the longitudinal distance, i.e.,
\begin{equation}\label{def:2} 
	  p = \gamma\beta mc,  ~~~  ~~~ 
          W =     \gamma mc^2, ~~~  ~~~ 
          K = (\gamma-1) mc^2,
\end{equation}
where $\gamma = 1/\sqrt{1-\beta^2}$ is the Lorentz factor.

	In the case of charged particle beams, we will see in Sec.~\ref{ben:0} that the paraxial approximation is equivalent to the statement that the \emph{effective beam current} $I_E$ generating the electromagnetic self-fields is small compared with the \emph{Alfv\'en current} $I_A$, a characteristic current defined as \cite{ALFVE1939-, MCCOR1982-}
\begin{equation}\label{def:3} 
	I_A = 4\pi \epsilon_0 c^2 \frac{p}{q}
           \approx 17000\frac{m_b}{m_e}\frac{\beta\gamma}{Z} ~ [\text{ampere}]
  \approx  \frac{pc~[\text{e-volt}]}{30 Z} ~ [\text{ampere}],
\end{equation}
where $q = Z |e|$ is the electric charge of the beam's particles, $m_b$ their mass, and $m_e$ the electron mass.

   In the discussion of problems like beam-plasma interaction and stability, the most convenient radial scale is not the RMS radius $\tilde{a} = \sqrt{<r^2>}$, but the \emph{scale radius} $a$ defined such that
\begin{equation}\label{def:4} 
	J_B(0) = \frac{I_B}{\pi a^2},
\end{equation}
where $J_B(r)$ is the areal current density and $I_B$ the total beam current.  This enables to write $I_B$ in terms of the on-axis beam particle density $n_b(0)$, i.e.,
\begin{equation}\label{def:5} 
	I_B = n_b(0) e \beta c \pi a^2,
\end{equation}
so that
\begin{equation}\label{def:6} 
    n_b(0) = \frac{1}{e\beta c} \frac{I_B}{\pi a^2}.
\end{equation}
   $J_b$ and $n_b$ given by equations \eqref{def:4} and \eqref{def:6} have the advantage to be equal to $J_b(r)$ and $n_b(r)$ for a beam with a constant particle density up to a radius $a$, a frequently used approximation.

   Apart from the \emph{volumic beam particle number density} $n_b(r)$, an often use parameter is the \emph{linear beam particle number density} $N_b$
\begin{equation}\label{def:6-1} 
	N_b= \frac{I_B}{e\beta c}.
\end{equation}

   We will use a cylindrical coordinate system with radial distance $r=\sqrt{x^2+y^2}$, azimuthal angle $\theta$, and longitudinal distance $z$.  Often, we will replace the time coordinate $t$ by a variable $\tau \DEF t - z/\beta ct$ (the `time within the pulse') which is zero at the beam head and equal to $\Delta \tau$ (the pulse duration) at the tail-end, or by $\zeta \DEF \beta c \tau = \beta c t - z$ which measures the `distance behind the beam head.'

   We will use the MKS system of units even though most papers in the bibliography use Gaussian units. To go from one to the other system, replace $4\pi e^2$ by $e^2/\epsilon_0$ for MKS.

   In order to avoid ambiguities we will express some of the more important quantities in terms of natural physical units, i.e., in terms of quantities such as the `{classical electron radius}' $r_e=e^2/4\pi\epsilon_0m_e c^2= 2.817 \times 10^{-15}$ m, or the `{electron rest energy}' $m_e c^2= 0.511$ MeV.  In these units, the Alfv\'en current \eqref{def:3} is
\begin{equation}\label{def:7} 
	I_A =  (\frac{ce}{r_e}) \frac{1}{Z} \frac{p}{m_e c}
            =  I_U \frac{m_b}{m_e}\frac{\beta\gamma}{Z},
\end{equation}
which shows that the natural `{unit of current}' is $I_U\DEF ce/r_e = 17.021$ kA, and that the linear density can be written as
\begin{equation}\label{def:7-1} 
          N_b = \frac{\nu_b}{r_e}, \text{~~~ ~~~ so that ~~~ ~~~}
        \nu_b = \frac{I_B}{\beta I_U},
\end{equation}
where the dimensionless number $\nu_b$ is the so-called \emph{Budker parameter} \cite{BUDKE1956A}.

   For a plasma with an {electron number density} $n_e$, the \emph{electron plasma frequency} is then
\begin{equation}\label{def:8} 
	\omega_p = e \sqrt{\frac{1}{\epsilon_0} \frac{n_e}{m_e}  }
                 = c \sqrt{4\pi r_e n_e},
\end{equation}
and the \emph{Debye length}
\begin{equation}\label{def:9} 
	\lambda_D = \sqrt{\frac{1}{4\pi r_e n_e} \frac{kT_e}{m_e c^2}}
                  = \frac{c}{\omega_p} \sqrt{\frac{kT_e}{m_e c^2}},
\end{equation}
where  $T_e$ is the {electron temperature}. 

  In the same spirit we define the \emph{beam plasma frequency} and the \emph{beam Debye length} by the expressions
\begin{equation}\label{def:10} 
	\omega_b  = c \sqrt{4\pi r_e n_b \frac{m_e}{\gamma m_b}}, ~~~  ~~~
        \lambda_B = \frac{c}{\omega_b} \sqrt{\frac{kT_b}{\gamma m_b c^2}},
\end{equation}
where $n_b$ is the beam-particle's number density in the reference frame observing the beam (e.g., the laboratory frame in which the beam current is measured), $m_b$ their mass, and $T_b$ their `temperature' to be defined below.  These are proper \emph{covariant} definitions, which like the covariant definitions of  momentum and energy, see \eqref{def:2}, contain a Lorentz factor $\gamma$ at the right place.

   While problems related to beam stability are best discussed in terms of  densities and frequencies such as $n_b$ and $\omega_b$, those related to beam propagation are generally discussed in terms of $I_B$ and $I_A$.  This implies that there is a frequent need for expressing the same quantity using either formalisms.  For example, according to equation \eqref{def:6} and \eqref{def:10}, the on-axis beam plasma frequency can be written as
\begin{equation}\label{def:11} 
       \omega_b(0) = 2 \frac{c}{a} \sqrt{\frac{I_B}{I_A}}
                   = 2 \frac{c}{a} \sqrt{Z\frac{m_e}{m_b}\frac{\nu}{\gamma}}.
\end{equation}

   A most important concept specifically related to particle beams is that of \emph{emittance}, which can be considered as a measure of the disorder in the  motion of the particles relative to the average motion of the beam \cite{LAWSO1973-}.\footnote{There are several, essentially equivalent, definitions of emittance. In this report we use the so-called `RMS emittance' which is both the most convenient and the most frequently used in our context.}   The emittance is an invariant of the motion if the properties of the beam acceleration and focusing system are linear.  On the other hand, nonlinear effects in the system increase the entropy and thus the emittance \cite{RHEE-1986-}.   In the general case it is necessary to distinguish between the transverse and longitudinal emittances, i.e., $\epsilon_{\tr}$ and $\epsilon_{\lo}$, which will be defined in the first section of the next chapter.  Since an emittance is basically the product of the extent of a spatial distribution along a given direction by an angular spread, i.e., a length times an angle, it is generally measured in units of `meter$\cdot$radian.'

   A concept that is related to beam emittance is that of beam \emph{brightness}, i.e., current divided by angular beam spread, which means that low-emittance implies high-brightness, and vice versa.  Both concepts are often used interchangeably to qualify a high-power directed beam, but we will use only the concept of emittance in the theoretical sections of this report.

   A third concept used to measure the quality of a directed particle beam is that of \emph{temperature}, see  \cite[p.207]{LAWSO1977-} and \cite{LAWSO1973-}. Indeed, if a non-relativistic beam pulse containing $N$ particles is pictured as a Maxwellian gas moving with a longitudinal velocity $\beta c$, it is natural to define a transverse and a longitudinal energy such that the total beam energy is equal to $\mathcal{E} = \mathcal{E}_{\tr} + \mathcal{E}_{\lo}$ with
\begin{equation}\label{def:12} %
	\mathcal{E}_{\tr} = N kT,     ~~~  ~~~ 
        \mathcal{E}_{\lo} = \tfrac{1}{2} N kT.
\end{equation}
However, a real beam is neither a `thermalized' (i.e., a Maxwellian) gas nor plasma so that the definitions of concepts such as temperature, pressure, entropy, etc., are depending upon the actual kinematical distribution functions, as well as of the beam shape and even the position within the beam. Nevertheless, it is possible to give sensible definitions for quantities such as the transverse and the longitudinal temperatures of a beam, and to related them to the respective emittances.  This is because the relation between temperature and pressure is such that a finite emittance can be interpreted as a pressure gradient tending to disperse the beam.

  In particular, one can use the word `temperature' in the sense of the mean kinetic energy spread of the beam, and define the \emph{longitudinal temperature} of a relativistic beam by
\begin{equation}\label{def:13} %
           kT_{\lo} = \delta W =  mc^2 \, \delta\gamma. 
\end{equation}
Using the differential identity $d\gamma = \tfrac{1}{2} \gamma^3 \, d \beta^2$ this can be approximated by
\begin{equation}\label{def:14} %
          kT_{\lo} \approx \tfrac{1}{2}  m \gamma^3 \tilde{v}_{\lo}^2,
\end{equation}
which apart from $m\gamma^3$, the so-called \emph{longitudinal mass}, has the form of the non-relativistic expression of the kinetic energy associated with the RMS longitudinal velocity spread $\tilde{v}_{\lo}$.  However, even for highly relativistic beams, this approximation is very good provided the fractional energy spread $\delta W/W$ is small. 

   The corresponding expression for the \emph{transverse temperature} is
\begin{equation}\label{def:15} %
  kT_{\tr} =  \tfrac{1}{2} m \gamma \tilde{v}_{\tr}^2, 
\end{equation}
where $m\gamma$ is the so-called \emph{transverse mass}. It applies to a beam for which transverse velocities are non-relativistic, again an excellent approximation, because for a highly focused beam with RMS angular spread $\tilde{\alpha}$ one normally has
\begin{equation}\label{def:16} %
     \tilde{v}_{\tr} \approx  \beta c \tilde{\alpha}.
\end{equation}

   Finally, an important concept to define is that of `high-intensity high-energy beam,' which appears in the title of report.  Here we will refer to the \emph{Wassenaar Arrangement}, an international agreement which controls the export of weapons and dual-use goods, that is, goods that can be used both for a military and a civilian purpose.  The lists of equipment, materials and related technologies which are part of the \emph{Arrangement} do not explicitly refer to particle beam weapons.  However, export of ``lasers of sufficient continuous wave or pulsed power to effect destruction similar to the manner of conventional ammunition,'' and ``particle accelerators which project a charged or neutral beam with destructive power'' are restricted in the \emph{Directed energy weapons systems} section of the \emph{Munition list}, document NF(96)DG ML/WP2 (16 March 1996) p.\,46.  Therefore, we will loosely define a \emph{high-energy high-intensity beam}, as a beam with energy and power comparable to those of a few kg of high-explosives, that is (according to the standard defined for nuclear weapons), a few $10^6$~calorie = 4.184 MJ, i.e., about 10~MJ.  Since the total energy in a beam pulse of duration $\Delta \tau$ is
\begin{equation}\label{def:17} %
     \Delta W_{\text{pulse}} = W \frac{I}{e} \Delta \tau,
\end{equation}
such an energy corresponds, for example, to a salvo of ten 10~kA, 1~GeV, 100~ns, endoatmospheric charged-particle beam pulses, or to a single 100~mA, 10~GeV, 10~ms, exoatmospheric neutral-particle beam pulse.

\newpage

\chapter{Particle beam propagation in vacuum or a negligible medium}

\section{Neutral beam in vacuum : ballistic propagation}
\label{neu:0}

In the absence of collisions between beam particles, the individual trajectories of neutral particles propagating in vacuum or a background of negligible density are all straight lines --- at least as long as ranges are sufficiently short for the effect of Earth's gravitational field to be ignored.   Neutral beam propagation is therefore `ballistic,' and the envelope equations can directly be derived from kinematics \cite{EMIGH1972-}.  For a beam pulse with axial symmetry, the RMS radius $\tilde{a}$, and the RMS half-length  $\tilde{\ell}$, are given of  by:\footnote{For an alternate derivation of this equation, which applies to any beam without beam-beam interactions in the absence of external forces, see \cite[p.187]{LAWSO1977-}.}
\begin{equation}\label{neu:1} 
      \tilde{a}''    = \frac{\epsilon_{\tr}^2}{\tilde{a   }^3}, ~~~~  ~~~~
      \tilde{\ell}'' = \frac{\epsilon_{\lo}^2}{\tilde{\ell}^3}.
\end{equation}

	In these envelope equations, the primes denote derivation with respect to the longitudinal coordinate $z$.  The constant $\epsilon_{\tr}$ and $\epsilon_{\lo}$ are the RMS transverse and longitudinal emittances which characterize the random distribution of the particles in the beam.  Specifically
\begin{equation}\label{neu:2} %
 \epsilon_{\tr}^2 = \tilde{a}^2 \frac{(\tilde{v}_{\tr})^2 - 
                                   v^2(\tilde{a}')^2}{v^2},  ~~~~  ~~~~
 \epsilon_{\lo}^2 = \tilde{\ell}^2 \frac{(\tilde{v}_{\lo})^2 - 
                                      v^2(\tilde{\ell}')^2}{v^2}.
\end{equation}
As $\epsilon_{\tr}$ and $\epsilon_{\lo}$ are constants of motion, they can be measured most conveniently at the point where the beam envelope forms a waist.  At such a point $\tilde{a}'=0$ or $\tilde{\ell}'=0$, and
\begin{equation}\label{neu:3} %
    \epsilon_{\tr} = \tilde{a} ~ \tilde{\alpha},  ~~~~  ~~~~
    \epsilon_{\lo} = \tilde{\ell} ~ \frac{\tilde{v}_{\lo}}{v}.
\end{equation}
where, in the paraxial limit, $\tilde{\alpha}$ is the RMS angular spread and $\tilde{v}_{\lo}/v \approx \gamma^{-2} \delta  p/p \approx \beta^{-2}\gamma^{-2} \delta W/W$ the RMS fractional longitudinal velocity spread of the beam.  The emittances $\epsilon_{\tr}$ and $\epsilon_{\lo}$ are thus characteristic of the beam quality: a small transverse emittance corresponds to a well collimated small angular divergence beam, and a small longitudinal emittance to a short pulse with a small energy spread.

  The general solution of equation \eqref{neu:1} for $\tilde{a}(z)$ is obtained by first multiplying it by the derivative $\tilde{a}'= d\tilde{a}/dz$ and integrating, which gives
\begin{equation}\label{neu:4} %
         (\tilde{a}'  )^2
       - (\tilde{a}_0')^2
       = (\frac{\epsilon_{\tr}}{\tilde{a}_0})^2
       - (\frac{\epsilon_{\tr}}{\tilde{a}  })^2.
\end{equation}
This can be solved by integrating $1/\tilde{a}' = dz/d\tilde{a}$, which yields 
\begin{equation}\label{neu:5} %
         \tilde{a}^2  = \tilde{a}_0^2
   + \Bigl( (\tilde{a}_0')^2 +(\frac{\epsilon_{\tr}}{\tilde{a}_0})^2 \Bigr) z^2
              - 2 z\tilde{a}_0 \tilde{a}_0'.
\end{equation}

	For possible applications in space-based beam weapon systems, this general solution has one essential feature: for a given emittance, and at a given target distance $z_{tar}$ from the accelerator generating the beam, the minimum beam spot size is inversely proportional to its initial radius $\tilde{a}_{0}$, i.e.,
\begin{equation}\label{neu:6} %
       \tilde{a}_{min} = z_{tar}\frac{\epsilon_{\tr}}{\tilde{a}_{0}}.
\end{equation}
This minimum is achieved for 
\begin{equation}\label{neu:7} %
                  \tilde{a}_{0}'=\tilde{a}_{0}/z_{tar},
\end{equation}
and is found by minimizing \eqref{neu:5} with respect to $\tilde{a}_{0}'$, the focusing angle of the beam.

  These remarkably simple results show that the critical parameter in focusing the beam is its emittance and not just its angular spread or radius. Similarly, the longitudinal emittance determines the minimum duration of beam pulses, and is thus an essential parameter for compressing the beam into short pulses at the target.

  According to \eqref{neu:6}, in order to focus a neutral beam into a 1~m radius at 1000~km, one would, for example, need a beam with an initial radius of 20 cm and an emittance of $2 \times 10^{-7}$ m$\cdot$rad.  A low-energy accelerator with such an emittance, and a focusing system for such a beam, have been developed at the Los Alamos and Argonne National Laboratories in the United States, as will be reviewed in Chap.~\ref{npb:0}.

\section{Charged beam in vacuum : space-charge-driven expansion}
\label{cha:0}

When a charged particle beam is launched into a vacuum, the beam tends to spread apart under the combined effects of its emittance and of the Coulomb repulsion between like charges, which is particularly strong if the beam energy is low.  However, this space-charge-driven expansion is only one of several effects which make that propagating a non-neutral, or non-neutralized, particle beam in a complete vacuum is very difficult.

For instance, when an unneutralized beam leaves the accelerator, the accelerator becomes charged with the opposite sign, pulling the beam particles back and causing them to decelerate.  Moreover, in the head of a beam pulse, where the beam current is generally a rapidly rising function of time, there are strong transient electromagnetic phenomena which tend to decelerate the particles. Space-charge-driven expansion can therefore be substantially faster in the head than in the body of a beam pulse.  As a result, the particles at the beam front spread apart radially, and the beam erodes becauses particles are continuously lost at the head.  These transient effects are particularly strong if the beam head is nearly flat rather than tapered along its length.  They also exist in the tail of the beam if it consists of a finite-length pulse rather then a semi-infinite one extending from the accelerator port.  All these processes happen simultaneously, which makes their analysis very difficult.  Nevertheless, to understand their main characteristics, it is useful to analyze them separately as will be done in the following sections.  We start with space-charge-driven expansion, first for an infinitely long beam, and then for a finite length beam pulse.

	In the paraxial limit the radial electrostatic force on a particle within a beam at a distance $r$ from the axis is
\begin{equation}\label{cha:1} 
     F_e(r) = 2 \frac{W}{r} \frac{I_B(r)}{I_A},
\end{equation}
where $I_B(r)$ is the total current flowing within the radius $r$.  This outward-directed force is partially compensated by the inward-directed pinch force due to the azimuthal magnetic field generated by the beam current within this radius:
\begin{equation}\label{cha:2} 
     F_m(r) = -2 \beta^2 \frac{W}{r} \frac{I_B(r)}{I_A}.
\end{equation}
As $\beta^2 < 1$, the net force $F_e + F_m$ is outward-directed, and causes the beam to spread apart.  By equating this force to the transverse acceleration force $m\gamma \ddot{r} = m \gamma \beta^2 c^2 r''$, we get the radial equation \cite{LAWSO1958-, LAWSO1959-}, \cite[p.134]{LAWSO1977-}
\begin{equation}\label{cha:3} 
             r'' =  \frac{2}{\beta^2\gamma^2}  \frac{I_B}{I_A} \frac{1}{r}
                 =  \frac{\kappa}{r},
\end{equation}
where $\kappa$ is called the \emph{perveance}, i.e., a measure of the extent to which a beam is influenced by its space-charge.

   The solution of Eq.\eqref{cha:3} can be expressed in terms of Dawson's integral \cite{LAWSO1958-,HARRI1958-}.  For small beam expansions, the approximate solution is \cite{HARRI1958-}
\begin{equation}\label{cha:4} 
             z(a_1 < 2a_0) 
                 =  \frac{a_1}{\sqrt{\kappa}} (1 - a_0^2/a_1^2)^{1/2},
\end{equation}
where $a_1 = r(z)$ and $a_0 = r(0)$, and for large beam expansions \cite[p.319]{ABRAM1968-}
\begin{equation}\label{cha:5} 
             z(a_1 > 2a_0) 
                 =  \frac{a_1}{\sqrt{\kappa}} ( \ln a_1^2/a_0^2)^{-1/2}.
\end{equation}

	While equation \eqref{cha:3} corresponds to the motion of an electron at the edge of a beam ($r=a$), the radial envelope equation combining the effects of the emittance as defined in \cite{EMIGH1972-}, i.e., equation \eqref{neu:1} for a neutral particle beam, with those of the electro-magnetic self-fields for a charged particle beam, is \cite{LEE--1976A}:
\begin{equation}\label{cha:6} 
 \tilde{a}'' = \frac{\epsilon_{\tr}^2}{\tilde{a}^3}
             + \frac{1}{\beta^2\gamma^2} \frac{I_B}{I_A} \frac{1}{\tilde{a}}.
\end{equation}
Compared to \eqref{cha:3}, we see that there is no factor `2' in the perveance term because the radial variable is the RMS radius $\tilde{a}$ rather then $r$. This is typical of the many essentially equivalent beam envelope equations which can be written down, and which differ by numerical factors on the order of 2.

   Neglecting beam emittance, and using Eq.~\eqref{cha:4} or \eqref{cha:5} to get a first estimate, the distance over which a beam has to propagate in order for it radius to double under the effect of its space-charge is approximately given by \cite{LAWSO1958-},  \cite[p.136]{LAWSO1977-}:
\begin{equation}\label{cha:7} 
         z_2 \approx a_0 \beta\gamma \sqrt{\frac{I_A}{I_B}}
         = 2 \gamma \frac{\beta c}{\omega_b},
\end{equation}
where equation \eqref{def:11} has been used to get the second form.  With an initial radius of $a_0 = 20$ cm, a kinetic energy of 500 MeV, and a current of 100 mA, this distance would be about 2600 km for electrons, but only about 5~km for a proton beam with the same characteristics.

  Therefore, if space-charge driven expansion was the only detrimental effect at work, a low-emittance electron beam with energies in the GeV-range could propagate quite far in vacuum.  As can be seen from the Lorentz-$\gamma$ appearing as factor in Eq.~\eqref{cha:7}, this advantage of electron over proton  beams is a relativistic effect, a feature that will reappear in many other occasions.

   This conclusion, as well as the radial expansion equation \eqref{cha:6}, have been obtained for a continuous beam, i.e., for an infinitely long pulse.  What about a beam pulse of radius $a$ and half-length $\ell$~?  In principle the answer is very difficult to obtain unless one considers a simplified model --- for example an ellipsoidal beam pulse with uniform charge density, in which case the self-forces are linear, and the problem is analytically solvable \cite[Sec.5.4.11]{REISE1994-}.

  For a finite beam pulse the self-charge forces lead to beam expansion in both the transverse and the longitudinal directions.  Therefore, the beam current is not constant, but rather decreases as the pulse expands longitudinally.  In the case of an ellipsoidal pulse, while the total number of particles $\mathcal{N}$ within the pulse remains constant, it is related to the maximum beam current (which we write $I_B$) and to $\ell$ according to the equation
\begin{equation}\label{cha:8} %
      \mathcal{N} = \frac{4}{3} I_B \frac{\ell}{e \beta c} .
\end{equation}

   The transverse envelope equation for the ellipsoidal pulse is then \cite[p.449]{REISE1994-}
\begin{equation}\label{cha:9} %
 a'' = \frac{\epsilon_{\tr}^2}{a^3}
        + 2\frac{1}{\beta^2\gamma^2} \frac{I_B}{I_A} \frac{1}{a}
          \Bigl(1 - \frac{1}{2} \frac{g_0}{\gamma^2} \frac{a^2}{\ell^2} \Bigr),
\end{equation}
and the corresponding longitudinal equation is \cite{KHOE-1977-}, \cite[p.449]{REISE1994-}
\begin{equation}\label{cha:10} %
 \ell'' = \frac{\epsilon_{\lo}^2}{\ell^3}
       + 2\frac{1}{\beta^2\gamma^2} \frac{g_0}{\gamma^2}\frac{I_B}{I_A} \frac{1}{\ell}.
\end{equation}

   The space-charge expansion terms in Eqs.~\eqref{cha:9} and \eqref{cha:10} contain a factor 2 as Eq.~\eqref{cha:3} because these equations give the motion of a particle at the edge of the ellipsoid.  Apart from this numerical factor, Eq.~\eqref{cha:9} is identical to Eq.~\eqref{cha:6} in the limit $\ell/a \rightarrow \infty$.  The bracketed factor which disappears in that limit is the coupling between the longitudinal and the transverse expansions.  The longitudinal envelope equation \eqref{cha:10} is also very similar to Eq.~\eqref{cha:6}, apart from the factor ${g_0}/{\gamma^2}$ where the Lorentz-$\gamma$ factor in the denominator implies that for highly relativistic beams the longitudinal expansion is much smaller that the transverse expansion.\footnote{The presence of this $\gamma^2$ factor could be expected because doing the derivation of Eq.~\eqref{cha:3} in the longitudinal rather then transverse direction would require to replace the transverse mass $m\gamma$ by the longitudinal mass $m\gamma^3$.  This simplified derivation would however neither yield the coupling between transverse and longitudinal expansion, nor the geometry factor $g_0$ in Eq.~\eqref{cha:10}.} 
 
  In both Eqs.~\eqref{cha:9} and \eqref{cha:10} the so-called geometry factor $g_0$ is given by the expression  \cite[p.405]{REISE1994-}
\begin{equation}\label{cha:11} %
      g_0 = \frac{2}{\xi^2}
                \Bigl(\frac{1}{2\xi} \ln \frac{1+\xi}{1-\xi} - 1 \Bigr),
\end{equation}
where $\xi$ is given by
\begin{equation}\label{cha:12} %
      \xi = \sqrt{ 1 - {a^2}/{\ell^2}}.
\end{equation}
For nearly spherical beam pulses ($a \approx \ell$) one has $g_0  \approx 2/3$, while for very elongated pulses ($a \ll \ell$) one as
$g_0 \approx \ln (4{\ell^2}/{a^2}) - 2$. Therefore, $g_0$ is always a number of order unity, and for a relativistic beam longitudinal space-charge driven expansion is always much smaller than transverse expansion.

  In the general case, Eqs.~\eqref{cha:9} and \eqref{cha:10} have to be solved simultaneously with Eqs.~\eqref{cha:8} and \eqref{cha:11} because the later two equations contain $\ell$.  But in the relativistic limit this does not affect the general behavior of the solution.  We therefore conclude this section by confirming that, for highly relativistic beams (i.e., $\gamma \gg 1$), space-charge-driven expansion as such is not a major problem for long range propagation.  In fact, while this expansion can be on the same order or larger than emittance-driven expansion in the transverse direction, it will in general be negligible compared to emittance-driven expansion in the longitudinal direction.

\section{Injection of a charged beam into vacuum : limiting current}
\label{inv:0}

As was just seen with space-charge-driven expansion, collective electromagnetic effects are important in high-intensity beams.  Another circumstance, in which the behavior of a high-current beam is different from that of a stream of non-interacting particles, is when a charged beam is launched from a source (e.g., an accelerator of a cathode) such that the total beam energy is initially fully characterized by the beam's current $I_0$ and  particle's kinetic-energy $K_0$, into a vacuum where the total beam energy consists of both kinetic and magnetic energy.

Assuming for simplicity that the beam has a constant current density up to a radius $a$, the magnetic energy density per unit length is readily calculated by integrating the azimuthal magnetic field
\begin{equation}\label{inv:1} 
       B_\theta (r \leq a) = \frac{\mu_o}{2\pi} I_B \frac{r}{a^2}, ~~~  ~~~
       B_\theta (r \geq a) = \frac{\mu_o}{2\pi} I_B \frac{1}{r},
\end{equation}
according to the standard formula
\begin{equation}\label{inv:2} 
        \frac{d W_{\text{mag}}}{d z}
            = \frac{1}{2\mu_0} \int_0^\infty B_\theta^2 ~ 2\pi r ~ dr
            = \frac{1}{2} \frac{d L}{d z} I_B^2,
\end{equation}
which yields the self-inductance per unit length
\begin{equation}\label{inv:3} 
     \frac{d L}{d z} 
   = \frac{\mu_0}{\pi} \bigl(\frac{1}{8} + \frac{1}{4} \ln\frac{b^2}{a^2} \bigr)
   \DEF \frac{\mu_0}{\pi} \mathcal{L}, 
\end{equation}
where the dimensionless inductance $\mathcal{L}$ is introduced for convenience, and the upper integration limit has been set the finite value $r=b$.  This cut-off is necessary to make the integral in \eqref{inv:2} finite, something that is not possible in principle for a beam propagating in free space, but is exact if the beam is sent into an evacuated conducting pipe of radius $b$. 

    On the other hand, the linear beam kinetic energy density is simply
\begin{equation}\label{inv:4} 
     \frac{d W_{\text{kin}}}{d z}
   = \frac{d W_{\text{kin}}}{d t} \frac{1}{\beta c}
   =  (\gamma -1) mc^2 \frac{I_B}{e\beta c},
\end{equation}
where the propagating beam parameters $\beta$, $\gamma$, and $I_B$ are related to the parameters $\beta_0$, $\gamma_0$, and $I_0=I_B\beta_0/\beta$ at injection by the energy rate conservation equation
\begin{equation}\label{inv:5} 
        \frac{d W_{\text{\text{inj}}}}{d t}
     =  \frac{d W_{\text{kin}}}{d t}
     +  \frac{d W_{\text{mag}}}{d t},
\end{equation}
which gives the identity
\begin{equation}\label{inv:6} 
       (\gamma_0 -1)\beta_0
        = (\gamma -1)\beta
        + 2 \mathcal{L} \frac{I_B}{I_U},
\end{equation}
where $I_U \approx 17$~kA is Alfv\'en's current unit.

   Since $I_B=e\beta c n_b \pi a^2$,  equation \eqref{inv:6}  contains three unknowns, and cannot therefore be solved without further hypotheses.  However, three important consequences can be derived from it. First, the kinetic energy of the propagating beam is always less than that of the beam upon injection, i.e., $\gamma < \gamma_0$.  Second, by comparing the two terms on the right hand side of  \eqref{inv:6}, the ratio of kinetic to magnetic energy is very poor for non-relativistic beams, i.e., 
\begin{equation}\label{inv:7} 
           \frac{d W_{\text{kin}}}{d W_{\text{mag}}}
       =   \frac{1}{2\mathcal{L}} \frac{\gamma-1}{\beta} \frac{I_U}{I_B}
   \approx \frac{1}{4\mathcal{L}}                 \beta  \frac{I_U}{I_B},
\end{equation} 
where the approximation corresponds to the limit $\beta \rightarrow 0$.  Finally, precisely because a low-velocity charged beam tends to have most of its energy in magnetic rather than in kinetic energy, there is a maximum critical current for injecting a beam into a vacuum.  An estimate of this limiting current is obtained by comparing the left hand side of \eqref{inv:6} to the magnetic term on the right, and by writing  $I_L$ for the compound $\beta^2I_B$.  This gives
\begin{equation}\label{inv:8} 
    I_L <  \frac{1}{2\mathcal{L}}   (\gamma_0 - 1)\beta_0   I_U
     \approx \frac{1}{4\mathcal{L}}                 \beta_0^3 I_U.
\end{equation} 
A more rigorous derivation, based on Poisson's equation and space-charge considerations, gives \cite{BOGDA1971-,THODE1979-}, \cite[p.90]{MILLE1982-}
\begin{equation}\label{inv:9} 
    I_L =  \frac{1}{4\mathcal{L}}   (\gamma_0^{2/3} - 1)^{3/2}  I_U
   \approx \frac{1}{6\sqrt{3}\mathcal{L}}  \beta_0^3  I_U,
\end{equation} 
which shows that \eqref{inv:8} overestimates $I_L$ by a factor of two in the ultra-relativistic limite, and by a factor of $\approx$2.6 in the non-relativistic limit.

  The conclusion of this section is that the injection of a low-velocity  high-intensity-beam into a vacuum, or extracting such a beam from a cathode in the initial stage of a particle accelerator, is very difficult.  This is illustrated by the factor $\beta_0^3$ in equations \eqref{inv:9}, which comes from that most of the energy goes into magnetic rather than kinetic energy.  The consequence is that generating a proton or heavy-ion beam, as well as accelerating and sending such a beam into vacuum at a relatively low energy, is much more difficult than doing the same with an electron beam.

\section{Inductive head erosion}
\label{ind:0}

At the head of a beam the current is rising and consequently the beam magnetic field as well as all quantities depending on the current are functions of time. This is in particular the case of the magnetic flux through any surface, which therefore by Faraday's law of induction
\begin{equation}\label{ind:1} 
       \oint \vec{E}\cdot\vec{dl}
        = - \frac{\partial}{\partial t}\iint \vec{B}\cdot\vec{dS}
        \DEF - \frac{\partial}{\partial t} \Phi,
\end{equation} 
induces a time varying electric field at the head of the beam.

  To get a good idea of the magnitude and impact of this induced electric field, it is sufficient to first consider a beam with a constant (i.e., independent of  $r$) current-density up to a constant (i.e., independent of $\tau$) radius $a$.  The magnetic field is then azimuthal, with intensity $B_\theta$ given by \eqref{inv:1}, and the flux per unit length along the beam path is 
\begin{equation}\label{ind:2} 
       \frac{d \Phi}{d z} (r \leq a)
     = \frac{\mu_o}{4\pi} I_B(\tau) \frac{r^2}{a^2}, ~~~ ~~  ~~ ~~~
       \frac{d \Phi}{d z} (r \geq a)
     = \frac{\mu_o}{4\pi} I_B(\tau) \ln \frac{r^2}{a^2}.
\end{equation}
The induced electric field is then longitudinal, i.e.,
\begin{equation}\label{ind:3} 
       E_z = - \frac{\mu_o}{\pi} \mathcal{L}(r)
               \frac{\partial}{\partial \tau} I_B(\tau),
\end{equation} 
where
\begin{equation}\label{ind:4} 
       \mathcal{L} (r \leq a)
     = \frac{1}{4} \frac{r^2}{a^2}, ~~~ ~~  ~~ ~~~
       \mathcal{L} (r \geq a)
     = \frac{1}{4} \ln \frac{r^2}{a^2}.
\end{equation}

   Therefore, the induced electric field is longitudinal, concentrated at the head where the current is varying most rapidly, and such that the beam's particles are subject to a drag
\begin{equation}\label{ind:5} 
       \frac{d K}{d z}  = q E_z (r,\tau),
 \end{equation}
which is greater on the edge than in the center of the beam.\footnote{Note that the electric field $<E_z>$ averaged over the beam cross-section satisfies Eq.~\eqref{ind:3} with $<\mathcal{L}>$ equal to $\mathcal{L}$ given by \eqref{inv:3}, because averaging the flux is equivalent to calculating the magnetic energy.}  Thus, particles at the head will have less energy than those in the body of the beam, so that according to Eq.~\eqref{cha:6} space-charge driven expansion will be faster at the beam front, where emittance driven expansion will also be faster because particles losing energy and deflecting from the beam will increase its emittance.  Since particles on the edge are experiencing a greater drag, and those leaving the beam keep experiencing a logarithmically increasing decelerating force,\footnote{Of course, there is a hypothetical cut-off $r=b$, the origin of which is not specified, because we are considering an idealized situation where the charged beam is propagating freely in vacuum.} all processes conspire to erode the beam front which will continuously regress into the pulse.  Simultaneously, the head of the beam takes upon a trumpet-like shape, which after sufficient propagation becomes a  slowly-varying self-similar function of time.

  The details of this erosion phenomenon are obviously very complicated, but the magnitude of its main effect --- a constant decrease of pulse length with propagation distance --- can nevertheless be obtained by an energy conservation argument similar to the one used in the previous section, under the assumption that all beam characteristics are slowly-varying self-similar functions of time.

   We therefore consider a semi-infinite beam pulse launched with initial velocity $\beta$, and with an eroding beam front moving forward at a velocity $\beta_F$.  We then assume that the processes at the beam head are such that the energy-rate conservation equation is
\begin{equation}\label{ind:6} 
        \frac{d W_{\text{inj}}}{d t}
     =  \frac{d W_{\text{kin}}}{d t}
     +  \frac{d W_{\text{loss}}}{d t}
     +  \frac{d W_{\text{mag}}}{d t},
\end{equation}
where
\begin{align}\label{ind:7} 
       \frac{d W_{\text{inj}}}{d t} & =  K \frac{I}{e},   \\
       \frac{d W_{\text{kin}}}{d t} & =  K \frac{I}{e}
                                              \frac{\beta_F}{\beta},  \\
       \frac{d W_{\text{loss}}}{d t}& =  K^*\frac{I}{e}
                                               (1-\frac{\beta_F}{\beta}), \\
       \frac{d W_{\text{mag}}}{d t} & =  \frac{\mu_0}{2\pi}
                                            \mathcal{L} I_E^2 \beta_F c  .
\end{align}
The first term is just the total amount of energy carried by the beam at injection and as it enters the head region. The second term is the kinetic energy moving forward at the beam front velocity $\beta_F$, assuming that most of the forward going particles have lost little energy by inductive drag.  (This is plausible since erosion affects more the particles close to the edge than those near to the center of the beam.  Moreover, consistent with self-similarity, this term is also the net kinetic energy moving forward through the beam head.)  The third term corresponds to the energy lost in the beam head through the erosion process:  The factor $(1-\beta_F/\beta)$ ensures that the number of particles is conserved, and $K^*$ is an effective kinetic energy at which particles are assumed to leave the beam. Finally, the last factor is the magnetic energy associated with the forward going particles, assuming some effective current $I_E < I$ taking into account the lower current-density and the finite rise-time characterizing the head region.

   Equation \eqref{ind:6} becomes then
\begin{equation}\label{ind:8} 
       \frac{\beta \, \beta_F}{\beta - \beta_F}
      = \frac{\mu_0}{2\pi} ec \mathcal{L} \frac{I_E^2}{I} \frac{1}{K-K^*},
\end{equation}
where the left hand side has a simple interpretation: During a time $\Delta t$, the front has moved a distance $\Delta z = \beta_F c \, \Delta t$, while the pulse has eroded by $\Delta x = (\beta -\beta_F) c \, \Delta t$.  Thus, if the pulse has a duration $\Delta \tau$, the beam has completely eroded when $\Delta x = \beta  c \, \Delta \tau$.  Therefore, we can define an \emph{erosion range} by 
\begin{equation}\label{ind:9} 
       z_{\Delta \tau} = \frac{\beta \, \beta_F}{\beta - \beta_F} c \Delta \tau,
\end{equation}
which using \eqref{ind:8} is thus
\begin{equation}\label{ind:10} 
       z_{\Delta \tau} = \frac{2\pi}{\mu_0}
                         \frac{1}{ec\mathcal{L}}
                         \frac{I}{I_E^2}(K-K^*) c\Delta \tau.
\end{equation}

   This remarquably simple result shows that inductive erosion depends essentially on three phenomenological parameters, $K^*$, $I_E$, and $\mathcal{L}$, for which one can make reasonable guesses.  For instance, if we assume that $K^* \approx K/2$ and $I_E \approx I = I_B$, we get
\begin{equation}\label{ind:11} 
       z_{\Delta \tau} \approx  \frac{\gamma-1}{\gamma\beta}
                                \frac{1}{4\mathcal{L}}
                                \frac{I_A}{I_B} c\Delta \tau.
\end{equation}
Then, assuming hypothetically $\mathcal{L}\approx 1$, and taking an electron beam characteristic of a low-atmospheric system, i.e., $K=1$~GeV, $I_B=10$~kA, and $\Delta \tau=10$~ns, we get $z_{\Delta \tau}=2.4$~km; and for an electron beam characteristic of a high-atmospheric system, i.e., $K>10$~GeV, $I_B<1$~kA, and $\Delta \tau>100$~ns, we get $z_{\Delta \tau}>2400$~km.

   The assumption $I_E \approx I_B$ is plausible when the beam rise time is short.  In the opposite case of a slowly rising beam one would have $I_E \ll I_B$, and \eqref{ind:10} shows that the beam erosion range may become very large.

\section{Injection into outer-space : spacecraft charging}
\label{ino:0}

{\bf NB:} This section should be expanded to discuss pulsed beams, as well as systems in which neutralized, or both-signs, beam pulses are launched.

   When a beam is sent into an infinite vacuum there is no return path for the beam current, and the beam's particles are slowed down under the action of the longitudinal restoring force due to charging up of the spacecraft launching the beam.  This effect puts a limit to the range of charged particles propagating in vacuum, which corresponds to the distance they can travel until they must turn back to the spacecraft.  This distance can be estimate by solving Poisson's equation for the potential energy, which leads to a non-linear equation that is not easy to integrate, except in the ultrarelativistic limit where
\begin{equation}\label{ino:1} 
      z_{max} \approx a_0  \sqrt{\frac{I_A}{2I_B}}.
\end{equation}
In the general case, a good approximation is provided by \cite{ROSIN1971-, WALLI1975-}\footnote{Note that this equation is formally similar to Eq.~\eqref{inv:9}. This is because the problem of launching a beam from an isolated platform is directly related to that of `limiting currents,' as can be seen by replacing $\pi R$ by $a_0$ in equation (3.53) of reference \cite[p.91]{MILLE1982-}.}
\begin{equation}\label{ino:2} 
      z_{max} \approx a_0 \frac{(\gamma^{2/3} -1)^{3/4}}{\gamma^{1/2}}
                          \sqrt{\frac{I_A}{2I_B}}.
\end{equation}

     If we take the same example as in Sec.\ref{cha:0}, where we compared a beam of protons to one of  electrons, each with a kinetic energy $K=500$~MeV, a current $I_B=100$~mA, and an initial radius $a_0 = 20$ cm, one finds that both beams would have a range of less than 1~km according to Eq.~\eqref{ino:2}.  In fact, looking at Eq.~\eqref{cha:7} we see that the Lorentz-$\gamma$ factor which gave a significant advantage to electron beams is missing in Eq.~\eqref{ino:1}, so that for relativistic beams the only significant parameter is $I_A \propto K$.

   This calculation alone would tend to rule out the use of charged particle streams as possible long-range beam weapons in vacuum.  However, as will be examined in the next chapter, if such beams were injected into outer-space, which is in fact a dilute plasma and not a vacuum (see Table~\ref{tab:atm}), the situation can be very different. This is illustrated, in particular, by studies of spacecraft charging in which the ability of the ionospheric plasma to return the current propagated by the beam back to the accelerator platform is taken into account.  In that case, provided the beams are of relatively low-energy (i.e., on the order of eV to keV) and low-intensity (i.e., such that beam current densities are comparable to ambient plasma densities), the ionosphere is able to return the current and spacecraft charging remains negligible \cite{OKUDA1987-}.  For high-intensity beams the situation is less clear, and either beams of neutral particles, or neutralized beams comprising an equal number of positive and negative charges, are preferable.  The later possibility will be discussed in Chap.~\ref{plb:0}, dedicated to plasmoid beams.

  This brings us to the end of this chapter, in which we have decomposed the complex processes which impede the propagation of non-neutral beams in a strict vacuum into hypothetically independent subprocesses.  To investigate how particle beams may propagate in outer-space, that is to understand how these subprocesses are modified by the outer-space environment, requires an understanding of the physics of propagating charged beams in a background gas or plasma of non-negligible density, which is the subject of the following chapters.

\chapter{Particle beam propagation in a gas or plasma}

\section{Beam charge and current neutralization}
\label{bcn:0}

Charged particle beams for use as directed energy weapons are in general injected either into the atmosphere for ground-  and aircraft-based systems, or into a plasma for space-based systems.  In the case of atmospheric systems, the beam will enter an initially neutral atmosphere and, by ionizing the air, turn it into a plasma along the beam path.  In the case of outer-space systems, the plasma will be the ionosphere for near-Earth orbiting systems, or the interstellar environment.  In all cases, the transient phenomena occurring at the head of a beam pulse are complicated.  We will thus concentrate first on infinitely long beams in the paraxial approximation.

   Similarly, we will start by assuming that the plasma can be described by a single fluid equation of motion, i.e., that the plasma ions are at rest, and that the equation of motion for the plasma electron fluid can be written 
\begin{equation}\label{bcn:1} 
  (\frac{\partial}{\partial t} + \vec{v_e}\cdot\vec{\nabla}) \vec{v_e}
   = - n_e\frac{e}{m_e}(\vec{E}+\vec{v_e}\times\vec{B})
     - \frac{\vec{\nabla} p_e}{m_e}  - \nu n_e \vec{v_e},
\end{equation}
where $e=|e|$, $m_e$, $n_e$, $p_e$, and $\vec{v}_e$ are the electron charge, mass, number density, pressure, and velocity respectively; $|\vec{v}_e| \ll c$, and $\nu$ is the effective (momentum transfer) collision frequency.    In order to make first order analytical calculations tractable we neglect the pressure term and the non-linear terms (for a justification see Appendix A of Ref.~\cite{KUPPE1973A}).  Thus
\begin{equation}\label{bcn:2} 
  \frac{\partial}{\partial t} \vec{v_e}
   = - n_e\frac{e}{m_e} \vec{E}  - \nu n_e \vec{v_e}.
\end{equation}
The plasma current density is by definition
\begin{equation}\label{bcn:3} 
     \vec{J}_P \DEF - e n_e \vec{v_e}.
\end{equation}
Therefore, the plasma equation of motion can be written \cite{COX--1970-}
\begin{equation}\label{bcn:4} 
       (\frac{\partial}{\partial t} + \nu) \vec{J}_P
                   = \epsilon_0 \omega_p^2 \vec{E},
\end{equation}
where we have introduced the plasma frequency \eqref{def:8}, and assumed that $n_e$ does not depend explicitly on time. (The first order disturbance of $n_e$ can be derived from the continuity equation, $\partial n_e/\partial t + \vec{\nabla}\cdot(n_e \vec{v_e}) =0$, if desired.)  This equation has two important limiting cases:
\begin{itemize}

\item \emph{Collisionless plasma}: $\nu \rightarrow 0$,
\begin{equation}\label{bcn:5} 
     \frac{\partial}{\partial t} \vec{J}_P = \epsilon_0 \omega_p^2 \vec{E}.
\end{equation}
This case corresponds to beam propagation in a tenuous gas or plasma, such as the high-atmosphere.  It corresponds also to the early stages of beam plasma interaction, i.e., to times that are small compared to $(a\omega_p/c)^2/\nu$, provided $\nu \ll \omega_p$, where $a$ is the beam radius \cite{KUPPE1973A}.

\item \emph{Collisional plasma}: $\nu \rightarrow \infty$,
\begin{equation}\label{bcn:6} 
     \vec{J}_P = \epsilon_0 \frac{\omega_p^2}{\nu} \vec{E}
         = \sigma  \vec{E}.
\end{equation}
This case corresponds to beam propagation in a dense gas or plasma, such as the low-atmosphere.

\end{itemize}
 Equation \eqref{bcn:6} is known as `Ohm's law' and
\begin{equation}\label{bcn:7} 
      \sigma \DEF  \frac{e^2}{\nu} \frac{n_e}{m_e}
             = \epsilon_0 \frac{\omega_p^2}{\nu},
\end{equation}
is by definition the \emph{scalar electric conductivity}.  If the magnetic force term is retained when going from \eqref{bcn:1} to \eqref{bcn:2}, the electric conductivity becomes a tensor \cite{SPITZ1956-}, \cite[p.500]{JACKS1975-}. This leads to various forms of `generalized Ohm's laws,' e.g.,
\begin{equation}\label{bcn:8} 
   \vec{J}_P = \sigma_{\lo} \vec{E}_{\lo}
             + \sigma_{\tr} \vec{E}_{\tr}
             + \sigma_{H} \frac{\vec{B}\times \vec{E}}{|\vec{B}|},
\end{equation}
where $\sigma_{\lo}$, $\sigma_{\tr}$, and $\sigma_{H}$ are called the \emph{longitudinal} (or \emph{direct}), \emph{transverse} (or \emph{Pedersen}), and \emph{Hall conductivities;} and $\vec{E}_{\lo}$ and $\vec{E}_{\tr}$ are the electric field components parallel and perpendicular to $\vec{B}$, respectively.

	In order to appreciate the relative importance of the plasma background for beam propagation, it is sufficient to compare the plasma density $n_e$ to the beam particle density $n_b$.  On the beam axis, according to \eqref{def:6},
\begin{equation}\label{bcn:9} 
    n_b(0) = \frac{1}{e\beta c} \frac{I_B}{\pi a^2}.
\end{equation}

	For example, in the case of a relativistic beam with $I_B= 100$ mA and $a = 20$ cm, $n_b = 1.6 \times 10^4$ cm$^{-3}$.  In comparison, in the ionosphere\footnote{The ionosphere is the region above $\approx 50$~km altitude where ultra-violet light from the Sun ionizes atoms and molecules in the atmosphere, to give free electrons and ions, albeit embedded in a dense neutral atmosphere except at great altitudes. The ionosphere is therefore a partially ionized plasma.   The magnetosphere is the region above $\approx 150$~km where the convection of the plasma and the motion of the free electrons and ions are predominantly controlled by the geomagnetic field.} between an altitude of 100 to 2000 km, the electron number density $n_e$, as well as the atomic number density $n_a$ of the residual atmosphere, are on the same order or larger (see Table~\ref{tab:atm}).  Therefore, plasma density effects cannot be ignored.

\begin{table}
\begin{center}
\hskip 0.0cm \begin{tabular}{|r|r|c|c|r|r|}
\hline
\multicolumn{6}{|c|}{\raisebox{+0.2em}{{\bf  \rule{0mm}{6mm} Typical ionospheric and magnetospheric data}}} \\ 
\hline
\raisebox{+0.2em}{altitude} \rule{0mm}{6mm} & 
\raisebox{+0.2em}{T~~                  } &  
\raisebox{+0.2em}{atomic density       } & 
\raisebox{+0.2em}{electron density     } & 
\raisebox{+0.2em}{$H_{\text{equator}}$ } & 
\raisebox{+0.2em}{$H_{\text{pole~~}}$  } \\ 
 
\rule{0mm}{0mm} [km]~~~ & 
[$^{\rm{o}}$K]          &
$n_a$ [m$^{-3}$]        & 
$n_e$ [m$^{-3}$]        &
[gauss]                 &
[gauss]                 \\
\hline
\rule{0mm}{5mm}    0~~~ &  300 & $5 \times 10^{25}$ &                 0  & 0.31~~  &  0.62~~ \\
                 100~~~ &  200 & $5 \times 10^{18}$ & $1 \times 10^{11}$ & 0.30~~  &  0.59~~ \\
                 300~~~ & 1000 & $5 \times 10^{15}$ & $5 \times 10^{12}$ & 0.27~~  &  0.54~~ \\
                1000~~~ & 1000 & $5 \times 10^{12}$ & $1 \times 10^{11}$ & 0.20~~  &  0.40~~ \\
                3000~~~ & 1100 & $5 \times 10^{10}$ & $1 \times 10^{10}$ & 0.10~~  &  0.19~~ \\

\hline
\end{tabular}
\end{center}
\caption[Typical ionospheric and magnetospheric data]{Typical time-averaged ionospheric and magnetospheric data as a function of altitude above ground.  The temperature and the atomic and free-electron densities are taken from reference \cite[Fig.1]{DUDEN1981-}. The horizontal and vertical components of the geomagnetic field at the magnetic equator and pole are from reference \cite[Sec.5h4]{SUGIU1972-}.  Throughout the report we take a representative value of 0.5 gauss, i.e.,  $5\times 10^{-5}$ tesla, for Earth's magnetic field.}    \label{tab:atm}
\end{table}

  Similary, the magnetic self-field on the edge of this beam, estimated by the elementary formula  $B = 2 \times 10^{-7} I/a$, is only 0.001 gauss, much less than the geomagnetic field (see Table~\ref{tab:atm}).  Therefore the effect of Earth's magnetic field on a charged beam, and on the magneto-plasma effects associated to its interaction with the atmosphere, cannot be ignored, as will be further discussed in Sec.~\ref{mag:0}.

	The first effect of a plasma background is that, because of quasi-neutrality, the excess charge locally introduced by the passage of the beam will tend to be neutralized.  This happens on a time-scale $\tau_q$ which, for a collisional plasma, is set by the plasma conductivity, i.e., $\tau_q \approx \tau_e$, where $\tau_e$ is by definition the \emph{electric diffusion time} of the plasma
\begin{equation}\label{bcn:10} 
            \tau_e \DEF \frac{\epsilon_0}{\sigma}
                      = \frac{1}{4\pi r_e c^2}\frac{\nu}{n_e}.
\end{equation}
In the case of a low-density gas or collisionless plasma the equation of motion is \eqref{bcn:5} provided the expelled plasma electrons are non-relativistic.  The charge neutralization time-scale is then set by the plasma frequency, i.e., $\tau_q \approx 1/\omega_p$.  The lowest possible charge neutralization time is obtained when the plasma electrons move radially in or out of the beam at relativistic velocities.  In that case $\tau_q \approx a/c$. See \cite[p.531]{OLSON1973-}.  In all cases the electric space-charge repulsion force \eqref{cha:1} will be reduced to a smaller value, i.e.,
\begin{equation}\label{bcn:11} 
     F_e(r) = 2 (1-f_e)\frac{W}{r} \frac{I_B(r)}{I_A},
\end{equation}
where $f_e$ is by definition the \emph{charge (or electric) neutralization fraction}, and where the difference $s_e = 1-f_e$ is called the \emph{electric screening factor}.  Equation \eqref{bcn:11} assumes that $f_e$ is independent of $r$, which is generally a good approximation since $f_e \approx 1$ in most practical situations.  When $n_i(r) \propto n_b(r)$ and $n_e(r) \propto n_b(r)$, e.g., when the beam and plasma distributions are similar, one has identically $f_e = |n_i-n_e|/n_b  = |N_i-N_e|/N_b$.

	The actual physical processes involved at the microscopic level are different depending upon the sign of the electric charge of the beam particles. To compare the main features of these two possibilities let us consider the case of a beam propagating in a preexisting quasineutral plasma, i.e., such that $n_e \approx n_i$ in the absence of the beam.  For an electron or antiproton beam, the electric field will quickly expel the plasma electrons and charge neutralization will be provided by the positive ions left within the beam.  Assuming these ions not to be able to move significantly during the passage of the beam pulse, the maximum charge neutralization fraction will be $f_e  = n_i/n_b < 1$ when $n_i < n_b$.  On the other hand, for a positron or proton beam, electrons from the plasma surrounding the beam will be attracted into the beam region.  The neutralizing fraction can be one even when $n_i < n_b$, which means that a positive beam can charge neutralize more easily than a negative beam.  This gives a significant advantage to positive beams compared to negative beams in some applications, especially for neutralization phenomena at the beam head \cite{LOTOV1996-}, and for non-relativistic positive ion beams \cite{KAGAN2001-}.  The fact that the charge neutralizing particles have very different masses implies that transient phenomena and stability conditions can be different depending on the sign of the beam particles.  However, in dense plasma, i.e., when $n_e \approx n_i \gg n_b$, most properties will be similar.  In particular, $f_e$ will essentially be one for beam pulses of duration longer than $\tau_e$ and radius  smaller than $c/\tau_e$.

	The second major effect of a plasma background is that a return current can flow through the plasma.  This current $I_P$ is driven by $E_z$, the longitudinal electric field induced by the variation of the effective beam current.  In a positive beam $E_z$ accelerates the plasma electrons forwards and ahead of the  beam pulse, and in a negative beam backwards into the beam pulse.  From Faraday's law of induction, Eq.~\eqref{ind:1},
\begin{equation}\label{bcn:12} 
   E_z = - \frac{\partial}{\partial t} \frac{\Delta \Phi}{\Delta z}
       = -\frac{1}{\pi\epsilon_0 c^2}
          \frac{\partial}{\partial t} (\mathcal{L}I_N),
\end{equation}
where $\Phi$ is the $\theta$-component of the magnetic flux, $\mathcal{L}$ a dimensionless inductance, and
\begin{equation}\label{bcn:13} %
                 I_N \DEF I_B + I_P,
\end{equation}
the \emph{net current} driving the magnetic self-fields \cite{BRIGG1974-}.

    For axially symmetric beams\footnote{$\mathcal{L} = \tfrac{1}{4} (\tfrac{1}{2}+\ln\tfrac{b^2}{a^2})$ for a beam with a constant density up to a radius $a$,  and $\mathcal{L} = \tfrac{1}{4} \ln(1+\tfrac{b^2}{a^2})$ for a beam with a Bennett density profile, Eq.~\eqref{ben:21}. }
\begin{equation}\label{bcn:14} 
       \mathcal{L} \approx \frac{1}{4} \ln \frac{b^2}{a^2},
\end{equation}
where $b$ is the maximum radius out to which the plasma background is significantly affected by the beam.  For instance, in beam-generated plasmas, $b$ is normally determined by the extent of induced breakdown around the beam head.  Typically, $b/a \approx 10$, i.e., $\mathcal{L} \approx 1$.  At the boundary of this region the conductivity becomes too small to ensure quasineutrality.  The charge imbalance from the beam is conducted to this surface, which is thus the path along which the beam current not neutralized by the plasma current is returned to the accelerator.

    From Ohm's law, Eq.~\eqref{bcn:5}, the longitudinal component of the plasma current is then
\begin{equation}\label{bcn:15} 
       I_P = \pi a^2 \sigma E_z
           = -\tau_m \frac{\partial}{\partial t} (\mathcal{L}I_N),
\end{equation}
where
\begin{equation}\label{bcn:16} 
      \tau_m \DEF \frac{\sigma}{\epsilon_0} \frac{a^2}{c^2} 
                = \mu_0 \sigma a^2
                = 4\pi r_e a^2 \frac{n_e}{\nu},
\end{equation}
is by definition the \emph{magnetic diffusion time} of the plasma.  From the sign in \eqref{bcn:15} it turns out that the effect of the plasma current is to decrease the magnetic field generated by the beam current, and thus to reduce the magnetic pinch force.  This is usually written as
\begin{equation}\label{bcn:17} 
     F_m(r) = -2 (1-f_m)\beta^2 \frac{W}{r} \frac{I_B(r)}{I_A},
\end{equation}
where $f_m \DEF -I_P/I_B$ is the \emph{current (or magnetic) neutralization fraction}, and where the difference $s_m = 1-f_m$ is called the \emph{magnetic screening factor}.\footnote{This assumes that $f_m$ is independent of $r$, which is generally not the case, but nevertheless a reasonable approximation when $f_m \approx 0$.}

   Because $E_z$ is a function of the beam current and shape variations, the plasma current, and thus $f_m$, will be largest at the beam head.  As in the case of charge neutralization, full current neutralization is easier to be achieved for a positive than a negative beam pulse \cite{LOTOV1996-,KAGAN2001-}.  This is because a positive beam can attract surrounding plasma electrons and create a forwards moving column which effectively neutralizes the beam charge and current even if the plasma density is low.  A negative beam, on the other hand, has to continuously expel plasma electrons in order to enable the plasma ions to charge neutralize the beam.  These electrons are concentrated in a narrow layer surrounding the beam where the electric field $E_z$ is less strong.  Their acceleration backwards (possibly together with plasma electrons from further away from the beam) in order to form the return current is therefore less efficient than in the case of a positive beam.

   If the plasma conductivity $\sigma$ is taken as a constant, and if some reasonable assumptions are made, it is possible to calculate $I_N$ as a function of the time $\tau$ measured from the beam front.  For example, for a pulse with an infinitely fast rise-time and a flat radial profile it is possible to derive complicated analytic expressions for $I_P$ by solving Eq.~\eqref{bcn:4} in combination with Maxwell's equations \cite{COX--1970-, HAMME1970-}.  But if the radial profile is approximated by a zeroth-order Bessel-function (which is much more realistic than a flat profile, and gives closed expressions for the fields), one finds the simple result \cite[p.147]{MILLE1982-}
\begin{equation}\label{bcn:18} %
     I_N = I_B \Bigl(1 - \exp(-\frac{\tau}{\mathcal{L}\tau_m})
                       + \exp(-\frac{\tau}{           \tau_e}) \Bigr).
\end{equation}
This expression shows that the beam current is quickly neutralized on a time scale given be $\tau_e$, but that $f_m$ decreases to zero within the pulse on a time scale set by the magnetic diffusion time.\footnote{This will be discussed in more details in Chap.~5. See also reference \cite{CARY-1980-}.}

  In the collisionless limit it is possible to derive a remarkably simple expression for the plasma current by combining equations \eqref{bcn:5} and \eqref{bcn:12}, i.e.,
\begin{equation}\label{bcn:19} 
    \frac{\partial}{\partial t} I_P
       = -\frac{a^2}{c^2} \omega_p^2
          \frac{\partial}{\partial t} \mathcal{L}(I_B + I_P).
\end{equation}

Therefore, provided $I_B$ is a varying function of time, $n_b < n_e$, and $I_B(0)=I_P(0)=0$,
\begin{equation}\label{bcn:20} 
               f_m =  \frac{\mathcal{L}\omega_p^2}
                           {\mathcal{L}\omega_p^2 + c^2/a^2}.
\end{equation}

From this expression, which was derived here from a one-dimensional model, one can calculate the magnetic screening factor $s_m $, and see that it is in good agreement with the two-dimensional analytical calculations done in Ref.~\cite{KUPPE1973A}, where for a beam with a Gaussian radial profile it is found that $\mathcal{L}=\sqrt{\e/8} \approx 0.58$.  This expression can also be compared to the detailed computer calculations of Ref.~\cite{LOTOV1996-}, in which the non-linear terms neglected when going from \eqref{bcn:1} to \eqref{bcn:2} are retained, and in which the plasma electrons are allowed to be relativistic. It can then be seen that, as a function of the dimensionless parameter $(a\omega_p/c)^2$, expression \eqref{bcn:20} underestimates current neutralization for positive beams, and overestimates it for negative beams \cite[Fig.7]{LOTOV1996-}.

   A two-dimensional generalization of Eq.~\eqref{bcn:19}, valid in the collisional and collisionless cases, is obtained by operating on Eq.~\eqref{bcn:4} with $\vec{\nabla}\times\vec{\nabla}\times$, using Maxwell's equations, and neglecting displacement currents, 
\begin{equation}\label{bcn:21} 
    (\frac{\partial}{\partial t} + \nu) \vec{\nabla}^2 J_P
       =  \frac{1}{c^2} \omega_p^2
          \frac{\partial}{\partial t}(J_B + J_P).
\end{equation}
Approximating the Laplacian by $-2/a^2$, and the currents by $\pi a^2$ times their corresponding current densities,  Eq.~\eqref{bcn:19} is recovered with $\mathcal{L}=0.5$ when $\nu=0$.

   From equation \eqref{bcn:21} it is clear that for $(a\omega_p/c)^2 \gg 1$ and short times we have $J_P+J_B\ \approx 0$, so that in this limit we have nearly full current neutralization, in agreement with expression \eqref{bcn:20}.  This condition can be written $a \gg \lambda_e$, where $\lambda_e = c/\omega_p$ is the so-called \emph{electromagnetic skin depth}, which is also the thickness of the sheath near the edge of the beam to which the net current is confined \cite{HAMME1970-}.

\section{Charged beam in a plasma : the Bennett pinch}
\label{ben:0}

   We now consider an infinitely long beam in the paraxial approximation and assume that the charge and current neutralization fractions $f_e$ and $f_m$ are given and independent of $r$ and $\tau$.  Such a model could be a first approximation to the body a pulse of the kind considered at the end of the previous section, between the times $\tau_e$ and $\tau_m$.   The total radial electromagntic force on a beam particle is then the sum of Eqs.~\eqref{bcn:11} and \eqref{bcn:17}
\begin{equation}\label{ben:1} 
          F_{em} = \Bigl( (1-f_e) - (1-f_m)\beta^2 \Bigr)
                   \frac{2}{r} \frac{W}{I_A} I_B(r),
\end{equation}
and the envelope equation including the plasma effects is therefore \cite{LEE--1976A}
\begin{equation}\label{ben:2} 
    \tilde{a}'' + \frac{I_E}{I_A}\frac{1}{\tilde{a}} 
                = \frac{\epsilon_{\tr}^2}{\tilde{a}^3},
\end{equation}
where $I_E$ is the \emph{effective current} associated with the total electromagnetic force on the beam's particles,
\begin{equation}\label{ben:3} 
   I_E \DEF I_B \Bigl( (1-f_m) -  \frac{1}{\beta^2} (1-f_e) \Bigr).
\end{equation}
When $f_e=f_m=0$, i.e., in a vacuum where $I_E = -I_B\beta^{-2}\gamma^{-2}$, equation \eqref{ben:3} is obviously equivalent to \eqref{cha:6}.

    The effective current $I_E$ contains charge imbalance as well as true beam and plasma currents.\footnote{The effective current $I_E$ should not be confused with the net current $I_N$, defined by Eq.~\eqref{bcn:13}, which is only equal to $I_E$ when $f_e=1$.}  It can have both signs, and the forces can either tend to separate or, on the contrary, to pinch the beam.  When $I_E$ is positive, a stationary solution with $\tilde{a}'' = \tilde{a}' = 0$ is possible.  When $f_m=0$, as is readily seen from Eq.~\eqref{ben:3}, this imposes the so-called \emph{Budker condition} \cite{BUDKE1956A}, i.e., $f_e > 1/\gamma^2$. The corresponding solution is called \emph{Bennett pinch}, and in that case \eqref{ben:2} gives the relation \cite{BENNE1934-,BENNE1955-,LAWSO1958-,LAWSO1959-}
\begin{equation}\label{ben:4} 
      \tilde{a} = \epsilon_{\tr} \sqrt{\frac{I_A}{I_E}} \DEF a_B.
\end{equation}
From \eqref{ben:3} we see that $a_B$, the \emph{Bennett pinch radius} defined by \eqref{ben:4}, is minimum for $f_e=1$ and $f_m=0$.  In that case $I_E=I_B$, the pinch force is maximum, and the beam is fully pinched.  The Bennett pinch solution exists, however, only in the paraxial limit \eqref{def:1}.  This can be seen from \eqref{ben:4} and \eqref{neu:2} which imply
\begin{equation}\label{ben:5} 
      \frac{I_E}{I_A}   <  \frac{\tilde{v}_{\tr}^2}{v^2}  \ll 1.
\end{equation}

  The Bennett pinch radius \eqref{ben:4} can be rewritten in a number of ways.  In particular, its relation to plasma physical parameters can be clarified by considering a beam with constant current density out to a radius $a$.  Introducing the beam plasma frequency \eqref{def:10} calculated for the effective beam particle density $n_b$ obtained by using the current $I_E$ rather than $I_B$ in \eqref{bcn:9}, we get
\begin{equation}\label{ben:6} 
      a_B = 2 \frac{\epsilon_{\tr}}{a} \frac{c}{\omega_b}.
\end{equation}
Using now the relation between the transverse emittance \eqref{neu:3} and the transverse velocity spread \eqref{def:16} we can introduce an effective beam Debye length according to \eqref{def:10}.  This gives
\begin{equation}\label{ben:7} 
      a_B = 2 \sqrt{2} \frac{1}{\beta} \lambda_{D\tr}.
\end{equation}
Writing the Bennett pinch radius in this way shows that a pinched relativistic beam is in fact a poor plasma \cite[p.41]{DAVID1974-}.  Indeed, it is only for a nonrelativistic beam such that $a_B \gg \lambda_D$ that the quasineutrality condition of ordinary plasma physics is satisfied.

\begin{figure}
\begin{center}
\resizebox{8cm}{!}{ \includegraphics{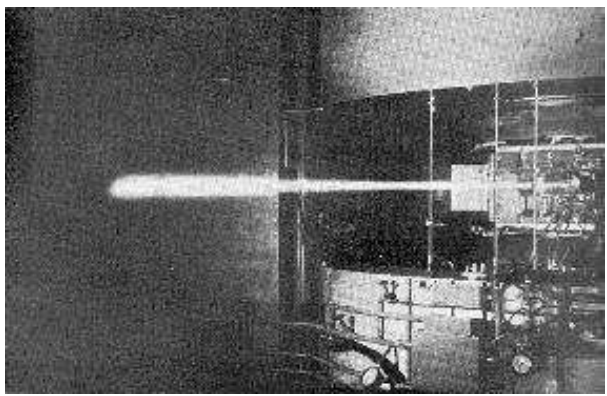}}
\caption[The pinch effect]{\emph{The pinch effect.} The photograph, taken in the early 1950s at the Argonne cyclotron (near Chicago), shows the glow produced when sending the full deuteron beam into the atmosphere.  Because of ionization the air near the beam is turned into a plasma which keeps the beam from expanding radially under the effect of Coulomb repulsion between like-charged particles: This is the `Bennett pinch effect,' first described by Willard H. Bennett in 1934 to explain focusing effects and breakdown in the residual gas of high-voltage electronic tubes, and later applied to the propagation of interstellar and interplanetary self-focussed beams of particles, such as proton streams traveling from the Sun towards the Earth.  As the beam loses energy and intensity because of interactions with atmospheric nitrogen and oxygen nuclei, plasma generation becomes less efficient and the beam progressively expands: This is the `Nordsieck effect,' after the name of Arnold Nordsieck who is generally credited for having first explained this expansion.  Ultimately, when the plasma effects become to weak to pinch the beam, it breaks-up.  This happens at a propagation distance on the order of the so-called `Nordsieck length.'  
\label{fig:pin}}
\end{center}
\end{figure}

	In a Bennett pinch, the beam particles perform harmonic motion around the beam axis.  The angular frequency $\omega_\beta(r)$ of the rotation is a function of $r$ and is called the \emph{betatron frequency}.\footnote{The betatron frequency $\omega_\beta(r)$ should not be confused with the beam plasma frequence $\omega_b(r)$, defined by \eqref{def:10}.} By equating the net force $F_e+F_m$ to the centrifugal force $\gamma m \omega_\beta^2 r$, and by averaging over the beam current density, one finds from \eqref{bcn:11} and \eqref{bcn:17} that the mean azimuthal velocity is given by
\begin{equation}\label{ben:8} 
 \tilde{v}_{\theta}^2  = \CON{\omega_\beta^2 r^2} = \beta^2 c^2 \frac{I_E}{I_A},
\end{equation}
which is independent of the beam profile.  In the general case, the betatron frequency  $\omega_\beta^2$ is distributed between zero and a maximum, the on-axis betatron frequency
\begin{equation}\label{ben:9} 
      \omega_{\beta m}^2 = \omega_\beta^2(0)
                         = 2\beta^2\frac{c^2}{a^2}\frac{I_E}{I_A},
\end{equation}
which is also independent of the beam profile.  In the special case of a beam with a constant current density out to a radius $a$, the betatron frequency is constant and equal to the maximum given by \eqref{ben:9}.

A quantitity directly related to the betatron frequency in the \emph{betatron wavelength}, whose minimum value
\begin{equation}\label{ben:20} 
      \lambda_{\beta m} = 2 \pi \frac{\beta c}{\omega_\beta}
                        = 2 \pi a \sqrt{ \frac{I_A}{2I_E} },
\end{equation}
enables to rewrite the paraxial limit condition \eqref{ben:5} as $a \ll \lambda_{\beta m}$, which may be taken as the postulate definining a beam such that the transverse velocity is much less than the longitudinal velocity \cite{LAWSO1958-}.

	For propagating self-pinched beams, the most natural equilibrium density profiles are those corresponding to a Maxwellian (i.e., Gaussian) transverse energy distribution \cite{BENFO1971-}.  Possible equilibria include filamentary and hollow current flows along the axis, with or without a return current \cite{IVANO1970-,BENFO1970-,KUPPE1973B,GRIGO1978-,HUBBA1988-}, but the simplest practical example is the so-called \emph{Bennett distribution} \cite{BENNE1934-,BENNE1955-}
\begin{equation}\label{ben:21} 
     J_B(r) = \frac{I_B}{\pi a^2} \bigr(1 + \frac{r^2}{a^2}\bigl)^{-2}.
\end{equation}
The RMS radius of this distribution diverges logarithmically.  However, both theory and experiment \cite{LEE--1976B,BRIGG1976-} that indicate a Bennett profile are not valid for $r \gg a$.  In practice, the current profile is often considered to be truncated at $r=2a$.  This yields $\tilde{a}= 1.006~a$.  In a beam with the Bennett profile \eqref{ben:21}, the betatron frequency $\omega_\beta^2$ is distributed between zero and the maximum given by \eqref{ben:9} according to
\begin{equation}\label{ben:22} 
 \omega_\beta^2(r) = \omega_{\beta m}^2 \bigr(1 + \frac{r^2}{a^2}\bigl)^{-1}.
\end{equation}

   The Bennett equilibrium is a particular case of Vlasov equilibria characterized by a constant axial macroscopic velocity for the beam particles \cite{DAVID1974-, GRATR1978-,DAVID1979-}.  Indeed, it is possible to impart an angular momentum to a beam by launching it from a source immersed in a magnetic field, which imparts  a component of angular velocity to the particles when they leave the field \cite[p.138]{LAWSO1977-}.   As will be seen in the discussion of beam propagation stability, an outwards centrifugal force can partially balance the inwards magnetic self-force, and therefore decrease the growth of filamentation instabilities.

   The Bennett pinch is important in many areas of science and technology.  In particular, it is important for understanding interplanetary particle streams \cite{ALFVE1939-, BENNE1955-} and for studying the ionosphere with beams launched from rockets \cite{GOUGH1980-} or the space shuttle \cite{KIWAM1977-, NEUPE1982-}.  It has many applications in thermonuclear fusion research, especially as a means for confining plasmas in devices such as the `Z-pinch.'  In this context the pinch condition is generally presented in the original form given by Bennett.  This form is obtained by first using \eqref{neu:3} to rewrite  \eqref{ben:4} as
\begin{equation}\label{ben:23} 
      \tilde{\alpha}^2 I_A \leqq I_E,
\end{equation}
where the equal sign has been replaced by the symbol `$\leqq$' to stress that Bennett's condition is actually a criterion for a beam to be self-focusing, i.e., to be able to pinch down (or expand) until the equilibrium implied be \eqref{ben:4} is reached.  Then, assuming a beam with constant current density out to a radius $a$ so that equation \eqref{bcn:10} can be used, one multiplies \eqref{ben:23} on both sides by $I_B= e\beta c \pi a^2 n_b$ to get Bennett's original form \cite[p.1589]{BENNE1955-}
\begin{equation}\label{ben:24} 
      2 N_B kT_{\tr} \leqq \frac{\mu_0}{4\pi} I_E I_B.
\end{equation}
In this formulation of the pinch condition the parameter $N_B=\pi a^2 n_b$ is the number of beam particles per unit length, and the right-hand side is generally written $I_B^2$ because for a fully pinched beam $I_E=I_B$.  This expression can also be written in the form
\begin{equation}\label{ben:25} 
          N_B kT_{\tr} \leqq \frac{1}{2} \mathcal{L}_B I_B^2,
\end{equation}
where $\mathcal{L}_B$ can be interpreted as a self-inductance per unit length so that the condition \eqref{ben:25} expresses the equality of two linear energy densities.

   In thermonuclear research and plasma physics a frequently used alternative expression of the pinch condition is obtained by dividing both sides of \eqref{ben:24} by $\pi a^2$ so that after introducing the magnetic field at the surface of the beam, i.e., $B(a)=\mu_0 I_B/2\pi a$, and a relative permeability $\mu_r = I_B / I_E$,  one gets
\begin{equation}\label{ben:26} 
      n_b kT_{\tr} \leqq \frac{1}{2\mu_0\mu_r} B^2(a).
\end{equation}
This remarkable expression simply means that in a Bennett equilibrium the outwards transversal thermal pressure of the beam's particles is equal to the inwards magnetic pressure at the beam's edge \cite[Sec.10.5]{JACKS1975-}.

    The seven equivalent forms of the Bennett pinch condition given in this section, equations $(\ref{ben:4},\ref{ben:6},\ref{ben:7},\ref{ben:23},\ref{ben:24},\ref{ben:25}, \text{and~} \ref{ben:26})$, illustrate the diversity of perspectives which can be used to discuss magnetic pinch phenomena, et partly explain the difficulty or reading and relating the numerous studies which have been published on this subject.\footnote{While equations \eqref{ben:24} to \eqref{ben:26} were here derived from \eqref{ben:4} assuming a beam with constant current density out to a radius $a$, they are valid for any beam profile $n_b(r)$.}

    The Bennett pinch has also been envisaged as a means for accumulating high energy particles in large rings in outer space.  Such rings could be used to store energy, or electrons to generate synchrotron radiation or free-electron laser optical beams \cite{SALTE1978-}.

     Finally, for endo-atmospheric beam weapons, the pinch effect provides the means for radially confining charged particle beams.  Its significance  is that when a beam is injected in a gas or plasma sufficiently dense to suppress the effect of the space-charge repulsion, the beam may pinch down to a minimum constant radius and propagate over large distances.  In particular, because the plasma background provides a means for carrying the return current, the range is no more strictly limited as it was in the case of charged beams in vacuum.  The limit to the propagation distance will now be set by scattering, energy loss, instabilities, etc., as will be seen in Sec.~\ref{nor:0}.

	The fact that the net charge transported by a beam pulse traveling through a plasma is equal to the charge of the beam itself, even when the beam is launched from a ground plane, is not an obvious result.  It is, however, correct, even when the plasma is generated within the pulse by beam-gas interactions \cite{LEE--1977A}.

	The Bennett pinch existence condition $I_E > 0$ requires Budker's condition $f_e > 1/\gamma^2$ to be satisfied when $f_m=0$.  For high-energy beams, i.e., $\gamma > 10$, this condition is easily satisfied, even in very low density plasmas.  This allows the transport of beams with current densities higher than the plasma density by a factor $\gamma^2$.  For low-energy beams, a more stringent condition is set by the paraxiality requirement, Eq.~\eqref{ben:5}, which is equivalent to the statement that the transverse velocity $v_{\tr}$ should be much less than the longitudinal velocity $v_{\lo}$.  To make this more precise, let us take Bennett's pinch condition in the form \eqref{ben:23}, in which we make the approximation $\tilde{\alpha} \approx \tilde{v}_{\tr} / \tilde{v}_{\lo}$, i.e.,
\begin{equation}\label{ben:27} 
      \frac{\tilde{v}_{\tr}^2}{\tilde{v}_{\lo}^2}  \approx \frac{I_E}{I_A}.
\end{equation}
For a monoenergetic beam we can write
\begin{equation}\label{ben:28} 
      \tilde{v}_{\tr}^2 + \tilde{v}_{\lo}^2 = \beta^2 c^2.
\end{equation}
Combining these two equation yields
\begin{equation}\label{ben:29} 
      \tilde{v}_{\lo} \approx \frac{\beta c}{\sqrt{1+ I_E/I_A}},
\end{equation}
which implies that for a non-relativistic pinched-beam the mean longitudinal propagation velocity can be substantially less than the mean particle's velocity $\tilde{v} \approx \beta c$.  For example, let us assume $\beta = 0.82$, which corresponds to a kinetic energy of 400~keV or 732~MeV, and to an Alfv\'en current of 25~kA or 46~MA, for an electron or a proton beam, respectively.   Then, if we further assume $I_E = 10$~kA, we find that $\tilde{\beta}_{\lo} \approx 0.69 < \beta$ for the electron beam, while $\tilde{\beta}_{\lo} \approx \beta$ for the proton beam.  Because of the lower mass of electrons, the paraxial approximation is therefore more difficult to satisfy for an electron beam than for an equal velocity non-relativistic proton or heavy-ion beam.

   The first laboratory experiments demonstrating stable propagation of a high power electron beam as a Bennett pinch through air were performed in 1965 in the United States \cite{GRAYB1966-}.  In these experiments  a 2.5~MeV, 17~kA, 20~ns beam pulse propagated over 3~m, with a loss of one-half in current density and total current, at a drift-tube pressure of 0.3~torr.  Subsequent experiments confirmed these results, showing that they agreed with theory, and demonstrating that considerable damage can be inflicted by such a beam on a thick metal target \cite[Fig.3]{ROBER1968-}.  Similar experiments were later performed in the Soviet Union \cite{RUDAK1972-, WALLI1975-}. 

   An overview of these early experiments is given in reference \cite{LINK-1968-}, together with excellent photographs illustrating the four characteristic behaviors of a 3~MeV, 50~kA, 30~ns electron beam injected in a 50~cm long beam chamber filled with air at various pressures, as summarized in Table~\ref{tab:elb}.

\begin{table}
\begin{center}
\hskip 0.0cm \begin{tabular}{|r|c|c|c|c|} 		\hline
\multicolumn{5}{|c|}{\raisebox{+0.2em}{{\bf  \rule{0mm}{6mm} Behavior of 3~MeV, 50~kA, 30~ns electron beam in air at various pressures}}} \\ 
\hline
\raisebox{+0.2em}{pressure [torr]}\rule{0mm}{6mm} & \raisebox{+0.2em}{observed behavior of beam} & \raisebox{+0.2em}{$f_e$} &   \raisebox{+0.2em}{$f_m$} & \raisebox{+0.2em}{force on beam electrons}  \\  
\hline
\rule{0mm}{5mm} $10^{-3}$~~~~~ & beam blows up            & 0 & 0 &   $1-\beta^2 \approx 0.02$ \\
                $10^{-1}$~~~~~ & beam pinches maximaly    & 1 & 0 &   $ -\beta^2 \approx -1.0$ \\
                  1~~~~~~~~~~~ & beam drifts force-free   & 1 & 1 &   $          \approx  0.0$ \\
                760~~~~~~~~~~~ & beam pinches and expands & 1 & 0 &   $ -\beta^2 \approx -1.0$ \\
\hline
\end{tabular}
\end{center}
\caption[Behavior of a 3~MeV, 50~kA, 30~ns electron beam in air]{At a fraction of a torr the pinch force is maximum and the beam diameter is a few millimeters: This is the `ion-focused regime.' At about 1 torr both the beam charge and current are neutralized and the beam drifts with nearly zero force.  At higher pressures the beam pinches again but expands under the effect of multiple scattering in air until it finally breaks up. \cite{LINK-1968-}.}    \label{tab:elb}
\end{table}

    According to these experiments, stable propagation conditions with $f_e=f_m=1$ exists at a pressure of about 1 torr in air, the current neutralization being provided by slowly counterstreaming plasma electrons.  The fact that this configuration is stable would make it attractive, in principle, for propagating a beam in a reduced density atmosphere or channel.  However, the beam is not pinched, and would thus spread apart because of collisions with the air molecules. 

  At a somewhat lower (generally sub-torr) pressure, where $f_e=1$ or even $f_e>1$, while $f_m\approx 0$, a more suitable propagation regime exists.  This propagation mode has been extensively studied in the United States \cite[p.10--11]{BEAL-1972-}, \cite{BRIGG1977-}, and in the Soviet Union \cite{DIDEN1976A, DIDEN1977-}, as well as in other countries, one of the earliest experiments being a Japanese-Dutch collaboration \cite{YAMAG1982-}.  In this regime the plasma electron are expelled from the beam so that $f_e=1$ and the ions in the resulting plasma channel produce a force which opposes that of the beam's space-charge --- hence the terminology \emph{ion-focused regime} (IFR) \cite{BRIGG1981-, BUCHA1987-, SWANE1993-}.  Additional to this electrostatic force is the self-magnetic pinching force which is not canceled by the current since $f_m\approx 0$.  Consequently the beam is fully pinched to the minimum radius consistent with the Bennett pinch condition \eqref{ben:4}.  For air, an estimate for the critical pressure for IFR propagation is given by the expression \cite{SMITH1985-}
\begin{equation}\label{ben:30} 
                  p\text{[torr]} < \frac{I_B~\text{[kA]}}{10~\text{[kA]}}
                                   \frac{1}{\tilde{a}~\text{[cm]}}.
\end{equation}

   The practicality of this regime for propagating electron beams over large distances in a low pressure atmosphere was demonstrated in the early 1970s \cite{BEAL-1972-, FLEIS1975-}.  For beams such as those just considered above, the pinch effect would provide an ultimate theoretical range of many thousands of kilometers.  However, the practical use of such beams in outer space will, in fact, be limited by the effect on them of Earth's magnetic field, as will be seen in Sec.~\ref{mag:0}.  Also, IFR propagation is more efficient in a preionized plasma background than in a neutral gas where the beam itself is used to create ionization \cite{SMITH1986-}.  This will be discussed in Sec.~\ref{dgc:0}, were propagation in prepared channels will be considered.

   On the other hand, in a full density atmosphere, the main obstacle to long range propagation is the collisions of the beam's particle with the air molecules, which result in a loss of beam energy and intensity, as well as to an increase of the beam emittance, as a function of propagation distance.  Because of the quantitative importance of these effects, which lead to the need of kA-intensity beams with several tens rather than just a few MeV-energies, the possibility of experimentally demonstrating the feasibility of using the pinch effect to propagate a charged-particle beam over more than a fraction of a meter in open air had to wait for the construction of large scale facilities at several laboratories \cite{MOIR-1981-,BARLE1981A}.  These experiments, as well as the related physics of the expansion of a beam propagating in a background gas of non-negligible density, will be examined in Sec.~\ref{nor:0}.

\section{Effect of internal forces : cohesion and coupling}
\label{coh:0}

   Propagation as a Bennett pinch allows a beam pulse to propagate without expanding from space-charge repulsion.  This means that the pinch force is somewhat similar to the molecular attraction by which the particles of a body are united to form a mass of liquid or solid.  The pinch force can therefore be interpreted as a \emph{cohesion} force, although of a much smaller strength than typical molecular forces.

   Moreover, when a beam pulse propagates through a gas, complex interactions at the head of the beam create a plasma and electromagnetic-field environment which greatly affect the subsequent parts of the pulse.  This is because, in general, the pinch force acting on a given beam slice results from the distribution of plasma charge and net current established by preceding beam slices.  Consequently, there is a causal relationship between the cohesion forces acting on subsequent beam segments, called \emph{coupling}, which implies for instance that the motion of the head of a beam determines to a large extent the motion of the remainder of the pulse.\footnote{Similarly, the electromagnetic fields generated by the beam's and plasma's charge and currents are causally connected.  There is therefore energy exchange between segments in beam pulses, as well as wake fields effects, that must be considered in the general case \cite{UHM--1991-}.}


   Thus, when external forces are acting on a beam, the existence of coupling has the important consequence that the behavior of a given beam slice is not determined by just the action of the external forces on that slice, but by that of a combination of the external with the coupling forces, which can either amplify, attenuate, or even compensate for the effect of the external forces.  In particular, at the head of a beam (where coupling forces are not yet established, and where the pulse is generaly expanding because of its self-charge) external forces will have a maximal effect, leading to deflection, tearing, and enhanced erosion in the direction of the external force.  But, behind the head, as soon as the coupling force becomes stronger than the external force, the beam will hold together.  This point, in the neck region of the beam where the restoring force is equal to the external force, is called the \emph{guiding point}.

   Coupling forces are therefore essential elements in the calculation of the net effect of external forces, and will be taken into account in Secs.~\ref{mag:0} on the effect of Earth's magnetic field and \ref{dgc:0} on the deflection and guiding by interfaces and channels; as well as in Secs.~\ref{dgc:0} on beam conditioning and \ref{mas:0} on mastering and damping beam instabilities.  Coupling forces are also essential in the study of beam stability, Chap.~\ref{sta:00}, where the notion of coupling is implicit in the way a beam responds to external perturbations.

   In practice, the analysis of coupling is complicated because there are both electric and magnetic forces, and because the coupling forces result from the mutual interactions of the beam charge and current distributions with those of the plasma conductivity and current, which all vary with propagation distance and from head to tail within the beam pulse.  Nevertheless, a first idea is obtained by considering these interactions for a sufficently thin beam slice, and by assuming that all distributions are axially symmetric and similar, and for ease of calculation well approximated by a Gaussian particle density profile, i.e.,
\begin{equation}\label{coh:1} %
     n(r) = \frac{N}{\pi a^2} \exp\bigl( -\frac{r^2}{a^2}\bigr),
\end{equation}
where $N$ is the linear particle density.  This distribution is normalized so that $\int d^2 r \, n(r) = N$ and has the advantage that the parameter $a$ is equal to both the scale radius and the RMS radius, i.e., $a = \tilde{a}$.

  We therefore need to calculate the interactions of a two Gaussian distributions, for example the electrostatic force between a beam of charge density $q_a n_a$ and radius $a$, and a non-neutral background plasma column of density $q_b n_b$ and radius $b$, with $a \neq b$ in general.  The radial electric field due to the beam is easily calculated
\begin{equation}\label{coh:2} %
     E_a(r) = \frac{1}{4\pi \epsilon_0} \frac {2}{r} q_a \int d^2 r \, n_a(r)
            = \frac{1}{4\pi \epsilon_0} \frac {2}{r} q_a N_a
               \Bigl( 1 - \exp \bigl( -\frac{r^2}{a^2}\bigr) \Bigr).
\end{equation}
The total force, projected on a transverse axis, between the beam and the background plasma is then obtained by multiplying this field by the plasma charge distribution and integrating
\begin{equation}\label{coh:3} %
   F_{\tr}(\vec{r}_a,\vec{r}_b) =  \int d^2 r \, 
              E_a(\vec{r}_a) \, q_b n_b(\vec{r}_b) \cos \theta,
\end{equation}
where $\vec{r}_a$ and $\vec{r}_b$ are the position of the center of the two distributions.  If these centers coincide the force is zero by symmetry.  However, if the two distributions are displaced by a transverse distance $\vec{x} = \vec{r}_a - \vec{r}_b$, there is a non-zero force 
\begin{multline}\label{coh:4} %
   F_{\tr}(x) =  
              \frac{q_a q_b}{4\pi \epsilon_0} \frac {2 N_a N_b}{\pi b^2}
              \int_0^{2\pi} d\theta \int_0^\infty dr \, r \,
              \frac{1 - \exp ( -{r^2}/{a^2} )}{r} \\
              \times
              \exp \Bigl( -\frac{r^2 + x^2 - 2 r x \cos \theta}{b^2}\Bigr)
              \cos \theta.
\end{multline}
This double integral yields a Bessel function which can be simplified in the end to give
\begin{equation}\label{coh:5} %
   F_{\tr}(x) =  \frac{q_a q_b}{4\pi \epsilon_0} \frac {2 N_a N_b}{x}
                \Bigl(1 - \exp \bigl( -\frac{x^2}{a^2+b^2} \bigr)  \Bigr).
\end{equation}
In the limit of large separation this expression gives the well known formula for the force between two charged wires, and in the limit $x^2 \ll a^2+b^2$ it gives the dipolar force between two slightly offset distributions of charge, i.e., 
\begin{equation}\label{coh:6} %
   F_{\tr}(x) \approx  \frac{q_a q_b}{4\pi \epsilon_0} 2 N_a N_b
                       \frac{x}{a^2+b^2}.
\end{equation}
Had we chosen distributions with a radial profile different from a Gaussian, we would have obtained expressions different from Eq.~\eqref{coh:5}, but all with the same limit when $x \rightarrow \infty$, and with  a limit differing from Eq.~\eqref{coh:6} only by a numerical factor of order unity (e.g., $2/3$ in the case of a Bennett profile) when $x \rightarrow 0$.  Moreover, had we made a multipolar expansion of the integrand in Eq.~\eqref{coh:4} before integrating, we would have obtained Eq.~\eqref{coh:6} as the first non zero term in the expansion, which is why it is qualified as dipolar. 

   We now turn to a specific application and consider a beam of current $I_B = e\beta c N_B$ propagating in a background such that the charge and current neutralization fractions equal $f_e = N_P/N_B$ and $f_m= -I_P/I_B$.  Since our intent is to understand the main features of coupling we focus on a thin beam segment labeled by the variable $\tau$. If the beam is not subject to any external forces and is perfectly aligned along a straight path all beam and plasma distributions are coaxial, so that the centroids of all charge and current distributions within a segment coincide.  On the other hand, if the beam is subject to external forces and moves along a bent trajectory, or if the beam tilts and makes an angle with the direction of propagation, the beam and plasma axes do not coincide anymore so that the corresponding beam and plasma distributions within a segment are displaced.\footnote{This is also the case if the beam performs an oscillatory motion about its main direction of propagation as a result of some perturbation, as will be seen in the study of beam instabilities.}  Consequently, when considering the cohesion forces acting on a beam segment, and when the analysis is restricted to electromagnetic forces (i.e., gravitational and centrifugal forces are considered separately), it is necessary to distinguish between three different centroids, which projected on the $x$ axis correspond to the following displacements from a common origin:
\begin{itemize}

\item $x_B$ : the \emph{beam centroid}, i.e., the centroid of the distributions $J_B(r)$ and $Q_B(r)=J_B(r)/\beta c$ of the beam current and charge densities;

\item $x_P$ : the \emph{plasma centroid}, i.e., the centroid of the plasma conductivity distribution which coincides with those of the plasma charge and current distribution ($x_P = x_B$ when the beam and plasma are not separated);

\item $x_N$ : the \emph{net current centroid}, i.e., the centroid of the net current distribution, which can be approximated by
\begin{equation}\label{coh:7} %
   x_N = \frac{I_B x_B + I_P x_P}{I_B + I_P} = \frac{x_B - f_m x_P}{1-f_m},
\end{equation}
($x_N=x_B$ when the beam and plasma are not separated).

\end{itemize}

   Let us now suppose that the beam centroid $x_B$ is suddenly displaced relative to the centroid $x_P$ of the electrostatic anti-pinch force by a small distance $x =x_B-x_P$, and calculate the electrostatic restoring force using Eq.~\eqref{coh:5}.  Dividing by $N_B$, the electrostatic coupling force per beam particles is then
\begin{equation}\label{coh:8} %
   F_{ce}(z,\tau) \approx  -\frac{2e}{4\pi \epsilon_0} f_e 
                      \frac{I_B}{\beta c} \, \frac{x_B-x_P}{a^2+b^2},
\end{equation}
where we have introduced the variable $z$ and $\tau$ as arguments to recall that all parameters on the right are possibly functions of the propagation distance and of the position of the beam slice within the pulse.

  Similarly, by analogy to the derivations of Eqs.~\eqref{coh:5} and \eqref{coh:8}, the magnetic coupling force can be calculated by considering a sudden displacement of the beam current centroid $x_B$ relative to the centroid $x_N$ of the net beam current $I_N=(1-f_m)I_B$.  This gives, as long as $f_m < 1$ so that $I_B$ and $I_N$ flow in the same direction, the magnetic restoring force
\begin{equation}\label{coh:9} %
   F_{cm}(z,\tau) \approx  -\frac{2e}{4\pi \epsilon_0} (1-f_m)
                      \beta \frac{I_B}{c} \, \frac{x_B-x_N}{a^2+b^2}.
\end{equation}
In deriving this expression we have assumed that the net current distribution,  just like previously the plasma charge distribution, is `frozen' for an instantaneous beam displacement. This is, because of Maxwell's equations, there is no immediate change in these distributions for such a displacement.

   The coupling forces \eqref{coh:8} and \eqref{coh:9} are both attractive and of similar strength when $f_e \approx 1$ and $f_m \approx 0$.  By comparison with Eq.~\eqref{bcn:17} it is seen that this strength becomes equal to the magnetic pinch force when $x$ approaches the beam radius $a$. The coupling forces are therefore strong and couple the beam longitudinally, causing the body to follow the head, because they persist for some time as the plasma retards their decay.  In first approximation, this decay is described by the relaxation equations
\begin{equation}\label{coh:10} %
   \frac{\partial}{\partial \tau} x_P =  \frac{x_B-x_P}{\tau_{ed}},
\end{equation}
and
\begin{equation}\label{coh:11} %
  \frac{\partial}{\partial \tau} x_N =  \frac{x_B-x_N}{\tau_{md}},
\end{equation}
which mean that after sufficient time the centroids of the plasma and net current distributions will realign with the beam centroid.\footnote{The formalism developed in this section can be applied to situations which are quite different from the one considered here.  For example, the role of the neutralizing plasma may be played by a comoving beam of oppositely charged particles (plasmoid beam).  The coupling forces persist then indefinitely, or at least as long as they are not affected by space-charge forces.}  The parameters $\tau_{ed}$ and $\tau_{md}$ are called the electric and magnetic dipolar diffusion times, and are on the order of the electric and magnetic diffusion times defined in Sec.~\ref{bcn:0}.

   The essence of coupling can now be stated in mathematical form by writing down the equation of motion of a beam particle subject to both external forces, $F_{ext}$, and internal coupling forces, i.e., \cite[p.75]{BUDKE1956A}, \cite[p.685]{BUDKE1956B}
\begin{equation}\label{coh:12} %
   m \gamma \frac{\partial^2}{\partial \tau^2} x = F_{ext} - F_{ce} - F_{cm}.
\end{equation}
Therefore, as announced at the beginning of this section, an external force may have a negligible effect on a beam slice within the body of a pulse, provided the coupling forces on this slice are larger than the external force.  If the beam and plasma distributions can adjust themselves in such as way that the external force is canceled from head to tail, the beam pulse will propagate undeflected.  This is typically the case when the electrostatic coupling force is equivalent to the electric dipolar polarization force which enables a fully neutralized beam to move undeflected across a magnetic field. (See Chap.~\ref{plb:0}.)  In general, however, the full effect of coupling is difficult to predict because of the complexity of the details at the beam head, and may have to be resolved by experimentation.

We are now going to exploit this notion of coupling in the following section devoted to the effect of Earth's magnetic field.

\section{Effect of external forces : Earth's magnetic field}
\label{mag:0}

   A beam of charged particles launched into the low atmosphere will necessarily be affected by Earth's magnetic field.  Similarly, the region above the atmosphere in which an orbiting charged particle beam weapon might be deployed contains plasma and magnetic fields of both solar and terrestrial origin.  The plasma may enable electron or proton beams to pinch and propagate over large distances, but the magnetic fields would strongly deflect the beam trajectory in most cases.

   A major issue in the use of a charged particle beam as a directed energy weapon is therefore the precision with which such a beam can be aimed at a target considering that a beam made of non-neutral particles will necessarily be affected, in a way or another, by Earth's magnetic field. 

   The main difficulty with this issue is not so much that a charged beam may be  significantly deflected when propagating over substantial distances in Earth's magnetic field, then the fact that the variations of the geomagnetic field are not known to such an accuracy that the beam can be aimed precisely enough to compensate for its deflection.  In particular, the geomagnetic field can significantly vary during a geomagnetic storm.  Moreover, the geomagnetic field can be disturbed in an unpredictable manner by nuclear explosions in and above the atmosphere, and even possibly by other means that may be sufficient to prevent the beam from being accurately or reliably aimed at a distant target.  In other words, the trajectory of a charged particle beam pulse is different from that of a ballistic missile because, contrary to Earth's gravitational field, Earth's geomagnetic field cannot be mapped with the requisite precision to ensure a direct hit of the beam on a relatively small object such as an ICBM reentry vehicle.  

   For these reasons, it is only for short-range endo-atmospheric applications (i.e., a few 100~m to a few km) that the effect of Earth's magnetic field can in principle be sufficiently reliably corrected for by precisely aiming a charged beam at the exit port of the accelerator.

  And, for the same reasons, it is likly that a practical long-range charged particle beam weapon will have to be coupled to a pointing and tracking system such that possible aiming corrections can be done using the information obtained by another (possibly much lower intensity) beam of comparable momentum.  Since this low-power beam could also be used for discriminating between targets and decoys, it could be an integral part of a high-lethality beam system, of which it would share many components, and therefore would not be an undesirable overhead.  

  In this context, one should keep in mind that other external forces than just Earth's magnetic field may interfere with beam propagation: image forces from the ground and obstacles, stray electromagnetic fields, unexpected inhomogeneities in the atmosphere, contermeasures, etc.  A balance should therefore be kept between mitigating the effects of external forces in order to decrease uncertainties, and the fact that even for neutral-beam and short-range systems, beam pointing and tracking is always an interactive process in the end, and that a function such as discrimination is a natural way of properly steering a beam before increasing its power and destroying the target.

  Nevertheless, in order to explore the full range of options, let us briefly recall that there are a number of particle beam concepts which avoid all together the problem of deflection by Earth's magnetic field:  Such beams may consist of neutral particles (e.g., neutral hydrogen beams), be guided by a laser-generated ion-plasma channel (e.g., electron beams propagating in the `ion-focused regime'), or have some mechanism for cancelling the effect of Earth's magnetic field (e.g., plasmoid beams, and fully neutralized positive ion beams).  All these options will be studied in a part or another of this book, and in particular the last one (neutralized proton beams) in the present section.

\subsection{Effect on charge and current neutralization}

   The possible deflection of a beam's trajectory is not the sole effect of an external magnetic field: of equal importance is that this field may interfere with plasma-physical processes such as charge and current neutralization which enable a beam to propagate in a pinched mode. 

   In order to study the full effect of an external magnetic field on a charged beam propagating through a gas or plasma it is necessary to  use a generalized Ohm's law such as \eqref{bcn:8}.  This makes the analysis difficult because one cannot use the paraxial approximation, and because some of the most important effects of the external magnetic field occur at the beam head, where the analysis is necessarily two- or three-dimensional even for a beam that would be axisymmetric in a non-magnetized plasma.  This is one reason why very few analytical studies have been published, i.e., references \cite{LEE--1971-, ROSIN1973-, CHU--1973-, BERK-1976-, CHRIE1986-}, of which only two explicitly deal with the case of an external magnetic field transverse to the motion of the beam, i.e., \cite{LEE--1971-} for electron beams, and \cite{CHRIE1986-} for proton beams, which causes the beam to be deflected sidewards. 

    To appreciate the impact of Earth's magnetic field it is sufficient to compare its intensity (which at sea level has a maximum value of $B_0 \approx 0.5$~gauss, i.e.,  $B_0 \approx 5\times 10^{-5}$~tesla, see Table~\ref{tab:atm}) to those of the beam's self-fields.  At the edge of a beam of intensity~$I_B$ and radius $a$, these are on the order of
\begin{equation}\label{mag:1} %
            B(a) \approx \frac{\beta}{c} E(a) 
                 \approx \frac{1}{4\pi\epsilon_0 c^2}\frac{2I_B}{a}.
\end{equation}
For a beam typical of an endo-atmospheric system ($I_B=10000$~A, $a=1$~cm) this gives $B \approx 2000$~gauss, but only  $B \approx 0.02$~gauss for a beam typical of an exo-atmospheric system ($I_B=10$~A, $a=100$~cm).  Therefore, while Earth's magnetic field may have only a second order influence on the complex plasma physical processes leading to charge and current neutralization of a typical endoatmospheric beam, it may completely prevent charge neutralization of a beam propagating through the Earth's ionosphere.  Indeed, if the beam's self-fields are small relative to the geomagnetic field, the only way to charge neutralize the beam is by motion of the plasma electrons along the ambient field lines, which requires tapering the beam density along its length in order to charge neutralize it \cite{CHRIE1986-}.  This means that propagating a charged beam through the magnetosphere is a very complex problem, which possibly may only be solved by extensive computer simulations coupled to actual full-scale experiments.

   In fact, even for beams such that the self-fields are not small compared to the external field, the effect of the external field $B_0$ can be such as to prevent full charge and/or current neutralization \cite{LEE--1971-, ROSIN1973-, CHU--1973-}.  Neglecting the effect of the plasma-ions \cite{ROSIN1973-, CHU--1973-}, this can be quantified in terms of the plasma-electron cyclotron and plasma frequencies, i.e., for non-relativistic plasma-electrons:
\begin{equation}\label{mag:2} 
         \omega_c = \frac{e B_0}{m_e}, ~~ ~~ ~~ ~~ 
          \omega_p = c \sqrt{4\pi r_e n_e}.
\end{equation}
First, if if the external magnetic field is parallel to the motion of the beam, charge neutralization cannot occur when $\omega_c \gg \omega_p$. This is because the plasma electrons are bound to move along the external field lines, and therefore cannot move in or out of the beam.  However, if the external magnetic field is transverse to the motion of the beam, charge neutralization is possible in principle, even if  $B < B_0$ as shown in reference \cite{CHRIE1986-}.  Second, in both cases, current neutralization is only possible if the following condition is met \cite{ROSIN1973-} 
\begin{equation}\label{mag:3} 
    a \gg \frac{c}{\omega_p}\Big(1 + \frac{\omega_c^2}{\omega_p^2} \Big)^{1/4}.
\end{equation}
When $\omega_c=0$ this is equivalent to the standard condition for the induction of a return current when a beam is injected into a plasma \cite{HAMME1970-}, i.e., $a \gg \lambda_e$, where $\lambda_e = c/\omega_p$ is the electromagnetic skin depth, as was shown at the end of Sec.~\ref{bcn:0}.

\subsection{Effect on beam trajectory}

     To discuss the question of beam deflection, let us first recall that in a homogeneous magnetic field $B_0$ the radius of curvature (or Larmor radius) of a particle of charge $q$ is
\begin{equation}\label{mag:4} 
	       R_L = \frac{p}{q B_{\tr}}
                   \approx \frac{p~[\text{GeV/c}]}
                          {0.3Z~B_{\tr}~[\text{tesla}]}~[\text{m}],
\end{equation}
where $B_{\tr}$ is the transverse component of the field.  Hence, for an electron or proton with momentum $p=1$ GeV/c,  $R_L \approx 66$~km in a transverse field of 0.5~gauss typical of Earth's geomagnetic field.  In the absence of any other effect, such a deflection would be a relatively small correction for an endo-atmospheric system with a range of a few kilometers.  However, for an outer-space system with a required range of several thousands of kilometers, the beam would spiral much as it did in low-energy beam experiments performed on the space shuttle \cite{KIWAM1977-}.

   For high current beams, whether in vacuum or a plasma, the single particle expression \eqref{mag:4} has to be corrected for the influence of the self-fields.  This can be done by calculating the transverse magnetic field required to close the trajectory of a uniform beam of current $I_B$ and radius $a$ so that it forms a toroidal ring of major radius $R$.  A first approximation for $R$ is then given by Budker's formula \cite[p.676]{BUDKE1956B}
\begin{equation}\label{mag:5} %
   R = R_L \Bigl(1 + \frac{I_B}{I_A} (2-f_e) \ln (\frac{8R}{a}) + ... \Bigr),
\end{equation}
which for $R \gg a$ is equivalent to a more precise formula, exact to $\mathcal{O}(I_B/I_A)^2$, derived in \cite{OTT--1971A, OTT--1971B}.  The leading correction term, with its logarithm, appears in many problems where the self-interaction of a bent beam is calculated (e.g., high-current beam stability or high-current betatron theory) and is sometimes called the `toroidal correction' \cite{OTT--1971B}.  Its effect is to resist bending, so that the radius of curvature of a high-current beam in a magnetic field is always greater than the Larmor radius.  This toroidal effect is maximum for a unneutralized beam ($f_e=f_m=0$), and vanishing in the limit of a fully neutralized beam ($f_e=f_m=1$), assuming that the beam and plasma currents do not separate, and that the effect of the plasma current is to replace $I_B$ by $I_N$ in Eq.~\eqref{mag:5}.  However, because of its dependence of the ratio $I_B/I_A$, the toroidal effect is small for a paraxial beam, e.g., a correction of only 0.5\% for a 10~kA, 1~GeV/c electron beam of 1~cm radius.  Moreover, if the beam does not make a full circle, or is just a short pulse, Eq.~\eqref{mag:5} tends to overestimate the impact of the self fields.

   It is therefore essential to investigate the potentially more important effect of an external field on the head of a beam pulse, which because of coupling determines the behavior of the body of a paraxial beam much more than the self-fields do.  In fact, the crucial importance of coupling on the deflection of a beam propagating in a magnetized gas was clearly demonstrated in a remarquable series of Russian experiments, in which the guiding effects of gas- and plasma-filled channels, as well as of metallic wires, were also demonstrated \cite{DIDEN1977-}.

   In these experiments a 35~kA, 1~MeV electron beam was propagated in the IFR mode through of 40~cm radius chamber filled with air at a pressure comprised between 0.1 and 0.6~torr, and subject to a transverse magnetic field of 2 to 200~gauss.  It was found that the radius of curvature was systematically smaller than the Larmor radius calculated in the single particle approximation for an electron energy of 1~MeV, i.e., the measured radius corresponded to an electron energy of about 0.4~MeV.  A careful analysis of the details of the experiment then showed that while the energy of the electrons in the body was 1~MeV, the average energy of the electrons at the head of the pulse (where the rising current was only a few kA) was less than about 0.5~MeV.  The behaviour of the pulse could therefore be explained by the fact that by ionizing the gas this lower-energy prepulse was able, despite its lower intensity, to produce a plasma environment such that the bulk of the electrons in the beam had to follow the trajectory of the beam front \cite[p.626]{DIDEN1977-}.\footnote{In this reference, as in many others, this `plasma environment' is termed `plasma channel,' even though it would be better to keep this term for a prepared plasma channel, because the effect of a self-generated plasma environment (or channel) is just that of coupling.}

   Consequently, in high-current mono-energetic beam propagation experiments where the toroidal correction \eqref{mag:5} is negligible, and where there is no current neutralization that could interfere with the effect of an external magnetic field on the beam head, i.e., $f_m = 0$, the trajectory of the beam should not be different from that of its individual particles.  In that case, \eqref{mag:4} applies directly to the beam as a whole, which has been verified experimentally \cite{HESTE1974-}.

\subsection{Effect on beam head}

   In the general case, i.e., when $f_e \neq 0$ and $f_m \neq 0$, the overall effect of an external magnetic field on a beam is more complex, and there is only one published paper related to this problem where the effect of a transverse field on the beam head is investigated in details, reference \cite{LEE--1971-}.  In that paper a simple plasma model is used to calculate the return current induced by an electron beam injected into a plasma that is magnetized by a field that is either parallel or transverse to the beam direction. The same model applies to the case of a positive beam, with the difference that in that case the plasma response would be a plasma-electron-current flowing forwards rather than backwards relative to the beam motion.  In the model used in this reference it is assumed that the beam particles move in straight line so that the calculation yields the first order plasma response corresponding to the induction of the return current, which implies that the question of the beam deflection as such is not explicitly discussed.

   The conclusion of reference \cite{LEE--1971-} is that in the presence of a transverse magnetic field the induced plasma-current is no more axisymmetric, but has (for a cylindrical beam) a typical $\sin \theta$ surface charge/current polarization density distribution such that the resultant electric field exactly cancels the external magnetic field's force on the plasma-current flowing within the beam.  This means that this distribution has just the proper character to enable the induced plasma-current to flow across the field in a force free region within the beam.\footnote{Incidentally, it is also found in reference \cite{LEE--1971-} that the decay of the induced plasma current is slower than exponential, so that equation \eqref{bcn:18} gives a somewhat pessimistic estimate of the duration of the current neutralized region at the head of a beam.}

   Thus, at the head of a beam pulse injected into a gas or plasma, there can be (under suitable conditions) a region of duration $\Delta \tau \approx \tau_m$ where $f_m \approx f_e \approx 1$ so that the beam current, as well beam charge, are fully neutralized.  In this region the plasma current is polarized in the plane perpendicular to the external magnetic field, which allows the plasma current to flow straight across this field,  whether the resulting configuration consists of a negative beam neutralized by a counterstreaming plasma-electron current, or a positive beam neutralized by a comoving plasma-electron current.  In order to find the implications of this polarization on the beam particles themselves, it is now necessary to carefully distinguish between these two configurations. 

  Suppose that the beam particles flow with a velocity $V=|\beta c|$ in the positive direction of the $z$ axis, and that the absolute value of the plasma-electron velocity along this axis is $v \ll c$.  Suppose also that Earth's magnetic field is directed in the positive direction of the $y$ axis and has intensity $B_{\tr}$.  Under these conditions the forces acting on the plasma and beam particles are along the $x$ axis, on which we now calculate their projections.

    First, consider a \emph{negative} (e.g., electron) beam.  The  cancellation of the force due to $\vec{B}_{\tr}$ by the force $\vec{F}_{pol}$ due to the polarization electric field means that
\begin{equation}\label{mag:6} %
          F_x(\text{plasma}) = -e v  B_{\tr} +  F_{pol} = 0.
\end{equation}
Thus, the total force (excluding the self-fields which are assumed to be canceled by the plasma charge and current) on a negative beam particle is
\begin{equation}\label{mag:7} 
       F_x(\text{beam}) = |-e \vec{V} \times \vec{B}_{\tr} +  \vec{F}_{pol} ~ |
                 = +  e (V + v) B_{\tr}.
\end{equation}
Therefore, the effect of the polarization electric field is to \emph{increase} the beam deflection caused by the external magnetic field, a small effect as long as $v \ll V$.

   Second, consider a \emph{positive} (e.g., proton) beam.  The  cancellation of the force due to $\vec{B}_{\tr}$ by the force $\vec{F}_{pol}$ due to the polarization electric field means that
\begin{equation}\label{mag:8} 
          F_x(\text{plasma}) = +e v  B_{\tr} -  F_{pol} = 0,
\end{equation}
where the direction of polarization force is reversed because the plasma electrons are induced to flow towards the head rather than the tail of the beam. Thus, the total force (excluding again the self-fields which are assumed to be canceled by the plasma charge and current) on a positive beam particle is
\begin{equation}\label{mag:9} 
       F_x(\text{beam}) = |+e \vec{V} \times \vec{B}_{\tr} +  \vec{F}_{pol} ~ |
                 = -  e (V - v) B_{\tr}.
\end{equation}
Therefore, the effect of the polarization electric field is now to \emph{decrease} the beam deflection caused by the external magnetic field, which implies that a positive beam may move straight ahead across a transverse magnetic field if $v=V$.

   Consequently there is a remarkable difference between a negative and a positive beam propagating in a gas or plasma, namely that the head of a positive beam may under some suitable conditions move undeflected across a transverse magnetic field. These conditions require in particular that the electrons of the plasma current induced by the positive beam have the same velocity as the beam, which can be interpreted as a `pick up' process by which the positive beam fully neutralizes itself by attracting electrons and taking them along.  Even if this process does not lead to a full cancellation of the transverse magnetic force it will nevertheless somewhat decrease the deflection by an external magnetic field, which implies that a positive beam has a considerable advantage over a negative one for use as a directed energy weapon in the Earth's magnetosphere.

   Indeed, when a beam propagates in a plasma and $f_e=f_m=0$ the monopolar pinch forces, Eqs.~\eqref{bcn:11} and \eqref{bcn:17}, are zero, i.e., $F_e=F_m=0$, and the beam expands freely.  On the other hand, the dipolar coupling forces, Eqs.~\eqref{coh:8} and \eqref{coh:9}, are such that $F_{cm}=0$ while $F_{ce} \neq 0$.  Therefore, a fully neutralized beam head can nevertheless guide a pinched beam body, and therefore lead a positive beam straight across a magnetized gas.

   However, it should be stressed right away that there are many practical difficulties in the way of taking this advantage fully into account.  In particular, as will be seen in the chapter on instabilities, the head of a fully current-neutralized beam is very much prone to instabilities of various kind, which may even lead to excessive ($f_m>1$) current neutralization \cite{SUDAN1976-}.  Moreover, fulfilling the condition $v=V$ is hampered by the fact that a high-power beam should preferably be highly relativistic (meaning that $v=V \approx c$), which has the inconvenience that relativistic effects (i.e., the magnetic field induced by the motion of the plasma current) tend to quench the polarization electric fields.  Finally, in order that the head can guide the body of a pulse, it is necessary that the coupling forces are sufficiently strong for the beam to remain intact, instead of tearing apart with the head going straight ahead and the body been deflected.\footnote{According to Eq.~\eqref{coh:12}, this may imply that the beam pulse will no more be straight and coaxial with its direction of propagation, but slanted in such a way that the coupling forces between neighbouring beam slices (calculated taking as in this subsection the non-axisymmetric response of the plasma into account) will compensate for the external magnetic force.}

   Nevertheless, it is possible that a favorable compromise may exist, and that a mildly relativistic beam such as a proton beam of about 2~GeV kinetic energy would have the requisite properties. (For such a beam, the condition $v=V$ implies that the comoving neutralizing plasma electrons would have an energy of about 1~MeV because 2~GeV~/~1~MeV $\approx 1836$, the proton to electron mass ratio.)  Besides of being only mildly relativistic, the key advantage of this beam energy is that it corresponds to the `minimum ionization window,' i.e., the velocity range in which both protons and electrons loose a minimum amount of energy when propagating through air at any pressure.\footnote{The minimum given by Bethe's stopping power, Eq.~\eqref{nor:5}, corresponds to $S(\beta)_{min} \approx 0.96$, where $S(\beta)_{min} \approx 0.22$~MeV/m in air at STP.} 

   In fact, while there is apparently little published discussion on the advantage of a positive over a negative beam with regards to its potential ability to move straight across a transverse magnetic field when propagating in a gas or plasma, there is no surprise in this property of a fully charge- and current-neutralized positive beam because such a beam is essentially equivalent to a \emph{neutral plasma beam} --- also called a \emph{plasmoid beam}.\footnote{The January 1980 \emph{Particle Beam Research Workshop} at the U.S.\ Air Force Academy strongly emphasized the potential advantages of plasmoid beams, which may not be ``significantly deflected by the Earth's magnetic field'' \cite[p.58]{GUENT1980-}; and of beam filaments that are ``current neutralized and thus are not deflected by kilogauss magnetic fields''  \cite[p.73]{GUENT1980-}.}

\subsection{Summary}

   In this section we have discussed, mostly in qualitative terms, the complicated effects that an external magnetic field has on a beam and its plasma environment.  These complications are such that the resulting behavior of the beam, and its trajectory, can only be determined by actual experiments or fully three-dimensional simulation programs \cite{HUI--1984A}.

   In the general case, when an external magnetic or electric force $F_{ext}$ acts on the beam, each beam slice will be subject to a different acceleration $F_{ext}/m\gamma$ if the velocity of the particles in that slice is a function of $\tau$.  If the shear in $F_{ext}(\tau)/m\gamma(\tau)$ is strong compared to the coupling force $F_{ce} + F_{cm}$ due to the beam self-pinch, the beam will tear.  The breakup will continue as $\tau$ increases until for some value $\tau_g$ the coupling force becomes stronger than the sheared external force; from this point on, i.e., $\tau > \tau_g$, the beam will hold together.  Once the guiding point $\tau_g$ is determined, one can find (a) the deflection due to the force $F_{ext}$, (b) what portion of the beam is torn out, and (c) the value of $\gamma(\tau_c)$, i.e., the energy of that part of the beam which actually guides the rest of the beam \cite{HUI--1984A}.  When the background plasma subjected to the external field responds in such a way that its distribution is no more axially symmetric the coupling forces are no more trivially determined by the pinch force, i.e., by Eqs.~\eqref{coh:8} and \eqref{coh:9}.  It it is then possible that the beam's trajectory is no more determined by just its energy at the guiding point, but by a combination of external and internal forces such that the beam as a whole is either more or less deflected than it would be in absence of any collective effects. 

   To conclude, let us summarize the main possibilities for propagating a particle beam across a magnetic field:

\begin{enumerate}

\item \emph{Neutral particle beams.}  A beam composed of electrically neutral particles is evidently not deflected by an electromagnetic field. This is the case of a beam made of intrinsically neutral particles such as neutrons, which are however very difficult to focus on a small or distant target, and of neutral atom beams.  In the later case, as is well known, the beam is not deflected even though the atoms are polarized by the external field:  This is because the electromagnetic field strength required to separate a bound electron from a nucleus is very large. As is easily estimated by calculating the electric field between two charges separated by a distance on the order of the Bohr radius, i.e., $r_e\alpha^{-2}$, this critical electric field is about $5 \times 10^{11}$~V/m, which corresponds to about $2 \times 10^{7}$~gauss.  On the other hand, an atomic beam propagating in the atmosphere is easily ionized (cross-section $\approx 10^{-18}$ cm$^2$) and this gives a lower altitude limit of about 150 km for using such a beam.

\item \emph{Plasmoid beams.}  If a neutral beam is composed of fully ionized atoms, e.g., a mixture of co-moving electrons and protons, the beam will polarize and under favorable circumstances generate a polarization sheath which may enable the plasmoid to move undeflected across the magnetized vacuum or plasma.   Indeed, since electrons and ions are deflected towards opposite directions, they create (on opposite sides of the beam) a pair of respectively negative and positive charge layers, which induce a polarization electric field.  This field yields an electric force which exactly cancels the magnetic force from the external magnetic field and the beam continues undeflected, loosing, however, particles from the polarization layers.  For such a beam, the effective current \eqref{ben:3} is evidently zero.  The envelope equation \eqref{ben:2} reduces then to that of neutral beam and, in the absence of the pinch effect, the neutral plasma beam expands through the influence of its emittance.  Moreover, as the electrons and ions move at the same velocity, a plasmoid beam can be seen as a ordinary plasma moving through a background gas or plasma.  Using standard concepts of plasma physics, and provided the background particle density is negligible compared to the beam particle density, it can be seen that beam expansion into the surrounding near-vacuum is described by a rarefaction wave propagating radially inwards at a velocity on the order of the ion sound velocity \cite{DENAV1979-}, i.e.,
\begin{equation}\label{mag:10} %
	          v_s \approx \sqrt{ \frac{kT_{\tr}}{\gamma m_i} }, 
\end{equation}
where $T_{\tr}$ is given by \eqref{def:15}.  The change in beam radius after propagating a distance $z$ is then approximated by the expression $\Delta \tilde{a} \approx z \epsilon_{\tr} /\tilde{a}_0$, which has the same form as \eqref{neu:6}, so that the `beam' and `plasma' pictures give compatible results.  A more precise  treatment of plasmoid beam expansion, including the emittance growth due to a non-negligible background, will be given in Sec.~\ref{ten:0}.  A plasmoid beam would thus be an alternative to a neutral particle beam for high-atmospheric or outer-space systems, an option that will be extensively discussed in Chap.~\ref{plb:0} where the possibility of both ion/electrons and matter/antimatter plasmoid beams will be considered.

\item \emph{Guided particle beams.} If a straight plasma channel is created by a laser beam in the atmosphere, the radial focusing force exerted by the ions on an electron moving in the channel may be larger than the lateral deflecting force due to an external field such as Earth's geomagnetic field.  This leads to the concept of `laser-guided ion-focused propagation' which may enable an electron beam to propagate undeflected through a low-density background such as the upper-atmosphere.  Creating a suitable channel over a very long distance may require a quite powerful laser, which by itself may not necessarily be very simple to build and operate, but which could be part of the pointing and tracking system of the weapon.  A basic assumption in this concept is that the massive ions of the channel are relatively immobile so that the lighter electrons are properly guided by the beam-channel tracking force.  Therefore, the guiding of proton or ion beams, or the guiding of electron beams in a dense background, will need other techniques, such as those discussed in Sec.~\ref{dgc:0} on beam deflection and guiding by interfaces and channels.

\item \emph{Neutralized proton beams.}  As we have seen in this section, a positive ion beam pulse may behave as a plasmoid in a gas or plasma, provided its head is able to fully charge and current neutralize, and its body to follow the head if the coupling forces are strong enough.  Thus, a sequence of proton beam pulses may propagate in straight line, creating suitable conditions for following pulses to move further ahead through a dense or relatively tenuous atmosphere.  To date, this mode of propagation has most probably still not been demonstrated, primarily because of the lack of sufficiently powerful sources and  suitable beams of positive particles. 

\end{enumerate}

\section{Charged beam in a dense gas or plasma : Nordsieck equation}
\label{nor:0}

When a beam propagates in a dense gas or plasma, the collisions with the electrons and nuclei result in a loss of energy by the beam particles, and eventually a loss in beam current when particles can effectively be removed from the beam by interactions with nuclei.  Furthermore, by increasing the angular and energy spread of the beam, the collisions lead to a continuous increase in both transverse and longitudinal emittance.  In the paraxial approximation the full radial envelope equation taking energy loss and scattering into account is as follows \cite{LEE--1976A,LEE--1976B}:\footnote{The theories developed in these two papers differ by some non-essential factors of `2.'  What matters is that they are used consistently when compared to experimental data.}
\begin{equation}\label{nor:1} 
    \tilde{a}'' + \frac{I_E}{I_A}\frac{1}{\tilde{a}} 
                + \frac{\tilde{a}'W'}{\beta^2W}
                = \frac{1}{\tilde{a}^3p^2}
   \Bigl(p_0^2\epsilon_{\tr 0}^2  + \int_0^z \tilde{a}^2p^2 ~ d\psi^2\Bigr).
\end{equation}
This result is obtained from kinetic theory which shows \cite{LEE--1976B} that a self-pinched beam subject to gas scattering evolves to a state in which its current density takes the form of the Bennett distribution \eqref{ben:21}, which is a similarity solution of the Fokker-Planck equation with $\tilde{a} \approx a$ for $r \leq \mathcal{O}(a)$.

	In \eqref{nor:1} all parameters are, in general, a function of the propagation distance $z$. The third term on the left hand side, for instance, is the contribution to radial expansion coming from the decrease of the beam's particle energy $W(z)$.  The right hand  side corresponds to the increase in transverse emittance because of Coulomb multiple scattering in the background gas.  In agreement with Liouville's theorem \cite{LAWSO1977-}, the emittance enters this formula through the product $p\epsilon$ which is the relativistic invariant, conserved emittance for a beam of varying energy.

  In first approximation, the multiple scattering angle is given by Rossi's formula \cite[p.67]{ROSSI1952-}\footnote{Rossi's formula provides a convenient first approximation for analytical calculations of the kind done in this section.  For more precise calculations, especially for beams of relatively low-momentum particles, it is better to use Moli\`ere's theory as formulated by Bethe \cite{BETHE1953-}.}
\begin{equation}\label{nor:2} 
       d\psi^2 = \bigl( \frac{E_s}{\beta c p}  \bigr)^2 \frac{dz}{X_0},
\end{equation}
where $E_s = \sqrt{4\pi/\alpha}\, m_ec^2 \approx 21.2$~MeV, and  $X_0 \approx 300.5$~m for air at standard temperature and pressure (STP) is the radiation length of the medium \cite[p.50]{BETHE1953-}\footnote{$\alpha=1/137$ is the fine-structure constant, and $r_e$ the classical electron radius.}
\begin{equation}\label{nor:3} 
       \frac{1}{X_0} = 4 \alpha r_e^2 N_A \frac{\rho}{A} Z^2 \ln(183 Z^{1/3}),
\end{equation}
where $A$, $\rho$, and $Z$ are the atomic mass, density, and atomic number of the medium.
	
The total energy of the beam particles as a function of $z$ is given by the equation
\begin{equation}\label{nor:4} 
          W' \DEF  \frac{dW}{dz}  =  -S(\beta) - \frac{W}{X_0}  + qE_z.
\end{equation}
The first term is Bethe's stopping power (i.e., the energy loss per unit path length) which corresponds to energy losses by ionization and excitation of the medium's molecules \cite[p.24]{ROSSI1952-}
\begin{equation}\label{nor:5} 
    S(\beta) = 4 \pi r_e^2 N_A \frac{\rho}{A} Z m_e c^2
               \frac{q^2}{\beta^2}
               \Bigl( \ln \frac{2\beta^2\gamma^2m_e c^2}{I(Z)} - \beta^2 \Bigr),
\end{equation}
where $q$ is the charge of the beam particles, and $I(Z) \approx 85$~eV for air is the mean excitation energy of the medium.  $S(\beta)$ is independent of the beam particles's mass and a slowly varying function of  their velocity when $\beta \rightarrow 1$.  It equals about 0.3~MeV/m in air at STP for single-charged particles with energies in the GeV range.  Between $\beta\gamma \approx 0.1$ and $\beta\gamma \approx 1000$, the stopping power calculated with Eq.~\eqref{nor:5} gives a result correct within a few percent.  $S(\beta)$ has a broad minimum at $\beta\gamma \approx 3.5$ (i.e., $\beta \approx 0.96$), which corresponds to a kinetic energy of about 1.3~MeV for electrons and about 2.3~GeV for protons. This is the so-called `minimum ionization energy' for these particles, at which $S(\beta) \approx 0.22$~MeV/m in air at STP. 

  The second term on the right of Eq.~\eqref{nor:4} is present only for electron beams and corresponds to bremsstrahlung radiation losses \cite[p.50]{ROSSI1952-}.  (Bremsstrahlung emission is negligible for particles of mass heavier than electrons).  The third term corresponds to the ohmic losses in the induced longitudinal electric field.  These losses which are associated with the charge and current neutralization process are concentrated in the head of the beam where they contribute to heating the plasma electrons.

\begin{figure}  
\begin{center}
\resizebox{12cm}{!}{ \includegraphics{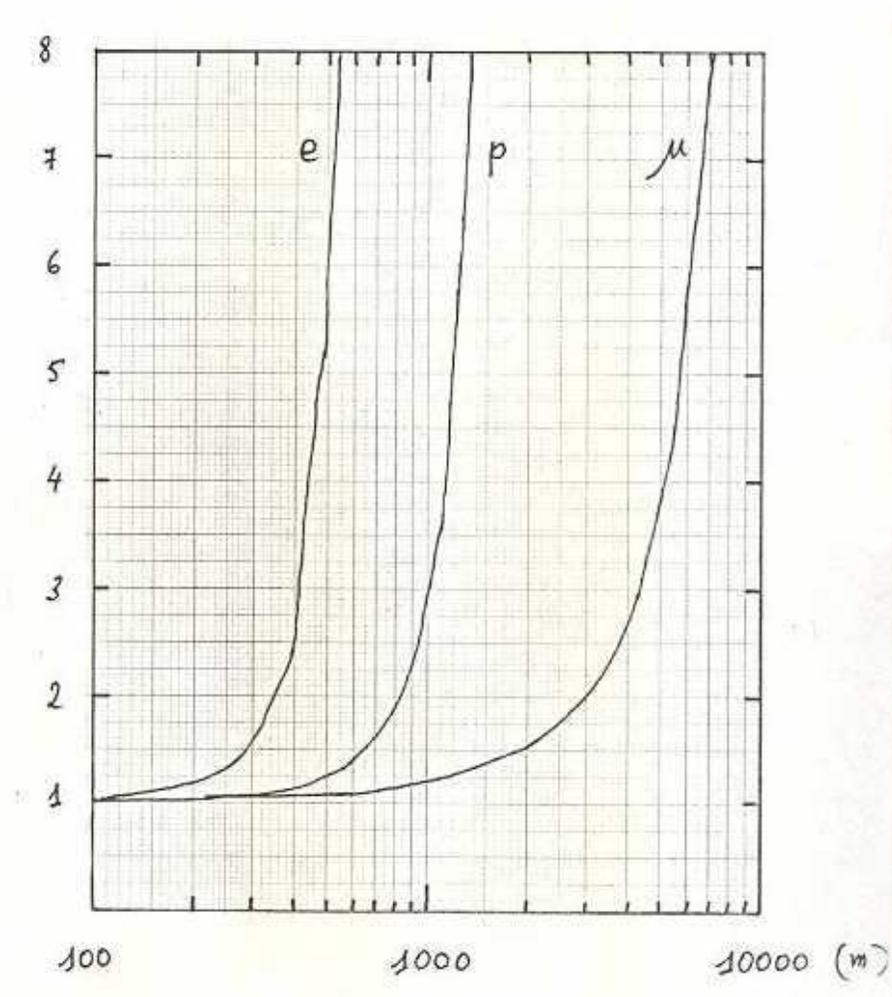}}
\caption[Beam expansion for various particles]{\emph{Beam expansion for various particles.} The radial expansion of a beam of intensity $I_B=10$ kA and momentum $P_B=10$ GeV/c, calculated by integrating numerically the full radial envelope equation taking all energy losses and scattering effects into account, is plotted as a function of propagation distance in air at STP.  The $\e$-folding range, also called the Nordsieck length, defined as the distance at which the beam radius has expanded by a factor of $\e \approx 2.718$, is of about about 400, 1000, or 4000 meters for an electron, proton, or muon beam, respectively. \label{fig:exp}}
\end{center}
\end{figure}
\begin{figure}  
\begin{center}
\resizebox{12cm}{!}{ \includegraphics{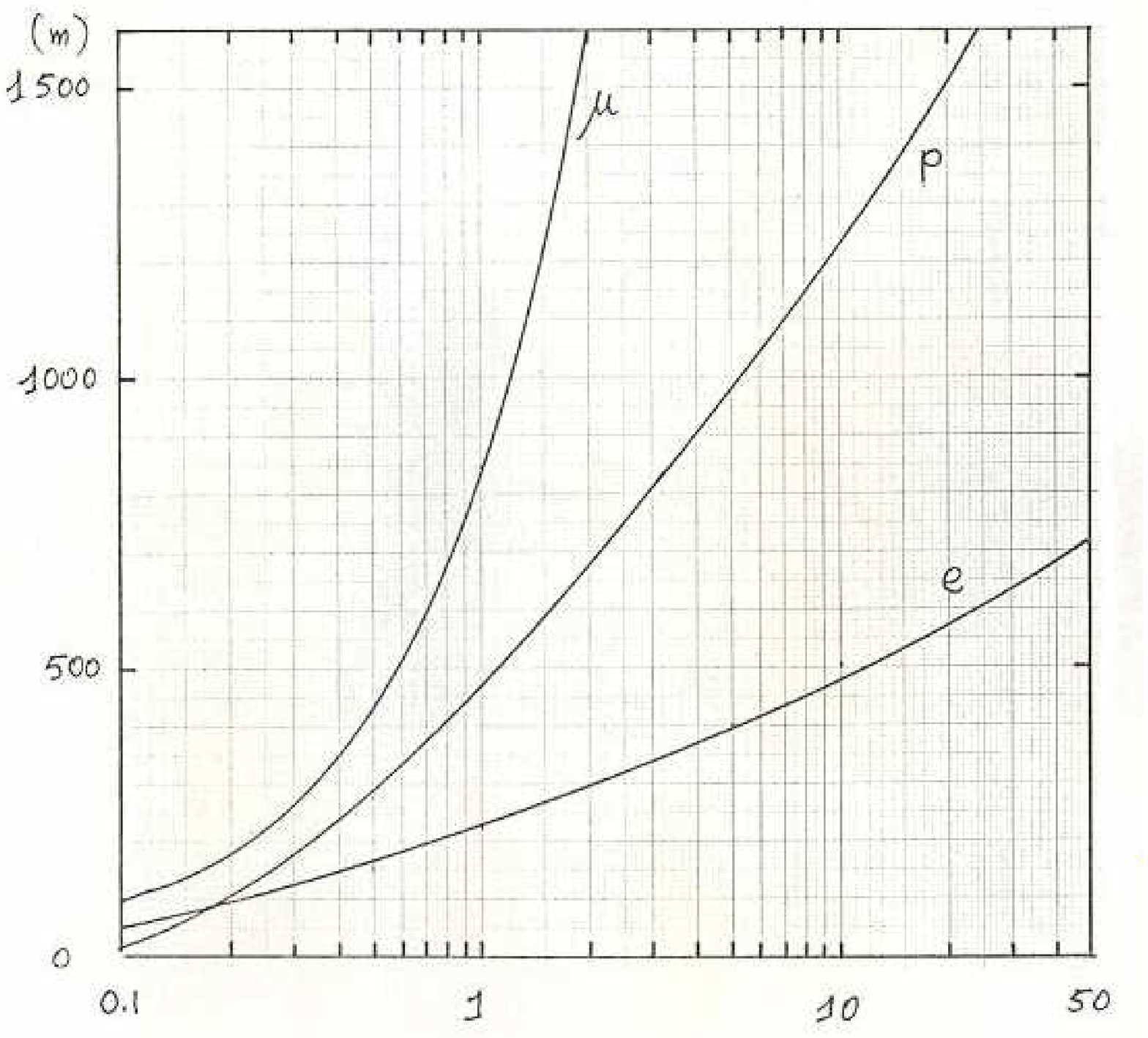}}
\caption[Effective beam range versus momentum for various particles]{\emph{Effective beam range versus momentum for various particles.} The effective range, defined as the distance at which the beam radius has expanded by a factor of $2e \approx 5$, is plotted as a function a beam momentum in GeV/c. For a beam intensity $I_B=10$ kA the range in air at STP of a single 1 GeV/c electron, proton, or muon beam pulse is of about 200, 500, or 800 m, respectively. For a 10 GeV/c pulse the effective range increases to about 500 or 1200 m for an electron or proton beam, and about 6000 m for a muon beam (see Figure~\ref{fig:exp}). 
\label{fig:ran}}
\end{center}
\end{figure}

	The beam current decreases because of collisions with nuclei.  In the case of proton beams, both inelastic and elastic nuclear collisions will effectively remove particles from the beam.  Therefore, in that case, the effective beam current \eqref{ben:3} will vary with $z$ as
\begin{equation}\label{nor:6} 
     I_E(z) = I_E(0) ~ \exp(-z/X_n),
\end{equation}
where $X_n$ is the nuclear collision length, about 500 m in air at STP for high energy proton beams.  In the case of electron beams, the pronounced statistical character of bremsstrahlung radiation losses will result in a wide energy spread for propagations over distances of the order of one radiation length.  The implication of such a large increase in longitudinal emittance is that particles with energies less than half the mean beam energy will in fact `evaporate' from the beam, thus leaving behind a reduced current beam \cite{HAFTE1979-}.  A calculation based on Bethe and Heitler's theory of straggling in bremsstrahlung emission \cite{ROSSI1952-} indicates that this effect would account for a 30\% loss in beam current for a high energy electron beam propagating over one radiation length.

	The solution of \eqref{nor:1} has generally to be found numerically.  However, considerable insight can be gained in the quasistatic limit  $\tilde{a}'' = \tilde{a}' \approx 0$,  where \eqref{nor:1} becomes
\begin{equation}\label{nor:7} 
    \tilde{a}^2p^2\frac{I_E}{I_A} 
   = \int_0^z \tilde{a}^2p^2 ~ d\psi^2,
\end{equation}
which can trivially be rewritten as 
\begin{equation}\label{nor:8} 
  \frac{d}{dz} \ln \Bigl( \tilde{a}^2p^2 \frac{I_E}{I_A} \Bigr)
                        = \tilde{a}^2p^2 \frac{I_A}{I_E} ~\frac{d\psi^2}{dz}.
\end{equation}
If one takes Rossi's multiple scattering formula \eqref{nor:2}, and uses the definition of the Alfv\'en current, this equation becomes the so-called \emph{Nordsieck equation} \cite{LEE--1973C,LEE--1976A,LEE--1976B}:
\begin{equation}\label{nor:9} 
  \frac{d}{dz} \ln (P_E\tilde{a}^2) = \frac{P_N}{P_E} \frac{1}{\beta^2 X_0}, 
\end{equation}
where $P_E = I_E(z) p(z) c/e$ is the effective beam power, and $P_N = c E_s^2/e^2 = (c/r_e)(4\pi/\alpha) m_ec^2 \approx 15\times 10^{12}$~W = 15~TW is a physical constant: the \emph{Nordsieck power}.  The remarkable property of this approximate envelope equation is that the radial expansion of a relativistic beam in a gas is a function of only two variables: the radiation length $X_0$ which characterizes the medium, and the effective power $P_E$ which characterizes the beam.

  The Nordsieck equation can be solved explicitly for several cases of interest.  For instance, for high-energy \emph{proton beams}, the dominant beam power loss comes from the decrease in beam current due to nuclear interactions: $P_E(z) \approx P_0 \exp(-z/X_n)$.  Then, for relativistic proton beams such that $\beta\approx 1$, 
\begin{equation}\label{nor:10} 
    a = a_0 \exp  \frac{1}{2}\Bigl( \frac{z}{X_n}
                 + \frac{P_NX_n}{P_0X_0} \bigl(\exp(z/X_n)-1\bigr) \Bigr).
\end{equation}
On the other hand, for high-energy \emph{electron beams}, the dominant effect is energy loss by bremsstrahlungs: $P_E(z) \approx P_0 \exp(-z/X_0)$.  Thus, as $\beta\approx 1$, 
\begin{equation}\label{nor:11} 
    a = a_0 \exp  \frac{1}{2}\Bigl( \frac{z}{X_0}
                 + \frac{P_N}{P_0} \bigl(\exp(z/X_0)-1\bigr) \Bigr).
\end{equation}
In both cases, the Nordsieck power $P_N$ plays an important role.  For $P_0 < P_N$, the radial expansion is very fast and the beam cannot propagate over sizable distances.

For $P_0 > P_N$, however, the $\e$-folding range, also called the \emph{Nordsieck length}, defined as the distance over which the beam radius expands by a factor of $e \approx 2.718$, becomes independent of initial beam power.  It is about $z_N \approx 2X_n \approx 1000$~m for protons and $z_N \approx 2X_0 \approx 600$~m for electrons in air at STP.  The implication is that for a beam to propagate, its total effective power has to be on the order of $P_N = 15$ TW, but any further increase in beam power would not substantially increase its range.

In the low-energy limit, i.e., $P_0 < P_N$, the beam will propagate less than one radiation or nuclear interaction length.  One can therefore assume that $z\ll X_0$ for electrons, and $z\ll X_n$ for protons, to find that in first approximation the {Nordsieck length} for both an electron or a proton beam is then given by
\begin{equation}\label{nor:12} 
    z_N \approx 2 \frac{P_0}{P_N+P_0} X_0.
\end{equation}
For example, the maximum power of the 10 kA, 50 MeV, ATA beam is $P_0 = 0.5$~TW.  Since the Nordsieck power is $P_N = 15$ TW, and the radiation length $X_0 \approx 300$~m in air at STP, the typical propagation length of a single ATA beam pulse should be on the order of $z_N \approx 20$~m.

	In order to check the validity of these conclusions we have written a computer program to solve \eqref{nor:1} in the general case, modeling all physical effects such as multiple scattering, energy losses, etc., as precisely as possible. \footnote{These computer simulations made in 1978 used Moli\`ere's theory of multiple-scattering and other refinments routinely used in the analysis of high-energy particle physics experiments.  It is only in 1984 that the results of a similar simulation and comparison to earlier experiments was published \cite{HUGHE1984-}. It should however be emphasized that the remarkable simplicity of the Nordsieck equation \eqref{nor:9}, which leads to closed form analytical expressions such as \eqref{nor:10} and \eqref{nor:11}, is due to the use of Rossi's formula \eqref{nor:2}.}

   First, we have solved \eqref{nor:1} for 10~GeV/c momentum, 10~kA beams of electrons, protons and muons in air at STP.  This corresponds to an initial beam power $P_0 = 6.6 P_N$, and the results are shown in Figure~\ref{fig:exp}.  Second, we have calculated the effective range, defined as the distance at which the beam radius has expanded by a factor of $2\e\approx 5$, as a function of momentum for 10~kA beams of these particles, see Figure 4.3.

   The $\e$-folding range of the \emph{proton beam} is about 1 km, as expected.  But it is only 400 m for the \emph{electron beam}.  (See Figure~\ref{fig:exp}.)  This discrepancy is the result of the approximation made in deriving the Nordsieck equation \eqref{nor:9} and of neglecting straggling.

   The range of proton beams in air is thus strictly limited by the nuclear interaction length, and that of electron beams by the radiation length.  However, as these lengths depend directly upon the atmospheric density, the corresponding ranges will increase in proportion to the decrease in atmospheric density.

   The \emph{muon beam}, which would be much more difficult to produce in practice than an electron or proton beam, is included for comparison.  As bremsstrahlungs and nuclear interactions are negligible for these particles, the muon beam range is of course much larger.  (See Figures~\ref{fig:exp} and~\ref{fig:ran}.)

   The solution to the Nordsieck equation for muon beams is easily obtained by letting $X_n \rightarrow \infty$ in equation \eqref{nor:10}, which becomes
\begin{equation}\label{nor:13} 
    a = a_0 \exp  \frac{1}{2} \Bigl( \frac{P_N}{P_0} \frac{z}{X_0} \Bigr).
\end{equation}
Therefore, the Nordsieck length for a muon beam is 
\begin{equation}\label{nor:14} 
    z_N = 2 \frac{P_0}{P_N} X_0,
\end{equation}
which gives an $\e$-fold range of 4 km for a 100 TW muon beam, in agreement with Figure~\ref{fig:exp}.

   Finally, in order to compare the beam expansion theory given in this section to the data, and therefore to validate the extrapolations given in Figures~\ref{fig:exp} and~\ref{fig:ran}, we have simulated and found good agreement with the beam expansion measurements given in reference \cite{BRIGG1976-}.  In these measurements, the ASTRON induction linear accelerator of the Lawrence Livermore National Laboratory was used to propagate a 250 ns, 0.85 kA, 5 MeV electron beam pulse over a distance up to 16 m in reduced density nitrogen.  This agreement can be also be used to predict with some confidence the propagation characteristics of the 10 kA, 50 MeV, ATA beam.  It is then found that one $\e$-folding beam-radius increase corresponds to a propagation distance of about 20 m in sea-level air, in excellent agreement with the analytical estimate given by equation \eqref{nor:12}, and that the beam radius increases to about ten times its initial value after propagating a distance of about 35 m.

  To conclude this section, in which the Nordsieck approximation was derived by neglecting the $\tilde{a}''$ and $\tilde{a}'$ terms relative to the $1/\tilde{a}$ term in Eq.~\eqref{nor:1}, it is important to clarify under which conditions this is possible.  To do this we restrict ourself to the case where the variation of the beam energy, i.e., the variations of $W$ and thus $P_0$, are small, so that Nordsieck's equation \eqref{nor:9} can be used to estimate the magnitude of the second derivative term that has been neglected.  This gives
\begin{equation}\label{nor:15} 
  \tilde{a}'' = \frac{\tilde{a}}{4}
                \Bigl (\frac{P_N}{P_0} \frac{1}{\beta^2 X_0} \Bigr)^2, 
\end{equation}
which is negligible in Eq.~\eqref{nor:1} if smaller than $(I_E/I_A)/\tilde{a}$, i.e., if
\begin{equation}\label{nor:16} 
  \frac{\tilde{a}}{2} \sqrt{ \frac{I_A}{I_E} }  <
               \frac{P_0}{P_N} \beta^2 X_0. 
\end{equation}
On the left hand side of this expression, apart from a non-essential numerical factor, we recognize the betatron wavelength \eqref{ben:20}, and similarly on the right hand side the Nordsieck length \eqref{nor:12}.  Therefore, the condition for Nordsieck's approximation to be valid, i.e., the condition for the pinch effect to be strong enough that beam spread due to scattering is constrained, can be written
\begin{equation}\label{nor:17} 
    \tilde{a}  \ll \lambda_{\beta m} \ll z_N, 
\end{equation}
where the first inequality recalls that the whole concept is only valid in the paraxial limit. 


\section{Particle beam in a gas or plasma : emittance-driven expansion}
\label{ten:0}

==>  suppress emphasis on plasmoid

There is a possibility that the concept of "diaxial beams" in pinch mode is more general than just outer-space:  a pair of positive/negative beams in the atmosphere will stay parallel much longer than the time to reach the target!


In Sec.~\ref{cha:0} we have considered the effect of space-charge-effects on the lateral and longitudianl expansions of a charged-particle beam pulse propagating in vacuum, independently of any other effects.  We have found that for a very relativistic beam, these effects were quite small

   However, for directed energy applications a much simpler plasmoid configuration is obtained by overlapping two co-moving beams of opposite charged particles in such a way that the combined beam is charge and current neutralized.  For example, a beam consisting of an equal number of electrons and positrons moving with the same velocity in the same direction.  For such a fully charge and current neutralized beam the effective current $I_E$ is zero, so that the envelope equation \eqref{nor:1} reduces to
\begin{equation}\label{ten:1} %
    \tilde{a}'' + \frac{\tilde{a}'W'}{\beta^2W}
                = \frac{1}{\tilde{a}^3p^2}
   \Bigl(p_0^2\epsilon_{\tr 0}^2  + \int_0^z \tilde{a}^2p^2 ~ d\psi^2\Bigr).
\end{equation}

   Obviously, this equation is the same as for a beam of neutral particles (e.g, un-ionized atoms, or neutrons) traveling through a background gas, provided the multiple scattering angle $d\psi^2$ is properly expressed in terms of the corresponding scattering processes.  For example, in the case of a neutral-hydrogen beam, multiple scattering is due to collisions between the hydrogen atoms of the beam with the molecules of the gas, while in the case of a beam of charged-particles multiple scattering is due to their Coulomb interactions with the electrons and nuclei of the gas's atoms.  In the latter case, a first approximation is provided by Rossi's formula \eqref{nor:2}, so that equation \eqref{ten:1} can be rewritten as
\begin{equation}\label{ten:2} %
    \tilde{a}'' + \frac{\tilde{a}'W'}{\beta^2W}
                = \frac{1}{\tilde{a}^3}
                  \frac{1}{X_0}
                 (\frac{E_s}{\beta c p})^2
                  \int_0^z \tilde{a}^2 ~ dz,
\end{equation}
where we have set $\epsilon_{\tr 0}=0$.

This envelope equation cannot be solved analytically in the general case.  However, if we assume that the energy $W$, and thus the momentum $p$, are nearly constant, \eqref{ten:2} simplifies to
\begin{equation}\label{ten:3} %
       3 \tilde{a}'\tilde{a}'' + \tilde{a} \tilde{a}'''
                   = \frac{1}{X_0}  (\frac{E_s}{\beta c p})^2.
\end{equation}
For an ideal beam with a zero initial emittance, and zero initial radius, the solution of this equation is
\begin{equation}\label{ten:4} %
      \tilde{a}(z) = \frac{E_s}{\beta c p} \frac{z^{3/2}}{\sqrt{3X_0}}.
\end{equation}
Therefore, the range at which the beam has expanded to a radius $\tilde{a}$ is given by
\begin{equation}\label{ten:5} %
    z({\tilde{a}}) = (\frac{\beta c p}{E_s})^{2/3} (3X_0)^{1/3}\tilde{a}^{2/3}.
\end{equation}

Taking for example a particle/antiparticle plasmoid beam (e.g., an electron/po\-si\-tron or proton/antiproton plasmoid) with a moment of 10 GeV/c per particle, the range at which the beam radius has increased to $\tilde{a}=1$ meter is given, as a function of altitude above ground, in Table~\ref{tab:plb}.

\begin{table}
\begin{center}
\hskip 0.0cm \begin{tabular}{|r|c|c|c|r|} 		\hline
\multicolumn{5}{|c|}{\raisebox{+0.2em}{{\bf  \rule{0mm}{6mm} 10 GeV/c plasmoid beam range in outer-space}}} \\ 
\hline
\raisebox{+0.2em}{altitude} \rule{0mm}{6mm} & 
\raisebox{+0.2em}{atomic density} & 
\raisebox{+0.2em}{$z_I$ } & 
\raisebox{+0.2em}{$X_0$ } & 
\raisebox{+0.2em}{$z({\tilde{a}})$~~~~~}  \\
\rule{0mm}{0mm}   
   [km]~~~         & 
$n_a$ [cm$^{-3}$]  & 
 [km]              &
 [km]              &
 [km]~~~~         \\
\hline
\rule{0mm}{5mm}    0~~~ & $5 \times 10^{19}$ &                30  &                 0.3 &         0.6~~ \\
                 100~~~ & $5 \times 10^{12}$ & $3 \times 10^{ 8}$ & $3 \times 10^{ 6}$  &      130~~~~~ \\
                 300~~~ & $5 \times 10^{ 9}$ & $3 \times 10^{11}$ & $3 \times 10^{ 9}$  &     1300~~~~~ \\
                1000~~~ & $5 \times 10^{ 6}$ & $3 \times 10^{14}$ & $3 \times 10^{12}$  &    13000~~~~~ \\
                3000~~~ & $5 \times 10^{ 4}$ & $3 \times 10^{16}$ & $3 \times 10^{14}$  &    60000~~~~~ \\

\hline
\end{tabular}
\end{center}
\caption[Range of a 10 GeV/c plasmoid beam propagating in outer-space]{The range $z({\tilde{a}})$ at which the RMS radius $\tilde{a}$ of a high-energy plasmoid beam has expanded to a radius of one meter because of multiple scattering in the atmosphere is given as a function of altitude above ground. The initial radius and angular spread are assumed to be zero. The ionization-energy-loss range $z_I$ and the radiation length $X_0$ are larger than $z({\tilde{a}})$ in the high-atmosphere.}    \label{tab:plb}
\end{table}

As can be seen, while the range is less than one kilometer in the low atmosphere, it is on the order of 1'000 to 50'000 km in the ionosphere, i.e., at altitudes between 300 and 3'000 km which correspond to the mid-course flight of ICBM reentry vehicles.  Moreover, while the range is on the order of $X_0$ (the radiation length) at sea-level, it becomes very much smaller than  $X_0$ in outer-space.  Since $X_0$ is on the order of $X_n$ (the nuclear interaction length), and since both $X_0$ and $X_n$ vary in inverse proportion to the atomic density, this means that the effect of energy losses on beam expansion can be neglected for high-energy electron and proton plasmoid beams propagating through the ionosphere.  Therefore, the approximations made in deriving \eqref{ten:3}, namely that $W$ and $p$ are constant, are correct for the purpose of calculating the ranges given in the table.

In summary, particle/antiparticle plasmoid beams should be considered as serious candidates for use as outer-space directed energy weapons.  Such beams are in principle not deflected by the Earth's magnetic field.  To assess the feasibility of this concept, many issues have to be addressed: overall stability, propagation across a magnetic field, beam losses at the boundary layers, beam losses by particle-antiparticle annihilations, effects of ionospheric plasma electrons and ions on beam propagation and stability, etc.  However, if the plasmoid beam is very relativistic, e.g., electrons/positrons with energies in the GeV to TeV range (i.e., $\gamma = 10^3$ to $10^6$), many effects which depend on powers of the Lorentz factor $\gamma$  will ensure that it will behave much more as a true neutral beam than as a charged-particle beam during its flight towards a target.

\section{Deflection and guiding by conductors and channels}
\label{dgc:0}

   In this chapter we have so far considered the propagation of charged particle beams in infinite media, except possibly for a cut-off such as in the dimensionless inductance $\mathcal{L}$.  We now examine some of the most important effects of finite-distance boundary conditions such as the proximity of a conductive plane, pipe, or wire; as well as the consequences of propagating the beam in a prepared channel, in which parameters such as the conductivity, temperature, and/or density have a predetermined profile.  While theses boundaries can have detrimental and/or beneficials consequences for the long-distance propagation of particle beams, they can also be used within accelerator systems to manipulate high-current beam pulses and therefore to prepare (or `condition') them before injection into another other section of the accelerator-system, or into the outside environment (as will be discussed in Sec.~\ref{bco:0} on beam conditioning).

   As the effects of the external forces deriving from these interfaces and channels have to be evaluated in relation to the beam's internal cohesion forces, frequent reference will be made to the concepts developed in Sec.~\ref{coh:0}.  In the absence of a universally accepted terminology, we will try to use the term \emph{tracking} for the effect of the guiding forces deriving from a pre-existing channel (i.e., a prepared plasma distribution), and keep the term \emph{coupling} (i.e., self-tracking) for the effect of the (self-)guiding forces deriving from a self-generated plasma distribution (or channel).

   Also, as the prototype external force is the magnetic force induced by the geomagnetic field, we will adhere to the convention of using Earth's magnetic field as a yardstick for measuring and qualifying the strength of external deflection and guiding forces.\footnote{On that scale, where the unit is $\approx 0.5$ gauss, the cohesion forces typical of high-current particle beams are in the range of few milligauss to a few kilogauss, while those of a solid body are measured in megagauss.}

\subsection{Deflection and guiding by conductors}

When an electrically charged particle passes nearby a conductor it induces currents and charges in the conductor that react on the particle and modify its motion.  These effects are well known in classical electrodynamics and particle accelerators physics where the method of `image charges and currents' is often used to quantify them.  As will be seen, these forces can be attractive or repulsive, depending on the magnetic permeability of the conductor, and on the ionization state of the medium through which the particle is moving.

For of a beam of radius $r$ propagating at a distance $d \gg r$ parallel to a  plane interface made of a highly conductive and diamagnetic material, such as a metallic plate or a ground surface, the reaction (or image) force on the beam can be estimated in good approximation by the image force \cite{LINK-1968-}
\begin{equation}\label{dgc:1} %
          F_{r} = -\Bigl( (1-f_e) - (1-f_m)\beta^2 \Bigr)
                   \frac{1}{d} \frac{W}{I_A} I_B,
\end{equation}
which is trivially obtained by replacing the radial distance $r$ in Eq.~\eqref{ben:1} by $-2d$, i.e., minus the distance between the beam and its image \cite{FERNS1991B}.\footnote{In this reference equations (11a) and (11b) for the image force are similar to our equation \eqref{dgc:1} with the beam current $I_B$ replaced by the net current $I_N$, and $f_e$ and $f_m$ interpreted as the `plasma shielding fractions' affecting the image charges and currents. Once these factors are calculated in section III of this reference, our equation \eqref{dgc:1} is recovered.}   This force is weaker than the radial pinch force \eqref{ben:1} on the beam by the factor $r/2d \ll 1$.

   Equation \eqref{dgc:1} shows that the reaction force can be attractive or repulsive depending on the values of $f_e$ and $f_m$.  In fact, because of radial symmetry, the sign of this force is the same whether the beam propagates parallel to a plane, or to the wall of a conductive pipe. This has importants consequencies, both for a beam propagating inside an accelerating system, as for a beam propagating in open air over a ground plane or near an obstacle.

   If the beam is propagating in a gas and $f_e \approx 1$ the image force should be repulsive. This effect was observed in 1967--1968 in some of the earliest published high-intensity pinched-state propagation experiments \cite{LINK-1968-}.  Open shutter photographs show how a 50~kA, 3~MeV beam is deflected by a conducting sheet put at 45-degrees in the way of the beam.  Had the conducting sheet been normal to the pinched beam a large hole in the sheet would have resulted from each shot.  These effects were systematically studied in the Soviet Union, showing, for instance, that a self-focusing relativistic electron beam propagating along the surface of a ferromagnetic plate is attracted to it, and reflected in the case of a diamagnetic (e.g., copper)  plate \cite{DIDEN1976B}. In fact, if the magnetic permeability (measured at an appropriate high-frequency) of the medium in which the neutralized beam propagates is $\mu_1$, and that of the conducting material $\mu_2$, the image force \eqref{dgc:1} should be multiplied by the factor $(\mu_1 -\mu_2)/(\mu_1 +\mu_2)$.

   If the beam is propagating in vacuum, e.g., in an evacuated beam pipe, the image force is attractive and proportional to $1/\gamma^2$.  This leads to a destabilizing force which tends to amplify transverse perturbations of the beam, especially if the restivity of the wall is finite \cite{CAPOR1980-}. However, if the beam is propagating in a conductive tube filled with neutral gas at pressures suitable for self-pinched propagation (e.g., in the ion-focused regime), the image force is repulsive and has a remarquable stabilizing effect on the transverse perturbations of a finite radius beam.  This effect can be used for centering and quieting a beam, and will be further discussed in the Sec.~\ref{bco:0} on beam conditioning.

   When the beam propagates in open air, a most important effect to consider is the influence of a conducting ground plane \cite{FERNS1991B}.  The potential seriousness of this deflection is illustrated by assuming that the only force acting on a beam pulse is given by Eq.~\eqref{dgc:1} and that $d$ represents the height above ground of a horizontally propagating beam.  At the head of the pulse, where the air is still not ionized by the beam, $f_e=f_m=0$ so that the ground-plane force is attractive (as in vacuum) but very small since proportional to $-1/\gamma^2$.  But in the body of the pulse (where $f_e\approx 1$ and $f_m\approx 0$) the ground-plane force $F_r$ is repulsive and much larger since proportional to $+\beta^2$.  Equating $F_r/ m \gamma$ to the transverse acceleration one gets an equation of motion showing that the body of a beam initially at height $d$ can reach targets at the same height at range $z$ only if \cite{FERNS1991B}
\begin{equation}\label{dgc:2} %
          z < d \frac{I_A}{I_B}.
\end{equation}
Since ${I_A}/{I_B}= 3.3$ for a 10~kA, 1~GeV electron beam, this limitation would be catastrophic for $d$ measured in meters, e.g., if the accelerator is on board a ship at sea, and the target a sea-skimming cruise missile.  

However, the bound \eqref{dgc:2} grossly underestimates the range because the beam coupling forces were neglected.  Using \eqref{coh:12}, i.e., the complete equation for the transverse motion of a beam slice, or simply the concept of guiding point, one reaches the conclusion that the image force from a ground plane has a negligible impact on the trajectory of a high-energy beam.  This is because the deflection force on the beam head is proportional to $1/(d\gamma^2)$, i.e., very small when $\gamma$ and $d$ are sufficiently large.  In quantitative terms, for a beam with a current in the kA range, this happens when  $\gamma > 5$, and the ground plane lies ten or more beam radii away \cite{FERNS1991B}.  Under these conditions the beam tilts in order that the coupling forces can compensate for the larger value of the image force on the beam body, but does not tear or alter its trajectory.

   A final effect related to those examined in this section is the interaction of beam with the thin conductive or resistive wire.  If the wire is conductive and grounded, it can be seen as the limit of a very thin plasma channel, and was therefore used in early experiments to show that such a wire (or thin plasma channel) was able to guide a beam across a magnetic field \cite{DIDEN1977-}. If the wire is resistive it will become charged in the presence of a beam, which is then strongly guided and focused by the oppositely charged wire \cite{PRONO1983-}.  More generally, wires can be used in various configurations, including multiple wire arrays, either passively as in the two previous examples, or actively as current-carrying wires in beam transport and conditioning systems as will be seen in Sec.~\ref{bco:0}.

\subsection{Magnetic tracking in discharge plasma channels}

The theoretical and experimental study of high-current plasma discharges has been undertaken very early on, often by people active in the study of long range propagation of particle beams \cite{MANHE1973-}, or in the construction of compact high-power particle accelerators \cite{PAVLO1975-}.  An important motivation for this was the potential of such discharges to create high-conductivity reduced density channels suitable to guide and propagate high-power particle beams over long distances, based on the observation that lightning discharges are stabilized by the presence of the background atmosphere \cite{MANHE1973-}.

   The most simple way to produce such discharges is by electrically exploding a thin wire, e.g., \cite{PAVLO1975-}.  Several experiments demonstrated the possibility of using the resulting plasma channel to propagate, over meter-long distances, MeV-energy, multi-kA-current electron \cite{MILLE1977-} and proton \cite{OTTIN1979-, OLSEN1980-} beams.  Another technique, more suitable for repetitively pulsed systems, is that of laser-initiated discharge channels \cite{GREIG1978}, in which a special technique is used to produce long paths of ionization by causing breakdown in a rapidly retracted focal region \cite{WEYL-1979-}.  It was applied to both electron, e.g., \cite{MURPH1987-}, and proton beams \cite{OLSEN1982-, OZAKI1985-, NERI-1993-}, and confirmed the efficient propagation of similar high power beams over distances of 2--5~meter.

   The beam focusing and guiding forces active in this propagation mode derive from the hot, highly ionized channel created by the discharge, which ensures that the beam is quickly charge and current neutralized, and that the total magnetic field is frozen and equal to that of the pre-formed channel at the moment of injection \cite{MILLE1977-, OTTIN1979-}.  Thus, while the self-pinch force is zero, the beam is pinched by that field, and beam cohesion is provided by the electrostatic coupling force, Eq.~\eqref{coh:8}, so that the beam is tracking the channel.  Moreover, if the channel and the beam are subject to a transverse magnetic field $B_{\tr}$, the channel can guide the beam across that field, provided its intensity is such that the corresponding deflecting force on a beam particle is equal to twice the maximum value of the tracking force, a condition that is written $B_{\tr} > 2 B_{\theta}$ in Ref.~\cite{RUDAK1973-}.  This is explained by the fact that a fully neutralized beam-plasma-channel system is easily polarized by an external field, which implies that in order to push the beam out of the channel the external force has to overcome both the coupling force and the restoring force from the electric polarization field, which are nearly equal when the beam and the channel separate.  This was shown in experiments where the plasma channel was simulated by a thin aluminum cylinder enclosing the beam \cite{RUDAK1973-, BARAN1976-}.\footnote{A gas-filled metallic tube with a wall thickness of a $\approx$10~$\mu$m is yet another technique for guiding a beam: The small thickness ensures the absence of image forces, and the interactions of the beam with the wall and the gas create a plasma that can be suitable for various applications. A thin-walled dielectric tube whose diameter is smaller than that of the beam may also be used:  The resulting smaller diameter plasma channel can still efficiently guide a 10 to 20~kA beam across a magnetic field, as is a thin wire which can be thought as a equivalent to a very narrow and immobile plasma channel \cite{DIDEN1977-,PRONO1983-}.}

   This mode of propagation has therefore the potential to propagate a high-power beam over distances that are compatible with applications such particle-beam driven inertial confinement fusion \cite{MILLE1977-, OTTIN1979-,OLSON2001-}.  On the other hand, when propagating over longer distances, the beam tends to be ejected from the channel as soon as the beam current separates from the plasma return current \cite{DIDEN1977-,OTTIN1979-, MURPH1987-}.  This means that while the plasma channel itself is relatively stable (on a time-scale defined by magnetohydrodynamic stability \cite{MANHE1973-, OLSEN1982-}) these channels quickly expel the beam pulse through beam-induced return currents, simply because currents flowing in opposite directions repel each other.

   Rarefied but highly conducting channels do not, therefore, enable long-distance beam guiding and transport.  Moreover, if a reduced density non-conducting channel is prepared, experiment shows that reduced density has little effect on beam propagation except from a decrease in scattering until sufficient conductivity is generated to trigger instabilities which cause the beam to be repelled or expelled from the channel \cite{MURPH1987-}.  Consequently, other means to guide and increase the propagation range of particle beams have to considered, as will be done in the next three subsections.

\subsection{Electrostatic tracking in ion-focusing channels}

  While the seminal papers of Willard Bennett \cite{BENNE1934-,BENNE1955-} had a strong and lasting influence on the US beam community, a similar influence on the Soviet community came form those of Gersh Budker \cite{BUDKE1956A, BUDKE1956B}.\footnote{Budker's first presentation outside the USSR, in 1956 at a symposium at CERN, Geneva, Switzerland, had an equally strong influence on European scientists, e.g., \cite{LAWSO1958-,  LINHA1959-, FINKE1961-}. The only two references given by Budker in the paper (presented by A.A. Naumov) published in the symposium's proceedings \cite{BUDKE1956A} were to Bennett's papers \cite{BENNE1934-,BENNE1955-}.}  In Bennett's papers a major emphasis was on `fully charge and current neutralized beams' and their potential ability to direct energy over large distances straight across magnetic fields \cite{GUENT1980-}, while in Budker's papers a major emphasis was on `relativistic stabilized high-current beams' that could find applications in linear and circular particle accelerators, as well as in thermonuclear energy devices.  As it turned out, both concepts were equal fruitful, although Bennett's original concept of magnetohydrodynamically stable self-focusing beams \cite{BENNE1955-}) proved to be more difficult to put into practice than Budker's idea, which directly led to what is now termed the ion-focused regime (IFR) propagation mode of high power electron beams \cite{BRIGG1981-, BUCHA1987-, SWANE1993-}.

   While this propagation mode was indirectly observed in some of the earliest high-power beam propagation experiments, e.g., \cite{BEAL-1972-}, it was only in 1976/1977 that it was first properly explained in published papers, i.e., \cite{DIDEN1976A, BRIGG1977-}.  The essence of this propagation mode is that either by is own effect, or as a result of some preparation, the electron beam of line density $N_e$ is propagating through an ion-channel of line density $N_i$ such that Budker's condition is satisfied, i.e., $1/\gamma^2 < f_n < 1$ with $f_n=N_i/N_e$, so that the net radial pinch force on the electron beam is inwards.\footnote{The charge neutralization fraction $f_e$, whose definition is somewhat ambiguous in Eq.~\eqref{bcn:11}, is written $f_n$ to make clear that it is here defined as $N_i/N_e$.}   Propagation is then stable, provided the beam region is free of secondary electrons, which is ensured by the condition $f_n < 1$, so that most instabilities are avoided, and there is no return current flowing within the beam region as in a discharge channel.

   If  the beam and ion channels are assumed to have radial Gaussian profiles with RMS radii $a$ and $b$, respectively, their electrostatic interaction force is simply given by expression \eqref{coh:5} with $N_a=N_e$ and $N_b=N_i$.  For short separations between the centroids $x_e$ and $x_i$ of the beam and channel distributions, this force reduces to expression \eqref{coh:6}, and the equations of motion of the centroids are simply \cite[p.75]{BUDKE1956A}, \cite[p.685]{BUDKE1956B}, \cite[p.232]{FINKE1961-}, \cite[p.226]{BUCHA1987-}
\begin{gather}
\label{dgc:3} %
   m \gamma N_e \frac{d^2}{d t^2} x_e = N_e F_{e,ext}
              - \frac{e^2}{4\pi \epsilon_0} 2 N_e N_i\frac{x_e-x_i}{a^2+b^2},\\
\label{dgc:4} %
   M        N_i \frac{d^2}{d t^2} x_i = N_i F_{i,ext}
              + \frac{e^2}{4\pi \epsilon_0} 2 N_e N_i\frac{x_e-x_i}{a^2+b^2},
\end{gather}
where $m$ and $M$ are the masses of the electrons and ions, respectively, and $F_{e,ext}$ and $F_{i,ext}$ some external forces.  In this equation we have supposed that the ions do not move in the longitudinal direction, i.e., $\beta_i =0$ so that $\gamma_i =1$.

   To take a concrete example, let us assume that the external force is due to a transverse magnetic field $B_0$.  The equations of motion are then
\begin{gather}
\label{dgc:5} %
   m \gamma N_e \frac{d^2}{d t^2} x_e = e N_e \beta c B_0
              - \frac{e^2}{4\pi \epsilon_0} 2 N_e N_i\frac{x_e-x_i}{a^2+b^2},\\
\label{dgc:6} %
   M        N_i \frac{d^2}{d t^2} x_i = 
              + \frac{e^2}{4\pi \epsilon_0} 2 N_e N_i\frac{x_e-x_i}{a^2+b^2},
\end{gather}
and can be interpreted as follows:
\begin{itemize}

\item When the mass of the ions is very large, i.e., $M \rightarrow \infty$, the ions are essentially immobile and the two equations decouple since $x_i$ is constant.  Eq.~\eqref{dgc:5} has then a stationary solution in which the difference $x_e-x_i$ is a constant such that the right hand side is zero: This is an illustration of the application of the concept of coupling that was introduced in Sec.~\ref{coh:0}, and which allows a beam-background system to respond in such a way that the effect of the external force is compensated for.  In the present case the beam and plasma distributions have different radii, and the stationary solution exists only as long as the approximation $(x_e-x_i)^2 \ll a^2+b^2$ that enabled to write Eqs.~\eqref{coh:6} and \eqref{dgc:5} is valid.  This gives a limit on the maximum value of $B_0$, i.e., 
\begin{equation}\label{dgc:7} %
       B_0 < B_{max} =  \frac{\mu_0}{4 \pi}
                        \frac{f_n}{\beta^2}
                        \frac{2 I_B}{\sqrt{a^2+b^2}}.
\end{equation}
Taking for example $f_n = 0.1$, $I_B = 10$~kA, $\beta\approx 1$, and $\sqrt{a^2+b^2} = 0.1$~m, we get $B_{max} < 20$~gauss.  Such a beam could therefore moved undeflected through Earth's magnetic field, the immobile ion channel providing a guide.  The quantity $B_{max}$ characterizes the tracking force provided by the ion channel when a beam propagates in the IFR mode.  It means that such a channel can guide the beam straight across an electromagnetic field, or along a bend if the channel is not straight, provided that the magnitude of the corresponding electromagnetic or centrifugal forces are less than that of a transverse magnetic field of strength $B_{max}$.  Finally, when $\beta\approx 1$, Eq.~\eqref{dgc:7} has a form similar to the magnetic field from a thin conductor at the distance $r=\sqrt{a^2+b^2}$.  This means that the guiding force provided by a charged or current-carrying wire is equivalent to that of a channel of infinitely heavy ions, and vice versa.

\item   Adding Eqs.~\eqref{dgc:5} and \eqref{dgc:6} yields an equation giving a linear combination of the transverse accelerations of the centroids of the electron and ion distributions in terms of the external field $B_0$.  If we postulate that due to coupling the beam follows its head (i.e., that the condition \eqref{dgc:7} is satisfied), and that the beam head continuously produces the ion channel so that the centroids of the ion and electron distributions follow parallel trajectories, we have $\ddot{x}_e(t,z)=\ddot{x}_i(t,z)$ in the body of the pulse.\footnote{Note that IFR propagation with the beam creating the required ionization can be problematic, see \cite{SMITH1986-}.}   The sum of Eqs.~\eqref{dgc:5} and \eqref{dgc:6} is then identical to the equation for the circular motion in a transverse magnetic field of a beam of `heavy particles' that would still have the velocity corresponding to $\gamma$, but an effective mass larger than $m$ and equal to
\begin{equation}\label{dgc:8} %
          m_{eff} = m (1 + f_n \frac{M}{\gamma m})  >  m.
\end{equation}
Thus, if the beam consists of electrons and the neutralizing channel of nitrogen ions, the radius of curvature of the beam body will be $ \approx 26'000 f_n/\gamma$ times larger than the Larmor radius of a single electron in the same magnetic field.\footnote{Since we neglect the self-fields except for their effect along the beam radius, this radius of curvature does not include the toroidal correction given by Eq.~\eqref{mag:5}.}  On the other hand, if the beam consists of protons and the neutralizing channel of electrons, the radius of curvature of the beam body will be only slightly different from the Larmor radius of a single proton.  This is because the tracking effect of the neutralizing channel in the ion-focused regime exists only if that channel can be considered as immobile, which is obviously not the case when an ion beam is neutralized by an electron background, whether the channel is self-generated by the beam or pre-formed by some other means.  Therefore, while we have seen in Sec.~\ref{mag:0} that the trajectory of a beam pulse is determined by the Larmor radius at its head, which is defined by $m$ rather than by $m_{eff}$, we can interpret the bound \eqref{dgc:8} as a condition for the body of a pulse to be tracking its head.  We can thus conclude that a beam propagating in the IFR mode and subject to an external force will be tracking the channel provide the conditions implied by Eqs.~\eqref{dgc:7} and \eqref{dgc:8} are both strongly satisfied.\footnote{A condition more general than \eqref{dgc:7}, in which $f_m$ is replaced by $\gamma M_{eff}/M$, is obtained by subtracting Eqs.~\eqref{dgc:5} and \eqref{dgc:6}, and requiring that $(x_e-x_i)^2 \ll a^2+b^2$.}

\end{itemize}

The guiding effect of IFR channels has been verified in a number of experiments.  Since the linear particle number density of a 10~kA relativistic beam is $N = I_B/(e\beta c) \approx 2 \times 10^{14}$~m$^{-1}$, the condition of partial neutralization $f_n < 1$ implies that for a 10~cm radius channel the required ion density corresponds to a partial pressure of less than $10^{-3}$~torr in a nitrogen atmosphere.  This means that IFR guiding is not very suitable for directing a beam through full-density air (in which it would be difficult to bore a wide and nearly fully evacuated channel), but rather for guiding a beam through an accelerator system or the ionosphere after preparing a low-density ionized channel with a laser beam.

   Possibly the first published paper on IFR propagation in a channel is a Soviet experiment in which a $p = 10^{-4}$ to $10^{-5}$~torr plasma channel was prepared by a 30~kW discharge in a 8~cm diameter and 1~m long silica tube.  A 1.5~MeV, 30~kA, 50~ns electron beam pulse was successfully transmitted, while pinching down from an initial 6~cm to a final 4~cm diameter radius \cite{DIDEN1976A}.  Optimum transmission was found for $f_n \approx 10$, i.e., substantially larger than one, which is not an obstacle as long as all secondary electrons are expelled by the beam's space charge field \cite{BRIGG1977-}.  In follow-up experiments, it was shown that the guiding effect of the channel was sufficient to enable straight motion across a magnetic field of intensity up to about 100~gauss, under conditions ($f_n \approx 1$) such that Eq.~\eqref{dgc:7} gives $B_{max} \approx 800$~gauss \cite{DIDEN1977-}.

   This opened the way to many applications, of which the most important is probably the guiding of high power beams through compact high-energy accelerators, see  Refs.~\cite{BRIGG1981-, PRONO1985B, CAPOR1986-, MILLE1987-} and Sec.~\ref{ata:0}, without which the possibility of sending lethal beams into the atmosphere would not exist.  In that application the radius of the guiding channel is generally intentionally smaller than the beam radius in order that the radial force be anharmonic and lead to damping of the transverse beam motion \cite{CAPOR1986-}.  This effect will be further discussed in Sec.~\ref{bco:0} on beam conditioning.

   Straight IFR channel can be used for injection and extraction of a relativistic electron beam in a high-current betatron \cite{HUI--1984B}, and bent IFR channels to deflect high-current electron beams.

  High-current electron beam bending using IFR channels and magnetic fields can been achieved using an number of techniques.  For example, two straight laser generated channels can be made to intersect at 45$^{\rm{o}}$ in the field of a relatively low intensity bending magnet.  The electron beam entering through one of the channels can then exit through the other one, the guiding forces from the channels, and the bending force from the magnet, providing enough steering for the beam to switch from one channel into the other \cite{FROST1985-}.  Two such 45$^{\rm{o}}$ bends can make a 90$^{\rm{o}}$ bend, which can be used to recirculate and accelerate a low-energy electron beam to a higher energy \cite{FROST1985-,SHOPE1986-}.  Another technique is to use a very-low energy, very-low current electron beam (e.g., 800~V, 250~mA produced by a hot tungsten filament) to form a bent channel in a reduced density gas by sending this beam through a low intensity (100~gauss) transverse magnetic field.  The resulting channel is sufficient to guide a 1~MeV, 18~kA, 1.5~cm radius electron beam through a 90$^{\rm{o}}$ bend \cite{SHOPE1985-}.

\subsection{Electrostatic channel-tracking in full-density air}

Due to the importance of conductivity in neutralizing a charged particle beam, and getting it to pinch, it was thought that a localized channel of higher conductivity could result in an attractive  electrostatic force to guide a beam by keeping it in the channel.  Such a conductivity channel in the atmosphere could be left behind by a previous beam pulse.  It could also be intentionally created by a laser pulse fired immediately preceding the beam pulse. 

    As a matter of fact, early calculations using a simplified model indicated that such an attractive electrostatic force exists, although usually small compared with the pinch (and thus coupling) force \cite[p.8]{LEE--1983A}.  More detailed computer simulations then showed that the electrostatic force produced by low-level channel preionization (in order to avoid problems with a possible return current) is weak, short lived, and partially repulsive \cite{HUI--1984A}.  These features were finally confirmed by analytical calculations, which clearly showed that depending on the relative radii of the beam and conductivity channels, the tracking force could be attractive or repulsive as a function of the position of the beam slice behind the beam head \cite{OBRIE1990A}.  In these calculations the strength of this force could also be accurately evaluated, and found to be equivalent to never much more than a few gauss near the guiding point in the head  region of a 10~kA pulse.

   It can therefore be concluded that `conductivity-channel tracking' is not a practical concept for guiding a beam.

\subsection{Magnetic channel-tracking in full-density air}
 
    An intense particle beam propagating through air leaves a channel that has a reduced density (after expansion) and a residual conductivity.  Subsequent pulses or even later portions of a long pulse may be guided by this channel and may propagate more easily in it.  This hole boring or density reduction effect has been noted as a means for increasing the propagation range of an intense beam in the atmosphere \cite{MURPH1987-}.  However, as explained in the previous subsections, reduced density on its own has little effect on beam guiding, too much conductivity leads to a return current which tends to expel the beam from the channel, and a low conductivity leads to an electrostatic force that is too weak and uncertain to be effective.

   Thus, it would seem that `hole boring' and `channel guiding' by a leading pulse in a train of pulses sent into the atmosphere could not work, unless a propagation regime in which a suitable electromagnetic tracking force exists could be found \cite{MURPH1987-}. 

   A magnetic guiding force of an initially un-ionized reduced density channel on an intense beam pulse was discovered in 1987, using three-dimensional simulations, and confirmed in 1990 in a double-pulse electron beam experiment \cite{WELCH1990-}.  The attraction between the low-density channel and the beam results from beam impact ionization and electrical conductivity dependence on the plasma electron temperature.  The higher  electron temperature in the rarefied channel depresses the conductivity and return current in the channel, thus shifting the centroid of the net current toward the channel axis.  Magnetic attraction between beam current and net current pulls the beam into the channel \cite{FERNS1991A}.  Contrary to electrostatic tracking in a very-low-density  ion-focusing channel, this magnetic tracking mode works for both electron and ion beams in relatively high-density channels. 

   In this tracking mode, also called `density-channel tracking,' the density reduction in the channel has to remain modest, i.e., on the order of a tenth to a third of ambient density, so that on-axis ionization by avalanching is weak and excessive return current is avoided.  A remarkable property of density tracking is its weak dependence on density reduction, with a guiding force equivalent to a few gauss per kA beam current, and only a logarithmic increase of that force below one tenth atmospheric density \cite[Fig.2]{FERNS1991A}.  Effects that degrade tracking include high-order chemistry effects and channel preionization, which both influence the mechanisms of conductivity generation \cite{FERNS1991A, MURPH1992-}.  Experiments showing that the magnetic tracking force is equivalent to at least 10~gauss for 10~kA beams, and exploring effects that degrade guidance, will be discussed in Sec.~\ref{usa:0}.

   As density tracking works for channels with on-axis reduced density on the order of 0.1 to 0.3 atmosphere, it enables multipulse propagation over distances equal to at least 3 to 10 Nordsieck lengths.  Multipulse hole-boring and propagation is therefore feasible for endoatmospheric applications such as point defense where a range of a few kilometers is sufficient.

\chapter{Injection of a high-power beam into the atmosphere}

\section{Plasma generation by a particle beam}
\label{pla:0}

In a endo-atmospheric system, except when the beam is launched into the ionosphere or a pre-formed plasma channel, the beam will, in general, be injected into initially un-ionized air.  At the very head of a beam pulse, there is thus no background plasma and therefore, both $f_e$ and $f_m$ are zero.  The beam head, therefore, expands at a rate governed by the net radial electromagnetic force, emittance and scattering.  However, as the beam ionizes the air, a plasma of increasing density builds up and, as plasma currents start flowing, the space-charge gets progressively neutralized.  When $I_E=0$, that is when $\gamma^2(f_m \beta^2 - f_e) = 1$ according to \eqref{ben:3},  the net electromagnetic force changes sign, and the beam stops expanding out and starts pinching in.

During the self-pinching of the beam, both the radius $a$ and the effective beam current $I_E$ change rapidly, and this happens while the beam current $I_B$, which in practice has a finite rise-time, increases.  As a result, a strong electric field is induced.  This field, given by \eqref{bcn:12}, accelerates the plasma electrons, which, as they gain sufficient energy, start ionizing the gas as well and therefore trigger an avalanche of secondary plasma electrons.  The rate of neutralization of the beam thus increases, until it levels off as the decreasing beam radius approaches an equilibrium radius given by the Bennett pinch relation \eqref{ben:4} with $f_e \approx 1$.  At this point, the induced electric field becomes very small and, the plasma current $I_P$, after having gone through a maximum, starts decreasing according to \eqref{bcn:15}.  Finally, for a sufficiently long pulse, after the plasma current has completely decayed, the beam becomes fully pinched.

\begin{figure}
\begin{center}
\resizebox{12cm}{!}{ \includegraphics{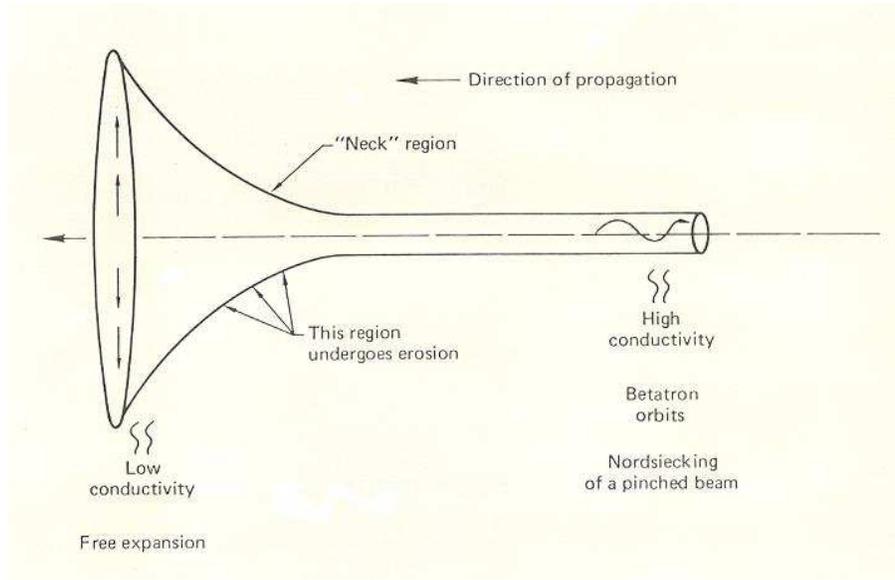}}
\caption[Propagation of a pinched beam]{\emph{Propagation of a pinched beam.} The diagram shows a propagating beam indicating the pinching, necking, and freely expanding regions. In the high-conductivity pinched (also called `Nordsiecking') region the particles perform betatron orbits with a slowly increasing radius.  In the neck region the beam expands and undergoes erosion as particles are lost as a result of weakening pinch forces.  Finally, in the low-conductivity free expansion region where the plasma effects are not sufficient to pinch the beam, the particles move away from each other because of Coulomb repulsion.   
\label{fig:pro}}
\end{center}
\end{figure}

	As a result of this process, the beam pulse takes on a characteristic `trumpet' shape, and can be divided into four distinct regions; the expanding beam head, the neck region in which $E_z$ and $f_m$ are maximum, the body in which $f_e \approx 1$ and Nordsieck's equation becomes a good approximation of \eqref{nor:1}, and finally the tail where $f_e=1$ and $f_m \approx 0$.  (See Figure~\ref{fig:pro}.)

	The description of this process requires equations for the plasma electron density $n_e$ and temperature $T_e$ so that the conductivity $\sigma$, and thus $\tau_e$ and $\tau_m$ can be calculated.  In many cases these equations can be be cast as a one-dimensional equation that it is usually called the \emph{conductivity equation}. Furthermore, in order that all quantities can be expressed self-consistently, an equation relating the various fields and currents is needed.  When this equation can be reduced to a one-dimensional equation it is called the \emph{circuit equation} because it basically yields the plasma current which returns the beam current to the accelerator.  In the general case this last equation has to be derived from Maxwell's equations in which the sources are the beam current density $J_B$ and the plasma current densities.

\subsection{Maxwell's equations and beam coordinate system}

   If a high-intensity beam is injected into full-density air, or into a reduced density channel in which the pressure in not much lower than 0.1~atmosphere, the plasma generated by the beam can be described by a simple scalar conductivity --- provided the external magnetic field is not too strong, e.g., on the order of Earth's magnetic field.  The plasma current density can then be related to the electric field by Ohm's law, Eq.~\eqref{bcn:6}, so that Maxwell's equations are
\begin{align}\label{pla:1} 
   \vec{\nabla} \times \vec{B} & = \mu_0 \Bigl( \vec{J_B} + \sigma\vec{E} \Bigr)
                     + \frac{1}{c^2} \frac{\partial \vec{E}}{\partial t} \\
   \vec{\nabla} \times \vec{E} & = - \frac{\partial \vec{B}}{\partial t}, \\
   \vec{\nabla} \cdot  \vec{B} & =  0,   \\
   \vec{\nabla} \cdot  \vec{E} & = \frac{1}{\epsilon_0} \rho.
\end{align}
In solving these equations it is often useful to introduce a scalar potential $\phi$ and a vector potential $\vec{A}$ such that
\begin{equation}\label{pla:2} 
            \vec{B} =  \vec{\nabla} \times \vec{A}, ~~~  ~~~
            \vec{E} = -\vec{\nabla} \phi - \frac{\partial \vec{A}}{\partial t}.
\end{equation}

\begin{figure}
\begin{center}
\resizebox{12cm}{!}{ \includegraphics{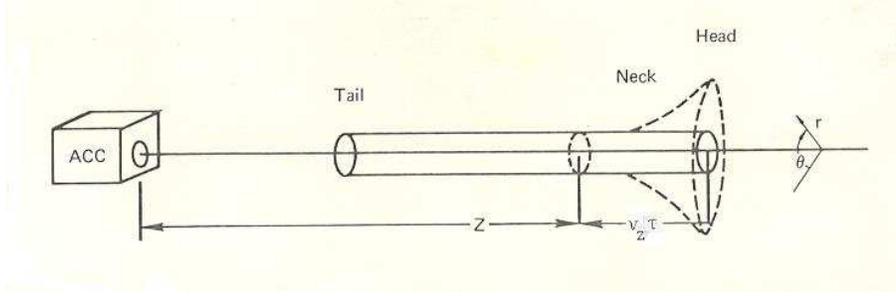}}
\caption[Coordinate system used in describing the propagating beam]{\emph{Coordinate system used in describing the propagating beam.} The variable $z$ is the distance from the accelerator to a point within a beam pulse, while the variable $\tau$ is the time such that the distance from this point to the beam head is equal to $v_z\tau$, where $v_z$ is the longitudinal velocity of the beam's particles.    
\label{fig:coo}}
\end{center}
\end{figure}

   The conductivity and circuit equations, which describe quantities directly related to the penetration distance of the beam into the gas, are best written down as a function of the variable
\begin{equation}\label{pla:3} 
    \tau \DEF t - z/\beta c.
\end{equation}
$\tau$ has the dimension of the time and is a fixed label for a particular beam particle (or beam slice) within the pulse, provided this particle (or slice) does not move relative to the beam head.  For the beam head $\tau=0$, and for the tail-end $\tau = \Delta \tau$, the beam pulse duration.  (See Figure~\ref{fig:coo}.)  For any function $f$, the partial time derivative transforms as 
\begin{equation}\label{pla:4} %
    \Bigl(\frac{\partial f(z,t)   }{\partial    t} \Bigr)_z  = 
    \Bigl(\frac{\partial f(z,\tau)}{\partial \tau} \Bigr)_z, 
\end{equation}
and the total time derivative as
\begin{equation}\label{pla:5} %
                  \frac{d        f        }{d        t}             =
            \Bigl(\frac{\partial f(z,t)   }{\partial t} \Bigr)_z    +
    \beta c \Bigl(\frac{\partial f(z,t)   }{\partial z} \Bigr)_t    = 
    \beta c \Bigl(\frac{\partial f(z,\tau)}{\partial z} \Bigr)_\tau. 
\end{equation}
In the $(z,\tau)$ variables $z$ plays therefore the role of time in the particle dynamics.  Consequently, the time derivative of a kinematical variable such as $\vec{r}$ is
\begin{equation}\label{pla:6} %
       \dot{\vec{r}}  \DEF \frac{d \vec{r}}{dt} 
     = \beta c \Bigl(\frac{\partial \vec{r}(z,\tau)}{\partial z} \Bigr)_\tau.
\end{equation}

    Instead of the standard cylindrical coordinates ($r,\theta, z, t$) we will now use as independent variables the coordinates ($r,\theta, z, \tau$) where $z$ is the propagation distance for a beam segment from its point of injection.  Dissipative processes such as energy losses and collision driven emittance growth will thus be calculated as functions of $z$, as was already done in Sec.~\ref{nor:0}.

   In this context it should be mentioned that instead of $\tau$ (the `time within the pulse') it is also possible to introduce a variable $\zeta \DEF \beta c t - z$ which measures the `distance behind the beam head.'  This has the advantages that apart from $\theta$ all coordinates have the dimension of a length, and that for an ultrarelativistic beam $\beta=1$ so that $\zeta = c t - z$.  This greatly simplifies some physical formulas and computer algorithms related to the propagation of very relativistic electron beams.  On the other hands, once an equation or formula is written using the $\zeta$ variable or the ultrarelativistic limit it can be very difficult to modify them for use in the normal-relativistic or non-relativistic cases.  For these reasons, the $\zeta$ variable and the ultrarelativistic limit should be used with great caution, e.g., only in specialized computer programs.

\subsection{Circuit equation}

	The circuit equation has to be derived from the full set of Maxwell's equations.  This is a very difficult task, even for relatively simple problems, because there are easily half a dozen variables and equations.  Indeed, it is not obvious which variable or equation, and under which circumstances, can be neglected or not.  It therefore took a number of years to find a satisfactory approach, which, as we shall see, provides an algorithm suitable for both detailed computer simulations and physically intuitive analytical calculations.

  For example, if we take a rigid beam propagating in a plasma described by a scalar conductivity $\sigma$, and if we suppose exact cylindrical symmetry so that the $\theta$ variable can be ignored, the whole set of Maxwell's equations can be reduced to a single equation for the $z$ component of the vector potential \cite{MCART1973-}
\begin{equation}\label{pla:7} %
        \frac{1}{r} \frac{\partial      }{\partial r}
           \Bigl( r \frac{\partial   A_z}{\partial r} \Bigr)
     +              \frac{\partial^2 A_z}{\partial z^2}
     - \frac{1}{c^2}\frac{\partial^2 A_z}{\partial \tau^2}
     = - \mu_0 \Bigl( J_B - \sigma \frac{\partial A_z}{\partial \tau} \Bigr),
\end{equation}
where the last term is the plasma current given by Ohm's law with $E_z= -{\partial A_z}/{\partial \tau}$. This equation is exact, but impossible to solve except for very simple boundary conditions.  In order to proceed it is necessary to make further assumptions.  For instance, the neglect of ${\partial^2 A_z}/{\partial z^2}$ is reasonable when the $z$ dependence of $J_B$ (the driving function here) is much weaker than the $r$ or the $\tau$ dependences.  On the other hand, the neglect of ${\partial^2 A_z}/{\partial \tau^2}$ (the displacement current) is not so obvious \cite{MCART1973-}.  The main reason for these difficulties is that while the potential $A_z$ implicitly contains all the information about the system and its evolution, nothing is explicitely known about how this information is compounded.

   A more constructive approach is to start from Maxwell's equations and to use an ordering scheme such that the higher order terms corresponding to a given problem can be identified, and the neglect of the smaller terms justified \cite{LEE--1973E}.  For instance, if the essential properties are the paraxial approximation, $I_E \ll I_A$, and the magnetostatic limit, $a \ll c\tau_m$, it is found that the large field components are $B_r$, $B_\theta$, and $E_z$, and that these components obey the equation
\begin{equation}\label{pla:8} %
      \frac{1}{r} \frac{\partial    }{\partial      r} \Bigl( r B_\theta \Bigr)
    - \frac{1}{c} \frac{\partial B_r}{\partial \theta}
  = - \mu_0 \Bigl( J_z - \sigma E_z \Bigr)
    + \frac{1}{c^2} \frac{\partial E_z}{\partial \tau},
\end{equation}
where the last term is on the order of $a/c\tau_m$ smaller than the previous terms.  Therefore, neglecting this last term corresponds to ignoring effects which are significant where the radius is large and the magnetic diffusion time (and thus the conductivity) small, i.e., effects that are typical of the beam head.  Consequently, by defining a potential $A$ such that
\begin{equation}\label{pla:9} %
       B_\theta =             -\frac{\partial A}{\partial      r}, ~~~  ~~~
       B_r      =  \frac{1}{r} \frac{\partial A}{\partial \theta}, ~~~  ~~~
       E_z      =             -\frac{\partial A}{\partial   \tau},
\end{equation}
and rewriting \eqref{pla:8} without the last term as \cite{LEE--1973E}, \cite[p.8]{LEARY1972-}
\begin{equation}\label{pla:10} %
        \frac{1}{r} \frac{\partial    }{\partial r}
           \Bigl( r \frac{\partial   A}{\partial r} \Bigr)
     - \frac{1}{r^2}\frac{\partial^2 A}{\partial \theta^2}
     = - \mu_0 \Bigl( J_z - \sigma \frac{\partial A}{\partial \tau} \Bigr),
\end{equation}
we obtain a field equation that is appropriate for the body of a paraxial beam pulse in cylindrical coordinates, or more generally such a beam provided $a \ll c\tau_m$.  Formally, this equation is similar to Eq.~\eqref{pla:7} without the second derivative and $\theta$ terms, but with a clear understanding of its range of applications. 

    Of particular interest is to solve the field equation \eqref{pla:10} under conditions which have a simple physical interpretation.   This is for instance the case when reasonable similarity assumption are made on the radial profiles of the various charge and current distributions.  In that case, as we shall see, the circuit equation takes a remarkably simple form.  First, it has to be noticed that because the plasma electrons are mostly generated directly by the beam, $n_e$, and thus $\sigma$, will have radial distributions close to that of $J_B$.  Second, as a consequence, the plasma current $J_P$ will also have a distribution similar to $J_B$ \cite{KINGS1973-} and $f_m$, just like $f_e$, will be independent of $r$.  The natural radial distribution function to take is that of a Bennett profile.  For $J_B$ it is given by Eq.~\eqref{ben:21}, where the radius is allowed to be a function $a(\tau)$, and for the conductivity we may write
\begin{equation}\label{pla:11} %
     \sigma(r,\tau) = \sigma_0(\tau) \bigr(1 + \frac{r^2}{a^2}\bigl)^{-2}.
\end{equation}
As for the potential $A$, which in Eq.~\eqref{pla:10} can be interpreted as an effective potential associated to the problem under consideration, we take \cite[p.66]{LEE--1976B}
\begin{equation}\label{pla:12} %
     A(r,\tau) = -\frac{\mu_0}{4\pi} I_E \ln \frac{a^2 + r^2}{a^2 + b^2},
\end{equation}
where $b$, which may possibly be function of $\tau$, is a cut-off measuring the maximum radial extent of the plasma generated by the beam.  As a matter of fact, if $B_\theta$ is calculated from \eqref{pla:12} using \eqref{pla:9}, the corresponding pinch force is that given by Eq.~\eqref{ben:1} because $I_E(\tau)$ is defined as the effective pinch current.  Then, inserting Eqs.~\eqref{ben:21}, \eqref{pla:11}, and \eqref{pla:12} in Eq.~\eqref{pla:9}, and using the definition of the magnetic diffusion time \eqref{bcn:16}, the field equation simplifies and becomes the circuit equation\footnote{It is remarkable that this equation is formally very similar to Eq.~\eqref{bcn:15} giving the plasma current in terms of the net current.} 
\begin{equation}\label{pla:13} %
    I_E = I_B  - \tau_m        \frac{\partial}{\partial\tau}(\mathcal{L}I_E),
\end{equation}
where $I_E$, $I_B$, and $\tau_m$ are the on-axis values of the corresponding distributions, and
\begin{equation}\label{pla:14} %
    \mathcal{L} = \tfrac{1}{4} \ln(1+\frac{b^2}{a^2}),
\end{equation}
the dimensionless inductance associated to a Bennett distribution.

  Equations \eqref{pla:10} and \eqref{pla:13} are strictly valid only in the body of a beam pulse.  In order to study phenomena such as beam head expansion and erosion, as well as plasma generation at the head of a beam penetrating an initially unionized gas, the difficulty is that the magnetostatic approximation $a \ll c\tau_m$ is strongly violated when the conductivity is very low.  However, in the paraxial approximation, it has been found that the three-dimensional Maxwell's equations expressed in cylindrical coordinates can  be greatly simplified \cite{CHAMB1979A, CHAMB1981-, HUI--1984A, KRALL1989B}, leading to a field equation and a circuit equation which are only slightly more complicated than Eqs.~\eqref{pla:10} and \eqref{pla:13}.

  This major advance is mainly due to Edward P.~Lee at the Lawrence Livermore National Laboratory, who suggested the use of a particular form of the Coulomb gauge \cite[Sec.6.5]{JACKS1975-} to take maximum advantage of the cylindrical coordinate system, i.e., $\vec{\nabla}_{\tr} \cdot \vec{A}_{\tr} = 0$.  Without entering into too much details, and restricting ourselves to axisymmetric beams so that only monopole fields have to be considered and all quantities are $\theta$ independent, the exact field equation in the very highly relativistic paraxial limit can be written as \cite{CHAMB1979A}
\begin{equation}\label{pla:15} %
        \frac{1}{r} \frac{\partial    }{\partial r} r
     \Bigl(1+ \frac{\epsilon_0}{\sigma} \frac{\partial }{\partial \tau} \Bigr) 
          \frac{\partial   A}{\partial r}
     - \frac{1}{c^2}\frac{\partial^2 A}{\partial \tau^2}
     = - \mu_0 \Bigl( J_z - \sigma \frac{\partial A}{\partial \tau} \Bigr).
\end{equation}
Here the scalar function $A=A_z - \phi$ is the effective potential.\footnote{As is well known, it is always possible to express the full Maxwell field in terms of two independent scalar functions.}  Because of the term containing the $\epsilon_0/\sigma=\tau_e$ factor the radial electric field is non-zero, and
\begin{equation}\label{pla:16} %
 E_z = -\frac{\partial A}{\partial \tau},  ~~~  ~~~
 E_r = -\frac{\epsilon_0}{\sigma} \frac{\partial^2 A}{\partial r \partial \tau}.
\end{equation}
This radial electric field gives a contribution to the pinch force that is given by
\begin{equation}\label{pla:17} %
  F_{em} = e (E_r - \beta B_\theta) = e \frac{\partial A}{\partial r}.
\end{equation}
Therefore, as with Eq.~\eqref{pla:10}, $B_E = -{\partial A}/{\partial r}$ can be interpreted as an effective magnetic field driving the pinch force, while the true magnetic field is given by
\begin{equation}\label{pla:18} %
 B_\theta = - \Bigl(1+ \frac{\epsilon_0}{\sigma}
                       \frac{\partial }{\partial \tau} \Bigr) 
                       \frac{\partial   A}{\partial r}.
\end{equation}

   Assuming again that both the beam current and the conductivity distributions have a Bennett profile, and taking for $A$ the Bennett potential \eqref{pla:12}, the complete circuit equation is found to be
\begin{equation}\label{pla:19} 
    I_E = I_B - \tau_e    a^2 \frac{\partial}{\partial\tau}(\frac{I_E}{a^2})
              - \tau_m        \frac{\partial}{\partial\tau}(\mathcal{L}I_E)
              - \tau_e \tau_m \frac{\partial^2}{\partial^2\tau}(\mathcal{L}I_E).
\end{equation}
The second term on the right is the radial current associated with the charge neutralization process and is thus most important at the beam head and neck.  The third term is the plasma return current $I_P$.  The last term is the displacement current which can be generally neglected, except possibly at the very head of the beam because $\tau_e \tau_m = a^2/c^2$.  In this equation, $\tau_e$ and $\tau_m$ are calculated from the on-axis values of $\sigma(\tau)$.

\subsection{Conductivity equation}

	The exact form of the equation giving the plasma conductivity depends considerably upon the chemical composition, the pressure, and the temperature of the gas, as well as on the intensity of the electric field $E$.  Moreover, while in a relatively dense gas the conductivity is given by Eq.~\eqref{bcn:6}, and is therefore a function of just $n_e$ and $\nu$, it depends in the general case on many more parameters and effects than just electron-neutral momentum transfer collisions.

   If the \emph{total conductivity} is defined in such a way that the plasma current density is given by Ohm's law $J_P=\sigma E_z$, a phenomenologically more complete expression is given by \cite{IYYEN1989-}
\begin{equation}\label{pla:20} %
   \frac{1}{\sigma} = \rho_{en} + \rho_{ei} + \rho_{ia} + \rho_{2s} + \rho_{2w},
\end{equation}
where $\rho_{en}$, $\rho_{ei}$, $\rho_{ia}$, $\rho_{2s}$, and $\rho_{2w}$ are the electron-neutral, electron-ion, ion-acoustic, two-stream, and two-stream-wave resistivities, respectively.  $\rho_{en}$ is given by Eq.~\eqref{bcn:7}, i.e.,
\begin{equation}\label{pla:21} %
        \rho_{en} =  \frac{m_e}{e^2} \frac{\nu_{en}}{n_e},
\end{equation}
where $\nu_{en}$ is the electron-neutral momentum transfer collision frequency.  $\rho_{ei}$ is the \emph{electron-ion resistivity of Spitzer} \cite{SPITZ1956-}
\begin{equation}\label{pla:22} %
        \rho_{ei} =  \frac{1}{2} \bigl( \frac{\pi}{2}       \bigr)^{1/2}
                           m_e c \bigl( \frac{r_e}{e}       \bigr)^{2}
                                 \bigl( \frac{m_e c^2}{T_e} \bigr)^{3/2}
                                 Z \ln{\Lambda}
   \approx 3 \times 10^{-5} \frac{Z \ln{\Lambda}}{T_e^{3/2}}
              ~~\Omega\cdot\text{m]},
\end{equation}
where $T_e$ is expressed in eV and the factor $\ln{\Lambda} \approx 10$ is called the Coulomb logarithm, a slowly varying function of $n_e$ and $T_e$.  The other resistivities in Eq.~\eqref{pla:20} are given in the Ref.~\cite{IYYEN1989-} for a beam of electrons, while the two last ones (which correspond to energy dissipation by means of two-stream instabilities) will be further discussed in Sec.~\ref{mic:0} when considering beam-plasma heating by instabilities.

   The general form of the plasma electron density equation is \cite{MCART1973-, IYYEN1989-}
\begin{equation}\label{pla:23} 
       \frac{\partial}{\partial\tau} n_e =  \beta c \frac{S}{w} n_b
                                         +  \nu_E n_e
                                         -  \nu_A n_e
                                         - \alpha_r n_e^2.
\end{equation}
The first term is the direct beam ionization rate where $S(\beta)$ is Bethe's stopping power, and $w = 33.7$ eV for air, the energy required to create one electron-ion pair.  The second source term corresponds to ionization by avalanche in the  induced electrical field at the beam head.  The third corresponds to conductivity electron loss by attachement to various molecules, and the last one to losses by molecular dissociative recombination.  In general, the avalanche and attachment rates, $\nu_E$ and $\nu_A$, and the recombination coefficient $\alpha_r$, depend on the gas pressure, the electric field, and the plasma temperature.  Therefore, for a real gas like air, a complete description would require a set of rate equations coupling the various populations of molecular and atomic nitrogen, oxygen, water, etc., in various states of ionization, to the beam source terms.  This leads to the development of complicated phenomenological models \cite{CARY-1980-, STRAS2003-}.

   The equation for the plasma electron temperature $T_e$ incorporating the same effects as those included in \eqref{pla:20} is \cite{IYYEN1989-}
\begin{equation}\label{pla:24} %
  \frac{\partial}{\partial\tau} \Bigl(\frac{3}{2}n_e T_e + n_e w_i \Bigr)
   = J_P^2(\rho_{en} + \rho_{ei} + \rho_{ia} + \rho_{2s})
   + J_B^2 \rho_{w} - P_B - P_C,
\end{equation}
where $w_i$ is the threshold ionization energy, $P_B$ the power lost because of brems\-strah\-lung and $P_C$ the power lost as a result of electron-conductivity cooling, both given in  Ref.~\cite{IYYEN1989-}.  As the average energy of a secondary electron is $w_s \approx 7.55$~eV for air, the average threshold ionization energy is $w_i = w - w_s \approx 26$~eV.  The first term on the right of Eq.~\eqref{pla:24} corresponds to heating by the return-current, to be discussed in Sec.~\ref{ohm:0}, and the second one to heating by wave resistivity, to be discussed in Sec.~\ref{mic:0}.

   As can be seen, the number of effects that may have to be included in the general case is quite large, which is why the expressions used in different contexts can considerably vary, see, e.g., \cite{HAMME1979-, HAAN-1982-, JANNS1984-, GLAZY1991-, OLIVE1996-}.  In practice, it is often sufficient to consider just the first two terms in the conductivity equation \eqref{pla:20}.  In that case $\sigma$ is given by Eq.~\eqref{bcn:7} where the effective collision frequency $\nu$ is the sum of two terms \cite{MCART1973-}, i.e.,
\begin{equation}\label{pla:25} 
       \nu =  \nu_{en}(T_e,n_m)  +  \nu_{ei}(T_e,n_m).
\end{equation}
The first term is the electron-neutral momentum transfer collision frequency, and the second one Spitzer's electron-ion collision frequency --- which is related to Spitzer's resistivity by Eq.~\eqref{pla:21} with the index $en$ replaced by $ei$.  For dilute plasmas, in which $n_e$ is less than a few per cent of the molecular gas density, $n_m = 0.5 n_a$ (for a diatomic gas), the second term can generally be neglected.  In the other extreme of a fully ionized gas, the first term can be neglected and the conductivity becomes Spitzer's conductivity, $\sigma_{ei} = 1/\rho_{ei}$ given by \eqref{pla:22}, which is nearly independent of $n_e$.

   Similarly, the equation giving the plasma temperature $T_e$ can also often be simplified.  This is generally done by phenomenologically relating $T_e$ to $E_z$ \cite{CARY-1980-,CHAMB1981-}, or by constructing an equation relating the conductivity $\sigma$ to Joule heating, thermal cooling by conduction and energy loss as free energy \cite{YU---1979-}.
 
\subsection{Current enhancement}

   Early high-intensity propagation experiments in low-pressure gas, in the United States \cite{BRIGG1977-} as well as in the Soviet Union \cite{BREJZ1977-}, showed that under some circumstances the net (or effective) current $I_E = I_B + I_P$ was exceeding the beam current $I_B$ by as much as a factor of three.  This phenomenon had been anticipated in the case of ion beams \cite{SUDAN1976-}, as well as in Tokamak discharges and the study of highly charged ions in plasmas \cite[Refs.~11-12]{SUDAN1976-}, and had even been observed earlier \cite[Refs.~1-2]{IPATO1984-}, but not well understood.

   The phenomenon of current enhancement (which is also called current multiplication, or current amplification) is potentially unfavorable to beam transport since a plasma return current above that necessary for current neutralization will eventually contribute to defocusing of the beam.  While several mechanisms may contribute to this phenomenon, it was found that classical Coulomb collisions alone are insufficient to drive the observed plasma current, but that the two-stream interaction between beam particles and plasma electrons, which can produce large-amplitude plasma oscillations, can drive such a plasma current, as much for ion-beams \cite{SUDAN1976-} than for electron-beams \cite{CHAMB1979B}.

   Current amplification was observed in a number of situations, e.g., at high pressures from near atmospheric \cite{IPATO1984-} down to about 100~torr \cite[Refs.~1-4]{LAU--1985-}, as well as at sub-torr pressures in the ion-focused regime \cite{PAITH1988-}.  It was also found in simulations showing that current multiplication can occur due to instabilities other than the two-stream mode, e.g., macro-instabilities such as the hose instability \cite{LAU--1985-, HUI--1985-, FERNS1986-}

   In summary, current enhancement is typically one of these undesirable ancillary effects which can occur due to beam instabilities and other processes that have to be controlled by carefully selecting the parameter regime most suitable for propagation.  This requires, in particular, a comprehensive understanding of instabilities, and of the techniques available for avoiding them.

\subsection{Summary}

	As was seen in this section, the complexity of plasma chemistry and the number of processes involved make the calculation of the conductivity rather difficult in general.  However, for short pulses in dense gases such as air in the pressure range of 0.1 to 1 atmosphere, a good first approximation is obtained by assuming an average plasma temperature of $T_e \approx 2$ eV and keeping dissociative recombination with oxygen as the main cause of electron loss.  In that case \cite{CHAMB1981-}\footnote{This reference gives the expression $\alpha_r(T_e) \approx 2.1 \times 10^{-7} (300/T_e$[$^{\rm{o}}$K]$)^{0.7}$.}
\begin{equation}\label{pla:26} 
     \alpha_r \approx 10^{-8} ~ \text{cm}^3\text{s}^{-1},
                                         ~~~ ~~~ \text{and} ~~~ ~~~
     \nu \approx \nu_{en} \approx 1.8  \times 10^{-7} n_m ~ \text{s}^{-1},
\end{equation}
which gives a collision frequency $\nu\approx 4.7 \times 10^{12} ~ \text{s}^{-1}$ for air at STP, which corresponds to $n_m \approx  2.6 \times 10^{19}$ cm$^{-3}$.

    A particular complication is that there are significant differencies in the details of the microscopic processes happening at the beam head for negative (electron or antiproton) and positive (proton or positron) beams.  These differences are exacerbated for low-energy (i.e., MeV range) beams where the non-relativistic behavior of the heavier particles (proton or ion) accentuates these differences.  Since the focus of this report is on high-energy beams, we refer to the literature for a discussion of these processes, e.g., to references \cite{OLSON1973-, IYYEN1989-} for electron beams, and \cite{OLIVE1996-, OTTIN2000-} for proton beams.

	The six equations \eqref{nor:1}, \eqref{nor:4}, \eqref{pla:19}, \eqref{pla:20}, \eqref{pla:23}, and \eqref{pla:24} constitute a full set of equations providing a complete one-dimensional description of the propagation of a beam pulse in a gas or plasma.  The main assumptions leading to these equations are the paraxial approximation, the description of the plasma by a scalar conductivity,\footnote{In more complete models a tensor conductivity is introduced to take into account decreased particle mobilities across magnetic field lines \cite{STRAS2003-}.  Similarly, because of the low collision frequencies and the presence of the geomagnetic field, the conductivity in the ionosphere above 70 km height is highly anisotropic, see, e.g., \cite{SUGIU1972-}.  While the consequencies for ionospheric beam propagation are discussed in some of the papers cited in the bibliogrpahy, they will not be developed in this report. } and the similarity of all radial distributions.  Whereas these coupled equations cannot be solved in general without a computer, their main features and implications for endo-atmospheric beam weapons can be derived by making some approximations.

\section{Charge neutralization : limitations due to atmospheric density}
\label{cnc:0}

In this section we derive a few simple consequencies of the essential requirement that a beam injected into the atmosphere should be able to sufficiently charge neutralize in order to propagate as a Bennett pinch.  This leads to strict limitations which are equivalent to altitude limits to effective beam propagation and related plasma effects, such as, in the case of a beam plasmoid, the ability to form a polarization sheath to propagate across Earth's magnetic field.

   In principle, these limitations should be derived taking numerous effects into account, something that requires solving a complicated analytical model, or using a computer simulation program.  However, a first order approximation is obtained by neglecting the influence of Earth's magnetic field, as well as of any return current, so that the criterion of `sufficient charge neutralization' becomes Budker's condition $f_e > 1/\gamma^2$.   We will therefore assume $f_m = 0$ throughout this section, and calculate under which conditions the charge neutralization fraction is equal to either one of the two limiting values, $f_e = 1$ and $f_e = 1/\gamma^2$, between which a beam pulse is able propagate as a Bennett pinch. In order to handle both cases at the same time, we will write where appropriate $\gamma^k$ instead of $\gamma$ or $\gamma^2$, and set $k$ equal to 0 or 2 when necessary.

  Leaving aside the option that a suitable plasma channel might be available, a beam injected into the low-atmosphere will necessarily have to produce its own plasma channel, while a beam injected in the high-atmosphere may take advantage of the ionospheric plasma background.  In both cases, however, there is an absolute limit to the areal beam current density $J_b$ which can be neutralized.  This is because the maximum beam single-charge number-density cannot be larger than the maximum background-plasma charge-pair number-density times $\gamma^2$.

   If we assume that the beam is made of single-charged particles with number density $n_b$, and that this beam is able to fully single-ionized a background gas of density $n_a$, the plasma charge-pair density is equal to $n_a$, and the maximum beam current will correspond to the Budker limit $n_b=n_a \gamma^2$, so that
\begin{equation}\label{cnc:1} %
                            J_{max}(n_a) \approx e \beta c n_a \gamma^2.
\end{equation}
The atomic number density of air and the corresponding value of $J_{max}/\gamma^2$ are listed in column 2 and 4 of Table~\ref{tab:alj}.  It can be seen that very large currents can in principle be propagate up to an altitude of 300 km, and still relatively large ones up to about 3000 km, especially is the beam is highly relativistic ($\gamma^2 \gg 1$).

  If instead of generating its own plasma the beam is neutralized by a ionospheric plasma with a charge-pair number density $n_e$, the corresponding maximum beam current density is then\footnote{This estimate is probably somewhat pessimistic for a positive beam because electrons from the surrounding plasma may be attracted into the beam, while for an electron beam charge neutralization is more strictly restricted by the plasma density because the ions are much less mobile.  The importance of these effects is difficult to estimate because they depend on the geomagnetic and beam self-fields.}
\begin{equation}\label{cnc:2} %
              J_{max}(n_e) \approx e \beta c n_e \gamma^2.
\end{equation}
The ionospheric plasma electrons density and the corresponding value of $J_{max}/\gamma^2$ are listed in column 3 and 5 of Table~\ref{tab:alj}.  Again, relatively large currents can in principle be propagated in the ionosphere at altitudes comprised between 100 and 1000 km.  However, if a beam propagates in straight line over large distances, i.e., on the order of several 1000 km, its current density will be limited to the lowest one that can be neutralized over its path.  This is why the  intensity of 3 A/m$^2$ considered in reference \cite{CHRIE1986-} is likely to be the highest one possible for a non-relativistic proton beam ($\gamma \approx 1$) propagating through the ionosphere using this mode of charge neutralization.  On the other hand, much higher currents can easily be transported by electron or positron beams, because their $\gamma$ becomes very large as soon as the beam energy exceeds a few tens of MeV ($\gamma \gg 10$).

\begin{table}
\begin{center}
\hskip 0.0cm \begin{tabular}{|r|c|c|c|r|} 		
\hline
\multicolumn{5}{|c|}{\raisebox{+0.2em}{{\bf  \rule{0mm}{6mm} Atmospheric density limitations}}} \\ 
\multicolumn{5}{|c|}{\raisebox{+0.2em}{{\bf  \rule{0mm}{0mm}  on beam current density}}} \\
\hline
\raisebox{+0.2em}{altitude} \rule{0mm}{6mm} & 
\raisebox{+0.2em}{$n_a$} & 
\raisebox{+0.2em}{$n_e$} & 
\raisebox{+0.2em}{$J(n_a)/\gamma^2$ } &
\raisebox{+0.2em}{$J(n_e)/\gamma^2$ }  \\  
\rule{0mm}{0mm}
    [km]~~~ &
[m$^{-3}$] &
[m$^{-3}$] &
[Am$^{-2}$] &
[Am$^{-2}$] ~ \\
\hline
\rule{0mm}{5mm}
    0~~~ & $5 \times 10^{25}$ &                 0  & $2 \times 10^{15}$ &    0~~~  \\
  100~~~ & $5 \times 10^{18}$ & $1 \times 10^{11}$ & $2 \times 10^{ 7}$ &    5~~~  \\
  300~~~ & $5 \times 10^{15}$ & $5 \times 10^{12}$ & $2 \times 10^{ 5}$ &  200~~~  \\
 1000~~~ & $5 \times 10^{12}$ & $1 \times 10^{11}$ & $             200$ &    5~~~  \\
 3000~~~ & $5 \times 10^{10}$ & $1 \times 10^{10}$ & $               2$ &    0.5~  \\

\hline
\end{tabular}
\end{center}
\caption[Atmospheric density limitations on beam current density]{ The maximum areal current density $J(n)/\gamma^2 = e\beta c n $ that can be transported by a beam is limited by the atmospheric number density if  charge-neutralization is achieved by beam-ionization of the atmosphere, $n \approx n_a$, or by the ionospheric plasma density if charge-neutralization is achieved by motion of ionospheric plasma electrons in or out of the beam, $n \approx n_e$.}    \label{tab:alj}
\end{table}

   Another essential requirement is that charge neutralization should be very fast in order that the head of the beam has only little time to expand under the effect of space-charge repulsion.  In the case of a beam producing its own plasma channel this means that the plasma-generation rate should be sufficiently high.  To investigate this point we consider equation \eqref{pla:20} under the assumption that ionization by avalanche and electron losses by attachment can be neglected.  We have therefore
\begin{equation}\label{cnc:3} %
       \frac{\partial}{\partial\tau} n_e =  \beta \alpha_c n_b n_a
                                                - \alpha_r n_e^2,
\end{equation}
where we have introduced the collisional ionization reaction rate $\alpha_c = cS/wn_a(0)$ to make the dependence on the atmospheric atomic number density $n_a(h)$ explicit.  For the stopping power we take the value $S = 0.37 \times 10^{6}$  eV/m appropriate to highly relativistic particles, so that $\alpha_c  \approx 4 \times 10^{-14}$ m$^3$s$^{-1}$, and for the attachment reaction rate the value $\alpha_r  \approx  \times 10^{-13}$ m$^3$s$^{-1}$ appropriate to the ionosphere assuming an electron temperature of about 1000 $^{\rm{o}}$K.

  To begin with we solve  \eqref{cnc:3} neglecting recombination. The solution (for an instantaneously rising beam pulse with constant $n_b$) is then the linear function
\begin{equation}\label{cnc:4} %
      n_e(\tau)  = n_b \frac{\tau}{\tau_C }, ~~ ~~ \text{where} ~~ ~~ 
         \tau_C  =     \frac{1}{\beta \alpha_c n_a}.
\end{equation}
Since we have neglected ionization by avalanche, $\tau_C$ is the minimum time taken by the beam to produce enough electron-ion pairs for $n_e=n_b$, which under ideal conditions is just sufficient to get $f_e=1$, i.e., full charge neutralization.\footnote{This assumes, in particular, that the plasma electrons can very quickly move in or out of the beam region, a process which in a collisional plasma happens on a time scale set by $\tau_e$.  Despite its crudeness, the estimation \eqref{cnc:4} for the charge neutralization as a function of time is often found to give good results.  See, e.g., \cite{NGUYE1987-}.} This time is listed in column~3 of Table~\ref{tab:aln}.  However, since we have neglected recombination, $\tau_C$ will be larger then the value given by equation \eqref{cnc:4}.  In fact, by solving \eqref{cnc:3} for $\alpha_r \neq 0$, it is easy to see that
\begin{equation}\label{cnc:5} %
             \frac{1}{\beta \alpha_c n_a}
                   < \tau_C(\alpha_c,n_b)
                 < \frac{1}{\alpha_c n_b},
\end{equation}
which shows that while the effect of recombination is not very large when $n_b \approx n_a$ (because for air $\alpha_c \approx \alpha_r$), the charge neutralization time $\tau_C$ can be much larger than given in Table~\ref{tab:aln} when $n_b \ll n_a$.

  If we consider a range of 1000 km, i.e., a time of flight of 3.3~ms, we see that $\tau_C$ is at least on that order at an altitude of 300 km.   Therefore, propagation as a fully charged-neutralized Bennett pinch over distances less than this range cannot be considered for altitudes much higher than 100 km (where $\tau_C = 3~\mu$s) if the beam its generating is own plasma.

\begin{table}
\begin{center}
\hskip 0.0cm \begin{tabular}{|r|c|c|r|r|r|r|}
\hline
\multicolumn{7}{|c|}{\raisebox{+0.2em}{{\bf  \rule{0mm}{6mm} Atmospheric density limitations}}} \\ 
\multicolumn{7}{|c|}{\raisebox{+0.2em}{{\bf  \rule{0mm}{0mm}   on charge neutralization}}} \\ 
\hline
\raisebox{+0.2em}{altitude} \rule{0mm}{6mm} & 
\raisebox{+0.2em}{$n_a$} & 
\raisebox{+0.2em}{$\tau_C$} & 
\raisebox{+0.2em}{$W_{\text{electron}}$ } &
\raisebox{+0.2em}{$W_{\text{proton}}  $~} &  
\raisebox{+0.2em}{$W_{\text{electron}}$~~} &
\raisebox{+0.2em}{$W_{\text{proton}}  $~~} \\ 
\rule{0mm}{0mm}
[km]~~~     &
[m$^{-3}$]  &
[s]         &
[GeV]~~     &
[GeV]~~     &
[GeV]~~~    &
[GeV]~~~    \\
\rule{0mm}{0mm}
    ~~ &
    ~~ &
    ~~ &
  $f_e = 1$~~ &
  $f_e = 1$~~ &
  $f_e > 1/\gamma^2$ &
  $f_e > 1/\gamma^2$ \\
\hline
\rule{0mm}{5mm}
    0~~~ & $5 \times 10^{25}$ & $3 \times 10^{-13}$&      0.008 &         1.2 & 0.003~~~   &     1.1~~ \\
  100~~~ & $5 \times 10^{18}$ & $3 \times 10^{-6}$ &    1.7~~~~~&      260~~~ & 0.07~~~~~  &   27~~~~~ \\
  300~~~ & $5 \times 10^{15}$ & $3 \times 10^{-3}$ &  17~~~~~~~~&     2600~~~ & 0.27~~~~~  &  110~~~~~ \\
 1000~~~ & $5 \times 10^{12}$ &          3         & 170~~~~~~~~&    26000~~~ & 1.1~~~~~~~ &  440~~~~~ \\
 3000~~~ & $5 \times 10^{10}$ &        300         & 800~~~~~~~~&   120000~~~ & 2.7~~~~~~~ & 1100~~~~~ \\

\hline
\end{tabular}
\end{center}
\caption[Atmospheric density limitations on charge neutralization]{ The minimum time $\tau_C$ taken by a beam to generate enough electron-ion pairs to fully charge-neutralize itself is inversely proportional to the atmospheric density $n_a$.  For beam particle densities $n_b$ comparable to $n_a$, the minimum beam particle energies, $W_{\text{electron}}$  and $W_{\text{proton}}$, compatible with that time (column~4 and~5), or with $\tau_C/\gamma^2$ for minimally pinched beams (column~6 and~7), become excessively large at high altitudes, especially for protons. For $n_b<n_a$, $W_{\text{electron}}$  and $W_{\text{proton}}$ have to be multiplied by  $(n_b/n_a)^{1/3}$.}    \label{tab:aln}
\end{table}

However, even for propagation over distances larger than $\beta c \tau_C$, the main consideration is that the beam head should not significantly expand during the time $\tau_C$, because otherwise the assumption of constant $n_b$ used in deriving Eq.~\eqref{cnc:4} would be violated.  If this expansion is characterized by the time $\tau_2$ taken by a non-charge-neutralized beam to double is radius,  this leads to the condition $\tau_2 > \tau_C/\gamma^k$, where $k=0$ or $2$ depending on whether we require that the beam density $n_b$ corresponds to a fully or a minimally pinched beam.  Therefore, using equation \eqref{cha:7} and developing, we get 
\begin{equation}\label{cnc:6} %
   \gamma^k\gamma^2\beta^2(\gamma mc^2)
     > \Bigl(m_ec^2 \frac{\pi r_e c^2}{\alpha_c^2 n_a(0)} \Bigr)
        \frac{n_a(0)}{n_a(h)} \frac{n_b}{n_a(h)}.
\end{equation}
This condition yields a very strong constraint on the beam particle energy, first because $\gamma$ appears to the third or fifth power on the left, and second because the numerical value of the constant factor inside the big brackets on the right is equal to 1.9 GeV.  As can be seen in column 4 and 5 of Table~\ref{tab:aln}, where $f_e=1$ is asked for, the particle's energies are all in the GeV range for $h> 100$~km, and become excessively large at higher altitudes, especially for proton beams.  In column 5 and 6 the corresponding numbers are given for minimally pinched beams:  The proton beam energies are still in the GeV to TeV range, while electron beam energies close to 1~GeV are sufficient between 300 and 3000 km altitude.  As the numbers given in the Table correspond to $n_b=n_a$, these energies would be $(n_b/n_a)^{1/3}$ smaller for beams such that $n_b<n_a$, but the beam power would then become negligible at high altitudes.  

  An immediate consequence of Eq.~\eqref{cnc:6} is that because of their small mass, beams of electrons are very much favored over beams of heavier particles.  This is readily seen in Table~\ref{tab:aln}, where the factors $W_{\text{electron}}/W_{\text{proton}} \approx (m_e/m_p)^{2/3} \approx 1/150$, and $W_{\text{electron}}/W_{\text{proton}} \approx (m_e/m_p)^{4/5} \approx 1/400$, lead to electron energies which are compatible with existing or near term technologies applicable to space-based systems, while  proton energies are much larger and the corresponding technologies much more demanding.

  In summary, despite the simplicity of the arguments leading to the numbers in Tables~\ref{tab:alj} and~\ref{tab:aln}, several important conclusions can be drawn from them with regards to pinched-beam propagation in the high-atmosphere, where charge neutralization is problematic.

\begin {itemize}

\item  For non-relativistic proton beams, charge neutralization by beam generated plasma is impossible at ionospheric altitudes.  The only option is by motion of ionospheric plasma electrons into the beam, which is complicated since they have to move along the Earth's magnetic field lines.  In reference \cite{CHRIE1986-} it is shown that it should nevertheless be possible to attain charge neutralization this way if the beam density is properly varied along itself. 

\item  For positron (or electron) beams, full charge neutralization by beam generated plasma is possible for altitudes up to 300 km, provided the beam energy is in the 1 to 10 GeV range.  At higher altitudes plasma generation tends to become to slow, so that partial charge neutralization, or charge neutralization by the ionospheric plasma, may become the only options.  In this context, it should be recalled that it can be easier for a positive than a negative beam to achieve complete charge neutralization \cite{LOTOV1996-,KAGAN2001-}.  This gives a considerable advantage to positron over electron beams.

\item For all type of charged particle beams, as well as for plasmoid beams, the residual atmospheric density and the ionospheric plasma density yield strong constraints on the maximum power that can be transported by the beam, as well on all plasma effects that rely on these densities.   In practice, when other considerations such as current neutralization and beam stability are taken into account, this implies that optimum Bennett pinch transport conditions often occur at beam particle densities $n_b$ comparable to $n_a$ or $n_e$.

\end{itemize}

\section{Conductivity generation and critical beam current}
\label{cri:0}

When a high-current beam pulse is injected into air at full atmospheric pressure, the avalanche ionization term can be neglected and \eqref{pla:20} can be solved explicitly to give $n_e$ as a function of $\tau$.  With the further approximations \eqref{pla:26}, one can calculate the plasma conductivity $\sigma$ and thus the magnetic diffusion time $\tau_m$:
\begin{equation}\label{cri:1} 
    \tau_m = \tau_M \frac{\exp(2\dfrac{I_B}{I_G}\dfrac{\tau}{\tau_M})-1}
                         {\exp(2\dfrac{I_B}{I_G}\dfrac{\tau}{\tau_M})+1},
\end{equation}
where
\begin{equation}\label{cri:2} 
    \tau_M = 2 a \sqrt{ \frac{\pi}{\nu} \frac{r_e}{\alpha_r} \frac{I_B}{I_G} },
\end{equation}
and
\begin{equation}\label{cri:3} 
     I_G = \frac{1}{4} \frac{e\nu w}{r_e S} \approx 7.2 ~~ \text{kA},
\end{equation}
with $r_e$ the classical radius of the electron and $e$ the electron charge.  $I_G$, the \emph{conductivity generation current}, which controls the rate of conductivity generation, is a critical current independent of pressure, and of beam parameters for $\beta \approx 1$, i.e., relativistic beams for which Bethe's stopping power $S(\beta)$ is nearly constant.

 The value of $I_G$ is typically on the order of 10 kA.  $\tau_M$, the asymptotic value of $\tau_m$, increases  with the beam radius and current, as well as with a reduction in atmospheric pressure.  For pulses that are long relative to $\tau_M I_G/I_B$, $\tau_m$ will thus approach $\tau_M$, whereas for short pulses, or at the beam head \eqref{cri:1} reduces to
\begin{equation}\label{cri:4} 
               \tau_m(\tau) \approx \frac{I_B}{I_G} \tau.
\end{equation}

	We will now make the drastic approximation that both $\tilde{a}'$ and $\tilde{a}''$ can be neglected in \eqref{nor:1}.  This approximation is only valid in the quasistatic limit leading to Bennett's or Nordsieck's equations.  However, this hypothesis corresponds approximately to the situation in which a beam head leaves the exit window of an accelerator to penetrate into the atmosphere.  Finally, we will assume a beam current $I_B$ with an infinitely short rise-time.  Under these conditions, neglecting the displacement current, the circuit equation \eqref{pla:19} can be solved.  With \eqref{cri:4}, the effective current is then
\begin{equation}\label{cri:5} 
   I_E = I_B \Bigl( 1 - \frac{1}{\beta^2}
             \bigl(1+\lambda\frac{\tau^2}{\tau_p^2}\bigr)^{-1/\lambda} \Bigr),
\end{equation}
where
\begin{equation}\label{cri:6} 
      \lambda = 2 \mathcal{L} \frac{I_B}{I_G},
\end{equation}
and
\begin{equation}\label{cri:7} 
     \tau_p = \frac{a}{c}  \sqrt{2 \frac{I_G}{I_B} }.
\end{equation}
It is now possible to calculate various quantities, and in particular the plasma current $I_P$ which turns out to be maximum at $\tau=\tau_p$.  The region where $I_P$ is maximum can be defined as the beam neck.  However, as $\tau_p$ is calculated for a beam with infinitely short rise-time, it does not give the absolute position of the beam neck.  Nevertheless, $\tau_p$ can be used to calculate relative quantities, and in particular the neutralization fractions at the point where the beam is pinching.  This is because plasma phenomena are dominant in the beam head, and to first approximation independent of rise-time: the conductivity becomes smaller as the rise-time is reduced, but the inductive electric fields becomes greater, thus resulting in the same plasma current when $I_B$ reaches its maximum \cite{BRIGG1974-}.

	For $\tau=\tau_p$, the charge neutralization is
\begin{equation}\label{cri:8} 
    f_{ep} =   1 - \bigl(1+\lambda\bigr)^{-1-1/\lambda}.
\end{equation}
At the pinch point, $f_e$ is thus equal to $1-1/e \approx 0.63$ for $\lambda \ll 1$ and about one for $\lambda \gg 1$.  The corresponding result for the current neutralization fraction is
\begin{equation}\label{cri:9} 
    f_{mp} = \frac{\lambda}{\beta^2}
             \bigl(1+\lambda\bigr)^{-1-1/\lambda}.
\end{equation}
The maximum current neutralization fraction is thus small as long $\lambda \ll 1$, but close to one for $\lambda \gg 1$.  Consequently, in order for a beam to be well pinched, which requires $f_m < 1$, one has to have $\lambda<1$, or, explicitly
\begin{equation}\label{cri:10} 
    I_B < \frac{1}{2\mathcal{L}} I_G.
\end{equation}
As $I_G \approx 10$ kA, this is a rather strong limitation on the beam current.  Indeed, for particle beam weapons, current in excess of $I_G$ are required for endo-atmospheric systems, and this creates some problems.

	These analytical results are in good agreement with detailed computer calculations which do not assume a constant beam radius \cite{SHARP1980-}.  In these calculations the radius of the beam at the neck is well approximated by
\begin{equation}\label{cri:11} 
   a_n \approx  a_\infty \bigl( 1 +  \frac{I_B}{I_G}  \bigr),
\end{equation}
which shows that for $I_B > I_G$, the neck radius is proportional to the beam current.

	The pinch condition is not the only one leading towards the requirement of small current neutralization fractions.  We will see, for instance, that a beam is also more likely to be unstable when $f_m > 0.5$.  Therefore, it turns out that $I_G$ corresponds to a \emph{critical beam current} setting an upper limit to the current of a beam pulse sent into the atmosphere.

	It is interesting to notice that the maximum current neutralization fraction calculated with \eqref{cri:9} is in reasonable agreement with the result of detailed computer calculations \cite{CHAMB1981-,SHARP1980-}.  For example, in the case of a 10 kA, 50 MeV beam pulse such as produced by the ATA accelerator at the Lawrence Livermore Laboratory, we find $f_{mp} = 0.50$, whereas the computer calculations \cite{CHAMB1981-} predicts $f_{mp} =0.4$.  In that case $a=0.5$ cm and $b \approx 10$ cm; therefore $\mathcal{L} = 1.5$ and $\lambda=4.2$.

	The maximum current neutralization fraction \eqref{cri:9} is independent of pressure.  But, as pressure is reduced, the ionization by avalanche can no longer be neglected.  The result \eqref{cri:9} is then a lower bound to the maximum current neutralization fraction.  For example, for the above mentioned ATA accelerator beam injected into air at 0.1 atmospheric pressure, the computer calculations \cite{CHAMB1981-} give $f_{mp} = 0.8$.

	The maximum value of the longitudinal electric field which appears during the self-pinching of the beam can be calculated from the circuit equation.  In first approximation one finds:
\begin{equation}\label{cri:12} 
    E_{z,max} = \frac{1}{4\pi\epsilon_0c^2} 
               2\frac{c}{a} \sqrt{\mathcal{L}} I_B.
\end{equation}
In the head region where $E_z$ is maximum, $a$ is of the order of $b$.  Thus:
\begin{equation}\label{cri:13} 
             E_{z,max} = 2500 \frac{I_B}{b} ~~ \text{[V/m]}.
\end{equation} 
For beam currents of 10 to 100 kA and head radii of 3 to 30 cm, we find fields of the order of 10 MV/m.  This maximum electric field is quite large and is approximately independent of the beam current.  This is because the beam head radius becomes proportional to the beam current when $I_B > I_G$, as it is shown by computer calculations \cite{SHARP1980-}.

	In the body of a beam pulse long enough for $\tau_m$  to reach its asymptotic value \eqref{cri:2}, the circuit equation becomes a linear equation with constant coefficients.  It can then be solved easily.  In particular, when neglecting $\tau_e$, the plasma current is
\begin{equation}\label{cri:14} 
     I_P = - I_B \exp(-\frac{\tau}{\mathcal{L}\tau_m}) f_{mp}.
\end{equation}
For the previous examples, the plasma current decay time $\mathcal{L}\tau_m$ would be of 10 ns and 35 ns at the respective pressures of 1 and 0.1 atmosphere.  Thus, if $f_m$ is large at the pinch point, the current neutralization fraction will remain almost constant throughout the pulse, provided its length is less than $\mathcal{L}\tau_m$.

	A last case of importance is that of a beam pulse injected into a plasma of such a density that its conductivity remains constant.  (This is in particular the case when $\nu_{ei} \gg \nu_{en}$ in \eqref{pla:25}, e.g., in a fully ionized gas, where $\sigma$ becomes the Spitzer conductivity $1/\rho_{ei}$ given by Eq.~\eqref{pla:22}.) In that situation, because the plasma density is high, plasma currents can start flowing right from the very beginning of the pulse.  The current neutralization is therefore almost complete at the beam head, and \eqref{cri:14} with $f_{mp} = 1$ will give the plasma current throughout the pulse.  Because of this large current neutralization, the beam head will not be pinched even though charge neutralization is complete.  Therefore, a pulse injected into a plasma or preionized background will also have an expanded head, and beam pinching then occurs as the longitudinal plasma current decays.

\section{Ohmic losses and return-current heating}
\label{ohm:0}

    High current pulses experience an ohmic drag $dW/dz=-|qE_z|$ which causes loss of energy even if ionization and other losses were negligible.  If the beam propagates in a dense gas or plasma, these losses are generally called `ohmic losses' because they are directly related to the plasma current by Ohm's law: $J_P = \sigma E_z$.  They are also maximum in the beam head where  $E_z$ is maximum, and are therefore strongly coupled to other effects such as beam head erosion.   Nevertheless, it is useful to consider ohmic losses on their own, for instance to derive some important first approximation formulas for the range and the overall ohmic heating effect of a beam pulse, because these effects do not depend very much on the details as they are obtained by integrating over the beam pulse. 

   For instance, the average longitudinal electric field is easily obtained by integrating Eq.~\eqref{bcn:12}
\begin{equation}\label{ohm:1} 
    <E_z> =   -\frac{\mathcal{L}}{\pi\epsilon_0 c^2}
               \frac{I_E(\Delta\tau) - I_E(0)}{\Delta\tau},
\end{equation}
where $I_E(0) = 0$ in a collisional plasma assuming full current neutralization at $\tau=0$.

  This average can be used to calculate $<dW/dz>$, and thus to provide an estimate of the ohmic range $z_{\Omega}$, defined as the distance a pulse could go if it only lost energy by this mechanism,
\begin{equation}\label{ohm:2} 
                z_{\Omega} = \frac{\gamma-1}{\gamma\beta}
                             \frac{I_A}{I_E(\Delta\tau)}
                             \frac{c\Delta\tau}{4\mathcal{L}}.
\end{equation}
Since $I_E(\tau) \approx I_B(1-\exp(-\tau/\mathcal{L}\tau_m))$, the range is maximum for pulses that are long, i.e., $\Delta\tau \gg \tau_m$, while for short pulses
\begin{equation}\label{ohm:3} 
          z_{\Omega} \approx \frac{\gamma-1}{\gamma\beta}
                             \frac{I_A}{I_B}
                             \frac{c\tau_m}{4}.
\end{equation}
A more precise two-dimensional calculation for a beam with a Gaussian current density profile, valid for pulses such that $\Delta\tau \ll \tau_m b^2/a^2$, gives \cite{LEE--1973D}
\begin{equation}\label{ohm:4} 
                z_{\Omega} = \frac{\gamma-1}{\gamma\beta}
                             \frac{I_A}{I_B}
                             \frac{c\Delta\tau}{\ln(1+2\Delta\tau/\tau_m)},
\end{equation}
which shows that \eqref{ohm:3} is underestimating the ohmic range by about a factor of two.    For a 10 kA, 1 GeV electron beam pulse with $\Delta\tau = \tau_m = 10$~ns, typical of an endoatmospheric beam, the ohmic range is therefore of about 10~km.  In comparison, the energy loss by ionization is $\approx 0.3$~MeV/m, which yields a range of about 3~km, i.e., on the same order.

	The energy given up by ohmic losses primarily heats the plasma electrons.  This `ohmic heating' process is due to the energy expended in driving the plasma return current, and is therefore often called `return-current heating.'  After the passage of the beam pulse, that energy contributes in forming a channel of hot air along the beam path.  On the other hand, the energy lost in the form of radiations (bremsstrahlung) or particles scattered out of  the beam do not contribute substantially into heating the air.  The total energy deposited per unit length along the beam trajectory can be calculated by integrating the collision and ohmic terms in \eqref{nor:4} over the length of the beam pulse,
\begin{equation}\label{ohm:5} 
    \frac{\Delta W}{\Delta z} = - \frac{I_B}{ce}
         \Bigl( Sc\Delta \tau + 4 cp\mathcal{L}\frac{I_E}{I_A} \Bigr),
\end{equation}
where $I_E$ is evaluated at the end of the pulse.  By replacing the dimensionless inductance $\mathcal{L}$ by the inductance per unit length, $L= \mu_0\mathcal{L}/{\pi}$, the ohmic term can be rewritten as
\begin{equation}\label{ohm:6} 
    \frac{\Delta W}{\Delta z}\Bigm|_{\Omega}  = - L I_B I_E.
\end{equation}
For a pulse long enough for $I_E=I_B$ this is twice the naive contribution $\frac{1}{2}L I_B^2$ that would be obtained by a simple magnetic circuit analysis.  In fact, the one-dimensional derivation leading to equations \eqref{ohm:5} and \eqref{ohm:6} neglects the fact that part of the energy lost by the beam is temporarily stored in the magnetic field.

   In fact, the beam energy loss due to the ohmic resistance of the plasma is the difference between the magnetic field energy and the work done by the beam, i.e.,
\begin{equation}\label{ohm:6A} 
       - \int_0^{\Delta \tau} d\tau~ R I_P^2 
       = \frac{1}{2} L I_B^2 
       - \int_0^{\Delta \tau} d\tau~ I_B \Phi,
\end{equation}
where $\Phi = L I_E$ is the effective flux linked by the beam.  As $I_B$ is constant throughout the pulse, except at the leading edge where we can assume $I_B = I_E = 0$,  $I_B$ can be taken out of the integral and we get \cite{LOVEL1971-}
\begin{equation}\label{ohm:7} 
        \frac{\Delta W}{\Delta z}\Bigm|_{\Omega}
      = - \frac{1}{2} L I_B^2 (2s_m-s_m^2),
\end{equation}
where $s_m=I_E/I_B$ is the magnetic shielding factor, noted $f$ in Ref.~\cite{LOVEL1971-}, calculated at the end of the pulse.  This expression gives $\frac{1}{2}L I_B^2$ for long pulses, and agrees with Eq.~\eqref{ohm:6} for short ones.

  Another general method of estimating return-current heating is by relating it to the ohmic range by the expression \cite{LEE--1973D}
\begin{equation}\label{ohm:8} 
        \frac{\Delta W}{\Delta z}\Bigm|_{\Omega}
      = - \frac{W_{\text{pulse}}}{z_\Omega},
\end{equation}
where $W_{\text{pulse}} = (\gamma-1)mc^2 (I_B/e) \Delta \tau$ is the total energy in the beam pulse.  This has the advantage to take the effects of finite beam and radius more easily into account than by explicit calculations \cite{GERWI1975-}.   For instance, with Eq.~\eqref{ohm:4}, one finds
\begin{equation}\label{ohm:9} 
          \frac{\Delta W}{\Delta z}\Bigm|_{\Omega}
      = - \frac{\mu_0}{4\pi} I_B^2 \ln(1+2\Delta\tau/\tau_m),
\end{equation}
which for $\Delta\tau \gg \tau_m$ agrees with Ref.~\cite{GERWI1975-}.

   Equations \eqref{ohm:5} and \eqref{ohm:7} show that, for a high current beam, the energy deposition by ohmic losses can be larger than the energy deposition by collisions.  This happens when the gas pressure is low enough for collisions to be negligible, or when the pulse is short.  Energy deposition can thus be enhanced by chopping a given pulse into a series of smaller pulses, because according to \eqref{ohm:7} each of these will deposit the same energy as the longer one provided $I_E$ remains about the same at the end of each smaller ones.  If the individual pulses of the pulse train propagate independently from one another, they will, of course, all be subject to erosion.  The most efficient configuration for energy deposition, for instance to bore a reduced density hole through the atmosphere in order to facilitate the propagation of subsequent pulses, will thus be a compromise.

\section{Beam head erosion}
\label{ero:0}

\begin{figure}
\begin{center}
\resizebox{12cm}{!}{ \includegraphics{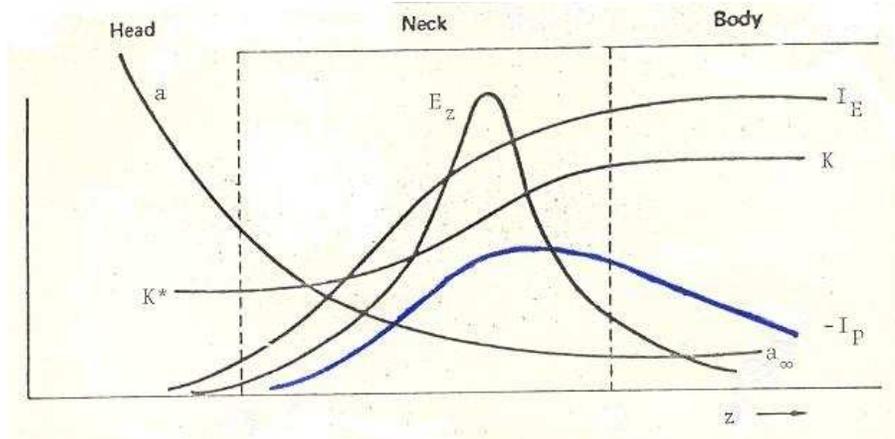}}
\caption[Beam neck profile]{\emph{Beam neck profile.} The variations of the effective beam current $I_E$, plasma return current $I_P$, kinetic energy $K$, longitudinal electric field $E_z$, and beam radius $a$, are shown in the neck region which separates the well-pinched body from the freely-expanding head of the beam pulse.   
\label{fig:nec}}
\end{center}
\end{figure}

In all situations, whether the beam is injected into vacuum, a neutral gas, or a plasma, the head of a pulse is not pinched and is followed by a neck in which there is a deep spike in the longitudinal electric field.  The beam particles at the very front of the pulse will thus spread out radially and get lost, and those in the neck will lose energy because of the longitudinal electric field.  As their energy decreases, the particles in the neck will also experience more scattering, and both effects will combine to increase the pinch radius.  Consequently, the beam head  expands continuously, and the neck region recesses progressively into the body of the pulse.  (See Figure~\ref{fig:nec}, as well as Reference \cite{BOUCH1988-} which gives the results of detailed simulations of `nose physics' for a 50 to 500~MeV, 10~kA, electron beam with a 1~ns rise-time propagating in air at full atmospheric pressure.)

	This beam head erosion process has been studied both theoretically and experimentally \cite{CHAMB1979A, LEE--1980A, SHARP1980-, GREEN1985-, BOUCH1988-, GLAZY1990-, ROSE-2002-}.  These studies show that after an initial transient period, the erosion process results in an almost constant rate of decrease in the length of the pulse as a function of the propagation distance.  Similar effects have also been observed with 50--200~eV electron beams, i.e., in a parameter-range comparable to that of 100--400~keV proton beams, where the axial propagation of the beam front is strongly modified by beam-plasma-particle interactions.  In particular, the injected cold beam front can erode during propagation and evolve into a warm front that propagates much slower than its injected speed \cite{CHAN-1991-,CHAN-1994-}.

	The erosion rate can be estimated in a simple way by assuming that the particles in the neck will get lost when their kinetic energy has dropped from its initial beam-front value $K=K(0)$ down to some critical value $K^*(\tau^*)$ because of energy losses in the longitudinal inductive electric field.  From \eqref{bcn:12} and \eqref{nor:4}, the erosion rate is 
\begin{equation}\label{ero:1} 
   \frac{d\tau}{dz} = -\frac{q}{\pi\epsilon_0c^2}
                       \frac{d}{dK} (\mathcal{L}I_E),
\end{equation}
which should be a constant if erosion proceeds at a steady state.  This equation can therefore be integrated between the times $\tau =0$ and $\tau =\tau^*$ to give the propagation range $z_{\Delta t}$ corresponding to the decrease $\Delta t$ in the pulse duration \cite{LEE--1980A}
\begin{equation}\label{ero:2} 
   z_{\Delta t} = \frac{\pi\epsilon_0c^2}{e\mathcal{L}}
                  \frac{K-K^*}{I_E^*-I_E(0)} \Delta \tau.
\end{equation}
Since $I_E(0) = 0$, and assuming $K^* = K/2$ as suggested by electron beam experiments \cite{LEE--1980A}, this gives  \cite{LEE--1980A,SHARP1980-}, 
\begin{equation}\label{ero:3} 
   z_{\Delta t} \approx \frac{1}{8\mathcal{L}}
                        \frac{I_U}{I_E^*}
                        \frac{K}{m_ec^2} c\Delta \tau,
\end{equation}
where $I_E^*$ is the effective current at the time $\tau^*$ corresponding to a point between the $E_z$ spike in the pulse neck and the non-expanded body of the pulse.

   For an electron beam with $K = 500$ MeV, $I_E^* \approx I_B = 10$ kA, and $\mathcal{L} = 1$ at the neck, a beam pulse length loss of $\Delta t = 10$ ns would allow for a range of about 600~m.  Beam head erosion is thus a severe limitation to the propagation of short pulses. 

    Comparing Eq.~\eqref{ero:2} and \eqref{ero:3} with Eq.~\eqref{ind:10} and  \eqref{ind:11}, we see that the erosion rates derived in this section agree within a factor of two with those derived in Sec.~\ref{ind:0} on the basis of an energy conservation argument for a beam propagating in vacuum.  But this difference is not significative, and has nothing to do with the fact that we consider here a beam propagating in a gas or plasma!  The discrepancy is only apparent since in both cases there are three phenomenological parameters (an energy, a current, and an inductance) which do not necessarily have the same numerical values since in one case we have a differential, and in the other an integral argumentation.  As a matter of fact, the effect of the resistive plasma response to $E_z$, i.e., the ohmic plasma current, is contained in the self-consistent value of the effective current $I_E=I_B+I_P$. 

   However, just like with Eq.~\eqref{ohm:6} in the previous section, the reasoning leading to Eq.~\eqref{ero:2} and \eqref{ero:3} did not included the contribution of magnetic energy into account, which is indeed the case since the estimate \eqref{ero:1} has been obtained under the implicit assumption that the particles in the beam neck do not move radially relative to the particles in the body of the pulse.  For this reason the energy conservation argument of Sec.~\ref{ind:0} is better, and in fact has recently been used to derive the following expression \cite{ROSE-2002-}\footnote{In that reference the dimensionless inductance is written $L=2\mathcal{L}$.}
\begin{equation}\label{ero:4} 
       z_{\Delta \tau} = \frac{\beta \, \beta_F}{\beta - \beta_F} c\Delta \tau
                       = \frac{1}{2\mathcal{L}}
                         \frac{I_U}{I_E}
                         \frac{I_B}{I_E}
                         \frac{K-K^*}{m_ec^2} c\Delta \tau,
\end{equation}
which is formally identical to Eq.~\eqref{ind:10} and where $K^*\DEF\alpha_F (\gamma_F -1) mc^2$ is assumed to be equal to a fraction $\alpha_F$ of the kinetic energy of a particle moving at the beam front velocity $\beta_F$.

     In Figure 2 of Reference \cite{ROSE-2002-} the erosion rate $(1- \beta_F/\beta)$ calculated with Eq.~\eqref{ero:4} is compared to computer simulations for a 0.1 to 2 GeV proton beam with $\mathcal{L}=2$, $I_B=10$~kA, and $I_E=5$~kA propagating in air at 760 torr.  When taking into account effects such as scattering and ionization energy loss, which are not explicitly included in \eqref{ero:4}, the value of $\alpha_F$ is found to be about 0.75 for $K \geq 1$~GeV.

   Considering that $\beta_F$ is very close to $\beta$ for a nearly relativistic beam, the factor $\alpha_F \approx 0.75$ in \eqref{ero:4} is equivalent to assuming $K^*/K \approx 0.5$ in \eqref{ero:2} when $I_E \approx  I_B$.  The latter assumption is a good approximation for high energy electron beams propagating in a high density gas, so that in that case Eqs.~\eqref{ero:3} and \eqref{ero:4} give comparable values for the erosion range.  However, in the case of proton beams, or of low-energy electron beams,  it is clear that Eq.~\eqref{ero:4} should be used, except for very non-relativistic beams (e.g., proton beams with $K \leq 0.1$~GeV), for which effects such as scattering and ionization energy-loss become overwhelming, so that different theoretical models, or computer simulations, should be used. 

   According to the simulations of Reference \cite{ROSE-2002-}, the erosion range for 0.1, 0.5, 1, and 2 GeV proton beam pulse with a 10~ns duration should be about 50, 600, 1700, or 4500 m, respectively.  This means that GeV-energy, kA-current proton beams have erosion ranges comparables to those of similar energy and current electron beams.

  Unfortunately, there is very little published data available on beam head erosion, and the only publication giving some significant data applies to a 400~kV electron beam with a rather short pulse duration, 3~ns, so that the average current of the beam decreases at the same time as the pulse is eroded \cite[Fig.11]{GREEN1985-}.  Nevertheless, this data can be compared to Eqs.~\eqref{ero:3} and \eqref{ero:4}, and reasonable agreement can be found with the eroison rates of about 14 to 18 cm/ns measured for propagation in air at pressures  between 1 and 8 torr \cite[Table~I\,]{GREEN1985-}.  For instance, from Eq.~\eqref{ben:29} the mean longitudinal velocity of a 400~keV electron beam (for which $I_A = 25$~kA) is 24~cm/s assuming $I_E = 1$~kA as in Ref.~\cite{GREEN1985-}.  This means that in this experiment the measured normalized erosion rate $(1- \beta_F/\beta)$ is comprized between 0.60 and 0.76.  Assuming $\mathcal{L}=1$, $I_E=1$~kA, $I_B=2$~kA, and $\alpha_F=7/8$, Eq.~\eqref{ero:4} gives an eroison rate of 0.49, which is also given by Eq.~\eqref{ero:4} with only two free parameters: $\mathcal{L}=1$ and $I_E = 1$~kA.  Therefore, with both equations one obtains a result suggesting that the erosion rate should be less than the measured value, in agreement with an analysis by French scientists \cite[p.202]{BOUCH1988-}. 

   The data of reference \cite{GREEN1985-} was also used by Russian scientists in an attempt to validate Eq.~\eqref{ero:3} and similar estimates based on other simplifying assumptions \cite{GLAZY1990-}.  In their conclusion they suggest an approximate expression which generalizes Eq.~\eqref{ero:3} by including the effect of scattering.  

   A significant implication of the derivations of Eqs.~\eqref{ind:10} and \eqref{ero:4} is that these equations are valid in both the non-relativistic and relativistic domains, something that is inherent in the derivation of \eqref{ero:2} which was taken from Ref.~\cite{LEE--1980A}, but that was questioned in Ref.~\cite{SHARP1980-}.  In fact, these formulas give good first order estimates as long as inductive losses are the dominant energy loss mechanism.  Moreover, the energy conservation argument leading to them is such that, with suitable redefinition of the parameters, they can be applied to other situations then propagation in free vacuum or open air:  For instance, they can be adapted to erosion of beams propagating in a pre-formed plasma channel.

    As we have seen in Sec.~\ref{dgc:0}, an electron beam propagating in the ion-focused regime in a pre-formed channel is subject to a strong radial centering force due to the electric field generated by the more massive background ions which populate the channel after the beam has ejected all free electrons.  If $f_n = N_i/N_b$ is the ratio of ions per unit length to beam electrons per unit length, beam propagation in the channel will be characterized by an effective current $I_E=f_nI_B$.   Inductive beam head erosion can then be estimated with Eq.~\eqref{ero:3} or \eqref{ero:4} in which $I_E^*$ is replaced by $f_nI_B$, as is confirmed by a direct derivation presented in Ref.~\cite[p.225]{BUCHA1987-}.  Since $f_n$ is typically a few percent, inductive erosion of a pulse propagating in a guiding channel is significantly less than for a similar pulse propagating in open air.

   Erosion rates for electron beams propagating in a preionized channel have also been estimated taking magnetic (or field potential) energy into account \cite{KRALL1989A}, or using an energy conservation argument \cite{MOSTR1996-}.  In both cases a formula similar to Eq.~\eqref{ero:4}, in which $I_E^2/I_B$ is replaced by $f_nI_B$, is obtained.  Obviously, just like with erosion of non-guided beams, these formulas only give order of magnitude estimates (or sensible interpolation formulas), and do not dispense of making detailed simulations and experiments \cite{KRALL1989A,WERNE1994-,MOSTR1996-}.

Finally, while the inductive-, emittance-, and scattering-driven erosion rates studied in this section are axisymmetric processes, there are also transverse erosion processes, due for example to an external magnetic field, or to a centrifugal force when the beam follows a curved channel.   These processes, which are referred to under the names of `magnetic (or centrifugal) erosion' \cite{WERNE1994-,MOSTR1996-}, as well as related beam injection losses (called  `evaporation' \cite{MOSTR1996-}), will not be further discussed here.

\section{Beam conditioning}
\label{bco:0}

While idealized models of beams (infinitely long pulses, or else instantaneous rise- and fall-times; constant radius; flat radial distributions;  constant energy, current, and emittance; etc.) were by necessity used in most early models of high-energy beam theory, it was realized long ago that in any practical system the actual spatial and temporal distributions of the beam's particles properties within a pulse would have a considerable effect on its propagation properties.

   This came as much from the traditional engineering experience telling, for example, the importance of shape and spin in order to extend and stabilize the propagation of a missile, than from early theoretical considerations.  For instance, in his papers of the mid 1950s,  Bennett anticipated that (contrary to a thunder bolt, or a thermonuclear Z-pinch) there would be no hydrodynamic-type instabilities in a self-pinched beam because of very rapid particle mixing in azimuth \cite{BENNE1955-}.  Thus, by properly adjusting a beam pulse's particles distributions, i.e.,  by preparing or \emph{conditioning} it, it may be possible to greatly extend its range and to significantly decrease the effects of detrimental plasma instabilities.

   Because a particle beam pulse is a non-neutral plasma, there are many more parameters that can be adjusted than, for example, for a solid bullet propagating through air.  This makes the systematic discussion of all possible beam conditioning techniques rather complicated, especially since their specific effects, and even more so their synergy if several of them are used at the same time, can only be fully assessed by performing actual experiments.

   Historically, one of the first beam conditioning technique to be experimentally tested was that of beam `pulsing' (or `chopping,' or `bunching'), in which the current of a relatively long pulse is modulated in order to suppress or decrease the effect of resistive hose instability \cite{BEAL-1972-}.  This same technique can be used to break down and compress a long beam pulse into a train of shorter ones, therefore enabling the burst to propagate over a longer range than the original single pulse \cite{NEIL-1980-}.

   In the 1980s many beam conditioning techniques were developed and tested, especially since such techniques became essential in order to accelerate beam pulses to higher energies, and subsequently to send them into open-air.  The experiments were mainly carried at Lawrence Livermore National Laboratory on the ETA \cite{CLARK1984}, and at Sandia National Laboratory on the RADLAC \cite{EKDAH1986-} accelerators.

   A basic technique for beam conditioning is to pass the pulse through a `conditioning cell' in which the beam propagates in the ion focused regime.  Due to the stability of this propagation regime, it is possible to act on the beam by external means to produce a desired effect (active conditioning), or else to let beam perturbations which may lead to unstable motion decay (passive conditioning).  Such applications were anticipated in the some of the earliest papers published on IFR propagation, e.g., \cite{DIDEN1976A}, and were emphasized as an important option for `quiteing' (or `cooling')\footnote{Cooling usually means decreasing the kinetic energy spread of a bunch of particles, but can also be interpreted as suppressing unwanted behaviors such as micro- and macro-instabilities.} a beam subject to the `beam break up' (BBU) instability \cite{BRIGG1981-}.

   One important application of IFR conditioning cells is to provide an interface between the near vacuum existing in accelerators and full-density open-air, in which the beam is to be sent \cite{EKDAH1986-}.  When entering such a cell the head and the tail of the beam rapidly expand to the wall within the first meter or two. This rapid expansion leads to scrape-off erosion of the low-current, low-energy segments of the beam, i.e., the head and tail where the effective emittance is large.  However, while the head and the tail erode in the first two meters of the cell, the residual, narrowed pulse is efficiently transported through the remaining of the cell to the open-air injection foil \cite[p.12--14]{EKDAH1986-}.  The importance of conditioning prior to open-air propagation was highlighted in multi-pulse tracking experiments, where the tracking effect was barely observable without beam conditioning \cite{MURPH1992-}.

   Another simple and important application of IFR conditioning cells is to produce a well centered beam, an application which compared to the traditional use of external magnetic focusing fields can simplify design and ease hardware requirements \cite{MYERS1995-}.  As seen in Sec.~\ref{dgc:0}, image currents driven in a conducting tube walls can provide a net restoring force which centers the beam.  Thus, a conducting tube filled with a neutral gas can be used to transport and center a beam, while at the same time reduce its transverse oscillations.  

   The beam conditioning effect of IFR transport can be enhanced by preparing a channel whose diameter is smaller than that of the beam: This has a stabilizing effect on beam macro-instabilities due to an-harmonic forces and phase-mix damping \cite{DIDEN1977-, CAPOR1986-}.  A similar effect is obtained by replacing the narrow channel by a thin wire \cite{DIDEN1977-,PRONO1983-}.
 
  During the 1990s, a major emphasis of research on beam conditioning was to find the best technique for preparing a beam prior to its injection into a dense gas, especially for combating the resistive hose instability \cite{FERNS1992-, HUBBA1993-, WEIDM1994-, MURPH1996-, ROSE-2001B}.  While this will be further discussed in Sec.~\ref{mas:0} on mastering and damping instabilities, two approaches can already be mentioned \cite{FERNS1992-}:  One strategy, that was just refered to, is to center the beam and reduce the transverse displacements that seed the instability.  A second strategy is to tailor the beam emittance so that it increases from head to tail.  Such tailoring detunes the instability and can reduce its growth substantially.  In general, both centering and tailoring are needed to propagate intense beams over long distances \cite{FERNS1992-, HUBBA1993-}.

   Due to the Bennett pinch relation, beam emittance and radius are related in such a way that tapering the radius can result into a desirable emittance tailoring of a beam pulse \cite{WEIDM1994-}. Indeed, if a pulse is `radius-tailored,' i.e., tapered from a large radius in the head to a small radius in the tail, then the growth rate of the resistive hose instability will be reduced. Such radius-tailoring can be produced using a fast rise-time focusing coil \cite{WEIDM1994-}, or active wire cells, i.e., current carrying wires inducing suitable magnetic fields affecting the beam shape in a controlled way \cite{MURPH1996-, ROSE-2001B}.

  Finally, high intensity beam conditioning can greatly benefit from the considerable experience gained with high-energy colliders and storage rings technology developed for fundamental research in particle physics.  In particular, the concept of stochastic beams has already been considered for the suppression of the ion-hose instability in the ion-focused regime, and for developing arrays of guide wires to cause exponentially fast phase decorrelation rates in wire-guided phase-mixing cells for beam conditioning \cite{OBRIE1990B}.

   More generally, the powerful methods developed for the `stochastic cooling' of ion beams in specially designed `conditioning rings' (rather than just single-pass conditioning cells) may find direct application for the conditioning of a high intensity proton beam pulse prior to its injection into the atmosphere. In such conditioning rings the beam pulse is treated as a random ensemble rather than a deterministic bunch of particles; and information gained by sampling the pulse at one point of the ring is sent over a secant (or a full diameter) to another point of the ring ahead of the pulse in order to correct for some unwanted perturbation at the moment when the pulse will pass at that point \cite{-MEER1980-}.\footnote{Of course, if the pulse does not fill the circumference of the ring, the information can be used to correct the beam pulse the next time it passes through the measuring point.}

\section{Propagation of a train of pulses}
\label{mul:0}

  This very brief section is included to remind that complex phenomena arise when a sequence of pulses is sent into the atmosphere.  These phenomena include the hydro- and plasmo-dynamic evolution of the perturbed atmosphere between  consecutive pulses, as well as the effect of electromagnetic wake fields that may connect distant pulses as well and their images in nearby obstacles.  Some of these phenomena and effects have been partly addressed in the previous sections and chapters, and will be further addressed in later ones.  However, a comprehensive discussion will not be attempted here, mainly because these issues are not fundamental, but rather technical in the sense that they will most probably be solved by trial and error through experiments.

\chapter{Stability of propagating high-power beams}
\label{sta:00}

\section{General considerations on beam stability}
\label{sta:0}

A crucial question concerning beam propagation is that of stability.  Because of the substantial source of free energy represented by the kinetic energy of the beam, a variety of instabilities could be excited and amplified during propagation.  For the purpose of using high energy particle beams as weapons, the problem is to find a set of beam parameters (energy, current, radius, pulse length, emittance, and energy spread) such that the beam can reach the target without being destroyed by the possible instabilities.  This is rather difficult, especially because the existing theoretical models predict stable  beam propagation in such a narrow range of conflicting parameters that only the actual testing of a beam will confirm whether these predictions were correct or not.  A further difficulty is that there are very few published studies in which a part or the full range of possible instabilities and their synergistic effects on electron (i.e., \cite{BUDKE1956A}, \cite{IVANO1970-}, \cite{BENFO1973-}) and/or ion (i.e., \cite{CALLA1977-}, \cite{JORNA1978-}) beam propagation are discussed in a somewhat systematic manner (see also, \cite{MILLE1982-}). 

	The stability of beam-plasma systems is investigated by the standard perturbation method.  If the initial perturbation of a stationary state of the system increases with time, the state is unstable under a perturbation of this type.  Usually one seeks a solution of the form\footnote{This form is conventional, but possibly the most frequently used.  However, all combinations of $\pm \vec{k}$ and $\pm \omega$ appear in the literature.}
\begin{equation}\label{sta:1} 
     f(\vec{r},t) = f(\vec{r}) \exp i(\vec{k}\cdot\vec{r} - \omega t),
\end{equation}
where $f$ is the deviation of any physical quantity from its stationary value.  The relation between the complex frequency $\omega$ and the complex wave number $\vec{k}$ is by definition the \emph{dispersion relation}
\begin{equation}\label{sta:2} 
                              D(\omega,\vec{k}) = 0.
\end{equation}
A wave is said to be \emph{unstable}, if for some real wave number $k$, a complex $\omega$ with a negative imaginary part is obtained from the dispersion relation, signifying growth in time of a spatially periodic disturbance:
\begin{equation}\label{sta:3} 
\left.
\begin{array}{c}
\IMA(k)      = 0 \\
\IMA(\omega) < 0
\end{array}
\quad \right\} \Longrightarrow \text{Instability}.
\end{equation}
The absolute value of the imaginary part of $\omega$ for an unstable wave, i.e., $\delta \DEF - \IMA(\omega)$ is called the \emph{increment} or \emph{growth rate} (or its inverse the $\e$-\emph{fold time}) of the perturbation, because
\begin{equation}\label{sta:4} 
         f(\vec{r},t)   \propto \exp (\delta~t).
\end{equation}

	Instabilities in a beam-plasma system are primarily the result of the interaction between the beam and the plasma.  The main parameters characterizing the plasma are its so-called \emph{plasma frequency} (or \emph{Langmuir frequency})  $\omega_p$  and $\nu$ the collision frequency of the plasma electrons.  The plasma frequency, already defined in \eqref{def:8}, is
\begin{equation}\label{sta:5} 
         \omega_p^2 = \frac{1}{\epsilon_0} \frac{e^2}{m_e}  n_e
                    = 4 \pi c^2 r_e n_e,
\end{equation}
where $n_e$ is the electron number density, $m_e$ the electron mass, and $r_e$ the classical electron radius.  The plasma dielectric constant is then
\begin{equation}\label{sta:6} 
              \epsilon_p = 1 - \frac{\omega_p^2}{\omega(\omega+i\nu)}.
\end{equation}
Similarly, as a plasma on its own, the beam is also characterized by its \emph{beam plasma-frequency} (or \emph{beam Langmuir-frequency})
\begin{equation}\label{sta:7} 
        \omega_b^2 = \frac{1}{\epsilon_0} \frac{e^2}{m\gamma}  n_b
                   = 4 \pi c^2 r_e \frac{m_e}{m\gamma} n_b,
\end{equation}
where $n_b$ is the beam particle number density, $m$ the beam particles rest mass, and $\gamma$ their Lorentz factor.  The on-axis beam-plasma frequency is related to the beam scale radius by \eqref{bcn:9}, and therefore
\begin{equation}\label{sta:8} 
       \omega_b^2(0) = 4 \frac{c^2}{a^2} \frac{I_B}{I_A}.
\end{equation}
Finally, with the definition of the plasma frequency, the charge neutralization time \eqref{bcn:10} can be written
\begin{equation}\label{sta:10} 
    \tau_e =   \frac{\nu}{\omega_p^2},
\end{equation}
and the magnetic diffusion time \eqref{bcn:16}
\begin{equation}\label{sta:9} 
    \tau_m = \frac{\omega_p^2}{\nu} \frac{a^2}{c^2}.
\end{equation}

	There are numerous kinds of instabilities and many methods of classifying them.  A rough phenomenological way is to divide them into two categories: macroscopic and microscopic.  The next step is then to distinguish between various characteristics, such as the fate of the perturbation (absolute or convective), the driving forces (electrostatic, electromagnetic, etc.), the relative importance of collisions (collisional/non-collisional, resistive/ non-resistive), the relative importance of spread in the velocity distributions or temperatures (hydrodynamic, kinetic), etc.  However, there is no sharp distinction between different possible categories.  In this paragraph we will simply introduce the main concepts involved.

\begin{itemize}

   \item {\bf Macroscopic instabilities} influence the spatial distribution of the beam.  They are usually classified according to the geometry of the distorsion.  For beams there are four main categories: (i) the \emph{sausage} or \emph{varicose} mode in which the beam contracts and dilates at regular intervals; (ii) the \emph{hose}, \emph{kink}, or \emph{sinuous} mode in which the beam oscillates sideways; (iii) the \emph{filamentation} mode in which the beam breaks up into several filaments;\footnote{In thin hollow beams filamentation corresponds to the \emph{tearing} modes in magnetized plasma sheets.} and (iv) the \emph{ripple} mode in which the beam is distorted by small-scale ripples on the surface.\footnote{This simple classification does not explicitly consider the \emph{hollowing} modes, see Sec.~\ref{mac:0}.  It illustrates, however, the extensive and often confusing terminology which makes the reading and correlation of publications related to instabilities very difficult.  As was observed by an early beam stability researcher:  
``The terms `sinuous' and `varicose' instability were applied by Lord Rayleigh, and there seems no need for the more recent coinings'' \cite[p.225]{FINKE1961-}.}

Macroinstabilities concern also other macroscopic degrees of freedom such as densities, hydrodynamic velocities, etc.  They are connected with the flow-out of a plasma as a whole from one region into another.  Macroscopic instabilities are also called \emph{magnetohydrodynamic}, \emph{hydrodynamic},\footnote{More precisely \emph{partly hydrodynamic} to distinguish them from the purely hydrodynamic instabilities in which electromagnetic forces are absent} or simply \emph{low frequency} instabilities.

   \item {\bf Microscopic instabilities} do not necessarily induce a macroscopic motion of the plasma as a whole, but they can excite local fluctuations of density and electromagnetic fields in the plasma.  These \emph{velocity-space} modes which do not appear to have direct effect on the beam will principally appear as an extra form of energy loss by which the energy of the beam can be transformed into powerful electromagnetic radiations.  The most important kind of microinstability is the so-called \emph{two-stream} instability.  This mode is found when one is studying the propagation of electromagnetic waves in a system consisting of two interpenetrating streams of particles.

Some kinds of microinstabilities are directly connected to macroinstabilities.  This is the case, for example, of filamentation which can be the macroscopic stage of a purely growing transverse electrostatic microinstability (the Weibel instability).  In general, the growth of microinstabilities may lead to the onset of macroinstabilities, which may eventually destroy the beam.

   \item {\bf Turbulent instabilities} should be interposed between microscopic and macroscopic instabilities, and should be included because when fully established and under control they could contribute to stabilize a nonconductive gas or plasma \cite{KOVAS1960-}.  The theoretical and practical problems with turbulence are notoriously difficult, however, and there have been only few studies on the possibility of suppressing beam-plasma instabilities by the introduction of turbulence \cite{HERSH1979-,FREUN1980-}.  Nevertheless, it is precisely in this domain dominated by non-linear effects that some of the recent progress with magnetically confined thermonuclear fusion has been made \cite{BURRE1998-}.

\end{itemize}

	In a given reference frame, two types of instabilities can be distinguished physically: \emph{convective} instabilities, and \emph{absolute} or non-convective instabilities \cite{BRIGG1971-}.  Briefly stated, the essential point is the distinction as to whether an initially localized disturbance (of an infinite system) grows exponentially with time locally (absolute instability), or ultimately decays because of the propagation of the growing disturbance away from the point of origin (convective instability).  For a beam propagating along the $z$ axis, the \emph{group velocity} of an unstable perturbation with longitudinal wave number $k_z$ is given by
\begin{equation}\label{sta:11} 
               v_g = \frac{\partial}{\partial k_z} \REA(\omega).
\end{equation}
If $v_g$ equals the beam velocity $v$, the perturbation is absolute in the beam frame; if  $v_g = 0$, the perturbation is absolute in the plasma; and if  $v_g < v$, the perturbation is convective.  Clearly, for beam propagation, the worst instabilities are the absolute ones in the beam frame.  Therefore, if an instability cannot be avoided, it should at least be convective in order not to completely hinder propagation.

	The effect on stability of beam particle collisions with the plasma can, in general, be neglected except when, as with bremsstrahlung losses of high energy electron beams, they contribute to broadening the beam velocity distribution.

	The collisions of the plasma electrons with the plasma molecules and ions often have an important effect on stability, and they can either increase or decrease the growth rates.   When collisions are negligible, the instabilities are in the \emph{collisionless} regime ($\nu \rightarrow 0$).  In the \emph{collisional} regime several distinctions can be made, in particular depending upon the relative values  of the magnetic diffusion time $\tau_m$  and the collision frequency $\nu$ \cite{BENFO1973-}.  For high-current beams propagating in air, one has $\tau_m \nu < 1$ for pressures below a few torr, and the instabilities are called \emph{non-resistive}.  For pressure above a few torr $\tau_m \nu > 1$, and the instabilities are of the \emph{resistive} kind.  In the former case, the key parameter is the dielectric constant \eqref{sta:6} and various time scale are possible for the growth rate.  In the resistive case the time-scale is $\tau_m$.  When $\tau_e > \tau_m$, the acting forces are predominantly electrostatic, whereas, in the resistive domain where $\tau_m > \tau_e$, they are primarily magnetic.

	In describing the electromagnetic oscillations associated with instabilities, especially in the case of microinstabilities, one often uses the standard conventions used for electromagnetic wave propagation in wave guides or plasmas.  If $\vec{E}_0$ and $\vec{B}_0$ are the unperturbed fields, and $\vec{E}_1$ and $\vec{B}_1$, their respective perturbations, one uses the following terminology:

\begin{itemize}
\item	$\vec{B}_1    = 0$	~~~~~~ \emph{electrostatic} wave,
\item	$\vec{B}_1 \neq 0$	~~~~~~ \emph{electromagnetic} wave,
\item	$\vec{k}~ \lo \vec{B}_0$	~~~~~~ parallel wave,
\item	$\vec{k} \tr \vec{B}_0$	~~~~~~ perpendicular wave,
\item	$\vec{k}~ \lo \vec{E}_1$	~~~~~~ longitudinal wave,
\item	$\vec{k} \tr \vec{E}_1$	~~~~~~ transverse wave.
\end{itemize}

	The hydrodynamic description of a beam-plasma system is strictly valid only in the limit of a monoenergetic beam penetrating a cold plasma.  The velocity distribution of the beam (also called the \emph{beam temperature}), and/or the velocity distributions of the plasma (the plasma ion and electron temperatures) can be taken into account by kinetic plasma theory.  Neglecting the plasma temperature effects, the \emph{hydrodynamic} regime of instability, as opposed to the \emph{kinetic} regime, is defined as follows:
\begin{equation}\label{sta:12} 
\begin{array}{l}
|v_p - v|  >  \Delta v  ~~~~ \text{hydrodynamic regime,} \\
|v_p - v|  <  \Delta v  ~~~~ \text{kinetic regime,~~~~~}
\end{array}
\end{equation}
where $v_p = \omega/k$ is the \emph{phase velocity} of the wave, $v$ the beam velocity, and $\Delta v$ the RMS beam velocity spread.  Introducing the \emph{Doppler shifted wave frequency}
\begin{equation}\label{sta:13} 
     \Omega \DEF  \omega - \vec{k} \cdot \vec{v},
\end{equation}
the kinetic regime is then defined as
\begin{equation}\label{sta:14} 
            |\Omega| < \vec{k} \cdot \Delta\vec{v}.
\end{equation}

	For propagating beams the velocity distribution is a function of both the spread in beam energy (longitudinal emittance) and in beam direction (transverse emittance).  For self-pinched beams the transverse velocity spread $\tilde{v}_{\tr}^2 = \Delta v_x^2 + \Delta v_y^2 = 2 \Delta v_{\tr}^2$ is related by \eqref{def:16} to the effective beam current corresponding to the Bennett pinch relation \eqref{ben:23}, i.e.,  $\tilde{v}_{\tr} = \beta c\sqrt{I_E/I_A}$.
The one-dimensional transverse and longitudinal components of the resulting velocity spread $\Delta\vec{v}$ are therefore \cite{LEE--1977B, NEWBE1982-}
\begin{equation}\label{sta:15} 
   \frac{\Delta v_{\tr}}{v} = \sqrt{\frac{I_E}{2I_A} },
\end{equation}
and \cite{BENFO1973-, THODE1976-}
\begin{equation}\label{sta:16} 
     \frac{\Delta v_{\lo}}{v} = \Oh (\frac{\Delta v_{\tr}}{v})^2
                            + \frac{1}{\gamma^2} \frac{\Delta W}{W}.
\end{equation}

	The combination of the effects of collisions with those of velocity spreads can considerably diminish the growth rate of instabilities.  However, the most serious instabilities cannot completely be suppressed.  Their effect will thus have to be minimized by the proper choice of beam parameters like current, shape, pulse length, deliberate velocity spreads, etc.  In the following sections the major instabilities will be discussed together with possibles remedies.

\section{Microinstabilities}
\label{mic:0}

	When a beam passes at high velocity through a plasma, its coupling to the electrons or ions can excite unstable electromagnetic oscillations.  For high energy beams propagating in ionized air, the coupling to the ions is in general negligible relative to to the coupling to the electrons.  The resulting microinstabilities are of several different kinds depending upon the type of electromagnetic wave excited.  The two most important ones are the so-called \emph{two-stream instability}, which refers to transverse or longitudinal electrostatic waves, and the \emph{Weibel instability} (or \emph{micro-filamentation}) which is a transverse electromagnetic mode.

	The complete analysis of the various types of streaming instabilities is rather complicated because the beam velocity distributions, the plasma temperatures, the beam's own magnetic field, etc., all have an effect on them.  The theoretical analysis by analytical models is thus restricted to the most simple cases for which a linear perturbation approach is possible.  The full analysis, including non-linear effects, requires computer simulations, and experiments will be needed to check the validity of the codes \cite{MOIR-1981-,BARLE1981A}.   The following discussion will present the main known results for the above-mentioned instabilities in the case of high energy beams launched into the air in the pressure range of interest for particle beam weapons.

\subsection{Two-stream instability}

	The dispersion relation of the \emph{two-stream instability} of a monoenergetic beam interacting with a cool plasma, neglecting the beam's own magnetic field, is \cite{BRIGG1971-,WATSO1960-,BLUDM1960-,LEE--1977B}
\begin{equation}\label{mic:1} 
   1 - \frac{\omega_p^2}{\omega(\omega+i\nu)}
     - \frac{1}{\gamma^2}\frac{\omega_b^2}{\Omega^2}
       (\gamma^2 \sin^2\phi + \cos^2\phi)
     = 0,
\end{equation}
where $\phi$ is the angle between the wave vector and the beam velocity, i.e.,
\begin{equation}\label{mic:2} 
    \vec{k} \cdot \vec{v} = k v \cos\phi = k_{\lo} v.
\end{equation}
The unstable waves which are solutions to this dispersion relation are obtained by resolving it for $\omega$ and finding the roots which have a negative imaginary part.  The peak growth occurs at $k_{\lo}v =  \omega_p$ and is given approximately as the lower of the collisionless ($\nu < \omega_p$) and collisional  ($\nu > \omega_p$) limits, given by
\begin{equation}\label{mic:3} %
   \delta_{NC} = \IMA(\omega)_{NC} = \frac{\sqrt{3}}{2}\omega_b
   \Biggl( \frac{\omega_p}{2\omega_b}
           \frac{(\gamma^2 \sin^2\phi + \cos^2\phi)}{\gamma^2} \Biggr)^{1/3},
\end{equation}
and
\begin{equation}\label{mic:4} 
   \delta_C = \IMA(\omega)_C = \omega_b
            \sqrt{ \frac{\omega_p}{2\nu}
                \frac{(\gamma^2 \sin^2\phi + \cos^2\phi)}{\gamma^2} }.
\end{equation}

   At relativistic energies, the fastest growing oscillations are those propagating almost perpendicular to the beam ($\sin\phi \approx 1$).  Such an angular dependence is due to the fact that when $\gamma \gg 1$, the transverse mass $m\gamma$ of the beam particles is much smaller than the longitudinal mass $m\gamma^3$, and the oscillations which are transverse to the beam are easier to build up.  The worst case will thus correspond to transverse waves for which
\begin{equation}\label{mic:5} 
   \delta \approx \omega_b \sqrt{ \frac{\omega_p}{2\nu} } \sin\phi.
\end{equation}
The most difficult mode to suppress corresponds to $k_{\tr} = 1/a$ because the beam has finite radius $a$.  Therefore
\begin{equation}\label{mic:6} 
       \sin\phi \approx k_{\tr}/k_{\lo} = \frac{v}{a\omega_p}.
\end{equation}
For air at atmospheric pressure, typical values for the plasma parameters of a 10~kA, 1~GeV, 0.5 cm radius electron beam are: $\omega_b = 5 \times 10^{10}$,  $\omega_p  = 6 \times 10^{12}$, and $\nu = 4.7 \times 10^{12}$ s$^{-1}$.  Thus $\sin\phi = 0.01$ and the $\e$-fold distance corresponding to the maximum growth \eqref{mic:5} is $v/\delta \approx 1$ m, showing that two-stream instability should be a major obstacle to the propagation of monoenergetic beams over significant distances.  However, taking beam velocity spreads --- which for a pinched beam are directly related to beam emittance by equations \eqref{sta:15} and \eqref{sta:16} --- into account, the two-stream instability can be totally suppressed, provided the parameters fall into the kinetic regime.

	The kinetic theory dispersion relation replacing \eqref{mic:1} in the case of transverse waves takes the form \cite{BRIGG1971-} 
\begin{equation}\label{mic:7} 
   1 - \frac{\omega_p^2}{\omega(\omega+i\nu)}
     - \omega_b^2 \int \frac{f(v)}{(\omega-kv)^2}dv
     \approx 0.
\end{equation}
The integral over the beam velocity distribution $f(v)$ has a singularity for $\omega = kv$, which was first studied by Landau for the collisionless damping of electromagnetic waves in warm plasmas (Landau damping effect).  In the case of the two-stream instability, a similar effect is associated with the beam velocity distribution and results in the suppression of the instability \cite{SINGH1964-}.  This suppression becomes effective in the kinetic domain, which for two-stream instabilities sets in when
\begin{equation}\label{mic:8} 
      |\Omega| \approx \IMA(\omega) < \vec{k} \cdot \Delta\vec{v}.
\end{equation}
The transverse velocity spread due to particle oscillations in the pinch field is given by \eqref{sta:15}.  The stability boundary, which is also the worst-case kinetic regime growth, is from \eqref{mic:5} and \eqref{mic:6}
\begin{equation}\label{mic:9} 
                \omega_b    \sqrt{ \frac{\omega_p}{2\nu} }
              < \frac{c}{a} \sqrt{ \frac{I_E}{2I_A} },
\end{equation}
which yields the condition \cite{LEE--1977B}
\begin{equation}\label{mic:10} 
          \nu  >  \frac{2}{\omega_p}   \frac{c^2}{a^2}.
\end{equation}

	This simple stability criterion predicts that the two-stream instability will be suppressed for high-current high-energy beams propagating in air at pressures above a few torr.  This has been verified experimentally \cite{BRIGG1977-,RUDAK1972-} and has been confirmed by extensive computer calculations for beam energies up to 1 GeV and currents on the order of 10 to 100 kA \cite{NEWBE1982-}.

	Below the critical pressure implied by the collision frequency bound \eqref{mic:10}, the two-stream instability is not suppressed, but its growth rate is considerably reduced by kinetic effects.  The growth rates of the longitudinal and transverse electrostatic waves, respectively, are then  \cite{THODE1976-,GRISH1973-}
\begin{equation}\label{mic:11} 
    \delta_{\lo} = \Oh \frac{1}{\gamma^2}
                       \frac{\omega_b^2}{\omega_p^2}
                   \nu \frac{v}{\Delta v_{\lo}},
\end{equation}
and
\begin{equation}\label{mic:12} 
    \delta_{\tr} = \Oh ~~
                       \frac{\omega_b^2}{\omega_p^2}
                   \nu \frac{v}{\Delta v_{\tr}},
\end{equation}
and can be considerably lower than \eqref{mic:5}.

\subsection{Weibel (or micro-filamentation) instability}

	The electromagnetic modes with $\vec{k}$ normal to  $\vec{v}$, $\vec{E}$ nearly parallel to $\vec{v}$, and $\vec{B}$ normal to both $\vec{v}$ and $\vec{E}$, is a variety of the \emph{Weibel instability} \cite{WEIBE1959-}.  This instability grows fastest in the absence of external or self-generated magnetic fields and is thus maximum for $f_m \approx 1$.  It is an absolute instability which grows at perturbation centers of enhanced beam density which magnetically attract nearby beam particles and repel plasma electrons.  Thus, the beam ultimately splits into filaments, each of which self-pinches.  The computer simulations of this process \cite{LEE--1973A} show that the filamentation stage in which the effect of this microinstability reaches the macroscopic level, eventually ends up with the beam breaking up into separate filaments, which may recombine into a single dense beam, from which the return current is expelled.

	For a charge- and current-neutralized beam, in which the self-fields are nearly suppressed, the dispersion relation of the Weibel instability for low frequencies ($\omega < \omega_p$ ) is \cite{LEE--1973A}
\begin{equation}\label{mic:13} 
       \omega^2 = -\beta^2 \frac{\omega_b^2}
                           { 1+\frac{\omega_p^2}{k_{\tr}^2 c^2} },
\end{equation}
where $k_{\tr}$ is the transverse wave number.  This is a purely growing mode, i.e., an absolute instability with maximum growth rate
\begin{equation}\label{mic:14} %
         \delta (k_{\tr} \rightarrow 0) = \beta\omega_b.
\end{equation}

   In the case of a collisional plasma the dispersion relation (\cite[Eq.8]{MOLVI1975-}) is obtained by multiplying  $\omega_p^2$ in \eqref{mic:13} by $\omega/(\omega+i\nu)$ according to \eqref{sta:6}.  Then, in the collisional limit($\omega < \nu$), the dispersion relation becomes approximately \cite[Eq.75]{JORNA1978-}
\begin{equation}\label{mic:15} %
           \omega^2 - i \frac{\omega_p^2\omega^3}{\nu k_{\tr}^2c^2}
                    + \beta^2\omega_b^2 = 0.
\end{equation}
For high frequencies ($\omega_b < \omega < \nu$) this reduces to
\begin{equation}\label{mic:16} %
             \omega = -i \frac{\nu}{\omega_p^2} k_{\tr}^2 c^2
                    = -i \frac{1}{\tau_m} k_{\tr}^2 a^2,
\end{equation}
where the definition of the magnetic diffusion has been used.  On the other hand, for low frequencies ($\omega < \omega_b$), equation \eqref{mic:15} reduces to \cite{MOLVI1977-}
\begin{equation}\label{mic:17} 
      \omega^3 = -i \frac{\nu}{\omega_p^2}\omega_b^2\beta^2 k_{\tr}^2 c^2,
\end{equation}
which also shows that collisions enhance this kind of instability.

  In the kinetic regime, the growth of microinstabilities is reduced, but kinetic effects alone are not sufficient to suppress them.  This is because for such a transverse instabilities $|\Omega| < |\omega|$ and there is no such effect as Landau damping in that case.  The kinetic regime boundary \eqref{sta:14} gives the maximum growth rate.  Thus, from \eqref{mic:17}, it can be seen that \cite{IVANO1970-}, \cite[Eq.14]{MOLVI1975-}
\begin{equation}\label{mic:18} 
         \delta = \nu \frac{\omega_b^2}{\omega_p^2}
           \bigl( \frac{v}{\Delta v_{\tr}} \bigr)^2.
\end{equation}
The same equation is obtained starting from \eqref{mic:16} by taking into account that $0 < k_{\tr} < \beta \omega_b/\Delta v_{\tr}$, see \cite{HUBBA1978-}.

To get a simple estimate, one can use \eqref{sta:8} and \eqref{sta:15} for $\omega_b$ and $\Delta v_{\tr}$.  The maximum $\e$-fold time in the kinetic regime is then on the order of
\begin{equation}\label{mic:19} 
    \frac{1}{\delta} \approx \frac{1}{8} \tau_m.
\end{equation}
This is very short and would, for this absolute instability, rapidly destroy the beam.  However, this is only the case for $f_m\approx 1$ and the presence of a sufficient non-zero magnetic field results in the stabilization of this mode.

   In fact, the stabilizing influence of velocity spreads and magnetic fields on the filamentation instability are known since the early days of thermonuclear research \cite{FURTH1963-}.  They can approximately be taken into account by rewriting \eqref{mic:13} as \cite{KAPET1974-}
\begin{equation}\label{mic:20} %
       \omega^2 = -\beta^2 \frac{\omega_b^2}
                           { 1+\frac{\omega_p^2}{k_{\tr}^2 c^2} }
                  + \omega_c^2 + (k_{\tr}\Delta v_{\tr})^2,
\end{equation}
where $\omega_c = qB/\gamma m$ is the beam particle's cyclotron frequency.  It then follows that filamentation is always suppressed when 
\begin{equation}\label{mic:21} %
         \omega_c > \beta\omega_b.
\end{equation}
Similarly, in the limit $k_{\tr} \rightarrow 0$, stability is ensured provided
\begin{equation}\label{mic:22} %
         \frac{\Delta v_{\tr}}{\beta c} > \frac{\omega_b}{\omega_p}.
\end{equation}

   In practice, for a beam propagating in open air, the magnetic field cannot come from an electromagnet as in laboratory experiments.  Nevertheless, if the beam is not fully current neutralized, there is a non-zero magnetic self-field and in the absence of any other magnetic field the stability criterion is \cite{DAVID1975-}
\begin{equation}\label{mic:23} 
      \Bigl( \beta^2(1-f_m) - (1-f_e)  \Bigr) > \frac{1}{2}\beta^2.
\end{equation}
Thus, for a charge neutral beam ($f_e=1$), a current neutralization fraction $f_m  < \Oh$ provides enough self-magnetic field to stabilize the micro-filamentation mode.  This criterion was derived for a collisionless plasma.  For a collisional plasma, where no analysis comparable to \cite{DAVID1975-} is available, the criterion \eqref{mic:23} is likely to be still applicable.  In fact, this is very plausible because the general criterion \eqref{mic:21} applied to a beam for which $\omega_b$ is given by equation \eqref{sta:8} shows that filamentation is suppressed if $B$ is at least equal to the maximum value of the self-magnetic field of the non-neutralized beam.

    It therefore remains the problem of current neutralized beams such that $f_m > 1/2$.  This has been clarified in a study which derived rigorously sufficient and approximaltely necessary conditions for the absence of the beam-Weibel microinstability \cite{CARY-1981-}.  This study concluded that even in the presence of collisions ($\nu \neq 0$) a beam satisfying the criterion \eqref{mic:21} is always stable, and that in the absence of collisions  ($\nu = 0$) a beam not satisfying \eqref{mic:21} is always unstable. Moreover, in the collisional case, a criterion weaker than \eqref{mic:21} was derived, i.e.,
\begin{equation}\label{mic:24} %
\omega_p^2 T_b/mc^2 + \omega_c^2 > \beta^2\omega_b^2 + \omega_p^2 T_p/mc^2,
\end{equation}
where $T_b=T_{\tr}$ and $T_p$ are the beam and plasma temperatures.  Therefore, if there is a small amount of magnetic field (e.g., Earth's magnetic field) such that
\begin{equation}\label{mic:25} %
         \omega_c^2 > \omega_p^2 T_p/mc^2,
\end{equation}
the criterion \eqref{mic:24} implies that the beam micro-filamentation mode can be stabilized primarily by beam temperature if $ \omega_p^2 T_b/mc^2 > \beta^2\omega_b^2$, i.e., if the criterion \eqref{mic:22} is satisfied.

   Finally, there is a last beam parameter which has a positive impact of controlling  filamentation and that we have not yet discussed: the anti-pinch action of a macroscopic beam rotation \cite{UHM--1983-}.  This can be done by realizing that the centrifugal effect of an axial rotation is similar to that of a transverse energy spread.  This enables to introduce a compound parameter, the transverse energy $\Delta W_{\tr}$, equal to the sum of the transverse energy spread $kT_{\tr}$ and the rotational energy $p_\theta^2/2m$.  In fact, using the notation of reference \cite{UHM--1983-}, one can define three dimensionless parameters ($\xi, \delta,$ and $\eta$) characterizing the influence of beam parameters on stability behavior:
\begin{equation}\label{mic:26} %
         \xi = k_{\tr} \beta c \omega_b \sqrt{2 \frac{\Delta W_{\tr}}{m c^2}}, 
\end{equation}
which represents the effect of the transverse energy $\Delta W_{\tr}$;
\begin{equation}\label{mic:27} %
      \delta = k_{\lo} \beta c \omega_b \frac{\Delta W_{\lo}}{\gamma^3 m c^2},
\end{equation}
which represents the effect of the longitudinal energy spread $\Delta W_{\lo}$; and
\begin{equation}\label{mic:28} %
         \eta = 2(1-f_m)  + \beta^2 \frac{\omega_c^2}{\omega_b^2},
\end{equation}
which represents the effect of the magnetic self-pinch and external field forces.  Using these parameters it is possible to present the results of the stability analysis in the form of two-dimensional contour plots (rather then just as lower or upper bounds) and see that the full transverse-energy and the longitudinal energy-spread can play major stabilizing roles \cite{UHM--1983-}.

\subsection{Beam-plasma heating by microinstabilities}

While microinstabilities are in general detrimental to long-range beam propagation, they are a welcomed mechanism when it comes to transfer energy from a beam to a plasma, for example to heat a plasma to thermonuclear temperatures.  For this reason microinstabilities and beam-plasma heating have been extensively studied in the context of thermonuclear fusion research.  A number of papers cited in the present report are therefore relevant to this subject, or motivated by it, e.g., \cite{IYYEN1989-,HAMME1979-,HAAN-1982-,JANNS1984-,VENUG1992-}.

\subsection{Discussion of microinstabilities}

	Concerning the stability of a beam with respect to microinstabilies, the situation can now be summarized as follows:

\begin{itemize}

	\item At very low gas pressure, the plasma generated by the beam can be so weak that the beam is only partially charge neutralized ($f_e<1$), and the conductivity so low that $f_m\approx 0$.  In that case, when $1 > f_e > 1/\gamma^2$, the beam is pinched but the plasma electron density may be insufficient for microinstabilities to develop.  With electron beams, the two-stream instability is avoided as long as the product of gas pressure and pulse length is low enough to keep secondary electrons from accumulating inside the beam \cite{BRIGG1981-}.  This is sometimes referred to as the \emph{low pressure propagation window}, in which the beam propagates in the so-called \emph{ion-focused regime} (IFR).  For air, it corresponds to about 1~torr for a 10 kA, 1 cm radius beam \cite{BEAL-1972-,BRIGG1981-}.  For such a beam, with an energy of 500 MeV, the range would be on the order of 200 km \cite{BRIGG1981-}.

	\item At intermediate pressures, the two-stream mode sets in, and the beam becomes unstable.  The range of the beams is then determined by the maximum growth rate given by \eqref{mic:11} or \eqref{mic:12}.

	\item At pressures above a few torr, for beam currents in the 10 to 100 kA range, the two-stream instability is suppressed.  This has been observed in many experiments --- for a review see \cite{WALLI1975-} and for actual experiments \cite{MOIR-1981-,BRIGG1977-}.  However, above the critical pressure, the rise in plasma conductivity enables a return current to flow and filamentation becomes possible as $f_m \rightarrow 1$.  Moreover, because of this increase in conductivity, other new instabilities, including macroinstabilities, also become possible.  Therefore, the narrow stability window observed in air near 1~torr is due to the suppression of microinstabilities by collisions and kinetic effects, and to the temporary absence of resistive macroinstabilities which set in at higher pressures.

\end{itemize}

\section{Macroinstabilities}
\label{mac:0}

	The small amplitude distorsion of the shape of a cylindrical beam can be described by giving the perturbed form of the beam surface, e.g.,
\begin{equation}\label{mac:1} 
      r = a + \Delta r(z,r,\theta,t).
\end{equation}
The standard method is to make a multipole expansion of the perturbation in a thin annulus as a function of the azimuthal angle $\theta$
\begin{equation}\label{mac:2} 
     \Delta r = \sum_m \Delta r_m(z,r,t) \cos m\theta.
\end{equation}
For small amplitude periodic perturbations, this can be Fourier analyzed as a superposition of modes such that \cite{FINKE1961-}
\begin{equation}\label{mac:3} 
    \Delta r = \sum_m A_m(r) \exp i (\vec{k}\cdot\vec{r} - \omega t + m\theta).
\end{equation}
The mode $m=0$ displays harmonic variations of beam radius with distance along the beam axis: this is \emph{sausage} instability.  The mode $m=1$ represents transverse displacements of the beam cross-section without change in the form or in a beam characteristics other than the position of its center of mass: this is called \emph{sinuous}, \emph{kink} or \emph{hose} instability.  Higher values of $m$ represent changes of cross-section from circular form: $m=2$ gives an elliptic cross-section, $m=3$ a pyriform cross-section, etc.  Modes with $m>1$ are referred to as \emph{filamentation} modes, because their growth leads towards the break-up of the beam into separate filaments.

	For a given azimuthal wave number $m$, various radial modes are possible.  They are usually classified according to the degree $2n$ of polynomial eigenfunctions of the dispersion equation.  For example, when $m=0$, $n=0$ corresponds to \emph{axial hollowing}, $n=1$ to standard \emph{sausaging}, $n=2$ to \emph{axial bunching}, etc., \cite{UHM--1981-}.  A systematic classification of the modes excited by resistive macroinstabilities in a simplified helical orbit beam model has been given by Steven Weinberg \cite{WEINB1967-}.

	The primary concern with macroinstabilities is to determine their growth rate and the conditions for which they are sufficiently convective to ensure that the growth will be of limited consequence for a sufficiently short beam pulse.  The analysis is then simpler in the Doppler-shifted frequency representation of the oscillations, i.e.,
\begin{equation}\label{mac:4} 
        \exp i(kz - \omega t) \DEF  \exp i(\Omega z/v - \omega \tau).
\end{equation}
In this representation, $\tau$ gives the position of a section of the beam pulse measured from the head of the beam \eqref{pla:3}, and $\Omega$ is the Doppler-shifted frequency \eqref{sta:13}.

	When $\Omega$ is real, and $\omega$ complex, the dispersion relation $D(\Omega,\omega) = 0$ yields  solutions giving the free growth of instabilities corresponding to initial value problems in the beam frame.  These are potential instabilities developing during the flight of the beam towards its target.  It is thus crucial that these instabilities are convective, i.e., that $\REA \omega(\Omega) \neq 0$.
	
	When $\omega$ is real, and $\Omega$ complex, the same dispersion relation yields solutions to initial value problems in the accelerator frame.  The convective nature of these instabilities, or the limited duration for the acceleration period for absolute instabilities, will ensure that their effect will disappear after the pulse has left the accelerator.  For practical reasons, one usually studies macroinstabilities in laboratory experiments by deliberately introducing perturbations at the end of the acceleration process \cite{LAUER1978-}.  In the following discussion, we will assume that the accelerator can be built in such a way that the pulse is not disturbed during acceleration.  We will thus concentrate on instabilities affecting the flight of the beam towards its target.

\subsection{Macro-filamentation}

   We begin the discussion of macroinstabilities by macro-filamentation because a partially or fully current neutralized beam may spontaneously break-up into filaments and blow apart without  having been subject to any significant external perturbation that could have initially disturbed the shape of the beam.  This is because macro-filamentation can result from the growth of micro-filamentation, which is always possible if there is some noise and some degree of current neutralization in the beam.

   Since macro-filamentation is  a complicated non-linear process it is best studied in computer simulations \cite{LEE--1973A}, although some analytical studies have been attempted \cite{BENFO1973-, MONTG1979-}.  Because it is a particularly dangerous form of instability in several proposed ion beam fusion  schemes \cite{OLSON1982-}, it has been extensively studied in that context, e.g., \cite{CALLA1977-, JORNA1978-, LEE--1980B}. 

   In its early (linear) stage, where the instability grows exponentially from noise, the micro-filaments are not actually formed and do not react on themselves as real current filaments would do \cite{MONTG1979-}.  However, as soon as the filaments are able to self-pinch, their development becomes non-linear and it is possible to qualitatively discuss their growth in terms of elementary concepts of electrodynamics --- such as the Biot and Savart law, which implies that parallel current attract, so that filaments grow by expelling currents of opposite sign, until the whole beam breaks up into separate filaments of minimum size given by the Bennett pinch condition.  Consequently, starting from an initial configuration that is partly or fully neutralized by two (or more) uniformly interpenetration streams (e.g., the beam current and the plasma return current), the final configuration may consist of many locally self-pinched `beamlets' of opposite signs, with a global degree of neutralization equal to that of the initial configuration.

   While such a bundle of self-pinched beamlets may possibly have interesting characteristics for some applications, we will use it here as a mental picture to derive two `marginal stability criteria' of the kind that were found in the study of stability problems related to early thermonuclear fusion devices \cite{FURTH1963-} (see also \cite{LEE--1980B}).   These criteria will be such that if they are met macro-filamentation should not occur, in which case micro-filamentation will not be suppressed but rather saturate and hopefully decay without impeding beam propagation.

    Let us rewrite the Bennett pinch condition \eqref{ben:4} of a fully charge neutralized beam as
\begin{equation}\label{mac:4a} %
      a_B(I_B,f_m) = \epsilon_{\tr} \sqrt{\frac{I_A}{(1-f_m)I_B}},
\end{equation}
in order emphasizes that for a an accelerator system producing a beam of given emittance $\epsilon_{\tr}$ and Alfv\'en current $I_A$, the pinch radius is a function of the variables $I_B$ and $f_m$.  Then, if a fixed radius is chosen for some particular reason, it is evident that an arbitrarily large current can be transported by the beam, provided $f_m \rightarrow 1$.   This possibility is of great interest in applications such as ion-beam fusion where a very large beam current is focused on a tiny target in order to compress it and hopefully reach thermonuclear ignition.  However, it is also evident that such a beam may split into several filaments, provided each of them carries a current satisfying the Bennett condition, which according to \eqref{mac:4a} may range from $(1-f_m)I_B$ to $I_B$.  The maximum number of possible filaments is therefore
\begin{equation}\label{mac:4b} %
      N_{fil(f_m<1)} = \frac{1}{1-f_m}.
\end{equation}
Since $N_{fil}$ must be an integer, one can derive a marginal stability criterion for a single beam not to decay into two filaments, i.e.,
\begin{equation}\label{mac:4c} %
      N_{fil} < 2  ~~ ~~ \Longrightarrow ~~ ~~ f_m < \Oh,
\end{equation}
which turns out to be consistent with the criterion \eqref{mic:23} for the absence of micro-filamentation.

   In the case where a beam is fully current neutralized, $f_m=1$, equation \eqref{mac:4a} does not apply any more and a separate analysis is required because $a(I_B)$ can in principle have any value.  However, if such a beam is prepared with an initial radius $a(I_B) = a_B(I_B,0)$, i.e., equal to that of a fully pinched beam with the same current, it is clear that it will not be able to break up into separate filaments if for some reason current separation occurs.  More generally, this is also the case if $a(I_B) < a_B(I_B,0)$, whereas when $a(I_B) > a_B(I_B,0)$ the beam may possibly break up into a number of filaments such that
\begin{equation}\label{mac:4d} %
      N_{fil(f_m=1)} = \frac{a^2(I_B)}{a_B^2(I_B,0)}.
\end{equation}
Again, this enables to derive a marginal stability criterion, i.e.,
\begin{equation}\label{mac:4e} %
    N_{fil} < 2 ~~ ~~ \Longrightarrow ~~ ~~ a(I_B) < \sqrt{2}a_B(I_B,0),
\end{equation}
which is compatible with \eqref{mac:4c}, in the sense that this criterion can also be written in the form $a_B(I_B,f_m) < \sqrt{2}a_B(I_B,0)$.

   In conclusion, we see that while beams that are not fully self-pinched are liable to macro-filamentation, they can nevertheless be marginally stable provided they have an initial radius less than $\sqrt{2}$ times the radius of a fully pinched beam of same non-neutralized current.  In principle, this discussion could be extended to plasmoid beams, in which the current neutralization is provided by comoving particles of opposite electrical charges.  We will leave that for the chapter explicitly dealing with plasmoids.

\subsection{Electrostatic kink (or `ion hose') instability}

	Analysis of the major macroinstabilities for propagation in air at pressure between 0.01 and 1 atmosphere shows that only short pulses may propagate over sizable distances.  Furthermore, because of the considerable difference in relativistic mass between transverse and longitudinal motions of the beam, instabilities involving transverse displacement will grow fastest.  As a result, the first order description of a beam pulse during the early stages of the growth of an instability will not be that of a thin and flexible thread but of an axially rigid rod, the cross-section of which is undergoing  distorsion.  This is the so-called \emph{rigid beam} approach.

	However, in the case of a beam propagating in a vacuum or a low density plasma, the \emph{thin thread} approximation can be satisfactory.  Let us consider, for example, the stability of a charged beam in a vacuum, and then the effects of a plasma on it \cite{JORNA1978-}.  In that case, the most serious instability is the $m=1$ electromagnetic kink mode in which the beam is distorted as a flat snake or a screw.  The growth rate can be determined by analyzing the lateral motion of a flexible cylinder of charged particles in a vacuum.  The calculation is easiest in the beam frame, and after Lorentz transformation into the accelerator frame the result is
\begin{equation}\label{mac:5} 
  \Omega = -\Oh i \omega_b \frac{ka}{\gamma} \sqrt{\ln (\frac{ka}{\gamma}) }.
\end{equation}
Because of the finite radius of the beam, the growth rate $\delta = -\IMA(\omega) = -\IMA(\Omega)$  for $k=k_{\lo}$ real, has a maximum at $ka = 0.6\gamma$, and a cut-off at $ka = \gamma$.  In the rest frame of the beam, the effect of a plasma environment is to reduce the electric field by a factor equal to the dielectric constant \eqref{sta:6} : hence to reduce the growth rate by $1/\sqrt{\epsilon_p}$ .  When $\nu > kv$, \eqref{mac:5} then becomes
\begin{equation}\label{mac:6} 
    \Omega = -\Oh i \frac{\omega_b}{\omega_p} \frac{ka}{\gamma}
                  \sqrt{\Oh \nu k v \ln (\frac{\gamma}{ka}) }.
\end{equation}
This is the dispersion relation of the electrostatic kink instability in the hydrodynamic limit for a thin beam.  It applies for beams propagating in collisional plasmas under such conditions that the magnetic forces can be neglected.  This is the case in air at pressures below 1 torr.  The maximum growth rate is at $ka = 0.72 \gamma$, where
\begin{equation}\label{mac:7} 
\delta = \frac{1}{4}\sqrt{\frac{\gamma}{\tau_m} \frac{v}{a} \frac{I_B}{I_A}}.
\end{equation}
In the kinetic regime, this instability is damped when  $\IMA(\Omega) < k \Delta v_{\lo}$.  This happens when
\begin{equation}\label{mac:8} 
   \tau_m  >  \frac{2}{\gamma} \frac{a}{v} \frac{I_A}{I_B}.
\end{equation}

	Let us consider now the stability problem of a rigid beam subject to transverse displacements.  This is the $m=1$ mode, and it is easiest to start by considering first a beam with a constant density profile and a sharp boundary, even though we will have to examine later the more realistic case of a beam with the Bennett profile.  We further assume that the beam is fully charge neutralized by a collisional plasma of constant density with radius larger than the beams radius.

	If a beam is slightly displaced in the transverse direction, a surface charge density distribution with a $\cos \theta$ azimuthal dependence will appear.  Such a charge distribution produces a homogeneous dipolar electric field which results in an electrostatic restoring force directly proportional to the displacement.  For a beam with a flat density profile, the equation of motion is simply \cite{LINHA1959-}
\begin{equation}\label{mac:9} 
   \frac{d^2}{dt^2} y_b = -\Oh \omega_b^2 (y_b - y_p),
\end{equation}
where $y_b$ is the position of the axis of the displaced beam, $y_p$ the position of the axis of the non-neutral plasma column which neutralizes the beam, and $\frac{d}{dt} = \frac{\partial}{\partial t} +v \frac{\partial}{\partial z}$  the total derivative.  From the equation of continuity and Gauss's law one can see that the plasma will move according to
\begin{equation}\label{mac:10} 
    \frac{\partial}{\partial t} y_p = \frac{1}{\tau_e} (y_b - y_p).
\end{equation}
Taking small perturbations of the form  $\exp i(kz-\omega t)$, this couple of equation yields the dispersion relation
\begin{equation}\label{mac:11} 
    -i\omega \tau_e = \frac{\Omega^2}{\Oh \omega_b^2- \Omega^2}.
\end{equation}
This is the electrostatic kink instability.  For $\Omega$ real, $\omega$ is purely imaginary and the growth is absolute in the beam frame.  Furthermore, the growth rate $\delta = -i\omega$ tends to infinity as $\Omega^2$ approaches $\Oh \omega_b^2$.  If we take for $\tau_e$ the complete expression
\begin{equation}\label{mac:12} 
     \tau_e = \frac{\nu - i\omega}{\omega_p^2},
\end{equation}
the dispersion relation \eqref{mac:11} can be put in the more familiar form \cite{KINO-1963-}
\begin{equation}\label{mac:13} 
           \frac{\omega_p^2}{\omega(\omega+i\nu)}
     + \Oh \frac{\omega_b^2}{\Omega^2}  = 1.
\end{equation}
Apart from the factor $\Oh$, it is identical to the dispersion relation for transverse (i.e., $\phi=\frac{\pi}{2}$) two-stream instability \eqref{mic:1}.  In the collisionless limit ($\nu=0$)  this result is well known to particle accelerators specialists \cite{BUDKE1956A}.

\subsection{Electromagnetic kink (or `resistive hose') instability}

	In a low conductivity plasma, $\tau_e$ is very large and \eqref{mac:11} reduces to stable oscillations at frequency $\Omega^2 = \Oh\omega_b^2$.   Physically, this corresponds to a beam neutralized by infinitely heavy ions, so that the beam's motion is simply harmonic oscillations about a cylinder of neutralizing charges which cannot move.  This is obvious from \eqref{mac:10} which shows that $y_p$ is constant when $\tau_e = \infty$.  At the other limit of high conductivity $\tau_e$ becomes very small.  In that case \eqref{mac:10} indicates that $y_b = y_p$, and this corresponds to the fact that when conductivity is very high the plasma neutralizing the beam can follow it exactly, and there are no electrostatic beam oscillations or instabilities.  However, when the conductivity is high, magnetic forces have to be taken into account:

	When a beam is displaced transversely to its direction of propagation, the motion of the magnetic field generated by the beam current $J_B$ induces a longitudinal electric field.  If  $\vec{v}_y = \frac{\partial}{\partial t} \vec{y}_b$ is the velocity of the sideways beam displacement and $\vec{B}_\theta$ the azimuthal magnetic field, the Lorentz transformation of this magnetic field gives $\vec{E}_z = \vec{v}_y\times \vec{B}_\theta$ in the limit of small $\vec{v}_y$.   This electric field generates a longitudinal plasma current $\vec{J}_z = \sigma \vec{E}_z$, which, if the plasma conductivity is high enough, generates an azimuthal magnetic field of sufficient strength to interfere with the beam-current-generated azimuthal magnetic field.  (This plasma current should not be confused with the plasma return current, which we assume to be negligible for the moment, $f_m = 0$.)  As the displacement of the plasma does not coincide with that of the beam when the conductivity is finite, the total magnetic field resulting from $J_B$ and $J_z$ will have an axis $y_m$ different from the beam axis $y_b$ or the plasma current axis $y_p$.  This displacement of that magnetic field axis is directly related to the magnetic diffusion time.  Thus
\begin{equation}\label{mac:14} 
  \frac{\partial}{\partial t} y_m = \frac{1}{\mathcal{L}\tau_m} (y_b - y_m).
\end{equation}

   When the conductivity is low (i.e., $\tau_m$ small) the plasma induced by the lateral displacement is negligible and $y_m = y_b$, i.e., the magnetic axis always corresponds to the beam axis.  On the contrary, when the conductivity is high (i.e., $\tau_m$ large), or the plasma infinite in extent (i.e.,  $\mathcal{L} \rightarrow \infty$), $y_m$ is constant and the magnetic field is `frozen.'  When the magnetic field axis does not correspond with the beam axis, the particles in the beam are subjected to a restoring magnetic force which is equivalent to the force needed to drive the plasma current $J_z$.  This force can be determined from the effect on the beam of the dipolar magnetic field resulting from the differences in position between the beam and the magnetic field axis.  This gives
\begin{equation}\label{mac:15} 
          \frac{d^2}{dt^2} y_b = -\Oh \beta^2 \omega_b^2 (y_b - y_m).
\end{equation}

  For periodic small amplitude oscillations, the system \eqref{mac:14}, \eqref{mac:15} gives the following dispersion relation \cite{LEE--1973B,LEE--1978-}
\begin{equation}\label{mac:16} 
 -i\omega\mathcal{L}\tau_m = \frac{\Omega^2}{\Oh\beta^2\omega_b^2- \Omega^2}.
\end{equation}
This is the dispersion equation of the \emph{electromagnetic kink} or \emph{hose instability} of a rigid beam in a resistive plasma.  It shows absolute instability in the beam frame and infinite growth for $\Omega^2 \rightarrow \Oh\beta^2\omega_b^2$.  This resistive hose instability has been extensively studied \cite{IVANO1970-}, \cite{UHM--1981-}, \cite{WEINB1967-}, \cite{ROSEN1960-}, \cite{LEE--1973B}, \cite{LEE--1975-}, \cite{LEE--1978-}, \cite{UHM--1980-} and \cite{LAMPE1981-}.

	The dispersion relation \eqref{mac:16} is similar to its electrostatic counter part \eqref{mac:11}, but corresponds to a different driving mechanism.  In fact, for $\tau_m<\tau_e$, the kink instability is of the electric kind and the dispersion relation is \eqref{mac:11}.  On the other hand, in high conductivity plasmas, where $\tau_m>\tau_e$, the instability is of the magnetic kind and the dispersion relation is \eqref{mac:16}.  In the limit of $\tau_m \propto \sigma \rightarrow 0$, the electric oscillations are stable with frequency $\Omega^2 =\Oh\omega_b^2$, and in the limit $\tau_m \propto \sigma \rightarrow \infty$, the magnetic oscillations are stable with frequency $\Omega^2 =\Oh\beta^2\omega_b^2$.

\subsection{Macrostability of a beam penetrating a neutral gas}

	When a beam pulse penetrates a neutral gas and generates its own  plasma, the electric and magnetic kink modes are encountered successively as the conductivity rises from zero to a maximum.  A model valid for arbitrary conductivity and combining the two instabilities is thus important.  Such a model is obtained by combining \eqref{mac:9}, \eqref{mac:10}, \eqref{mac:14}, and \eqref{mac:15}.  This leads to the system of equations:
\begin{equation}\label{mac:17} 
   \frac{d^2}{dt^2} y_b + \Oh        \omega_b^2 (y_b - y_p)
                        + \Oh \beta^2\omega_b^2 (y_b - y_m)  = 0,
\end{equation}
\begin{equation}\label{mac:18} 
   \frac{\partial}{\partial t} y_p + \frac{1}{\tau_e} (y_p - y_b) = 0,
\end{equation}
\begin{equation}\label{mac:19} 
\frac{\partial}{\partial t} y_m + \frac{1}{\mathcal{L}\tau_m} (y_m - y_b) = 0.
\end{equation}
The resulting dispersion relation is \cite{UHM--1981-,LEE--1975-}
\begin{equation}\label{mac:20} 
   (1-i\omega\mathcal{L}\tau_m)(1-i\omega\tau_e)
 = \frac{  \Oh\omega_b^2 (\beta^2
         + i\omega\tau_e\gamma^{-2}
         + \omega^2\tau_e\mathcal{L}\tau_m) }
        {  \Oh\beta^2\omega_b^2  - \Omega^2 }.
\end{equation}
As the $m=1$ oscillations are stable in the limits $\sigma \rightarrow 0$ and $\sigma \rightarrow \infty$, the instability growth rate has a maximum at some finite value of the conductivity.  The dispersion relation \eqref{mac:20} shows that this maximum is obtained when
\begin{equation}\label{mac:21} 
      \tau_e = \mathcal{L}\tau_m = \frac{a}{c} \sqrt{\mathcal{L}}.
\end{equation}
By comparison with \eqref{cri:4} and \eqref{cri:7}, one sees that this happens at a point very early in the beam pulse, and that $m=1$ instabilities will have maximum growth in the neck region.

	In the head region of the beam, $\tau_m$ and $\tau_e$ are directly function of the beam current \eqref{cri:4}.  Therefore, in order to minimize the length of the neck region over which the growth is largest, the conductivity generation by the beam should be as large as possible, requiring
\begin{equation}\label{mac:22} 
                     I_B > I_G.
\end{equation}
However, this condition implies high plasma return currents: the hose instability should therefore be examined in the $f_m \neq 0$ case as well.

	A non-zero plasma return current has essentially no effect on the electrostatic forces, and thus on the electric kink instability.  On the other hand, the magnetic forces are directly affected, and the first effect of a return current is to diminish the magnetic force in \eqref{mac:15} by the factor $(1-f_m)$ which is equivalent to replacing by $I_E$ the current $I_B$ used to calculate $\omega_b^2$  with \eqref{sta:8}.  But there is also an additional effect: as anti-parallel currents tend to repel each other, when the beam is displaced relative to the plasma channel in which the return current flows, the interaction between the beam current and the return current tends to increase the displacement.  This is the so-called \emph{self-hose} effect which results in a supplementary force on the beam proportional to $I_BI_P = f_m I_B^2$.  Including those two effects, the equation of motion \eqref{mac:15} becomes
\begin{equation}\label{mac:23} 
\begin{array}{r}
    \dfrac{d^2}{dt^2} y_b = - \Oh \beta^2 \omega_b^2 (y_b - y_m) (1-f_m)
                            + \Oh \beta^2 \omega_b^2        y_m     f_m \\
                         ~~\\
                          = - \Oh \beta^2 \omega_b^2 
                              \bigl(y_b(1-f_m) - y_m\bigr).
\end{array}
\end{equation}
Similarly, they result into the replacement of $\omega_b^2$  by $\omega_b^2(1-f_m)$ in the denominator on the right hand side of  equation \eqref{mac:20}, \cite{UHM--1981-}.  Thus, for $\tau_m>\tau_e$,
\begin{equation}\label{mac:24} 
      -i\omega\mathcal{L}\tau_m = 
        \frac{\Oh\beta^2\omega_b^2f_m +\Omega^2}
             {\Oh\beta^2\omega_b^2(1-f_m) - \Omega^2}.
\end{equation}
With the effect of the self-hose included, the growth rate of the magnetic hose instability is clearly worse.  In particular contrary to \eqref{mac:16}, the growth rate at $\Omega=0$ is now non-zero.  This implies that when $f_m \neq 0$, even a very slow transverse displacement may result in the disruption of the beam.  Furthermore, as $f_m \rightarrow 1$, the growth rate increases and the position of the pole moves towards smaller frequencies.

	The electric and magnetic hose modes for rigid oscillations of flat profile cylindrical beams show very bad stability properties.  The consequences of various damping effects on these instabilities will be discussed below.  Before that, we will examine the $m\neq 1$ modes with the same flat profile rigid beam model, but only for resistive modes in the limit of $\tau_m>\tau_e$.

	For $m=0$, the sausage mode dispersion relation is as follows \cite{UHM--1981-,LEE--1981-} :
\begin{equation}\label{mac:25} 
      -i\omega\frac{1}{6}\tau_m = 
        \frac{-\beta^2\omega_b^2(1-2f_m) +\Omega^2}
             {2\beta^2\omega_b^2(1-f_m) - \Omega^2}.
\end{equation}
The main difference between this dispersion relation and the hose dispersion relation \eqref{mac:24} is that the sausage mode is unstable in a limited frequency range only: for $f_m\approx 0$, between $\omega_b$ and $2\omega_b$.  This will enable the kinetic effects to stabilize this instability provided that, as can be seen from \eqref{mac:25},
\begin{equation}\label{mac:26} 
                          f_m < \Oh.
\end{equation}

	For $m>1$, there are numerous unstable modes possible.  However, in the limit of small $f_m$, the main modes obey the approximate dispertion relation \cite{WEINB1967-}
\begin{equation}\label{mac:27} 
      -i\omega\tau_m = 
        \frac{-(m^2-1)\Oh\beta^2\omega_b^2 +\Omega^2}
             {m^2\Oh\beta^2\omega_b^2 - \Omega^2}.
\end{equation}
In the case $m=1$ one recognizes the hose mode, and for $m>1$, similarly to the sausage mode, instabilities exist only in a limited frequency range, which becomes narrower in proportion when $m^2$ increases.

	The essential difference between the hose mode and the $m\neq 1$ modes is that when $m=1$ a low frequency disturbance can be produced without internal compression or distorsion: only a simple transverse displacement is required \cite{WEINB1967-,LEE--1978-}.  Because there is no change in internal pressure to produce a restoring force, hose instability appears at an arbitrary low (Doppler shifted) frequency.  Furthermore, the stabilizing effects of possible spreads in the beam velocities vanish in that limit.  For the $m\neq 1$ modes, instability potentially appears at finite frequency, but is strongly suppressed by kinetic effects.  We will thus concentrate on these effects on the hose mode only, assuming that coping with this worst case mode will be sufficient.

\subsection{Macrostability of beams with rounded radial profiles}

	In examining now the stability of more realistic beam plasma models we will see that the instabilities are somehow not so bad as for idealized models such as the sharp boundary, flat profile, cylindrical beam studied so far.  We will first look at the effect of rounded beam profiles --- specifically of the Bennett profile \eqref{ben:21} --- and then at the effect of spreads in velocity distribution, mainly arising from the particle's oscillations in the pinch field.

	Let us rewrite the hose dispersion relation for $f_m=0$ in the following form
\begin{equation}\label{mac:28} 
      -i\omega\tau_b = 
        \frac{\Omega^2}
             {\beta^2\omega_{\beta}^2 - \Omega^2}.
\end{equation}
$\omega_\beta$ is the \emph{betatron frequency}, equation \eqref{ben:9}, which for a beam with a flat current profile is simply
\begin{equation}\label{mac:29} 
        \omega_\beta^2 = \Oh\beta^2\omega_b^2 \DEF \omega_{\beta m}^2.
\end{equation}
In the case of a beam with the Bennett profile \eqref{ben:21}, the betatron frequency is not constant but uniformly distributed between zero and the maximum $\omega_{\beta m}$.  The force equation \eqref{mac:15} has then to be averaged over the Bennett distribution and the resulting effective betatron frequency is \cite{LEE--1978-}
\begin{equation}\label{mac:30} 
    \omega_\beta^2  = \frac{1}{6}\beta^2\omega_b^2.
\end{equation}
Similarly, for the magnetic axis diffusion equation \eqref{mac:14}, the dipole magnetic diffusion time must be recalculated by properly averaging over the beam and plasma conductivity profiles.  Assuming the beam current and the plasma conductivity to both have Bennett profiles with scale radius $a$ for the beam and $b$ for the plasma, one finds
\begin{equation}\label{mac:31} 
    \tau_d = \frac{3}{2} \frac{\eta^4}{(\eta^2-1)^2}
             \Bigl(\frac{\eta^2+1}{\eta^2-1} \log(\eta) -1 \Bigr)\tau_m,
\end{equation}
where $\eta=b/a$.  In the case where $a=b$, this is
\begin{equation}\label{mac:32} 
                                \tau_d = \frac{1}{8} \tau_m,
\end{equation}
and by comparison with \eqref{mac:31} one sees that $\tau_d > \frac{1}{8}\tau_m$ when $b>a$, so that the beam is more stable in the case where the conductivity profile is broader than the beam profile \cite{LAMPE1981-}.

	The dispersion relation \eqref{mac:28} has been derived in the hydrodynamic limit.  However, from \eqref{mac:28} and \eqref{sta:15}, 
\begin{equation}\label{mac:33} 
    |\Omega| = \omega_\beta \frac{|\omega|\tau}{\sqrt{1+|\omega|^2\tau^2}}
             < k_{min}\Delta v_{\tr} = \omega_\beta.
\end{equation}
Thus, according to \eqref{sta:14}, kinetic effects cannot be ignored for the hose instability.  The simple dispersion relation \eqref{mac:28} should therefore be replaced by the expression
\begin{equation}\label{mac:34} 
      -i\omega\tau_d =  \int_{0}^{\omega_{\beta m}^2}
        f(\omega_\beta^2) \frac{\Omega^2}
                               {\beta^2\omega_\beta^2 - \Omega^2}
                          ~d\omega_{\beta}^2,
\end{equation}
where $f(\omega_\beta^2)$ is a suitable distribution function.  In principle this distribution function could be derived from the Vlasov kinetic plasma theory.  But an exact solution to this problem has still not been found.  However, two different phenomenological approaches \cite{LEE--1978-,UHM--1980-}, which agree with computer simulations \cite{LAMPE1981-}, give plausible results with the following function
\begin{equation}\label{mac:35} 
   f(\omega_\beta^2)  = 6 \frac{\omega_{\beta}^2}{\omega_{\beta m}^2}
                \bigl(1 - \frac{\omega_{\beta}^2}{\omega_{\beta m}^2}\bigr).
\end{equation}
For $f_m=0$, the resulting dispersion relation is \cite{LEE--1978-,UHM--1980-}
\begin{equation}\label{mac:36} 
  -i\omega\tau_d = 6 \frac{\omega_{\beta}^2}{\omega_{\beta m}^2}
               \Bigl(\frac{1}{2}
                   - \frac{\Omega^2}{\omega_{\beta m}^2}
                   + \frac{\Omega^2(\omega_{\beta m}^2-\Omega^2)}
                          {\omega_{\beta m}^4}
                          \log\frac{\omega_{\beta m}^2-\Omega^2}{\Omega^2}
               \Bigr).
\end{equation}
In comparison with \eqref{mac:28}, this dispersion relation shows much less serious instability problems:

\begin{itemize}

	\item The growth rate has a cut-off at $\Omega=\omega_{\beta m}$ and there is no pole at $\Omega=\omega_\beta$.  Instead, for $\Omega=0.52\,\omega_{\beta m}$, the growth rate has a maximum
\begin{equation}\label{mac:37} 
            \delta_{max} =  \frac{0.69}{\tau_d}.
\end{equation}

	\item The instability, from absolute in the beam frame, becomes convective.  Therefore, as a perturbation of the beam grows, it will at the same time move backwards into the pulse.  For a perturbation of amplitude $y(0)$ generated at the head of the beam, a saddle point analysis of the dispersion relation \cite{LEE--1978-} shows that the growth of the hose is such that at the tail of the beam pulse its amplitude will be 
\begin{equation}\label{mac:38} 
     y(\Delta t) = y(0) \bigl(1 + \frac{\Delta t}{\tau_d}  \bigr).
\end{equation}
\end{itemize}

	The existence of a maximum and of a cut-off in \eqref{mac:36} have been verified experimentally \cite{LAUER1978-}.  Similarly, other experiments \cite{BRIGG1977-} have also demonstrated the convective nature of the hose instability.  In these experiments, the hose instability leads to an erosion of the beam tail.  This is because, for a beam of finite duration, the disturbance is maximum at the end of the pulse.

	The significance of \eqref{mac:38} is that for a pulse of finite duration, a disturbance ultimately disappears.  Thus, if a beam pulse has a duration of the order of $\tau_d$, it will be able to propagate over large distances. In practice, the problem is that $\tau_d$ is only of a few nanoseconds for a high current beam injected into the atmosphere at normal pressures.  Therefore, stable beam propagation is restricted to very short pulses.

	A further difficulty with the hose instability is that its maximum growth and its convective nature depend on the extent of current neutralization of the beam.  It is found, for instance, that even a small amount of current neutralization (i.e., $f_m\approx 0.1$) is strongly destabilizing, particularly at low frequencies, and leads to absolute instability as seen by the beam \cite{UHM--1980-}.   The hose instability in the presence of current neutralization has been studied on a computer in the frame work of the Vlasov theory \cite{LAMPE1981-}.  The results show that for a current neutralization fraction larger than 0.5 the return current driven self-hose becomes the dominant destabilizing mechanism.

	The general properties of the hose instability can be transposed to the case of the $m\neq 1$ macroinstabilities and this has been verified in a number of cases.  For example, it has been found that the sausage mode is stable provided that \cite{LAMPE1981-}
\begin{equation}\label{mac:39} 
                 \frac{f_m}{1-f_m} < 2 \frac{I_B}{I_G}.
\end{equation}

\subsection{Discussion of macroinstabilities}

	The stability with respect to macroinstabilities seems to be ensured for beams of pulse length of the order ot $\tau_m$, provided that $f_m$ is small.  In practical cases, these requirements may be somewhat less stringent.  For instance, the effect of various damping factors (such as variations of beam current during the pulse, energy spreads, smooth radial and longitudinal beam profiles, finite pulse length, etc., \cite{BEAL-1972-}) cannot be calculated easily.  Experiments such as those that are possible with the high-current high-energy accelerators which have been proposed or built at Los Alamos \cite{MOIR-1981-}, Livermore, \cite{BARLE1981A} and in the Soviet Union \cite{PAVLO1980-} are thus crucial.

\section{Mastering and damping instabilities}
\label{mas:0}

        A direct application of the study of instabilities is the design of methods to suppress or (at least) to control instabilities in actual systems.  For beams propagating within accelerators there are a number of classical techniques such as
\begin{itemize}

\item longitudinal magnetic fields,

\item focusing magnetic fields,

\item conductive walls,

\item cooling,

\item etc.,

\end{itemize}
which have the effect of mastering and damping instabilities, as well as of shaping and conditioning beam pulses.  When the goal is to extract the beam and to inject it into an external medium in which the pulses freely propagate, further beam shaping and conditioning techniques are required.  As will be explained when discussing beam propagation experiments, these techniques comprise various empirical methods such as
\begin{itemize}

\item  beam chopping, e.g., \cite{BEAL-1972-,HESTE1974-},

\item  radius tailoring, e.g.,  \cite{FERNS1992-,WEIDM1994-},

\item  emittance tailoring, e.g., \cite{WEIDM1994-},

\item  beam quieting,

\item  etc.

\end{itemize}

	Because of the need for short pulses, actual beam weapon systems will have to use trains of small pulses in order to send sufficient energy towards the target.  This creates additional stability problems, because pulses within a train will have to propagate in the plasma background generated by the previous pulses.  The analysis of these complicated problems, together with that of the boring of reduced density plasma channels, will be part of the conceptual design of practical particle beam weapons.

\chapter{Plasmoid beam propagation}
\label{plb:0}

\section{Plasmoids in fundamental and applied sciences}
\label{pfa:0}

In conventional plasma physics a plasmoid is defined as an isolated plasma which holds together for a duration much longer than the collision times for the constituent particles.  The term plasmoid was suggested by  Winston Bostick in relation to early experiments which showed that toroidal shaped plasmas, i.e., `plasma rings,' could be created and projected across magnetic fields \cite{BOSTI1956-}.  The original concept of directed plasmoids, also due to Bostik \cite{BOSTI1956-}, i.e., of localized clumps of plasma projected by a `plasma gun,' e.g., \cite{MARSH1960-,ARETO1965-}, is therefore discussed since the 1950s, as is explained in the review \cite{WESSE1990-} covering theoretical descriptions, experimental observations, and computational results up to 1990. 

   Starting from these early experiments, theoretical and experimental studies related to the acceleration of \emph{compact plasmoids}, e.g., doughnut shaped toroids, over distances many times their own dimensions and to directed kinetic energies much greater than their stored magnetic and thermal energies proceeded slowly.  As a new type of accelerator able to accelerate such rings to high energy was proposed in 1982, i.e., \cite{HARTM1982-}, some speculations were made on their possible use as weapons \cite{BRADN1982-}.
 Experiments performed in 1983 demonstrated the gross stability and self-contained structure of compact toroids \cite{TURNE1983-}.  This lead to the proposal of a dedicated experimental facility to demonstrate the formation, compression, and acceleration of compact toroid plasmoids \cite{HARTM1984-}, which confirmed that such plasmoids could be routinely produced and accelerated to velocities of $>10^3$ km/s, e.g., \cite{PARKS1988-}. However, despite further progress on these and related systems based on rotating ion rings (e.g., \cite{KAPET1980-,SCHAM1993-,OLIVE1994-}), or on related activities such as `ball lighting' research (e.g., \cite{GILMA2003-, SHMAT2004-}), it seems that the prospect for practical applications of ring type plasmoids is not sufficiently high to justify claims such as those made in the early 1980's, e.g., \cite{BRADN1982-}.

   Similarly, the concept of directed plasmoid beams (or \emph{beam plasmoids}), i.e., of fully charge- and current-neutralized accelerated beams, is implicitly contained in the seminal works of Alfv\'en \cite{ALFVE1939-,ALFVE1981-}, and explicitly in those of Bennett \cite{BENNE1955-, BENNE1954-}, because such beams are naturally occurring as cosmic streams of particles and plasmas.\footnote{In his 1954 paper, reference \cite{BENNE1954-}, Bennett considered nearly charge and current neutralized proton beams streaming between the Sun and the Earth, with currents on the order of 15~A up to 150~kA, and energies of 50~MeV down to 50~keV, respectively.}  It is therefore quite natural that the surprising 1966-discovery (i.e., reference \cite{GRAYB1966-}) of the unsuspected low-pressure propagation window for high-intensity electron beam in air near 1~torr was quickly interpreted by Bennett as a confirmation of his theory, which strongly emphasized the importance of full charge and current neutralization, see \cite{ROBER1968-, COX--1970-, HAMME1970-}.\footnote{As stressed by H.L.~Buchanan of the \emph{Defense Advanced Research Projects Agency}, ``theoretical understanding of beam physics in this pressure regime, now called the ion-focused regime (IFR), has evolved slowly'' \cite[p.221]{BUCHA1987-}:  Indeed, it took many years to fully appreciate that propagation in this mode is most advantageous when the plasma is less dense than the beam, and its conductivity sufficiently low that beam-induced return currents and thus beam-current-neutralization are negligible.}   Possibly for this reason, the January 1980 \emph{Particle Beam Research Workshop} at the U.S.\ Air Force Academy similarly emphasized the importance of plasmoid beams, stressing in particular that ``An intense high-energy plasmoid beam has several operational advantages in exoatmospheric military applications'' \cite[p.54]{GUENT1980-}. 

   In fact, according to astrophysical observations, and in agreement with the original ideas of Alfv\`en and Bennett, the cosmos is full of directed energy phenomena such as jets \cite{VAUCO1979-}, flares, and bursts of gases in various states of ionization, as well as of very energetic particles (cosmic rays) and photons (gamma ray bursts), which can be highly collimated and able to propagate over thousands of light-years \cite{SHIBA2003-}.  Some cosmic streams even appear to consist of matter-antimatter plasmoids, i.e., jets of electron-positron pairs, and there is considerable debate on their origin and the mechanisms responsible for their acceleration \cite{GALLA1992-,HUILI1996-}.  Since the physics of the propagation of such beams through the interstellar plasma is closely related to the subject of this report, there are many publications in astrophysical journals of direct interest to it.  This is especially the case when considering long-range propagation of particle beams under conditions in which streaming microinstabilities cannot be avoided, e.g., \cite{ROSE-1984-, ROSE-1987-, SCHLI2002-}.

   On the interplanetary rather than interstellar scale there are many phenomena such as solar flares, geomagnetic storms, solar wind, etc., as well as many plasma effects in the Earth's iono- and magneto-spheres, which have become accessible to direct observation by rockets, artificial satellites and deep-space probes.  Moreover, many near-Earth phenomena of such type have become accessible to direct manipulation, e.g., by means of particle beams \cite{GOUGH1980-,KIWAM1977-,NEUPE1982-,GRAND1982-, OBAYA1984-, WINCK1984-, ARNOL1985-,  WINGL1987A, WINGL1987B, WINGL1987C, LIVES1989-} and materials \cite{HAERD1986-} injected into the near-Earth environment from rocket-borne platforms or orbiting space laboratories.   Historically, it is through the study of geomagnetic storms \cite{CHAPM1960-}, which are caused by streams of neutral ionized gas ejected from the sun, that Chapman and Ferrero discovered many basic plasma properties which were independently obtained under laboratory conditions by Langmuire (who originally gave the name \emph{plasma} to such neutral ionized gases).  Of crucial importance to the subject of this report is that these studies are directly related to the problem of propagating streams of energetic particles across magnetic fields --- hopefully in straight line and with minimum losses.

   In effect, basic physical processes such as plasmoid electric-polarization and/or paramagnetic/diamagnetic-magnetization in response to external electromagnetic fields, are the same in astrophysics \cite{CHAPM1960-, ABE--2001-}, in the motion of spacecrafts across the magnetosphere \cite{DRELL1965-}, or in heating/fueling magnetically confined thermonuclear fusion plasmas \cite{TUCK-1959-,SCHMI1960-,MANHE1977-,PARKS1988-,BRENN2005-}, as in `strategic defense' \cite[p.3554]{HEIDB1992-}.  However, despite many similarities, there are important differences between these domains, essentially because their characteristic energy and length scales can be very different \cite[Table.II]{WESSE1990-}.  In particular, astrophysical plasmoids are generally extremely energetic and wide, while solar and magnetospheric plasmoids are significantly less.  But in both cases the energy density and the directionality are very low in comparison to those required for either compact or beam plasmoid weapons.  Similarly, in thermonuclear fusion devices (which may be of the magnetic or inertial confinement type), the velocities of the plasmoid's particles are typically non-relativistic, while in plasmoids for strategic defense the particles may have to be highly relativistic in order to enable propagation over large distances.

    These differencies are clarified in the next two subsections, where the distinctive characteristics of compact and beam plasmoids are discussed in quantitative terms in view of their possible strategic applications.  When considering the propagation of such plasmoids across an external magnetic field $B_0$ \emph{transverse} to the direction of motion, the key parameters (see, e.g., the review papers \cite{WESSE1990-, BRENN2005-}) are:
\begin{itemize}
\item The so-called \emph{kinetic-beta}, defined as the ratio of the plasmoid kinetic energy density to the magnetic field energy density, i.e.,
\begin{equation} \label{pfa:1} 
      \beta_K = \frac{\Oh m (\beta c)^2 n_\pm}{\Oh \epsilon_0 c^2 B_0^2},
\end{equation}
where $n_\pm$ is the number density of the plasmoid particle of either sign,  $m = m_i + m_e \approx m_i$ the sum of the electron and ion masses, and $\beta c = v$ their velocity.

\item The ratio of the plasmoid radius $a$ to the Larmor gyroradius $R$ of the ions, equation~\eqref{mag:4}, i.e.,
\begin{equation} \label{pfa:2} 
      a/R = a\frac{e B_0}{\beta c\gamma m_i},
\end{equation}
so that a plasmoid such that  $a/R < 1$ will be called \emph{narrow}, and \emph{wide} if $a/R > 1$.

\end{itemize}
In both criteria the magnetic field intensity appears explicitly because the Lorentz force is maximal for a transverse field.  In the case of a \emph{longitudinal} magnetic field, the key parameter (see, e.g., the review paper \cite{ROBER1983-}, and for a comparison between the transverse and longitudinal geometries \cite{ROBER1981-}) is: 
\begin{itemize}

\item The ratio of the plasmoid radius $a$ to the electromagnetic skin depth $\lambda_S$ of the electrons,
\begin{equation} \label{pfa:2-} 
      a/\lambda_S = a\frac{c}{\omega_e}
                 = a\sqrt{ \frac{\epsilon_0 \gamma m_e}{e^2 n_\pm} },
\end{equation}
so that when $a/\lambda_S < 1$ the magnetic field penetrates into the plasmoid and all of the particles trajectories are deflected towards the plasmoid axis; while for $a/\lambda_S > 1$ the magnetic force is concentrated at the surface of the plasmoid and the plasmoid is compressed radially until the internal pressure balances the magnetic pressure.

\end{itemize}
Therefore, as is well known, the main effect of an axial magnetic field is to focus and compress a beam or plasmoid along its axis, while that of a transverse magnetic field is to deflect it and spread it apart, which is the problem studied in the remainder of this Chapter.

\subsection{Compact plasmoids}


Compact plasmoids have properties that are more closely related to those of ordinary plasmas than to those of particle beams.  The theoretical methods can thus be borrowed to the fields of standard magnetodynamics and plasma physics \cite{TUCK-1959-}, provided the usual conditions such as quasi-neutrality on a length scale characterized by the Debye length are satisfied.  The distinctive characteristic of these plasmoids is to be diamagnetic to the extent of completely excluding any external magnetic field, which make them attractive for applications such as refueling \cite{PARKS1988-} or heating magnetic confinement fusion devices \cite{MANHE1977-}.  The theoretical feasibility of this cross-field propagation mode has been proved on general grounds \cite[Appendix]{MANHE1977-}, and by constructing explicit solutions \cite{POUKE1967-}.  The properties and formation of compact plasmoids are also linked to the problem of magnetic-field-line reconnection, an unsolved fundamental issue in magnetohydrodynamics \cite{BRUNE1982-}, geophysics \cite{ABE--2001-}, and astrophysics \cite{BLACK1994-}.

   According to ideal magnetohydrodynamics, a compact plasmoid such that $\beta_K \gg 1$ and $a/R \gg 1$ should exhibit a strong diamagnetic behavior, i.e., behave as a perfect conductor, so that the exclusion of the magnetic field should result in simple ballistic propagation.  However, in a series of experiments with a 4 kA neutralized beam of ions (composed for about 75\% of protons) with peak ion energy in the range of 100--200~keV, it was found that relatively narrow plasmoids with $\beta_K$ between 0.01 and 300 were able to propagate over a magnetized vacuum, but showed very little diamagnetic behavior \cite{HONG-1988-, WESSE1988-}.  This was explained by anomalously fast penetration of the transverse magnetic field into these plasmoids \cite{WESSE1990-, WESSE1988-, SONG1990-}, a phenomenon which has also been observed under different experimental conditions in several active space experiments.  Conversely, while these experiments showed that self-polarization can occur for narrow plasmoids with $\beta_K$ as large as 300, magnetic expulsion as been reported \cite{SONG1990-} for wide plasmoids with $\beta_K$ down to 1.3. This means that ideal magnetohydrodynamics is insufficient to understand these phenomena, and that one should rely on detailed experimental and computer simulation investigations \cite{BRENN2005-}.

   The way $\beta_K$ is defined emphasizes that it is a quantity which is mostly used in the context of non-relativistic magnetohydrodynamics.  Nevertheless, since protons with energies up to about 200 MeV (which have a relativistic-beta $\beta=v/c\approx 0.56$) may still be considered as non-relativistic in first approximation, the concept of magnetic field exclusion may still correctly apply to hydrogen plasmoid beams with proton energies in the 50 to 100 MeV range.  

   For example, using computer simulations, 50-MeV-proton plasmoid configurations with $\beta_K \gg 1$ have been shown to be able to propagate uninhibited through a magnetized plasma corresponding to the Earth's ionosphere at an altitude of approximately 300 km, provided the plasmoid-to-background mass-density ratio is large \cite{PAPAD1988-}.\footnote{The size of this ratio ensures that the electromagnetic skin depth is much smaller than the plasmoid radius.}  More precisely, such simulations have been able to demonstrate that ``ion beams in the mega-electron-volt range with current density of the order of $10^{-2}$ to $10^{4}$ A/cm$^2$ will be able to propagate ballistically over distances of 500 to 2000 km'' \cite[p.1090]{PAPAD1991-}.

   The adjective `ballistic' used by the authors of these studies suggests that in the high-kinetic-beta mode of propagation strong diamagnetic effects imply that the background magnetic field and plasma are excluded from the plasmoid, which therefore behaves almost like a solid conductor moving across a magnetic field.  This means that the behavior of the plasmoid and its interactions with the ambient plasma correspond to what is expected from ideal magnetohydrodynamics, which is known to be applicable to relativistic plasma beams provided \cite[Appendix]{FINKE1961-} 
\begin{equation}\label{pfa:3} 
        \frac{I_B}{I_A} \gg 1.
\end{equation}
In the case of the simulations \cite{PAPAD1991-}, where $I_A \sim 100$~kA and $I_B \sim 40$~MA, this criterion is clearly satisfied.  On the other hand, in the previously mentioned experiments \cite{HONG-1988-, WESSE1988-}, where $I_A \sim 500$~kA and $I_B \sim 4$~kA, it is not satisfied, which may explain why diamagnetisme was not observed.

   The crucial idea leading to the Finkelstein-Surrock's criterion \eqref{pfa:3} is that the particles should be confined by the plasmoid's self-magnetic field to remain within the transverse radius $a$.  It then turns out that this is a sufficient condition for ensuring charge neutrality, i.e., the condition $\lambda_B \ll a$ (provided the particles's temperature $kT_b$ is less than their directed energy $W$), as well as for neglecting the plasma electron inertia, which enables writing Ohm's law in the form customary in magnetohydrodynamics.  However, as stressed by Finkelstein and Surrock, the  criterion \eqref{pfa:3} is only a necessary condition for the applicability of magnetohydrodynamics to relativistic plasma streams, so that detailed experiments and simulations cannot be dispensed of.

   In conclusion, the simulations \cite{PAPAD1988-,PAPAD1991-} confirm that there exist a parameter range in which the old idea of `plasma bullets' hurling across magnetized plasmas \cite{TUCK-1959-} is essentially correct, and demonstrate that this parameter range may correspond to relatively narrow compact plasmoids.  In practice, however, this range corresponds to very-high-current but rather low-energy-density plasmoids, such as those considered in reference \cite{PAPAD1991-}, which are about 6~km long and 2~km wide at half-density.\footnote{While these dimensions could correspond to the propagation-size of some initially narrower plasmoids, there is at present no practical concept for designing a `plasma gun' suitable to generate and fire them into the ionosphere.  On the other hand, the beam plasmoids discussed in the next sub-section are directly related to extrapolations of existing or near-term particle beam technology.}   These simulations therefore show the limits of that idea, because (unless the plasmoids are made of antimatter, e.g., antiprotons neutralized by positrons, or of relatively high-energy, i.e., 200 to 400~MeV, ions) their lethal effect will only come from the amount of kinetic energy intercepted by the target, which may turn out to be quite small, even if the plasmoid's velocity is a non-negligible fraction of the speed of light.  For this reason, the concept of compact plasmoids propagating in the ballistic mode is in direct competition with the so-called `kinetic-energy interceptors,' i.e., solid bullets launched by electromagnetic guns or other means, and will not be further discussed in this report.

\subsection{Beam plasmoids}

   Beam  plasmoids have properties that are more closely related to those of particle beams than to those of ordinary plasmas, so that their properties are in many ways similar to those of the particle beams studied in the previous chapters of this report.  They can be considered as the limiting case of very high kinetic-beta and narrow plasmoids, i.e.,  $\beta_K \gg 1$ and  $a/R \ll 1$, with the added constraint of being highly directional, i.e., of low-emittance with angular divergences typically measured in $\mu$rad.  In contrast to compact plasmoids (where the currents and magnetic field lines are essentially `closed') they are more `open' structures, and contrary to the Finkelstein-Surrock magnetohydrodynamicity criterion \eqref{pfa:3} they generally satisfy the Bennett-Alfv\`en paraxiality criterion \eqref{ben:25}
\begin{equation}\label{pfa:4} 
        \frac{I_B}{I_A} \ll 1.
\end{equation}

   For example, at the January 1980 U.S.\ Air Force Academy \emph{Particle Beam Research Workshop} the following significant parameters were defined for a beam plasmoid accelerator: ``100 to 200~A/cm$^2$, 5 to 10~MA, 20~ns, 200 to 400~MeV, 20 to 30~$\mu$rad, hadron plasmoid; Option:  25 to 50~A, 1~ms, 200 to 400 MeV, 1~$\mu$rad'' \cite[p.55]{GUENT1980-}.  In this concept the idea was to accelerate (within a common structure) a number of hadron (i.e., protons, ions, or other strongly interacting particles) beams, and to neutralize them with co-moving electrons in order to form a $\sim$~1~m radius multi-beam plasmoid at the exit of the accelerator.\footnote{Several accelerator designs, based on both radio-frequency and induction linac technology, in which multiple beams thread common components, have been studied during the 1980s as possible drivers for inertial confinement fusion.}

In practice, there are numerous possibilities, all based on some variant of the concept of fully charge and current neutralized high-energy beams.  The most promising designs are not necessarily those based on hadronic beams, or a mixture of hadrons and leptons, but possibly on purely leptonic beams because the lethality of high-energy (i.e., multi GeV) electrons and positrons interacting with heavy materials is comparable or higher to that of hadronic beams.  

To illustrate a few of these possibilities, and to relate them to single-species particle-beam configurations, it is interesting to consider a classification that has been suggested by researchers of the University of Maryland  for an initially pure electron beam injected into a possibly ionized and magnetized gas  \cite{DESTL1988-}:
\begin{enumerate}

\item The intense beam is injected into an initially neutral gas and ionization at the beamfront results in the creation of a neutralizing plasma channel that allows for effective beam propagation in the ion-focused regime.

\item A pre-formed channel is produced by various methods.  In this manner beamfront erosion can be significantly reduced and transverse instabilities can be suppressed.  Moreover, the channel may act as a guide, and therefore enable the beam to propagate undeflected across a magnetic field.

\item The electron beam is injected through a localized plasma into vacuum. Beam space-charge effects accelerate ions downstream where they provide a co-moving channel of neutralization that permits effective electron beam propagation into the vacuum region.

\end{enumerate}

   The third class obviously corresponds to a basic method for creating beam plasmoids because particles from the initially non-neutral beam carry along oppositely-charged neutralizing particles picked-up from the localized plasma.  (Another method would be to merge two separately accelerated streams of oppositely charged particles).

   When the initial beam is made of positively charged particles (e.g., protons or heavy ions) the neutralizing plasma may simply consist of a localized electron source such as a glowing tungsten filament.  It is well known that such a neutralization technique can be very efficient, and should have negligible effect on the ability to focus beams, \cite{DOLIQ1979-, HUMPH1981-, LEMON1981-}.  In the jargon of light- or heavy-ion inertial confinement fusion research this mode of plasmoid formation and propagation is named `neutralized ballistic transport' \cite{ROSE-2001A}, which is experimentaly studied in the Neutralized Transport Experiment (NTX), a joint venture of the Lawrence Berkeley and Lawrence Livermore National Laboratories \cite{ROY--2004-}.

   When the initial beam is made of electrons, as in the above third class, the method has been tested and shown to lead to Bennett-type pinch equilibria for plasmoids in which the co-moving ions and electrons do not necessarily have the same longitudinal velocity, a possibility that was first contemplated by Bennett \cite{BENNE1955-}.  For example, in an experiment where electrons had a velocity of about $0.84~c$, and the ions about $0.05~c$, the plasmoid was able to propagate across a transverse magnetic field of up to 200 gauss \cite{ZHANG1989-}.  Therefore, it is technically possible to envisage beam plasmoids propagating in a pinched mode \cite[p.188]{DESTL1988-}.  However, if the particles have different velocities the plasmoid will not remain localized for long propagation distances.  (Going to the plasmoid rest frame, it is obvious that all species have to move at the same velocity in order to stay together.)  Moreover, as different particle species are likely to experience different type of losses during propagation, the ideal beam plasmoids for long-range applications are most likely to be `co-moving pair plasmoids,' e.g., particle-antiparticle beams in which both species move at the same velocity.

    Conversely, it is possible to envisage non-neutral beams in which different numbers of electrons and ions move at the same velocity:  In that case the plasmoid is not charge neutral, and its deflection by an external magnetic field can be such that the ions are bent in the same direction as the electrons.  Such configurations may find applications in endo- or exo-atmospheric accelerator systems for steering or bending high-current high-energy beams using much less heavy and bulky electromagnets then conventional methods \cite{NGUYE1985-}.

   There is therefore a large spectrum of beam plasmoid configurations, with various species of particles moving in the same direction with possibly different velocities, with two or more overlapping or radially separated streams, and some special configurations such as co-moving particle-antiparticle streams.  For many reasons it is not necessary to examine all these possibilities in details.  In particular, it is intuitively clear that many important properties can be derived from the study of the most promising and simple configurations, such as beams of co-moving H$^\pm$ ions, proton-electron, proton-antiproton, and positron-electron pairs.

   For definiteness, we will focus on particle-antiparticle plasmoids, and refere to other possibilities when appropriate.  We will also use the word \emph{stream} for each of two single-species particle-beams composing them, and the word \emph{beam plasmoid} for their combined configuration, even though the streams may only partially overlap, or even be radially separated.  Finally, we will define the \emph{current} of a beam plasmoid as the current corresponding to each of these streams, and give it the plus sign if the positively charged one is moving forward. Therefore, we will write $J_+ = +e\beta c n_+$ and  $J_- = -e\beta c n_-$ for the stream's areal current densities, where $n_+=n_-=n_\pm$ is the beam-plasmoid's particle number density.

\subsection{Time-scales for beam plasmoid propagation}

In the following sections we are going to consider the feasibility of striking distant targets with beam plasmoids sent across the magnetosphere at altitudes above 100~km and over distances of up to 10'000~km.  This means that while the plasmoids do not have to be absolutely stable, their `life-time' should nevertheless be somewhat larger than the time required to reach the target, i.e., a `time of flight' of
\begin{equation}\label{pfa:5} 
        \tau_f = \beta c \Delta z \sim 33 ~\text{ms},
\end{equation}
in the ultrarelativistic limit.

   A second time-scale is set by the deflection of the plasmoid's particles in the geomagnetic field, which is characterized by their cyclotron (or Larmor) frequency $\omega_c$ , i.e.,
\begin{equation}\label{pfa:6} 
    \tau_c = \frac{1}{\omega_c} =  \frac{p}{e B_0 \beta c} \sim 2.2 ~\text{ms},
\end{equation}
if we assume a magnetic field of $B_0=0.5 \text{~gauss}=5 \times 10^{-5} \text{~tesla}$, and a particle's momentum of $p=10$~GeV/c.  In the case of a narrow plasmoid, as defined by Eq.\ \eqref{pfa:2}, this time is much larger than the time taken by two oppositely charged particles to move apart by a transverse distance equal to twice the streams's radius $a$. A more stringent time-scale is therefore set by the time required for the streams composing a plasmoid to fully separate under the effect of the geomagnetic field, which, provided nothing is opposing such a separation, is on the order of
\begin{equation}\label{pfa:7} 
   \tau_s \approx \sqrt{ \frac{2a}{\beta c} \frac{1}{\omega_c} }
   \sim 3.8~\mu\text{s},
\end{equation}
assuming $a=1$~m and a momentum $p=10$~GeV/c.

   A third time scale is set by the particle density of the streams composing the plasmoid,  which is characterized by their beam plasma frequency $\omega_b$, i.e.,
\begin{equation}\label{pfa:8} 
    \tau_b = \frac{1}{\omega_b} 
           = \frac{a}{2c} \sqrt{\frac{17\text{kA}}{I} \frac{\gamma m_b}{m_e}}
          \sim 0.3~\mu\text{s},
\end{equation}
where we assume an energy $W=\gamma m_b c^2 = 10$~GeV, a current $I = 10$ kA, and a stream radius $a = 1$~m.  This time defines the scale on which   instabilities develop, as well as self-field effects such as space-charge expansion, which according to \eqref{cha:7} corresponds to a radius doubling time of
\begin{equation}\label{pfa:9} 
    \tau_2 \approx 2 \gamma \frac{1}{\omega_b}
          \sim 6.6~\mu\text{s},
\end{equation}
for 10 GeV proton streams.

\section{Propagation across a magnetized vacuum}
\label{pmv:0}

While a plasmoid propagating in the high-atmosphere will be greatly affected by its interactions with the ambient ionospheric plasma, and possibly with the self-generated plasma due to its particles's collisions with the residual atmosphere, it is useful to start by considering the simpler case of a plasmoid propagating in vacuum.  Also, in order to identify the key physical processes which enable a plasmoid to move in straight line across a magnetic field, we begin with a very simple configuration, the well-known `capacitor model' \cite{SCHMI1960-, CHAPM1960-, DOLIQ1963-, SINEL1967-, PETER1982-, BOROV1987-, CAI-1992-}.

   This plane-symmetric capacitor model has been much studied and many of its limitations have been identified and discussed in details, at least in the non-relativistic limit, e.g., \cite{BOROV1987-}.   We will not go into all the details of this model, but rather focus on the main features of a special case, namely that of an electron-positron or proton-antiproton plasmoid (which for the moment we assume to be stable with regards to annihilation), and examine the implications of its motion being relativistic rather than non-relativistic.  In the next section, still using this model, we will examine the main impact of an ambient plasma on propagation across a magnetic field.  Then only, in the following section, we will move to a more realistic model: the axially symmetric `Gaussian-profile beam plasmoid model.'

   In the capacitor model the plasmoid is supposed to be very long in the direction of propagation, and of oblong rectangular cross-section with the external field $\vec{B}_0$ perpendicular to the smaller side.  In our case we suppose that this plasmoid consists in first approximation of two overlapping and incompressible oppositely-charged flat particle-streams which may move rigidly relative to one another.  We also focus on the transverse motion and neglect possible variations in the longitudinal direction.  In the absence of any external field such a configuration is stable: the only possible effect of a temporary perturbation is that the negative and positive streams may oscillate about their equilibrium position.  When a non-zero external field $B_0$ is applied, the negative and positive streams are displaced towards opposite directions so that two polarization layers appear, one on each side  orthogonal to the directions of both the magnetic field and the velocity.\footnote{As this phenomenon is rather common in electrodynamics and plasma physics, it is associated with a number of more or less equivalent names such as polarization layers, polarization sheets, charge layers, double layers, polarization sheath, etc.  A similar terminology is used for polarization current layers, sheets, etc.}  If $\Sigma$ is the surface charge density of the polarization layers, assumed to be very thin, the `capacitor model' implies that the polarization electric field $E_p=\Sigma/\epsilon_0$ will be nearly constant within the plasmoid, so that the transverse equation of motion of an electron (or of a positron by changing the sign of the electric charge) located between the polarization layers will be
\begin{equation}\label{pmv:1} 
    \gamma m \ddot x = e \beta c B_0 - \frac{e}{\epsilon_0} \Sigma.
\end{equation}
In writing this equation we have assumed that the motion is non-relativistic, i.e., that $\beta \ll 1$, so that the self-magnetic field $B_p=-\beta \Sigma/\epsilon_0$ due to the motion of the polarization layers could be neglected. Since the streams's particles's density is $n_{+} = n_{-} = n_\pm$, the surface charge density can be expressed in terms of the polarization layer thickness $\lambda$, i.e., $\Sigma = e \lambda n_\pm$.  The equation of motion becomes then
\begin{equation}\label{pmv:2} 
     \gamma m \ddot x = e \beta c B_0 - \frac{1}{\epsilon_0} e^2 \lambda n_\pm,
\end{equation}
or
\begin{equation}\label{pmv:3} 
         \ddot x = \beta \omega_c - \lambda \omega_b^2,
\end{equation}
where we have introduced the beam cyclotron and plasma frequencies according to \eqref{pfa:6} and \eqref{pfa:8} with $n_b=n_\pm$.

    The meaning of Eqs.~\eqref{pmv:2} or \eqref{pmv:3} is that under the effect of the external field the polarization layer thickness adjusts itself until the right-hand side is zero, so that the particles inside the plasmoid (i.e., `between the capacitor plates') move undeflected by the magnetic field.  This can be interpreted as an `$\vec E \times \vec B$ drift' effect in which a self-polarization electric field $\vec E_p$ is self-consistently compensating for the effect of an external magnetic field $\vec B_0$, i.e.,
\begin{equation}\label{pmv:4} 
        \beta c B_0 = E_p,
\end{equation}
which from Eqs.~\eqref{pmv:2} and \eqref{pmv:3} implies that $\lambda$ satisfies the equation
\begin{equation}\label{pmv:5} 
        \lambda_{\text{n.r.}} = \beta c \frac{\epsilon_0 B_0}{e n_\pm}
                              = \beta \frac{\omega_c}{\omega_b^2}, 
\end{equation}
where the label `n.r.' recalls that it corresponds to a non-relativistic plasmoid.

   An important quantity which can easily be calculated from $\lambda$ is the polarization current, i.e., the current $I_\lambda = e \beta c \lambda w n_\pm$ which flows in the polarization layers assuming that the streams have a width $w$.  In practice, e.g., to facilitate comparison with axially symmetric models, we will agree that each stream has an effective cross-sectional area $\pi a^2$, so that $a$ is the usual scale radius, and that $w=\pi a$.  Then, 
\begin{equation}\label{pmv:6} 
         I_{\lambda,\text{n.r.}} = \pi \frac{B_0}{\mu_0} \beta^2 a,
\end{equation}
where we have used the identity $\epsilon_0\mu_0c^2=1$, and where the constant $\pi {B_0}/{\mu_0}$ equals 125 A/m in a typical geomagnetic field.  With this convention, the condition $\lambda \ll a$ which means that the boundary layers must be thin in comparison to the plasmoid radius $a$ for the capacitor model to be valid, is equivalent to the condition
\begin{equation}\label{pmv:7} 
         I_\lambda \ll I_\pm,
\end{equation}
which implies that the current in the polarization layers must be small compared to the current in the streams.

   Equation \eqref{pmv:6} has been derived under the assumption that the particle's density is constant within the polarization layers.  In reality, the streams do not fully neutralize each other in these layers, and the external magnetic field is only partially compensated by the polarization electric field. Consequently, space-charge repulsion and deflection by the external field imply that the polarization layers of a plasmoid sent into a magnetized vacuum expand as it propagates.

  In principle, for an very long plasmoid, this expansion stops when the particles deflected by the magnetic field have made half a gyration and start moving backwards relative to the streams until they come back close to them, and then continue a cycloidal motion at an average velocity smaller than that of the streams.  This suggests that there may exist self-similar solutions such that, after a period of expansion, the boundary layer thickness remains constant.  Indeed, writing down the corresponding equations of motion, it is not difficult to see that there is always a self-similar solution to the `capacitor model' such that a non-relativistic beam plasmoid can in principle propagate over very long distances, provided it has enough energy and current to overcome the losses in beam energy and current at the head and tail, e.g., \cite{BOROV1987-}.  However, these steady state solutions are such that the polarization layers have a particle density much less than $n_\pm$ (so that they are very wide, i.e., $\lambda$ of the order of the particle's gyroradius $R=\beta c /\omega_c$), and have a transverse longitudinal-velocity profile such that $\beta c$ goes to zero at the outer edge. Finally, they exist only for simple (e.g., plane symmetric) configurations.\footnote{In particular, if the plasmoid boundary is less simple than in the capacitor model, a steady state may not be possible.  An example is any cylindrical boundary with its generators parallel to the beam velocity: the particles in the surface transition layers are only in quasi-equilibrium \cite{CHAPM1960-}. The charge in the layers is driven away from the streams by electrostatic repulsion, resulting in a continual repolarization of the plasmoid \cite{GALVE1991-}.} Thus, these non-relativistic solutions are more appropriate to the motion of geomagnetospheric plasmas \cite{CHAPM1960-}, than to the long-distance propagation of narrow plasmoid beams. 

   Let us return to the question of the response of a beam plasmoid to an external field $B_0$ assuming that its boundary layers may have expanded so that the condition $\lambda \ll a$ is not true anymore.  Their particle density will then be less than that of the streams, i.e., $< n_\pm$, implying that $\lambda$ will be larger than its  capacitor model value given by \eqref{pmv:5}.  However, despite of this, the polarization electric field $E_p$ and the polarization current $I_\lambda$ will still in good approximation be equal to their values calculated according to Eqs.~\eqref{pmv:4} and \eqref{pmv:6}, because these quantities depend on the charge integrated over the polarization layers's thickness. Therefore, for propagation to be possible, the main necessary condition remains that given by Eq.~\eqref{pmv:7}, which states that there should be enough current (or charge) in the beam for a pair of boundary layers to form and be able to cancel the deflection due to $B_0$.

  For \emph{non-relativistic} plasmoids, where $\beta \ll 1$, the condition \eqref{pmv:7} may easily be satisfied, even for magnetic fields much larger than the geomagnetic field, and consequently a number of experiments have demonstrated successful transport of neutralized low-energy ion beams across magnetic fields in vacuum, e.g., \cite{ROBER1981-,ISHIZ1982-, ZHANG1989-, BRENN2005-}.  For \emph{relativistic} plasmoids, however, equation \eqref{pmv:1} has to be modified to include the magnetic force due to the self-magnetic field $B_p=-\beta \Sigma/\epsilon_0$ induced by the currents corresponding to the longitudinal motion of the polarization layers.  This leads to replace Eq.~\eqref{pmv:3} by
\begin{equation}\label{pmv:8} 
         \ddot x = \beta \omega_c - (1-\beta^2) \omega_b^2 \lambda,
\end{equation}
which shows that the self-magnetic field tends to quench the polarization electric field, i.e., to reduce its effectiveness in compensating the external magnetic field.  Indeed, as $1-\beta^2 = 1/\gamma^2$, the corresponding relativistic expression for the boundary layer thickness is
\begin{equation}\label{pmv:9} 
    \lambda = \gamma^2\beta c \frac{\omega_c}{\omega_b^2},
\end{equation}
which can be much larger than $\lambda_{\text{n.r.}}$ since $\gamma > 1$ for a relativistic beam.

   Similarly, the relativistic version of the polarization current \eqref{pmv:6} is
\begin{equation}\label{pmv:10} 
         I_{\lambda} = \pi \frac{B_0}{\mu_0} \beta^2 \gamma^2 a,
\end{equation}
which shows that in the non-relativistic limit ($\beta^2 \ll 1$ and $\gamma^2 \approx 1$) the polarization current may be small, while in the ultra-relativistic limit ($\beta^2 \approx 1$ and $\gamma^2 \gg 1$) it may be very large.  For example, let us consider a plasmoid composed of particle/antiparticle pairs having 1 GeV of kinetic energy, i.e., $\beta=0.87$ and $\gamma=2.07$ for protons, and $\beta=0.99$ and $\gamma=1950$ for electrons. Requiring that the plasmoid radius should be $a =1$~m when reaching the target, the polarization current would have to be about 400~A for the proton/antiproton plasmoid, and about $5 \times 10^8$~A for the electron/positron plasmoid.  Since, beam plasmoids should be significantly relativistic, preferably with energies on the order of at least 10~GeV according to Table \ref{tab:plb}, we see that propagating an electron/positron plasmoid through magnetized vacuum requires truly enormous currents,\footnote{I.e., characteristic of relativistic inter-stellar particle streams.} while for a proton/antiproton plasmoid $(m_p/m_e)^2\approx 3 \times 10^6$ times smaller currents may suffice.

   However, satisfying condition \eqref{pmv:10} insures propagation only as long as there is enough particles in the streams to replace those in the polarization layers which are lost during propagation as a result of their deflection by the external magnetic field, or because of space-charge repulsion.  Supposing that after some initial phase the boundary layer thickness remains constant, this requires calculating the `life time' of the polarization layer, i.e., estimating the time that a particle spends in the boundary layer between the moment it emerges from the fully neutralized stream (where the electric field is equal to $E_p$) and the moment it reaches the edge of the charge layer (where the polarization electric field is nearly equal to zero).

   The rate of boundary layer's particle loss due to deflection by the external magnetic field can be estimated by assuming that the electric field decreases linearly within the charge layers.\footnote{This corresponds to a constant charge density within the polarization layer, which is actually the case in the non-relativistic capacitor model.}  An elementary calculation, requiring that the particle's trajectories and velocities are continuous, leads to a non-linear equation which can be solved numerically, i.e., 
\begin{equation}\label{pmv:11} 
   \tau_{\text{strip}} \approx
                  1.16 \sqrt{ \frac{2\lambda}{\beta c} \frac{1}{\omega_c} },
\end{equation}
which (as could be expected) has the same form as Eq.~\eqref{pfa:7}.  We call $\tau_{\text{strip}}$ the magnetic stripping time because it corresponds to the time taken by the force exerted by the magnetic field to drive a particle out of the boundary layer. (In the axially symmetric Gaussian-profile model we will encounter an electric stripping effect such that polarization layer particles are lost along the magnetic field lines.)

   The rate of boundary layer's particle loss due to space-charge expansion can be estimated by assuming that the particles are lost on a time scale equal to that required for the charge layer to double its thickness under the effect of space-charge alone.  This requires deriving the one-dimensional counter-part of Eq.~\eqref{cha:7}, which leads to the closed form expression
\begin{equation}\label{pmv:12} %
         \tau_2 = \frac{\gamma}{c} 
                  \sqrt{\frac{w\lambda}{\pi} \frac{I_A}{I_\lambda}}.
\end{equation}
Writing $w=\pi a$, and replacing $\lambda$ by its value given by \eqref{pmv:10}, one finds a remarkable result, namely that $\tau_2=\tau_{\text{strip}}$, provided $\lambda = a$ and the numerical factor 1.16 is ignored in Eq.~\eqref{pmv:11}.

  Therefore, $\tau_2$ and $\tau_{\text{strip}}$ are on the same order, and both nearly equal to the separation time scale $\tau_s$ defined by Eq.~\eqref{pfa:7}, which can therefore be taken as a measure of the the boundary layer life-time, provided one takes for its thickness during propagation the value $\lambda=a$, which is also a very plausible estimate.  For our example of a 1 GeV proton/antiproton plasmoid, which corresponds to a particle's momentum of 1.7 GeV/c, the life-time of the boundary layer is therefore of about 1.5 $\mu$s, which is equivalent to a propagation distance of only 400 m.  Therefore, if the plasmoid is to propagate over a distance suitable for an outer-space system, e.g., at least 400 km, the initial current has to be at least 1000 times larger than the 400 A required in the polarization layers, i.e., of at least 400 kA. 

   Consequently, mainly because of boundary layer losses due to magnetic stripping, the current of a relativistic beam plasmoid propagating across a typical geomagnetic field of 0.5 gauss has to be very large, possibly so large that the required high-current accelerator could be far too heavy and bulky to be used in outer-space.  However, this conclusion was reached using a one-dimensional model --- the capacitor model in its relativistic form --- while a number of published two-dimensional computer simulations, e.g.,  \cite{GALVE1988-, GALVE1989A, GALVE1991-}, show that the propagation of a plasmoid in vacuum is more complicated than suggested by the analytical studies mentioned at the beginning of this section. In particular, it would be important to confirm whether or not the relativistic effects which are particularly detrimental to electron/positron-plasmoids propagation also affect their motion across a magnetized plasma.  Moreover, the effects of several possible instabilities and further erosion mechanisms should also to be taken into account \cite{PETER1983B, GALVE1986-, GALVE1988-, CAI-1992-}.  It is therefore essential to see how a plasmoid would propagate in a plasma background before going into further details.

\section{Propagation across a magnetized plasma}
\label{pmp:0}

The difficulty of propagating a plasmoid across a magnetized plasma was anticipated long ago when it was realised that the presence of a ionized background gas tends to neutralize and reduce (i.e., to `quench' or `short circuit') the polarization surface charges of the moving plasmoid \cite[p.921]{CHAPM1960-}.  These quenching effects have been extensively studied in a number of computer simulations \cite{WINGL1987A, WINGL1987B, WINGL1987C},\footnote{The third of these papers explicitly describes the beam propagation physics to be studied with the U.S.\ Air Force \emph{Beams on Rockets} (BEAR) experiment, prior to its launching into space, on 13 July 1989. See Sec.~\ref{gta:0}.} and \cite{LIVES1989-, GALVE1989B, GALVE1990-}, as well as experimentally \cite{HONG-1988-, WESSE1988-}.  The complexity of these effects, especially if the plasmoid is made of non-equal mass particles, and if the induction electric-field effects become important, are also described in simple terms with reference to the `capacitor model' in reference \cite[Sec.III]{BOROV1987-} (see also \cite[p.1896]{GALVE1991-}).

   In references \cite{LIVES1989-, GALVE1989B, GALVE1990-} two-dimensional simulations are made for the convection of a finite slab-shaped plasmoid across a magnetized vacuum or a magnetized plasma: this enable to clearly understand what happens when the background gas is of non-negligible density.  For instance, as recalled in reference \cite{GALVE1989B}, a plasmoid propagating in vacuum loses momentum with distance due to three erosion effects: (i) the erosion of the charge layers due to the velocity shear (magnetic stripping), (ii) the erosion of the charge layers due to their expansion along the magnetic field lines (electric stripping), and (iii) the erosion of the head of the stream due to Larmor-radius effects.  When the plasmoid streams across a magnetized plasma, these erosion effects are diminished due to the short circuiting of the electric field by the background plasma.  However, the convection velocity decreases with distance until the plasmoid is stopped.


   Similar effects are observed in experiments where the propagation of a plasmoid beam through a magnetized vacuum is compared to its propagation through a magnetized plasma  \cite{HONG-1988-, WESSE1988-}.  While these experiments were not the first to demonstrate that a plasma background tends reduce the polarization electric field and inhibit $\vec E \times \vec B$ drift, see, e.g., \cite[Fig.20]{BOSTI1956-}, they were the first to study that effect under conditions closely related to the ``propagation of a neutralized ion beam (plasmoid) in and above the ionosphere in a background magnetic field and a low density partially ionized plasma'' \cite[p.3778]{WESSE1988-}.

  However, just like in the previously mentioned computer simulations, these experiments made with a 5 kA, 0.15 MeV beam do not correspond to the conditions of an actual directed plasmoid beam weapon, because such a device would require beam energies of at least 100~MeV to several GeV or more to be effective, and that under these circumstances self-magnetic field effects become very important, especially for electron beams.  Therefore, instead of these relatively low-energy simulations and experiments, one has to investigate the effect of a background plasma on relativistic plasmoids of the type considered in the previous section. 

   In this perspective, we will restrict our discussion to very narrow plasmoids, and use the paraxial approximation in a way similar to that used in Sec.~\ref{ben:0} to introduce the effective current $I_E$ defined by Eq.~\eqref{ben:3}.   This is because in that approximation the main effect of a plasma background is to modify the relativistic equation of motion \eqref{pmv:8} according to the substitution
\begin{equation}\label{pmp:1} 
    \frac{1}{\gamma^2} = \Bigl(1-\beta^2\Bigr) ~~~ \rightarrow ~~~
                         \Bigl((1-f_e) - \beta^2(1-f_m)\Bigr),
\end{equation}
which is equivalent to defining an \emph{effective gamma} $\gamma_E$
\begin{equation}\label{pmp:2} 
    \frac{1}{\gamma^2_E}  = \frac{1}{\gamma^2} + \beta^2 f_m -f_e,
\end{equation}
so that the transverse equation of motion \eqref{pmv:8} becomes
\begin{equation}\label{pmp:3} 
         \ddot x = \beta \omega_c - \frac{1}{\gamma^2_E} \omega_b^2 \lambda.
\end{equation}
   Consequently, the same substitution has to be made in Eqs.~\eqref{pmv:9} and \eqref{pmv:10}, i.e., 
\begin{equation}\label{pmp:4} 
     \lambda = \pi \frac{B_0}{\mu_0} \beta^2\gamma^2_E \frac{a^2}{I},
\end{equation}
\begin{equation}\label{pmp:5} 
         I_{\lambda} = \pi \frac{B_0}{\mu_0} \beta^2 \gamma^2_E a.
\end{equation}

   Qualitatively, the main difference between the magnetized-vacuum equation \eqref{pmv:8} and the magnetized-plasma equation \eqref{pmp:3} is that  $\gamma^2_E$ can take any positive or negative value, while $\gamma^2$ was restricted to $\gamma^2 \geq 1$.  In particular, when $f_m=0$ and $f_e > 1/\gamma^2$, the effective gamma is imaginary, $\gamma^2_E < 0$, and we get  quenching phenomena similar to those mentioned at the beginning of this section for a non-relativistic plasmoid injected into a plasma rather then into vacuum: the external magnetic field is no more compensated by the polarization electric field so that the whole plasmoid tends to charge-separate; erosion decreases because the boundary layers pinch rather than expand, but the $\vec E \times \vec B$ drift stops as soon as $E_p=0$, etc.

    Moreover, when $\gamma^2_E < 0$, and the initially overlapping positive and negative streams are separated transversely by a small distance $\lambda$, Eq.~\eqref{pmp:3} shows that the particles within the plasmoid are submitted to an electromagnetic force which adds to the Larmor gyration force, and therefore tends to increase the separation. In fact, this additional force tends to increase the effect of any transverse perturbation even if $B_0 = 0$~: Contrary to the vacuum case, a beam plasmoid is absolutely unstable when propagating through a charge-neutralizing plasma.  This is of course a consequence of the fact that opposite currents repel, and a simple example of the filamentation process by which the oppositely charged streams forming a beam plasmoid tend to separate into independent filaments as soon as magnetic repulsion becomes more important than electric attraction.

  Therefore, in order to maximize the benefits of the polarization electric field, and to avoid gross instability, the charge neutralization fraction $f_e$ should be as small as possible, and the current neutralization fraction $f_m$ such that $\gamma^2_E$ is positive, i.e.,
\begin{equation}\label{pmp:6} 
        \beta^2 f_m  > f_e  - \frac{1}{\gamma^2}.
\end{equation}
Thus, if $f_m$ is not too small, it is possible to have an effective gamma $\gamma_E$ that is much smaller than the kinematic gamma, i.e.,
\begin{equation}\label{pmp:7} 
        \gamma_E \approx \frac{1}{\beta \sqrt{f_m}}  \ll \gamma,
\end{equation}
so that (even for an electron beam) the polarization current $I_{\lambda}$ given by Eq.~\eqref{pmp:5} may be much smaller for a relativistic plasmoid propagating in a plasma than for the same beam plasmoid propagating in vacuum.

   The question is therefore whether it is possible or not to satisfy Eq.~\eqref{pmp:6} for a highly relativistic plasmoid propagating in the upper atmosphere.  In other words, when $\beta\approx 1$ and $\gamma^{-2} \ll 1$, whether it is possible to satisfy the condition $f_m > f_e$, which means that the plasma electrons should stay within the plasmoid streams, rather than been expelled or attracted into them.  This implies that the longitudinal electric field $E_z$ should be non-zero while the transverse (or radial) electric field $E_r$ should negligible, i.e., that $|E_r| < |E_z|$.  If we take an axially symmetric plasmoid and for $E_r$ its maximum value $E_r(a)$,  this leads to the condition
\begin{equation}\label{pmp:8}
   \Bigm| \frac{\mu_0}{4\pi} \frac{2I}{\beta a/c} \Bigm| ~ < ~
   \Bigm| \frac{\mu_0}{\pi} \mathcal{L} \frac{\partial}{\partial \tau} I \Bigm|,
\end{equation}
which implies that the rise-time of the plasmoid current should be less than $2 \mathcal{L} a/c \approx 6.6$~ns for $a=1$~m.  Therefore, the condition $f_m > f_e$, which is opposite to the usual conditions required in order to have a fully charge-neutralized beam, e.g., \cite[p.1998]{MCART1973-}, may be satisfied provided the plasmoid current variation is sufficiently fast.\footnote{In reference \cite{CHRIE1986-}, where the goal is to propagate a 200 MeV proton beam as a Bennett pinch in the high-atmosphere, the solution is to carefully tapper the beam current density in order to avoid a fast rise-time which would result in little charge neutralization (see, in particular, the discussion on page 1676).}

   The magnetic neutralization fraction of a beam propagating in the high-atmosphere can be estimated by means of Eq.~\eqref{bcn:19} which applies 
here because collisions can be neglected. Thus, according to Eq.~\eqref{bcn:20}, and provided $n_b < n_e$,
\begin{equation}\label{pmp:9} 
               f_m =  \frac{\mathcal{L}\omega_p^2}
                           {\mathcal{L}\omega_p^2 + c^2/a^2}
                   \approx \mathcal{L}\frac{a^2}{c^2} \omega_p^2.
\end{equation}
where the approximate value corresponds to the limit $f_m \ll 1$.  Taking for the ionospheric plasma density $n_e$ the numbers given in Table~\ref{tab:atm}, we obtain for $a=1$~m a maximum current neutralization of $f_m \approx 15$\% at an altitude of 300 km, and smaller values, with a minimum of $f_m \approx 0.3$\%, between 100 and 1000 km.  As $f_e \approx 0$ and $f_m \neq 0$, the boundary layers will of course expand rather than pinch.  But that was also the case in vacuum, and since $I_{\lambda,\text{plasma}} < I_{\lambda,\text{vacuum}}$, boundary layer expansion will actually be slower in a plasma than in vacuum.

   The main difficulty with Eq.~\eqref{pmp:9} is that it is strictly valid only when the charge density $n_b$ in the boundary layers is less than the ambient plasma density $n_e$, i.e., when $n_b < n_e$ so that effect of the longitudinal electric field in the boundary layers is simply to put the ionospheric plasma electrons into motion.  This means that there are sever restrictions to the applicability of Eq.~\eqref{pmp:7}, namely that the boundary layers's charge density $n_b$ should at most be equal to $f_m n_e$, or if the plasmoid is able to ionize the background gas molecules to $f_m n_a$, where $n_a$ is the atomic density of the  residual atmosphere.  These limitations are similar to those given in Table \ref{tab:alj} for charge-neutralized beams, i.e., that if we take $f_m =10\%$ for example, the absolute limit to the current density in a boundary layer is about 2 MA/m$^2$ at an altitude of 300~km, and rapidly decreasing when going to higher altitudes; and that if the plasmoid's current neutralization is to rely on the ionospheric plasma alone, the current density in a boundary layer would be limited to a maximum of 2 kA/m$^2$, which occurs at an altitude of 300~km.
 
    In conclusion, the boundary layers of a relativistic beam plasmoid propagating in the ionosphere between 100 and 600 km can be current neutralized at a level of about 0.1 to 10\% under suitable conditions, so that the polarization current \eqref{pmp:5} may be calculated with $\gamma_E$ given by Eq.~\eqref{pmp:7}.  This leads to a drastic reduction in beam current requirement, especially if the plasmoid's trajectory is such that the lowest beam current requirement corresponds to the end of its range.  Since propagation as a Bennett pinch corresponds to the case $f_e \neq 0$ and $f_m \approx 0$, the present case where $f_e \approx 0$ and $f_m \neq 0$ may be qualifed as an `anti-pinch' propagation mode.

\section{Gaussian-profile beam plasmoid model}
\label{gau:0}

{\bf NB:}  The normalization is wrong in this chapter $\frac{1}{\sqrt{2\pi}}$ should not be there in Eq.~\eqref{gau:1}. 

   To finalize our assessment of the theoretical prospect of relativistic beam plasmoids it is necessary to use a simple two-dimensional model, because such a model enables to discuss a number of effects which are absent or cannot be properly estimated in a plane symmetric model, and to better normalize major parameters such as the current required for successful long-distance propagation.  For this purpose, the most simple model would be to consider a plasmoid consisting of two constant density streams of equal radii.  However, it is not much more difficult to consider a more realistic charge/current distribution, such as a Bennett profile \cite{ZHANG1989-}, or a Gaussian profile \cite{NGUYE1985-}, which we will assume here, i.e.,
\begin{equation}\label{gau:1} %
     n(r) = \frac{N}{\pi \tilde{a}^2}
            \frac{1}{\sqrt{2\pi}} \exp\bigl( -\frac{r^2}{2\tilde{a}^2}\bigr),
\end{equation}
where $N=I/(e\beta c)$ is the linear charge density and $\tilde{a}$ the RMS stream radius.  Using this distribution we will therefore repeat a number of steps already made with the capacitor model, making all calculations as for propagation in vacuum, and replacing where appropriate $\gamma$ by $\gamma_E$ to enable the transition to  propagation in a plasma.

  The starting point is to calculate the average (electric and magnetic) force exerted by one Gaussian stream on the other one, assuming that they maintain fixed density and current profiles when they are laterally displaced by a distance $x$.  This leads to calculating a double integral similar to Eq.~\eqref{coh:4}, so that the resulting equation of motion is simply \cite{ZHANG1989-}
\begin{equation}\label{gau:2} %
    m\gamma \ddot{x} = \beta c e B_0   -
            \frac{1}{\gamma^2_E} \frac{e^2 N}{4\pi \epsilon_0}
            \frac{2}{x}\Bigl(1-\exp\bigl( -\frac{x^2}{4\tilde{a}^2}\bigr)\Bigr).
\end{equation}
The corresponding total stream-stream force per unit length is easily found to be
\begin{equation}\label{gau:3} %
            \frac{\Delta F}{\Delta z} =  I B_0  -
            \frac{\mu_0}{4\pi} \frac{1}{\beta^2\gamma^2_E} I^2
            \frac{2}{x}\Bigl(1-\exp\bigl( -\frac{x^2}{4\tilde{a}^2}\bigr)\Bigr),
\end{equation}
where, in the limit of large separation ($x \gg \tilde{a}$), which corresponds to ignoring the exponential term, we recognize on the right the standard expression for the force between two thin electrical streams separated by a distance $x$.

  The next step would be to investigate under which conditions the equation of motion, \eqref{gau:2} or \eqref{gau:3}, leads to a bound state by setting the left hand side to zero.  However, it is immediately seen this would lead to two solutions, one with $x$ small and the two streams closely overlapping, and one with $x > \tilde{a}$ and the two stream nearly separated.  This calls for a proper analysis, which implies integrating the equation of motion.  We therefore rewrite Eq.~\eqref{gau:1} in non-dimensional form as
\begin{equation}\label{gau:4} %
            \ddot{\xi} = A   -
            B\frac{1}{\xi}\bigl(1-\exp( -\xi^2)\bigr),
\end{equation}
where $\xi=x/(2\tilde{a})$, and the parameters $A$ and $D$ are independent of $\xi$.  After multiplication by $\dot\xi$ this equation can be integrated with the initial condition $\dot\xi(0)=\xi(0)=0$, and we obtain \cite{ZHANG1989-}
\begin{equation}\label{gau:5} %
       \frac{1}{2}(\dot\xi)^2 = A \xi - \frac{D}{2} \operatorname{Ein}(\xi^2),
\end{equation}
where the function $\operatorname{Ein}(\xi)$ is a particular form of the exponential integral \cite[p.228]{ABRAM1968-}.

   Equation \eqref{gau:5} is the kinetic energy of the interacting streams, so that the zeros of the equation $(\dot\xi)^2=0$ correspond to stationary separations of the streams.  Solving numerically this equation, one finds that these solutions (which come in pairs) exist only if $2A/D < 1$.  Replacing $A$ and $D$ by their values this translates into the condition
\begin{equation}\label{gau:6} %
      I > I_{\text{min}}
          \DEF 8 \pi \frac{B_0}{\mu_0} \beta^2\gamma^2_E \tilde{a}
            =  1~ \text{[kA]} ~~\beta^2\gamma^2_E \tilde{a}.
\end{equation}
The limiting case, $2A/D = 1$, which corresponds in very good approximation to $\xi = 2$, defines the domain of stability. The solutions such that $\xi > 2$ are therefore unstable, so that the stable solutions are confined to the interval
\begin{equation}\label{gau:7} %
         x < 4 \tilde{a}.
\end{equation}
In the limit $x \ll 4 \tilde{a}$ these solutions are easily evaluated from Eq.~\eqref{gau:2} or \eqref{gau:3} by setting the force equal to zero and keeping only the first order term in the exponential, i.e., 
\begin{equation}\label{gau:8} %
     x \approx \tilde{\lambda}
       \DEF  8 \pi \frac{B_0}{\mu_0} \beta^2\gamma^2_E \frac{\tilde{a}^2}{I}.
\end{equation}

  Equation \eqref{gau:7} shows that bound solutions exist for stream separations as large as four root-mean square radii, i.e., Gaussian streams that are almost 90\% separated.  Equation \eqref{gau:6} gives the minimum current $I_{\text{min}}$ necessary for the streams to move undeflected across a transverse magnetic field: This current is therefore the Gaussian-stream model counterpart of the polarization current given by equations \eqref{pmv:10} or \eqref{pmp:5} in the capacitor model.  This enables to notice that that model was in fact underestimating by a factor of about five (if we take $a\approx 1.5\tilde{a}$) the minimum polarization current required by an axially symmetric plasmoid to propagate across a given magnetic field.   The same discrepancy is noticed by comparing Eq.~\eqref{gau:8} with its capacitor model counterpart, Eqs.~\eqref{pmv:9} or \eqref{pmp:4}.

    An important quantity that we did not calculated with the capacitor model is the energy required to separate the two streams in order to polarize in response to an external magnetic field.  This energy corresponds to a loss which affects both streams every time that the plasmoid has to repolarize.   It is evaluated by calculating the work done by the streams to separate, which for a small separation $\delta x \ll \tilde{a}$ is obtained by integrating the stream-stream force term on the right of Eq.~\eqref{gau:3}, i.e.,
\begin{equation}\label{gau:9} %
                   \frac{\Delta W_{\text{pol}}}{\Delta z} =
 \int_0^{\delta x} \frac{\Delta F}{\Delta z} dx = 
                   \frac{\mu_0}{4\pi} \frac{1}{\beta^2\gamma^2_E} I^2
                   \Bigl(\frac{\delta x}{2\tilde{a}}\Bigr)^2.
\end{equation}
Taking for $\delta x$ the separation $\tilde{\lambda}$ given by Eq.~\eqref{gau:8} we get
\begin{equation}\label{gau:10} %
                  \frac{\Delta W_{\text{pol}}}{\Delta z} =
                  \frac{B_0^2}{2\mu_0} \beta^2\gamma^2_E
                  8 \pi \tilde{a}^2,
\end{equation}
which is equal to the linear electromagnetic energy density stored in the polarization electric field.

  In the capacitor model the planar symmetry implied that boundary layer expansion and losses could only proceed in the direction perpendicular to both the beam velocity and the external magnetic field, i.e., the $x$-direction along the magnetic force $\vec{v} \times \vec{B}_0$.  In an axially symmetric model it becomes possible to study the expansion and losses of the streams's polarization charge layers in the other transverse dimension, i.e., the $y$-direction along the magnetic-field lines (parallel to either directions of $\vec{B}_0$).  This will be done qualitatively by referring to the most simple model, the `cylindrical dipole model,' in which the plasmoid consists of two solid streams or radius $a$ displaced by an infinitesimally small transverse distance  \cite{GALVE1991-}.  In that model, the external magnetic field induces a polarization electric characterized by a maximum value such that $E_0 = \beta c B_0$.  This field is uniform within the plasmoid
\begin{equation}\label{gau:11} %
   E_{x,\text{inside}} = E_0, \hspace{4cm} E_{y,\text{inside}} = 0, \hspace{2cm}
\end{equation}
and has a typical dipolar angular dependence outside
\begin{equation}\label{gau:12} %
    E_{x,\text{outside}} = E_0 \cos(2\theta)\frac{a^2}{r^2}, \hspace{2cm}
    E_{y,\text{outside}} = E_0 \sin(2\theta)\frac{a^2}{r^2}.
\end{equation}
Consequently, within a polarization layer, there is an electric field component $E_y$ along the magnetic lines which has a maximum value $E_{y,\text{max}} = \pm E_0/\sqrt{2}$ at $\theta=\pm\pi/4$.  This field, which is fully uncompensated by $B_0$, therefore leads to a continuous stripping of the boundary layer along the magnetic field lines.  In fact, since the average driving force of this electric-stripping mechanism is about half as large as the magnetic-stripping force, the net effect of both mechanisms are comparable because a particle entering the boundary layer at any point on the surface of the plasmoid will be approximately at the same distance away from it after a time on the order of that required to move over a distance of about $a$.

   From the perspective of the effectiveness of the polarization layers in shielding the inside of a plasmoid from deflection by the geomagnetic field, what matters most is the average effective charge that remains in these layers during propagation.   It is therefore clear that the particles which have moved transversally beyond the edge of the plasmoid because of the $E_y$ component of the polarization field should be considered as lost.  Consequently, during propagation, the effective charge contributing to the generation of the polarization field will be concentrated in a region of approximate thickness $a$ and width $2a$ on both sides of the plasmoid, in qualitative agreement with the behavior of the ions in computer simulations of these effects, e.g., \cite[Fig.4]{GALVE1991-}.\footnote{The fact that these simulations are non-relativistic does not matter here since we consider the effect of forces that are transverse to the streams.}

   This qualitative analysis enables to confirm the reasoning which in the capacitor model led to the conclusion that in a relativistic plasmoid the magnetic striping time \eqref{pmv:11}, and self-charge expansion doubling time \eqref{pmv:12}, are nearly equal to the magnetic separation time scale $\tau_s$ defined by Eq.~\eqref{pfa:7} where the length scale is precisely $a$. This allows us to continue using Eq.~\eqref{pfa:7} for the life-time of the boundary layer in the Gaussian-stream model, where to be on the conservative side we take $\tilde{a}$ instead of $a$ for the length scale, i.e., . 
\begin{equation}\label{gau:13} 
   \tau_s \approx \sqrt{ \frac{2\tilde{a}}{\beta c} \frac{1}{\omega_c} }
           =      \sqrt{ \frac{2\tilde{a}}{\beta c} \frac{\gamma m}{eB_0} }.
\end{equation}
This life-time is related to the propagation distance over which the charge layer has to be replaced in order to continue moving across the magnetic field, i.e.,
\begin{equation}\label{gau:14} 
   z_s = \beta c \tau_s   =  \sqrt{ 2\tilde{a} \frac{p c}{eB_0} },
\end{equation}
which is remarkable since it depends only on the plasmoid radius and on the momentum of the particles.

  For example,  with $\tilde{a}=1$~m and $p=100$~GeV/c, one finds e.g., $z_s=1.14$~km, a relatively small distance.  This implies that if such a plasmoid were to propagate over 1000 km, it would have to repolarize about 900 times, and that the corresponding particle current and energy would be lost.  We therefore define a \emph{repolarization range}, that is the maximum range a plasmoid can travel until its repolarization current supply is exhausted, i.e., using Eqs.~\eqref{gau:6} and \eqref{gau:14}
\begin{equation}\label{gau:15} 
   z_{\text{max}}(I) = z_s I/I_{\text{min}}  =
              \frac{I}{8 \pi} \frac{\mu_0}{B_0}
              \Bigl( \frac{2m c}{eB_0} \Bigr)^{1/2}
              \tilde{a}^{-1/2}
              \beta^{-3/2}\gamma^{1/2}\gamma_E^{-2}.
\end{equation}

   In order to get a meaningful result, the repolarization range has to match the collision driven expansion of the plasmoid, which can be estimated according to Eq.~\eqref{ten:5}, that is the maximum range at which the plasmoid streams have expanded to a radius $\tilde{a}$, i.e.,
\begin{equation}\label{gau:16} %
           z_{\text{max}}({\tilde{a}})
                   = \Bigl(\frac{mc^2}{E_s}\Bigr)^{2/3} \bigl(3X_0\bigr)^{1/3}
                     \tilde{a}^{2/3}
                     \beta^{4/3}\gamma^{2/3},
\end{equation}
which at an altitude of 300~km, where $X_0=3 \times 10^{12}$~m, gives
\begin{equation}\label{gau:17} %
    z_{\text{max}}({\tilde{a}}) =  273 ~ \text{[km]} ~~ p^{2/3}
                     \tilde{a}^{2/3}
                     \beta^{2/3}.
\end{equation}
when the momentum $p$ is expressed in units of GeV/c.

Setting $\gamma_E^{-2}=\beta^2f_m$ according to Eq.~\eqref{pmp:7}, and equating $z_{\text{max}}(I)$ with $z_{\text{max}}({\tilde{a}})$, we get an equation for the initial plasmoid current, i.e.,
\begin{equation}\label{gau:18} %
    I(z_{\text{max}}) =
              8 \pi \frac{B_0}{\mu_0}
              \Bigl( \frac{eB_0}{2mc}  \Bigr)^{1/2}
              \Bigl( \frac{mc^2}{E_s}\Bigr)^{2/3}
              \bigl( 3X_0            \bigr)^{1/3}
              \bigl( \beta\gamma     \bigr)^{1/6}
                     \tilde{a}^{7/6}
                     \beta^{2/3}
                      f_m^{-1}.
\end{equation}
which at an altitude of 300~km, and for $p$ expressed in  GeV/c, gives
\begin{equation}\label{gau:19} %
    I(z_{\text{max}}) = 760~ \text{[kA]} ~~ p^{1/6}
                     \tilde{a}^{7/6}
                     \beta^{2/3}
                      f_m^{-1}.
\end{equation}
A related quantity is the number of repolarizations, i.e., the ratio
\begin{equation}\label{gau:20} %
    I(z_{\text{max}}) /I_{\text{min}} =
              \Bigl( \frac{eB_0}{2mc}  \Bigr)^{1/2}
              \Bigl( \frac{mc^2}{E_s}\Bigr)^{2/3}
              \bigl( 3X_0            \bigr)^{1/3}
              \bigl( \beta\gamma     \bigr)^{1/6}
                     \tilde{a}^{1/6},
\end{equation}
which is seen to be independent of $\gamma_E^{-2}$ and only weakly dependent on $p$ and $\tilde{a}$, and which (again at $h=300$~km and $p$ in GeV/c) gives
\begin{equation}\label{gau:21} %
    I(z_{\text{max}}) /I_{\text{min}} = 760~  p^{1/6}
                     \tilde{a}^{1/6}
                     \beta^{2/3}.
\end{equation}
A final quantity of interest, which also leads itself to a simple expression, is the fractional energy loss due to repolarization.  Combining Eqs.~\eqref{gau:10}, \eqref{gau:21}, and the linear energy content of each plasmoid stream, i.e., ${\Delta W_{\text{kin}}}/{\Delta z}=mc^2(\gamma-1)$ $\times I/(e\beta c)$, this is
\begin{equation}\label{gau:22} %
     \frac{\Delta W_{\text{pol}}}{\Delta z}
     \frac{\Delta z}{\Delta W_{\text{kin}}}
     \frac{I(z_{\text{max}})}{I_{\text{min}}}
     \frac{z}{z_{\text{max}}} = 
     \frac{\tilde{a}}{2R}
     \frac{\beta^2\gamma}{\gamma-1}
     \frac{z}{z_{\text{max}}},
\end{equation}
where $R$ is the Larmor gyroradius, and which for a relativistic plasmoid is a loss of negligible magnitude.

   Equations \eqref{gau:16} to \eqref{gau:22} are the main results of this section:  They provide a consistent and realistic first order estimate of the key parameters of a beam plasmoid propagating in the ionosphere at altitudes comprised between 100 and 600~km. It is seen that if the $\beta^{2/3}$ factor is set equal to one (which is an excellent approximation for relativistic streams), these equations depend only on $p$ and $\tilde{a}$, so that they are independent of the particle's mass, and therefore directly apply to both proton and electron plasmoids.  Moreover, the $p$-dependence of $I(z_{\text{max}})$ is very weak, which means that the plasmoid current is almost independent of the momentum, which itself is essentially defined by the range through Eq.~\eqref{gau:17}.

   For example, if the ultimate range is set to $z_{\text{max}}=6000$~km (which means that the effective range would be something like 5000~km since the plasmoid would already have lost 80\% of its current after propagating that distance), Eq.~\eqref{gau:17} implies that the momentum is 100 GeV/c for a focus of 1~m radius on the target when $f_m=15\%$.  The initial current in each stream is then given by Eq.~\eqref{gau:19}, i.e., $I(z_{\text{max}})=11$~MA, which according to \eqref{gau:21} is enough for 1600 repolarizations.  Assuming that the pulse length is 3~ns, the energy and power delivered on a 5000~km distant target would be about 1000~MJ and 300~PW. These parameters --- a momentum of 100 GeV/c, a pulse length of 3~ns, and a current of 10~MA in both plasmoid's streams --- are technologically very demanding.  However, they are not very much different from those required by other daunting technological enterprises such as laser- or particle-driven inertial confinement fusion.\footnote{Moreover, a current of 10 MA and in pulse of 3 ns represent $2 \times 10^{17}$ antiprotons, i.e., only 0.3 ng of antiprotons.}

   Nevertheless, if we return to the discussion following Eq.~\eqref{pmp:9} in the previous section, we see that the present boundary layer current density of 11~MAm$^{-2}$~/~1600 $\approx$ 7~kA/m$^2$ is larger than the 2~kA/m$^2$ which can be supported by the ionospheric plasma alone.  This implies that for a range of 6000~km the full set of plasmoid/residual-atmosphere plasma-physical effects should be taken into account, which can only be done with a full-fledged two- or three-dimensional computer simulation program.  For a smaller range, the constraints will of course be reduced, but due to the relatively mild scaling dependencies appearing in Eqs.~\eqref{gau:16} to \eqref{gau:21}, the plasmoid currents will remain on the order of kA to MA, and the momenta in the multi-GeV/c range.

   There are also other considerations and potential difficulties to take into account.  For instance, because of the very large current and momentum, a pulse duration $\Delta \tau$  of 3~ns (i.e., a plasmoid beam-length of about 1~m) is enough to deliver a sufficiently large energy on the target. This has number of advantages: For example, the condition that the current is rapidly varying, which is a prerequisite for the derivation of Eq.~\eqref{pmp:9}, is satisfied; and, most important, the pulse length is sufficiently short for most instabilities not to have enough time to develop.  On the other hand, a short pulse means that inductive and any other form of erosion at the head and tail should be very small.  If for example Eq.~\eqref{ind:11} is blindly applied to a boundary layer with the above characteristics (100~GeV, 7~kA, 3~ns), the inductive erosion range would be of only 100~km.  However, Eq.~\eqref{ind:11} was derived for a narrow beam while here we have ended up with a plasmoid which comprising the boundary layers has a transverse width larger than its length.  Therefore, most particles that would be lost according to the standard picture of inductive erosion find themselves simply displaced within the boundary layers, so that the would-be head erosion process becomes part of boundary layer stripping, and that Eq.~\eqref{ind:11} does not apply.  Thus we reach again the conclusion that beyond the calculations presented in this section, a two- or three-dimensional computer simulation program is most probably the only way to reliably study these complicated effects.

\section{Feasibility of matter-antimatter beam plasmoids}
\label{epp:0}

A matter-antimatter plasmoid beam propagating across a magnetized plasma is possibly the most complex system considered in this review.  But this system is also an important contender for a possible long-range outer-space particle beam weapon, and despite its complexity it is entirely based on concepts and processes which have been considered in previous parts of this review --- with the exception of the annihilation processes which have so far been neglected, and which may take place when antiparticles interact with ordinary particles.

We have chosen to discuss this system because it has a number of obvious advantages over asymmetrical ion-electron beam plasmoid systems (to which many of the results obtained in this chapter apply after suitable modifications), and because on a time-scale compatible with the full-scale deployment of high-power long-range particle beam weapons in outer-space, availability of antimatter in the form of positrons (and at a greater cost of antiprotons) should not be the main obstacle.

Moreover, apart from the advantages which come from having a plasmoid composed of two charge-symmetric streams, there are advantages coming from the fact that a particle-antiparticle plasmoid is a pair plasma so that several complications present with ordinary plasmas are absent.  For example, the conductivity of a pair plasma is purely scalar, and unlike the Ohm's law for an ion-electron plasma there is no Hall effect or pressure contributions \cite{BLACK1993-}.  This has no impact on the fact the background atmosphere is an electron-ion plasma, but can drastically change the character of various beam-beam and beam-plasma interactions.  For example, some instabilities such as the dicotron mode are less likely to arise in a pair plasma than in an electron-ion plasma \cite{GALVE1988-}.

Finally, there are some aspects of beam-plasmoid physics, such as the importance of transverse and longitudinal emittances (or equivalently transverse and longitudinal temperatures), which have not been addressed in the previous sections, and which become essential in the context of possible beam-beam annihilation effects, so that they find their place in this section.   We are therefore going to examine the magnitude of these effects, in order to see to what extend they may prevent propagation across the magnetosphere towards a distant target.

The first major difference between a particle-antiparticle beam plasmoid and either a same-charge-particle beam, or a neutral-particle beam, is that interactions with the background gas or plasma may lead to annihilation reactions, which will remove some antiparticles from the plasmoid.  Since this creates a charge imbalance, an equal number of particles will be ejected from the plasmoid so that this \emph{stream-background annihilation} effect leads to an equal decrease in the current of both the matter and the antimatter streams as a function of propagation distance.

The second major difference is that particles of one stream may annihilate with their antiparticles in the other stream.  This \emph{stream-stream annihilation} effect is potentially important since oppositely charged particles attract rather than repel, so that while a same-charged-particle beam is characterized by a blow-up due to space-charge repulsion, as seen in a referential moving with the beam, a particle-antiparticle plasmoid is characterized by a drive towards matter-antimatter explosion \cite{GSPON1987A}.

Let us take an electron-positron beam plasmoid.  The first consideration is therefore to evaluate the positrons's rate of loss due to their annihilations with electrons in the outer-space medium, which at altitudes above 100~km is a partially ionized plasma with a total electron number density approximatively given by the product $Zn_a$, where $Z\approx 7.2$ is the average atomic-number of air, and $n_a$ the number density of the atoms in the residual atmosphere (see Table~\ref{tab:atm}).  If $\sigma_{ep}$ is the positron-electron annihilation cross-section, the annihilation rate is then given by
\begin{equation} \label{epp:1} 
      - \frac{dN}{dt}  = Zn_a \sigma_{ep} \beta c = 1/\tau_{ep},
\end{equation}
so that the number of positrons decreases as
\begin{equation} \label{epp:2} 
   N(t) = N_0 \exp (-t / \tau_{ep}).
\end{equation}
The annihilation cross-section is given by Dirac's formula \cite{DIRAC1930-}\footnote{In fact, this formula is valid for positron velocities $\beta > \alpha$, i.e., for energies greater than $\approx 6.3$ eV, the binding energy of positronium.  Since formation of positronium atoms tends to slow down the annihilation process, the cross-section \eqref{epp:3} leads to a conservative estimate of positron losses by annihilations.}
\begin{equation} \label{epp:3} 
   \sigma_{ep} = \pi r_e^2 \frac{1}{\gamma+1} \Bigl[
    \frac{\gamma^2+4\gamma+1}{\gamma^2-1} \ln(\gamma + \sqrt{\gamma^2-1})
                  - \frac{\gamma+3}{\sqrt{\gamma^2-1}}   \Bigr].
\end{equation}
In the low-energy limit, this cross-section reduces to
\begin{equation} \label{epp:4} 
   \sigma_{ep}(\gamma \rightarrow 1) = \pi r_e^2 \frac{1}{\beta},
\end{equation}
and, in the high-energy limit, to
\begin{equation} \label{epp:5} 
   \sigma_{ep}(\gamma \rightarrow \infty) = \pi r_e^2 \frac{1}{\gamma} 
                 \bigl[ \ln (2\gamma) -1 \bigr].
\end{equation}

   Because the positrons in the plasmoids considered in this review are highly relativistic, the high-energy limit applies and the stream-background annihilation life-time \eqref{epp:1} becomes
\begin{equation} \label{epp:6} 
   1/\tau_{ep} = \pi r_e^2 c Zn_a\frac{1}{\gamma} 
                 \bigl[ \ln (2\gamma) -1 \bigr].
\end{equation}
Using $\pi r_e^2 c = 7.5 \times 10^{-9}$ m$^{3}$s$^{-1}$, this gives $\tau_{ep} \approx 10^{3}$ seconds for a 1~GeV plasmoid (i.e., $\gamma \approx 2000$) propagating at an altitude of 100 km where $n_a \approx 5 \times 10^{18}$ m$^{-3}$. Therefore, already at this altitude the plasmoid current loss due to annihilations with background electrons is very small, even to reach targets located many 1000 kilometers away, i.e., corresponding to propagation times of several tens of milliseconds.  At a somewhat greater altitude, or for a higher energy beam, this effect would become even more negligible.

The second consideration is to evaluate the stream-stream annihilation rate, i.e., the rate of positron loss due to their annihilations with electrons in the overlapping part of the plasmoid.  In that case, while the cross-section is still given by Dirac's formula \eqref{epp:3}, the annihilation rate is not determined by the beam's velocity as a whole, but by the average velocity of the positrons relative to that of the electrons in a frame of reference moving with the streams.  In that frame, the electron-positron plasmoid is a non-thermalized plasma characterized by two temperatures, $T_{\tr}$ and $T_{\lo}$, which for a actual beam are in general different.  For example, in a typical neutral hydrogen beam system envisaged for outer-space particle beam weapon use, the initial emittance is $2 \times 10^{-7}$ m$\cdot$rad, and the beam energy spread $\Delta W/W = 0.1\%$ \cite[p.24-77]{KNAPP1980-}.  Using equations \eqref{def:15} and  \eqref{def:16} this give  $kT_{\tr m} \approx 2$ eV and $kT_{\lo m} \approx 0.25$ eV, respectively.  In the case of a kA to MA current, multi-GeV beam plasmoid of the kind considered in the previous chapter, the temperatures will also have to be in that range.  Therefore, as is the case for high-quality (so-called `cold')\footnote{As a `temperature' of 1 eV corresponds to about 11600 $^{\rm{o}}$K, this terminology is of course not related to the everyday notion of temperature.} beams, both the transverse and longitudinal temperatures are relatively low and non-relativistic (i.e., $kT \ll mc^2 = 0.5$ MeV).  Consequently the cross-section to use is given by \eqref{epp:4} rather than \eqref{epp:5}, as was the case in the stream-background annihilation effect previously investigated.

In the frame moving with the beam, writing $n$ for the electron or positron beam number density, the stream-stream annihilation rate is given by the equation \cite{GSPON1986A}
\begin{equation} \label{epp:7} 
      - \frac{dn}{dt}  = n^2 \langle \sigma v \rangle,
\end{equation}
where $\langle \sigma v \rangle$ denotes some average over the effective cross-section obtained by multiplying \eqref{epp:4} with the probability of finding an electron at the position of the positron in a Coulomb field, and $v$ the relative electron-positron velocity.  Assuming a Maxwellian velocity distribution, this average is
\begin{equation} \label{epp:8} 
  \langle \sigma v \rangle = \pi r_e^2 c \alpha \sqrt{\frac{4\pi mc^2}{kT}}.
\end{equation}

   The solution of equation \eqref{epp:7} is very simple, namely
\begin{equation} \label{epp:9} 
  n = \frac{n_0}{1 + t/\tau_{bm}},
\end{equation}
where $\tau_{bm}$ is the beam half-life in the frame moving with the beam
\begin{equation} \label{epp:10} 
  1/\tau_{bm} = \frac{1}{2} n_0 \langle \sigma v \rangle,
\end{equation}
while the beam half-life in the accelerator frame is $\tau_{1/2} = \gamma\tau_{bm}$ because of Lorentz time dilatation.

Taking again a plasmoid stream energy of 1~GeV, and a very intense current of 10~MA in a radius of 1~m, we have $n_0 = 6.6 \times 10^{16}$ m$^{-3}$, and from the calculated beam temperatures we have $\langle \sigma v \rangle \approx 10^{-19}$ m$^{3}$s$^{-1}$.  Therefore, using either the transverse or longitudinal temperature, we find from \eqref{epp:10} that $\tau_{1/2} = \gamma\tau_{bm}$ is on the order of $10^{5}$ seconds, i.e., very large.  The main reason for this is that the beam density is comparatively low, even for a current of 10~MA, which ensures that contrary to the case of a high-density matter-antimatter plasma \cite{GSPON1986A}, the life-time of a particle-antiparticle plasmoid beam of the type to be used in an outer-space system is very long.

   In the case of proton-antiproton annihilations, there is no simple theory leading to an analytical formula similar to Dirac's result \eqref{epp:3}.  However, in the relatively low beam-temperature domain characteristic of beam plasmoids, the proton-antiproton annihilation rates are about nine times larger than the corresponding electron-positron rates, as can be seen by looking at the respective reaction rates shown in Figure~1 of reference \cite{GSPON1986A}. The second difference between electron-positron and proton-antiproton plasmoids of a given energy is their Lorentz factor $\gamma$, which for relativistic energies differ by their mass ratio, $m_p/m_e=1836$.  Therefore, the times calculated in this section for positrons should be divided by $\approx$17'000 to get the corresponding numbers for antiprotons, which brings them down into the range of a fraction to a few seconds, which still means that proton-antiproton plasmoid life-times are significantly larger than any conceivable time of flight to a target orbiting the Earth.  This concludes this section, in which we have shown that matter-antimatter annihilation during propagation is not a major effect for relativistic plasmoids propagating in the high-atmosphere.

\chapter{Scientific and technical prospect}
\label{sci:0}

\section{Discussion of theoretical prospect}
\label{the:0}

	We have examined the main physical problems involved in the propagation of particle beams for possible exo-atmospheric or endo-atmospheric beam weapon systems.

	There is no problem of principle with the propagation of neutral particle beams in outer space.  Systems with adequate characteristics are under development since more than twenty years \cite{KEEFE1981-}.  For short range exo-atmospheric systems, i.e., on the order of a few hundreds of kilometers, charged beams or plasmoid beams might eventually be used.

	For land-based systems, the main problem is to cope with the conflicting nature of the following set of conditions:

\begin{enumerate}

\item The beam power has to be larger than the Nordsieck power \eqref{nor:9} in order for the beam to propagate over sizable distances;

\item the beam current should be less than the critical current \eqref{cri:3},  and the current neutralization fraction as small as possible \eqref{cri:10}, so that the beam is well pinched;

\item the beam pulses have to be long enough so that they will not be completely eroded before reaching their target \eqref{ero:3}.  Furthermore, the beam should be relativistic enough ($\gamma > 10$) for the steady state erosion hypothesis to be valid;

\item the pulse duration has to be of the order of the dipole magnetic diffusion time so that propagation is not disrupted by macroinstabilities \eqref{mac:38};

\item in order to keep the growth of instabilities at a minimum, the current neutralization should be minimized \eqref{mic:23}, \eqref{mac:24}, \eqref{mac:26}, and \eqref{mac:39}, and the conductivity  should rise fast enough to avoid excessive growth in the head \eqref{mac:22};

\item the beam radius, emittance, energy, and current should match the Bennett pinch radius \eqref{ben:4} at the exit of the accelerator system, and should be compatible with the requirements of beam stability and hole boring.

\end{enumerate}

	These conditions cannot easily be simultaneously satisfied.  This is primarily because the pulse length is set by magnetic diffusion time.  For a given beam radius, this time can only be increased by increasing the beam current, and thus producing an undesirable large current-neutralization fraction.

	\emph{However, within rather tight limits, a set of acceptable beam parameters seems to exist.  In order to find out if these conditions are really satisfactory for beam weapon applications, careful experiments are needed.} This concluding sentence was written in November 1982, together with the additional remark:  \emph{These crucial experiments will be done with accelerators currently under construction in the United States \cite{MOIR-1981-,BARLE1981A} and probably in the USSR \cite{PAVLO1980-}.  Within a couple of years, at most, the final answer should be known.}  What can we add to this conclusion, more than twenty years after it was written? 

\begin{figure}
\begin{center}
\resizebox{4cm}{!}{ \includegraphics{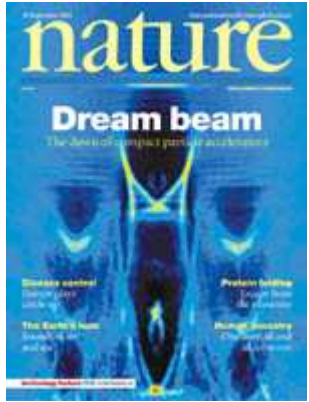}}
\caption[`Dream Beam:' Ultra-high-energy laser-driven beam]{\emph{Ultra-high-energy laser-driven beam.} Under the heading `Dream beam' the journal \emph{nature} published in  September 2004 the papers of three groups (\cite{MANGL2004-} at Rutherford Appleton Laboratory, U.K., \cite{GEDDE2004-} at Lawrence Berkeley National Laboratory, U.S.A., and \cite{FAURE2004-} at Ecole Polytechnique, France) experimentally confirming that GeV per centimeter accelerating-gradients (i.e., thousands of times stronger than in conventional radio-frequency accelerators) were possible, as predicted by T.\ Tajima and J.\ Dawson in 1979.  
\label{fig:db}}
\end{center}
\end{figure}

\begin{itemize}

\item First:  That the laws of physics have not changed, and that the theoretical analysis done in the previous chapters remains valid.  In fact, having followed the literature over the past twenty years, it appears that nothing really new and important has been found and published.  In particular, the most difficult theoretical problems addressed in this report --- the questions related to the stability of propagating a charged-particle beam through a pre-existing or beam-generated plasma --- are still discussed in recent papers by referring to the seminal work done during the 1960s to the 1980s.\footnote{See, for example, references number 20 to 25 cited in a recent paper where the resistive hose instability is reinvestigated for low-collisional plasmas \cite{UHM--2003-}.}  This is not to say that no progress has been made, but that the basic physical processes at work in high-intensity high-energy beam propagation in plasmas appear to have been properly recognized and analyzed during these decades.

\item Second: That a major step in the theoretical understanding of beam-propa\-ga\-tion stability was made between the time of the early analytical calculations, such a those of Steven Weinberg \cite{WEINB1967-}, and the time of  the more sophisticated studies, such as those of Edward P.\ Lee \cite{LEE--1978-}, which showed that oversimplified models were likely to predict far too pessimistic behaviors in comparison to more realistic ones.  This means that an essential step after accelerating a beam pulse is that of `conditioning,' that is of shaping and smoothing its spatial an possibly temporal extent in such a way to avoid the onset and growth of instabilities during propagation.\footnote{This is the plasma-physical counterpart of the hydrodynamical need to smooth boundaries and remove all unwanted discontinuities in order to avoid drag, turbulence, and instabilities in the motion of ships, aircrafts, or missiles.}

\item Third: That some experimentally well-established observations, such as stable long-distance propagation in low-density gas (i.e., in the ion-focused regime \cite{BRIGG1981-}), are now also much better understood theoretically than some years ago \cite{BUCHA1987-}.  This means that one can have much greater confidence in the theory, both for free-propagation in high- or low-density air, as for controlled transport and conditioning in the accelerator system generating the beam. In particular, the interaction of intense relativistic electron beams with pre-formed laser-generated channels in the atmosphere \cite{MURPH1987-, MURPH1992-}, as well as the stability of such beams in beam-induced channels \cite{NGUYE1987-,WHITT1991-}, appear to confirm present understanding of such phenomena.

\item Fourth: That conceptual and technological advances are constantly changing the context in which specific systems have to be assessed, both from a theoretical and a practical point of view \cite{GSPON2000-}.  For example:

\emph{Antimatter and muon beams.} The possibility of the use of antihydrogen beams in exo-atmospheric systems came to the author's mind in 1983 when a former CERN colleague told him about Los Alamos management's interest to hire him for antimatter work at their laboratory \cite{GRINE1984-}.  This lead him to realize that antimatter was the sole portable source of muons, which are the main byproducts of the annihilation of antiprotons with matter \cite{GSPON1987B}.  This means that if muons from such a source could be `cooled'\footnote{`Cooling' means decreasing the kinetic energy dispersion of a bunch of particles.} and accelerated, the possibility of exploiting the high-range of muon beams (see Figures~\ref{fig:exp} and~\ref{fig:ran}) would become theoretically possible.  In fact, as will be seen in Sec.~\ref{him:0}, the problem of generating and cooling intense beams of muons is the subject of vigorous research since a few years.  While this effort may not succeed in producing muon-beams sufficiently intense to propagate in self-pinched mode over significant distances, the development of muon technology for fundamental research will help assessing whether muon beams, and beams of their related neutrino decay products, may have other practical applications \cite{AMALD1999-,DETWI2002-}.

\emph{Laser-driven particle accelerators.}  Radio-frequency-based accelerators are limited to relatively low accelerating electric fields, i.e., 10 to 100 MV/m, requiring tens to hundreds of meters to produce a multi-GeV beam.  On the other hand, ultra-high-intensity lasers can produce accelerating fields in the 10 TV/m range (1 TV = 10$^{12}$ V), surpassing those in radio-frequency accelerators by six orders of magnitude.  It had therefore been anticipated for a long time, and demonstrated in computer simulations, that GeV of electron energy per centimeter of acceleration distance was theoretically possible \cite{TAJIM1979-}.  The experimental confirmation of this prediction came in 2004, when three groups were independently able to generate beams of 80--170 MeV electrons with low divergence and a small energy spread (less than three per cent) \cite{MANGL2004-,GEDDE2004-,FAURE2004-}.  This achievement was made possible by the application of a number of techniques, including some that have  already been discussed in this report, e.g., the use of a preformed plasma-channel to guide the laser. As the practical implications for the design of much cheaper and smaller particle accelerators were immediately recognized, the publication of these results was the highlight of the September 30, 2004, issue of the journal \emph{nature}, see Fig.\,\ref{fig:db}

\end{itemize}

\section{Discussion of beam propagation experiments}
\label{exp:0}

The discussion of the technical prospect of particle beam technology for  exo-atmospheric applications is quite different from that of endo-atmospheric applications.\footnote{A summary of the early efforts on particle beam weapons research, from World War II to 1980, as well as some comments on Soviet particle beam research, are given in a comprehensive fact-sheet published by the U.S.\ Department of Defense \cite{USDOD1980-}.}  Indeed:

\begin{itemize}

\item Exo-atmospheric beam technology is very similar to the relatively low-intensity (mA), relatively high-energy (GeV), beam technology used in numerous, officially non-military, national and international nuclear and elementary-particle research laboratories.  Moreover, the typical accelerator and beam steering technologies required for outer-space beam weapons is very similar to those of emerging applications of accelerators such as cancer therapy, breeding of fissile and fusion materials, transmutation of nuclear waste, breeding of antimatter \cite{GSPON1987A}, or proton-radiography of the detonation of conventional explosives and nuclear weapon's primaries, \cite{HARTO2000-} and \cite[p.84--89]{CUNNI2003-}.  Finally, the more advanced technology for beam weapons is significantly overlapping with that of powerful research accelerators such as the Large hadron Collider (LHC) at CERN and the future linear-collider (which will most probably be a laser-driven accelerator \cite{KATSO2004-}) to be built (most probably in the US) as a fully international effort \cite{CERN-2002-}.  This means that most components of space-based beam weapons can be developed and tested in civilian laboratories  (Sec.~\ref{nhb:0}), and that the role of military laboratories is to develop and test systems which integrate these components, and merge them together with the application specific technologies related to their use in an outer-space battle-field environment, rather than as research tools in a laboratory (Sec.~\ref{gta:0}).  Finally, most experiments on exo-atmospheric systems can be done on the ground, e.g., using evacuated pipes for propagating the beam, and at reduced beam power, i.e., without requiring safety precautions that would be dramatically more stringent than those required by the operation of typical accelerators used for scientific research.

\item Endo-atmospheric beam technology is very similar to the relatively high-intensity (kA), relatively low-energy (MeV), beam technology used in military laboratories for the flash x-ray radiography of the detonation of conventional explosives and nuclear weapon's primaries \cite[p.81--84]{CUNNI2003-}, the generation of high-power radio-frequency or free-electron-laser beams, and the simulation of nuclear weapons effects \cite{FLEIS1975-}.  On the other hand, this technology has only few civilian applications, such as industrial radiography, electron-beam welding, and possibly future inertial confinement fusion drivers and very high-energy particle accelerators \cite{BARLE1990-}.  Therefore, the development and demonstration of this technology is mostly done in military and national laboratories, as will be seen in Secs.~\ref{ata:0} to \ref{usa:0}.  Nevertheless, there is significant overlap between these developments and those of conventional research accelerators, as is shown by the fact that both type of technologies are generally discussed at the same national and international scientific conferences, and published in the same technical journals.  Finally, an additional reason for locating these experiments in military laboratories is that even relatively modest endo-atmospheric systems create special hazards because the total energy in typical beam pulses can easily be equivalent to the energy content of gram- to kilogram-amounts of high-explosives.

\end{itemize}

The purpose of this report is to review the physics of high-intensity high-energy particle beam propagation.  The discussion of exo-atmospheric systems could therefore be very brief since (as noticed in Sec.~\ref{neu:0}) there is in principle no scientific obstacle to propagating a neutral beam over long distances in the near-vacuum of outer-space.  We will nevertheless address some of the critical issues related to these systems, even though they are more of a technological than physical nature.  This is because the development of exo-atmospheric beam weapons is in many ways linked to those of endo-atmospheric ones, and because their development illustrates the very strong interdependence which characterizes civilian and military particle accelerator research, as well as the importance of informal international collaborations and scientific exchanges (including between `enemies') in relation to the development of advanced weapons systems.

\chapter{Neutral particle beams propagation experiments}
\label{npb:0}

\section{Neutral hydrogen beam technology development}
\label{nhb:0}

According to Sec.~\ref{neu:0}, what is required to focus a stream of neutral particles on a 1~m target at a distance of 1000 km is a beam with an initial radius of 20 cm and an emittance of $2 \times 10^{-7}$ m$\cdot$rad.\footnote{Chapter VII and VIII of the Los Alamos Accelerator Technology Division 1978 progress report is describing a number of features associated with a system based on a 20--100 mA, 50--500 MeV beam with such an emittance \cite[p.24-77]{KNAPP1980-}.} As a charged particle is needed in order to be accelerated by electromagnetic means, the general technique is to start from a negative-ion source producing a low-energy,  low-emittance beam of H$^-$ or D$^-$ atoms, which is then injected into an accelerator whose main quality is to increase the energy of the ions to about 100 MeV without unduly increasing the emittance of the beam.  At the exit of the accelerator this high-brightness beam is very precisely focused on a target by a magnetic beam-steering optics, and then passed through a neutralizing cell in order to remove the extra electron (without increasing the beam emittance) so that the final beam is neutral.

Most of the basic technologies necessary for building such a system have been invented, built, and tested between the late 1970s and the early 1990s.  This is what will be summarized in the next subsections:

\subsection{The ion source}

 The first advance advance which made the concept of a neutral particle beam weapon scientifically feasible was the invention of a sufficient low-emittance negative ion source by the group of V.G. Dudnikov in the USSR \cite{DUDNI1974-, DIMOV1977-}.  This pulsed surface-plasma ion source was quickly adopted world-wide, and Los Alamos soon demonstrated its superior performances relative to previously existing sources \cite[p.117]{KNAPP1982-}, \cite{SMITH1981-}.

\subsection{The injector}

The second major advance was again a Soviet invention. It solved the problem of injecting into a linear accelerator a low-energy low-emittance beam while maintaining its brightness.  Moreover, this new injector was very compact and rugged.  It is so-called `radiofrequency quadrupole' (RFQ), which, in combination with the Dudnikov high-brightness H$^-$ source, led to speculations in the years 1978-1980 that the USSR might have had a considerable advance in particle beam weapons technology.  However, while development of RFQ technology started in Russia in 1970 already, it only reached maturity around 1977, about the time when Los Alamos started its own RFQ program \cite{WELLS1985-}.

In fact, it is in 1978 that a new \emph{Accelerator Technology} (AT) division, headed by Edward A.\ Knapp, was formed at Los Alamos. The first step in the development of a neutral particle beam system for possible deployment in space, code-named `White Horse,' was to build a 100 mA, 5 MeV accelerator test stand, which could be scaled up to about 50 MeV for outer-space re-entry-vehicle/decoy discrimination, and later to about 500 MeV for boost-phase or mid-course intercept of ballistic missiles. This accelerator test stand (ATS) was to integrate in a test-bed a Dudnikov H$^-$ source, a RFQ to reach an energy of about 2 MeV, a drift-tube linear accelerator section to increase the beam energy to 5 MeV, and beam diagnostic equipments.  This relatively small and low-cost system was to be supplemented by several collaborative undertaking on related military or civilian projects.  This is why Los Alamos was glad to accept a proposal form CERN, reported in the 1980 AT division progress report \cite[p.36]{KNAPP1982-}:

\begin{quote}
``(CERN) in Geneva, Switzerland, has asked the AT division to cooperate in the design and construction of an RFQ linac.  This linac would replace the [...] injector [...] used with the `old' linac for the CERN proton synchrotron. [...] The plan is to carry out the mechanical design and assemblies at CERN, and then send them to Los Alamos for machining of the pole tips.''
\end{quote} 

The reason why CERN needed Los Alamos for building an RFQ is that the precise machining of its pole tips required a computer-controlled milling-machine of a type that was only available in a nuclear-weapons laboratory such as Los Alamos.  Reciprocally, Los Alamos needed CERN as it provided a test-bed for coupling an advanced RFQ injector to a high-energy linac, which, moreover, happened to be the most recent and advanced in the world at the time.

Therefore, while the first successful high-power operation of a prototype RFQ with beam was obtained in February 1980, the first operational RFQ built at Los Alamos was shipped to CERN and installed on its new linac, while the second to become operational was installed on the Los Alamos accelerator test stand in November 1982~ \cite[p.53]{JAMES1984-}.

\subsection{The accelerator}

  A major achievement of the Los Alamos accelerator technology division has been to reach, in 1979, a quantitative understanding of emittance growth in linear accelerators \cite{JAMES1979-}, showing that there is a lower limit to the emittance of such accelerators \cite{STAPL1979-}.  This understanding had been made possible by numerous exchanges between Los Alamos and many other institutions, including the CERN laboratory in Geneva. As an example, we cite a few sentences of the section on emittance growth of the 1978 Los Alamos accelerator technology division progress report \cite[p.63,64,71]{KNAPP1980-}:\footnote{We stress that these quotes are from a 47-page long chapter entirely dedicated to the discussion of the design of an accelerator and of a beam-steering system for possible use as a particle beam weapon in the context of the `White Horse' project.}

\begin{quote}
``The new 50-MeV injector linac of CERN is the latest of this type proton linear accelerator to be commissioned in the world. [...] Preliminary results from CERN during the early summer 1978, using the first tank at 10 MeV, indicated larger than anticipated emittance growth.  Knowing our urgent need for verification of the design approach for the FMIT project, the CERN Linac Group graciously invited our participation in the commissioning of the full machine.  An on-site collaboration afforded us the opportunity to reach a level of understanding impossible to achieve at conferences or by letter. [...] We very much appreciate the complete openness, candor, and hospitality of the whole CERN linac group.  Their willingness to let us use their raw data, before they had time to publish their results, was the key that enabled us to answer our basic questions in the urgent time scale imposed by current projects.''
\end{quote}

\subsection{The beam focusing and steering optics}  

A major issue in a space-based neutral beam system is the ability to focus the beam into an acceptably small spot at some distance, and to provide a steering capability for aiming the beam.  Using the ray-tracing computer program TURTLE (a typical particle beam optics development tool devised by a collaboration of accelerator physicists from FNAL, SLAC, and CERN), the Los Alamos accelerator group was able to show in 1978 that a 50-meter long system of bending magnets and solenoids was able to focus 91\% of a few 100 MeV hydrogen beam within the desired radius, starting with an initial radius of 0.2 m and an emittance of $2 \times 10^{-7}$ m$\cdot$rad \cite[p.52--62]{KNAPP1980-}.  An optical system with these characteristics was ultimately built and successfully tested in 1990 at the Argonne National Laboratory near Chicago. 

In this context it is important to recall that the beam focusing and steering technology for particle-beam weapons does not overlap just with beam technology for high-energy-physics, but also with the more directly related technolgy of ion-beam focusing and steering for particle-beam driven inertial confinement fusion.  In the United States, this technology is developed as a collaboration between the Lawrence Berkeley and Lawrence Livermore National Laboratories, and has led to significant results, including recent final focus experiments \cite{MACLA2001-}.

\subsection{The neutralizing cell}

The final part of a neutral particle beam accelerator system, the neutralizing cell in which the extra electron is detached from the accelerated ion to form a beam of neutral atoms, is possibly the most controversial component of the whole system.  This is because simple techniques (such as passing the beam through a foil or a gas to strip-off the extra electron) are likely to significantly increase the emittance of the beam and therefore to unduly decrease its brightness, at least for the most demanding applications of outer-space weapon systems.  The most promising technique is therefore that of laser photodetachment, which is potentially 100\% efficient rather than limited to about 55\% as with collisional processes.  High efficiency in outer-space is very important since any degree of incomplete neutralization of the beam could result in charging the orbiting accelerator, ultimately causing arcs that could lead to self-destruction.  Moreover, it could be that by means of a multiphoton detachment process the excess electron could be removed without increasing beam emittance \cite{STINT1995-}.  However, the kind of laser required for this purpose would add a considerable level of complexity to a space-based particle beam device, so that the whole subject of neutralization is still an open issue, despite constant progress \cite{ZHAO-1997-,KUAN-1999-}.

\subsection{Summary}

The developments presented in this section clearly show that the kind of accelerator and beam optics required for a neutral particle beam weapon can be built. Therefore, they established the scientific and technical feasibility of such systems.  However, they does not prove their practical engineering feasibility, namely they do not demonstrate that a complete system (including power supplies, cooling, and many ancillary equipments) can be put in orbit and operated in outer-space.  This is where the specific developments described in the next two sections come in.

\section{BEAR and GTA at Los Alamos National Laboratory}
\label{gta:0}

Until 1983 the development of neutral particle beam weapon's technology and systems was made in the form of small, relatively low-cost, and mostly `paper' research programs.  This was because (as shown in the previous section) the much more costly civilian research programs related to fundamental research could be used to develop and build all the key components of such systems, and that even international laboratories such as CERN, located in a neutral non-nuclear-weapon state, could be used for testing key component in full-scale accelerators.\footnote{The same will happened in the mid-1980s, and still continue today, with the development of antimatter technology where U.S. scientists from weapons-laboratories or working on defence-contracts have essentially free, unlimited access to all of CERN's antimatter facilities \cite{GSPON1987A,GSPON1988-,THEE-1988-}. }

For this reason, the accelerator test stand (ATS), extensively referred to in the previous section, was sufficient to provide first hand experience on the most crucial components of a neutral particle beam system, and was even able to demonstrate reliable operation at 170 mA and 5 MeV, making it the brightest high-current H$^-$ beam in the world \cite[p.3]{BURIC1986-}.


Nevertheless, with the advent of President Reagan's Strategic Defense Initiative, neutral particle beam weapons were to be evaluate in competition with other directed-energy systems such as high-energy lasers and nuclear-pumped x-ray lasers.   This lead to the definition of a comprehensive, national neutral particle beam research and development program lead by the Los Alamos laboratory, described in a September-1986 brochure \cite{BURIC1986-}, and related technical reports \cite{LAGTA1986-}.  The objectives of this program were: (i) to provide the necessary basis for a decision by 1992 to build a space-based neutral particle beam system to be used as a decoy/warhead discriminator (near term goal); (ii) to develop the technology in stages to ultimately build a neutral particle beam weapon capable to provide a rapid hard kill of enemy warheads (far term goal).  To meet these objectives and more directly the first one the essential elements of the program were:

\begin{enumerate}

\item The \emph{Integrated Space Experiment 1} (ISE-1) and\\ the \emph{Beam Experiment Aboard Rocket} (BEAR);

\item The \emph{Ground Test Accelerator} (GTA) and\\ the \emph{Technology Program}.

\end{enumerate}

The near-term goal was therefore to use the space shuttle to launch, in 1991, a 50-MeV-accelerator-based system into space (ISE-1), a quite ambitious objective.  This was to be preceded by the suborbital launch, in 1987, of a 1-MeV accelerator by a single-stage Aries rocket.  In this context, the GTA was to be representative of what was actually to be placed into orbit in ISE-1, while the BEAR payload was basically to be a ruggedized and less powerful version of the ATS, packed into a cylindrical volume of 1 m diameter and 7 meter length, together with diagnostic instrumentation.

However, as a consequence of the January 28, 1986, \emph{Challenger} space shuttle disaster and funding constraints, the Integrated Space Experiment (ISE-1) was cancelled, and Los Alamos was asked in December 1987 to complete BEAR on an accelerated schedule with limited funding \cite[p.25]{LARHL1989-}.  The BEAR payload was launched 200 km into space on July 13, 1989, and several accounts, e.g., reference \cite{LARHL1989-}, as well as the final report, i.e., reference \cite{NUNZ1990-}, described the experiment as a success:

\begin{quote}
``The US Department of Defense's Strategic Defense Initiative Organization is sponsoring the development of neutral particle beam (NPB) technology for strategic defense applications. The first step in demonstrating the functioning of an NPB in space was the development and launch of the Beam Experiments Aboard a Rocket (BEAR) in New Mexico in July 1989. A government, laboratory, and industrial team, under the technical coordination of Los Alamos National Laboratory, designed, developed, and tested the BEAR payload. The primary objective of BEAR was the operation of an NPB accelerator in space. The payload was also designed to study (1) the effects on the space vehicle of emitting an NPB and associated charged beams into the space environment; (2) the propagation and attenuation characteristics of an NPB in space; (3) the dynamics of the charged particle components of the beam in the geomagnetic field; (4) the effects of neutral effluents from the vehicle; and (5) any anomalous or unanticipated phenomena associated with operating an NPB in the space environment. The BEAR experiment successfully demonstrated operation of an NPB accelerator and propagation of the neutral beam as predicted in space, obtained first-of-a-kind NPB physics data, and demonstrated the ability of the BEAR accelerator to survive recovery and to continue operating normally. No unanticipated phenomena were encountered that would significantly delay further development of NPB technology for defensive, space-based weapon systems'' \cite{NUNZ1990-}.
\end{quote}

On the ground, despite the cancellation of the first (and thus subsequent) integrated space experiment(s), construction of the GTA proceeded more or less on schedule, and the accelerator was commissioned in 1992, producing a 24 MeV, 50 mA beam with a 2\% duty factor \cite{SAND1992-}.  Consistent with the design report, \cite{LAGTA1986-}, and other reports, e.g., \cite{BURIC1986-,LARHL1990-}, work on GTA and related equipments, such as the magnetic optics, proceeded in collaboration with numerous laboratories and universities (e.g., the Oak Ridge, Lawrence Berkeley, and Argonne national laboratories, the Northeastern University and the University of Texas) as well as with the industrial contractors associated to the project.  In particular, a state-of-the-art RFQ was built at Los Alamos and delivered to serve as a proton source in the Superconducting Super Collider (SSC) expected to be built in Texas \cite{LARHL1990-, MARSH1994-}.  Similarly, development and testing of the magnetic optics components were done at the Argonne and Lawrence Berkeley laboratories.

However, although all its components had been built, and many of them successfully tested, the GTA was never fully assembled, and its construction abandoned at the end of 1993 \cite{MARSH1994-}.  The floor space was later used for assembling and testing the Low Energy Demonstration Accelerator (LEDA) of the Accelerator Production of Tritium (APT) project \cite{SCHNE1996-}; various equipments and spare parts were reused in other projects;  and finally the GTA accelerator itself was donated in 2004 to the University of Indiana for its Low Energy Neutron Source (LENS) facility, to be used as a training ground for scientists who will later work at the \$1.5 billion neutron source to be completed in 2006 at Oak Ridge \cite{RINCK2004-}...

\section{Emerging neutral beam technologies}
\label{eme:0}

As recalled in the previous section, the years immediately following the collapse of the Soviet Union coincided with the abandonment of the most visible components of the U.S.\ national neutral particle beam research and development program: the ISE experiment and the GTA accelerator. Nevertheless, this did not mean that interest in neutral particle beam was lost, neither that development stopped there.  In particular, the numerous told and untold reasons for terminating an attempt to demonstrate the viability of a neutral particle beam system in outer-space were much more of a technical and political nature than of a fundamental one.  With a successful proof of principle experiment such as BEAR, there was no really compelling scientific reason to make another space-based experiment which would not have added very much to the understanding of the underlying physical issues.  In fact, problems such as developing an appropriate kW to MW class energy source for the whole system, a suitable cooling system for the accelerator, or a highly efficient beam neutralizer, would not have much benefited from a crash-program to put GTA into outer-space.

The points to be stressed are therefore:
\begin{enumerate}

\item That the accelerators and associated technologies for neutral beam weapon systems are so closely related to those of accelerators and detectors used for fundamental nuclear and elementary particle research that their development does not need to be done in military laboratories;

\item  That the specific characteristics of neutral particle beams for military applications such as ballistic missile defense should be discussed in relation to the full range of technologies available for accomplishing similar decoy/weapon discrimination and/or target destruction objectives \cite{GSPON2000-}.

\end{enumerate}

Indeed, the most important and almost unique advantages of high-energy particles  are their ability to penetrate deeply into any target and to interact strongly (that is by inducing nuclear reactions) with any substance.  This implies that the potential damage can be considerable, and that even for very low-intensity beams the secondary particles emitted in the nuclear interactions with the materials in the targets provide a signal giving a lot of information on their composition.  This is illustrated by the current method which uses background cosmic-ray particles as a natural beam for the remote analysis of the elemental composition of artificial satellites \cite{HARRI1995-}.  While this method requires times on the order of months or years to integrate sufficient events to achieve useful data, directing a modest beam of accelerated particles at a given spacecraft would achieve the same result in a fraction of a second.\footnote{This requires, of course, the use of very sensitive and clever techniques for detecting the secondary particles emitted by the target, and for discriminating them against the cosmic-ray background, something that can only be compared to the skill required to design the detectors and analyze the results of sophisticated high-sensitivity nuclear or elementary-particle physics experiments.}

Considering that it is most probably easier and less expensive to destroy a ballistic missile or a spacecraft by means of some kind of a kinetic interceptor than by either a high-energy laser or particle beam, the most potent application of a space-based particle beam is quite certainly that of warhead/decoy discrimination.  This is possibly why this application had been assigned in priority to space-based particle beam systems starting 1986 in the United States (see, e.g., references \cite{BURIC1986-} and \cite{LAGTA1986-}, where it was stated that a discriminator/weapon decision was to be made by 1992).  The remaining question to answer before designing and possibly deploying such discriminators is therefore whether there could be any strong competitor on the design horizon, of which three can easily be identified:

\subsection{Antihydrogen beams}

  While a neutral hydrogen beam containing 5 to 10~MeV protons would be sufficiently energetic to strongly react with the surface of a target, a beam with an energy of at least 50 to 100~MeV is necessary to penetrate inside, or to generate sufficiently many neutrons on the surface, in order to determine whether there could be a nuclear warhead within.  If instead the beam would consists of low-energy antihydrogen atoms, the antiprotons would spontaneously annihilate on the surface, generating several high-energy pions for every antiproton hits, which would deeply penetrate into the target and strongly interact with the materials.  An antihydrogen beam would therefore enormously simplify the design of the accelerator, which could operate at a much lower beam energy and current than for conventional hydrogen beam.  Therefore, the technological burden would be transferred to the antimatter technology, which, however, is under intensive development since more than ten years \cite{GSPON1987B}.

\subsection{Positronium beams}

Positronium (i.e., atomic bound states consisting of an electron and a positron) can be formed in two states: parapositronium with a life-time of $1.2 \times 10^{-10}$~s and orthopositronium with a life-time of $1.4 \times 10^{-7}$~s. Thus, even for a beam of orthopositronium atoms, a very high kinetic energy per atom is required for long-range propagation to be possible.  However, as with matter-antimatter plasmoids, propagation may not be the main problem:  forming a positronium beam pulse of the required energy might be the most difficult step \cite{OTTWI1988-,WEBER1988-,SURKO1986-}.

\subsection{Ultra-high-energy laser beams}

Comparisons of the relative ability of neutral particle beam and laser systems for discriminating between reentry vehicle and decoys show that particle beams can typically discriminate about hundred times as many objects as can lasers, and do so with significantly greater certainty \cite{CANAV1989-}. This applies, however, to lasers with relatively low peak power.  The recent invention of `chirped pulse amplification,' which provided a factor of one million (i.e., $10^6$) increase in the instantaneous power of lasers, enabled tabletop lasers to produce nuclear reactions directly \cite{MOURO1999-}. Therefore, such superlaser beams combine the ease in steering and focusing of optical laser beams with the capacity of particle beams to generate high-energy secondary particles in distant targets, which gives them the ability to `x-ray' remote objects and discriminate whether they are warheads or decoys \cite{GSPON2000-}.

\subsection{Summary}

One can therefore conclude that neutral particle beams provide a credible option for discrimination,\footnote{In Sec.~\ref{him:0} it will be shown that charged \emph{muon beams} also have a potential for warhead/decoy discrimination.  This could be used to discriminated objects over at least part of their ballistic flight, and therefore become a competition or a complement to neutral particle/antiparticle-beam systems, provided muon-beam systems could operate in the high atmosphere.}  but that this option should be constantly compared to emerging alternatives which arise as technology advances.  In particular, it may happen that the laser system required to neutralize a negative hydrogen beam could in fact be of a complexity comparable to that of a superlaser able to do the same discrimination task on its own...  But it would then be necessary to assess that further advances in antimatter technology would not require a complete reevaluation of the possible military uses of antimatter, both in offensive and in defensive weapons (see, \cite{GSPON1987A,GSPON1988-,THEE-1988-,GSPON1987B} and references therein).

 Conversely, if some major scientific or technical advance is made in accelerator technology, such as very-high-efficiency superconductive acceleration and radio-frequency generation, very-high-brightness laser-acceleration of neutral-particles, etc., neutral particle beams may find again their leading position as a potential rapid hard-kill system, simply because of the intrinsic strongly-interacting nature of high-energy particles.\footnote{The abstract of a typical comparative study (possibly biased in favor of particle beams) summarized this fact as follows: ``This report explores the role of directed energy weapons (DEWs) in theater defenses. For ranges shorter than $200 - 300$ km they are much cheaper than space-based interceptors (SBIs); they are competitive with ground-based interceptors (GBIs). For inter-theater ranges of $\approx 1000$ km, lasers are competitive with the SBIs, but NPBs are significantly cheaper than either. For nominal laser and space-based interceptor (SBI) costs, lasers are strongly preferred for ranges under $300 - 500$ km. For ranges 700 km, SBIs have a slight advantage. Neutral particle beams (NPBs) appear dominant for ranges over $400 - 1000$ km'' \cite{CANAV1991-}.}

\chapter{Charged particle beams propagation experiments}
\label{cpb:0}

\section{ATA at Lawrence Livermore National Laboratory}
\label{ata:0}

The purpose of the construction and operation of the Advanced Test Accelerator facility (the ATA) at the LLNL, as well as its basic characteristics, have been described in a number of informal, e.g., \cite{LLNL1978-, BARLE1981A, BARLE1981B, B.M.S1982-}, and more technical, e.g.,  \cite{COOK-1983-, JACKS1983-, REGIN1983-}, papers and reports.  Summarizing from these publications:

\begin{quote}

The main uncertainty in the concept of charge-particle beam weapons is whether it is feasible to propagate an intense self-focused electron beam through the atmosphere.  That is, an electron beam held in tight focus by its own magnetic field.  To conduct a comprehensive program of electron-beam propagation experiments, LLNL has constructed between 1978 and 1982 a 50 MeV, 10 kA, linear accelerator (the ATA) at its high-explosive test location, \emph{Site 300}, which is well equipped for managing experiments with unusual hazards. 

Together with its associated program of beam propagation physics, the ATA represents the largest single component of the Defense Advanced Research Agency (DARPA) particle-beam technology program, whose aim is to establish the feasibility of particle beam weapons.  The prime goal of the Department of Defense particle-beam technology program is to resolve what is and is not possible in beam propagation.   Accordingly, the goal of the ATA is to develop an experimental capability that can resolve critical questions about beam propagation physics in a timely and cost-effective fashion.

A first generation of particle-beam weapons will emphasize short-range applications.  Potentially first applications  of particle-beam weapons may be for the defense of large ships against cruise missiles.  Another early use may be for terminal ballistic missile defense of hardened sites such as missile silos or national command authority centers.  The short range, which reduces the sensor burden on search and fire-control radars, demands the high lethality that particle-beams possess.  Boring its way out to targets at a rate of a kilometer per millisecond, the beam can deposit megajoule of energy almost instantaneously.  Consistent with fire-control system considerations, beam weapons can have the capability of engaging tens of targets per second.  With such characteristics, charged-particle beams are particularly well suited to counter small, very fast, highly maneuverable threats.

Photographs of the completed 200-meter-long ATA facility, as well as drawings of the 80-meter-long underground experimental tank are shown in reference \cite{BARLE1981B}.  Also visible is the 4-meter-thick shielded door which can be moved aside for beam experiments in open air.  For such open air experiments the beam is directed towards a staging area where it may interact with various targets after propagating in free air over distances which are only limited by the topography of the ATA site, located in a shallow valley at \emph{Site 300}.

\end{quote}

The main characteristics of the ATA in relation to his primary purpose, the study of endo-atmospheric beam propagation and interaction with military targets, are as follows:

\begin{itemize}

\item  Beam energy : 50 MeV

An energy of 50 MeV means that the electron beam is fully relativistic but of an energy still substantially below the 1'000 to 10'000 MeV that are needed to have a beam power on the order of the Nordsieck power, which is the required for propagating a distance of about one radiation length in open-air in a single pulse (Figure 4.3).  This means that the experiments at ATA will be done under rather difficult conditions, which has the advantage that the results will easily extrapolate to higher energies.  In this respect, it is important to recall that the main difficulties with particle acceleration is in the low energy section: as soon as an electron beam has reached an energy of 10 to 100~MeV it is easy to inject it into a betatron and further accelerate it to higher energies.  An example of such a betatron, designed to accelerate the 10~kA ATA beam from 50 to 250 MeV, is given in reference \cite{PETER1983-}.

\item Beam current : 10 kA 

A current of 10 kA is on the order of the critical current given by equation \eqref{cri:3}, which corresponds to the maximum current for self-pinched propagation through the atmosphere a sea-level pressure.

\item Beam transverse emittance : $10^{-4}$ m$\cdot$rad

For a 50 MeV electron beam the Alfv\'en current is $I_A \approx 1.7$~MA. Assuming that $f_e=1$ and $f_m=0$ when the beam is injected into the atmosphere, the effective current is $I_E=I_B=10$~kA.  The Bennett pinch relation, Eq.~\eqref{ben:4}, gives therefore a minimum initial beam radius $\tilde{a}=1.5$~mm.\footnote{The published measurements of the ATA injector brightness, Ref.~\cite{WEIR-1985-}, yield the value $\epsilon_n \approx 0.75$~rad$\cdot$cm for the average `normalized emittance,' which is related to the transverse emittance by the equation $\epsilon_n = \gamma\epsilon_{\tr}$.}

\item Beam radius : 0.5~cm

The transverse emittance at the exit of the accelerator yields a minimum beam radius of 0.15~cm for propagation in full density air.  However, since beam conditioning before injection into the atmosphere results is some emittance growth, the nominal beam radius is generally taken as $\tilde{a}=0.5$~cm in  standard test cases of beam propagation \cite{CHAMB1981-} or hose instability \cite{LAMPE1984-} simulation programs.

\item Pulse length : 70 ns

A pulse length of 70 ns is adequate for studying nose erosion and tail losses during propagation, as well as the convective nature of major instabilities such the hose instability.

\item Burst rate : 1 kHz --- Average rate : 5 Hz

A maximum repetition rate of 1'000 Herz, and an average rate of 5 Herz, imply that every second five successive pulses separated by a time-delay between 1 and 200 milliseconds can be sent into the experimental tank or into the atmosphere. This enables to study hole boring and channel evolution, as well as tracking and stability of subsequent pulses through the low-density plasma generate by preceding pulses in reduced-density or ground-level air.

\item Lethality : 35 kJ/pulse

The is only 3.5 percent of the 1 MJ energy which would be delivered by a 100 ns, 10 kA, 1 GeV beam pulse, i.e., the energy equivalent of about 0.25 kg of TNT, which is considered to be on the order necessary to destroy a typical target.  Nevertheless, this is sufficient to test the lethality of the beam on numerous targets without completely destroying them and the sensors used in the measurements.

\end{itemize}


The accelerating principle used in the ATA is \emph{magnetic induction}, a technology traditionally used in circular accelerators such as the betatron, which was pioneered and developed for use in linear accelerators at the Lawrence Livermore National Laboratory by N.C. Christofilos \cite{CHRIS1964-}. For a general discussion of this technology, and its applications to both linear and circular accelerators in the United States and in Russia, see the review \cite{KAPET1985-}. For a more technical discussion of solid core induction accelerators, such as ASTRON, ETA, and ATA, as well as of next generation concepts such as HBTS (High Brightness Test Stand), see the report \cite{SWING1985-}.

\begin{quote}

``The successful completion and operation of the ASTRON accelerator \cite{CHRIS1964-} provided a new tool to test particle beam weapon propagation ideas under the auspices of ARPA.  This program, named SEESAW, was centered at LLNL from 1958 to 1972'' \cite[p.3]{SWING1985-}.

``The Beam Research Program at Livermore was dormant from 1972 to 1974 when it was revived by the Navy under the name CHAIR HERITAGE.  Under Navy sponsorship, LLNL built the Experimental Test Accelerator (ETA) which produced currents an order of magnitude higher than had been previously acheived. (...) Beginning with the construction of the ATA, the project was placed under the auspices of the Defense Advanced Project Agency (DARPA), which provided funding for the particle beam weapons research from 1978 to the present'' \cite[p.4]{SWING1985-}.

\end{quote}

The construction of the ATA started in 1980. The first tests of the electron gun (injector) began in November of 1982, and its full 10 kA, 2.5 MeV, beam was delivered to the ATA main accelerator in January 1983.

The ATA started operating at its full design specifications of 10~kA and 50~MeV in July 1984.  However, the beam current was found to be strictly limited by instabilities within the accelerator so that only very uniform beam pulses could be accelerated to full energy when injected into the 85-m-long main accelerator structure.


The most serious such instability, termed \emph{beam break-up} (BBU), is a very rapid growth of any beam's transverse displacement to a disruptive amplitude.  A radical cure to this problem was found and tested in the early 1985 using another lower-energy induction accelerator, the Experimental Test Accelerator (ETA), \cite{MARTI1985-, NEWSC1985-}.  The idea was to fill the accelerator with benzene gas at a pressure of $10^{-2}$ Pa, corresponding to an altitude of 80 to 120 kilometer above the Earth, and to created a plasma channel by sending a low-energy laser pulse through it.  This proved to be very successful, the channel providing an electrostatic guide for the beam all the way through the accelerator \cite{PRONO1985A}.  Moreover, this breakthrough provided a way to greatly simplify the construction, and to reduce the weight, of future linear induction accelerator.  Indeed, quoting from an October 1985 review of the ATA progress:

\begin{quote}

``Clearly, the laser guiding technology gives  a tremendous improvement in accelerator performances as well as simplifies accelerator operation and future construction (i.e., no longer needed are transport solenoid or steering magnets)'' \cite[p.3145]{PRONO1985B}.

\end{quote}

The final confirmation of the full and reliable operation of ATA by the use of the laser guiding transport technique was given on September 1986, \cite{CAPOR1986-, MILLE1987-}.  From then on the ATA could be used for what it had been built for: to investigate the feasibility of an intense charged-particle beam as an endo-atmospheric point-defense weapon.  It turned out that the answer to one basic question of principle, namely that the propagation of the beam over long distances in open air is possible, came very soon.


Indeed, in his State of the Laboratory statement of 1987, the director of LLNL was able able to highlight:

\begin{quote}

``A major accomplishment in the laboratory's beam research program was the first demonstration of open-air propagation of an electron-beam.  Using the Advanced Test Accelerator (ATA), researchers were able to tailor the electron beam to permit stable propagation in the open air'' \cite[p.3]{LLNL1987A}.

\end{quote}

In the Beam Research section of the same annual report, the following details were given:

\begin{quote}

``Recently we successfully transported a high-intensity, 50-MeV electron beam from our Advanced Test Accelerator (ATA) into conditioning and diagnostic cells and then into free air.  This test of stable beam propagation in free air is the first of its kind at this energy, level of beam current (5 to 10 kA), and pulse rate (1 Hz).  We have studied carefully the effects of beam parameters upon stability in air and have begun measurements of ancillary phenomena that will be important in assessing the practicality of using high-intensity electron beams as tactical weapons'' \cite[p.54]{LLNL1987B}.

\end{quote}

However none of the experimental details concerning these propagation experiments appear to have been published, except for the abstract of a presentation at the April 1990 meeting of the American Physical Society:

\begin{quote}

``The Advanced Test Accelerator (ATA) was designed and built by LLNL under the auspices of DARPA to examine the feasibility of stably propagating high current ($\approx 10$ kA), moderate energy ($\approx 50$ MeV) electron beams in the atmosphere.  We report on a number of experiments conducted at ATA over the past five years that studied propagation characteristics of beams at pressures ranging from 20 millitorr to full atmosphere.  At pressures above 20 torr, the most serious problem with propagation stems from the resistive hose instability.  We discuss various techniques of beam conditioning and quiteing to reduce the growth of this instability.  With a proper tailoring of the radial and emittance profile at the beam head, propagation without catastrophic hose disruption was possible over 20 meters in an experimental tank.  We also discuss the benefits and drawbacks of "laser guiding" on a photoionized benzene channel within the actual accelerator and the effects this transport scheme had upon beam conditioning and propagation'' \cite{FAWLE1990-}.

\end{quote}

From then on there will be no publications any more.  In particular, there is still no information on whether or not the beam generated by the ATA has been injected into another, most probably circular, accelerator to increase its energy from 50~MeV to 500~MeV or more \cite{PETER1983-, HUI--1984B, PETIL1987-, KAPET1991-}.  The only subsequent open publications related to the ATA are those concerned with its use as a driver for a free-electron laser, e.g., \cite{BARLE1990-}, or publications related to technological developments which may be used to upgrade the ATA or to build highly reliable and efficient components for a new generation of high-current high-energy accelerators, e.g., \cite{HELLE1999-}.

In conclusion, if we take the above statements for granted, we have to assume that the ATA project has succeeded in meeting its stated goals.  This means, in particular, that single pulses must have propagated in a stable manner in free air over a distance of at least one Nordsieck length, equation \eqref{nor:12},  i.e., about 20 meters according to the calculations made at the end of Sec.~\ref{nor:0}.  

   Morover, since ATA is able to fire five closely spaced pulses in a single burst, operation in this burst-mode must have allowed to verify that the stable propagation of a tightly focused beam is possible up to a distance of about 100~m.  According to the available drawings of the ATA facility, e.g., the cover page of the brochure \cite{LLNL1978-}, this range is about the distance between the exit port of the ATA and the focal point of the out-doors staging area.  As a matter of fact, a private communication to the authors of a paper published in 1993 claimed that ``In this particular series of ATA experiments, known as the multipulse propagation experiment (MPPE), a train of up to five 10~MeV pulses, separated by several msec, was conditioned and propagated'' \cite[p.4184]{HUBBA1993-}.

\section{RADLAC at Sandia National Laboratory}
\label{snl:0}

The main advantages of linear induction accelerators such as the ATA used for beam research, or such as the FXR, DARHT, or AIRIX used for flash x-ray radiography \cite{YU---1996-}, are their intrinsic simplicity and capability to produce high quality beam pulses under reliable conditions.  However, while such accelerators are well suited for research applications, they are very heavy and bulky, and therefore not suitable for applications in which relatively compact and light-weight accelerators are required.

One alternative technology has been successfully developed by A.I.\ Pavlovskii in the Soviet Union,  with possible applications to \emph{linear induction accelerator without iron} \cite{PAVLO1974-, PAVLO1980-}, and \emph{pulsed air-cored betatrons} \cite{PAVLO1976-,PAVLO1996-}.  This technology, which does not use ferrite- or iron-loaded cavities, was subsequently developed in the United States, and the first device based on this principle was built and successfully operated at the Sandia National Laboratory \cite{MILLE1981-}. The current and energy achieved in this accelerator, called RADLAC-I, were of 25 kA and 9 MeV with an average accelerating gradient of 3 MV/m.  This accelerating gradient is substantially larger than that of the ATA (about 0.5 MV/m) which implies that the RADLAC is a promising candidate for a compact high-power accelerator.  Moreover, the RADLAC can operate at high repetition rates since it is not affected by the classical beam breakup (BBU) instability, so that laser guiding of the beam in a laser-generated channel within the accelerator is not necessary \cite[p.1185]{MILLE1981-}.\footnote{For a general discussion of this technology, and its applications to both linear and circular accelerators in the United States and in Russia, see the review \cite{KAPET1985-}. For a more technical discussion of RADLAC, with many references, see \cite{PREST1986-}.  And for an up to date review, including the discussion of recent progress \cite{SMITH2004-}.}

   The RADLAC technology was substantially improved during the 1980s and the construction of RADLAC-II initiated. This accelerator consists of two accelerating modules called RIIM.  Using several such modules in succession, or recirculating a given pulse several times through them, a beam can be accelerated to higher and higher energies. In 1985 RIIM was capable of reliable operation at output levels of 40 kA and 9 MeV \cite{MILLE1985-}.

  With the RIIM operating, RADLAC-II could be assembled and beam propagation experiments using its 40 kA, 18 MeV beam were soon successful \cite{MAZAR1986-}.  In early 1986 the beam was extracted, without significant losses, and propagated into a magnetic-field-free, air-filled experimental tank. At a pressure of 1 atm the beam propagated straight without oscillations, and the radius was measured to be about 0.75 cm, somewhat smaller than the 0.9 cm beam radius within the accelerator.  The RADLAC-II beam was then conditioned in a 16-m-long, ion-focused region, and was allowed to propagate outside the accelerator building in open air where it propagated in a stable manner for quite a distance.   As a matter of fact, figure 9 of reference \cite{MAZAR1986-} is an open-shutter photograph of the RADLAC-II beam propagating outdoors at night.  The Manzano mountains of New Mexico are visible in the background.

Subsequent publications gave no further details on out-doors propagation experiments. Nevertheless, research and development did not stop there, as is indicated by the summary presented at the subsequent DARPA conference \cite{HASTI1986-}~:

\begin{quote}
``RADLAC program activities are reviewed. The work is broadly categorized under lead pulse stability (LPS), channel tracking, and Recirculating Linear Accelerator (RLA) activities. In LPS activities, stable, open-air propagation of the RADLAC-II beam was demonstrated over ranges longer than a Nordsieck length. These shots were coordinated with the activities of other experimenters measuring beam induced emissions, and demonstrated that RADLAC-II could be fired on a predetermined schedule to allow numerous, coordinated, and geographically widespread measurements to be made. Since those experiments, improvements in the RADLAC-II accelerator, ion-focus regime (IFR) beam conditioning cells, and matching of the accelerator beam to those cells have produced a beam which should allow greater than 20 betatron wavelengths in a Nordsieck length and saturation of hose growth to be observed. Channel tracking activities have included continued hardware development on the RADLAC-II Module (RIIM) for pulse-to-pulse channel tracking, the design of a laser for conductivity channel tracking, and demonstration of a crude beam director for a high current beam. Codes which allow channel tracking simulations to be done have also been developed. Pulsed power and beam transport experiments on the Recirculating Linac have led to hardware and techniques which will allow demonstration of beam recirculation of a high current beam this year and a recirculating linear accelerator next year. These transport schemes and pulsed power developments can be extended to higher energies and a conceptual RLA for Navy charged particle beam weapon (CPBW) applications has been developed.''
\end{quote}

\noindent As well as at the 1987 SDIO/DARPA Services Annual Propagation Review \cite{HASTI1987-}~:

\begin{quote}
``The RADLAC program encompasses high power electron beam propagation experiments and accelerator development, both for advanced propagation experiments and to develop compact accelerator options for future charged particle beam weapons (CPBW). Propagation experiments include conditioning cell and lead pulse stability (LPS) experiments on RADLAC-II, and channel-tracking experiments on IBEX. The RIIM accelerator was used for two-pulse accelerator experiments to explore two-pulse configurations for RADLAC-II. The ion-focused regime (IFR) transported, recirculating linear accelerator (RLA) experiment is aimed at future CPBW compact accelerator development. This paper briefly outlines recent work in these areas.''
\end{quote}

\noindent Indeed, a considerable emphasis of research at Sandia National Laboratory is on developing high-energy accelerators suitable for defending high-value force-projection assets such as aircraft carriers \cite[p.5]{HASTI1986-}~:

\begin{quote}
``IFR guiding is not sufficient for energies higher than the order of 10~MeV. For higher energies, a combination of IFR and strong focus sector magnets has been theoretically shown to produce the required transport. CPBW accelerator concepts based on coaxial cavities and IFR/sector-magnet beam transport have been developed.  With stable propagation, these conceptual designs can satisfy Navy mission requirements and meet volume and weight constraints (...) for a fleet defense CPBW.''
\end{quote}

   Concerning the stable open-air propagation of the RADLAC-II beam, it is remarquable that it was demonstrated in 1986 already, \cite{MAZAR1986-}, that is possibly one year earlier than the same experiment with the ATA beam, \cite[p.3]{LLNL1987A} and \cite[p.54]{LLNL1987B}.  This illustrates both the maturity of the RADLAC technology, and the level of the inter-US competition between national laboratories.

    However, as is explicitly recognized in Ref.~\cite{HASTI1986-}, and further explained in the report~\cite{EKDAH1986-} where RADLAC-II beam conditioning is discussed, stable propagation of the relatively low-energy RADLAC-II beam is not sufficient to guarantee the stability of the first pulse of a high-energy beam: ``the critical physics issue facing the use of intense relativistic beams for weapons in the lower atmosphere'' \cite[p.1]{EKDAH1986-}.  This is because the effect of the low net-current (i.e., $f_m \approx 30$--50\%) and large beam radius (i.e., $a \approx 1$--5~cm) in these experiments severely limit the number of betatron wavelengths that the beam propagates before expanding by scattering.  According to Eqs.~\eqref{ben:20} and \eqref{nor:12},  the number $N(z_N) = z_N/\lambda_{\beta m}$ of wavelengths within the first Nordsieck length of propagation scales as $N(z_N) \propto I_E^{3/2}/a$ when $P_0 < P_N$. This scaling clearly shows that large radius, low effective-current beams will expand before propagating many betatron wavelengths.\footnote{In relation to this, see the discussion at the end of Sec.~\ref{nor:0}.}  This has the effect of suppressing hose-instability growth by temporal betatron de-tuning, ``an effect that will be minimal for weapon grade beams'' \cite[p.1]{EKDAH1986-}.  Therefore, subsequent work on RADLAC II has been aimed at increasing the effective current and decreasing the beam radius by properly conditioning the beam before injection into open-air \cite{EKDAH1986-}. The goal of 20 betatron wavelengths per Nordsieck lengths should be achieved, which means that the beam will truly propagate as a Bennett pinch, and that collective effects such hose-instability can be studied in order to assess lead pulse stability.

As for the significance of the RADLAC-II open-air experiments in comparison to the corresponding ATA experiments, it should be remembered that what matters most in first order is beam power, as is shown by the elementary solution \eqref{nor:11} to the Nordsieck equation \eqref{nor:9}. For ATA the initial beam power is $P_0 \approx 0.5$~TW, and for RADLAC-II $P_0 \approx 0.72$~TW.  Therefore, according to Nordsieck length's approximation, Eq.~\eqref{nor:12}, RADLAC-II should have a single-pulse range on the order of about 30~m, instead of about 20~m for ATA.  However, if the calculations are made by computer-integrating the complete beam envelope equation \eqref{nor:1} using Moli\`ere's theory of multiple scattering \cite{BETHE1953-} and detailed energy-loss models, it turns out that both RADLAC-II and ATA have about the same $\e$-fold range of approximately 22~m.  In other words, while Nordsieck's equation provides a good first approximation (especially for beams of high-energy particles), computer simulations are indispensable in the relatively low-energy domain in which both ATA and RADLAC-II are operating.

A final important conclusion deriving from the RADLAC-II stable propagation experiments is that the plasma generation current \eqref{cri:3}, which should theoretically limit stable propagation to currents less then about 10 kA, does not appear to be so critical, since the RADLAC-II beam intensity is of 40 kA.

\section{LIA-10 and LIA-30 at Arzamas-16}
\label{lia:0}

The development of high-current electron-beam technology in the Soviet Union and Russia parallels in many ways the corresponding development in the Western and other major countries such as Japan and China.  There are even several instances of importants discoveries which were first made in the Soviet Union \cite{KASSE1975-, WELLS1986-, BYSTR1995B}.  For example, in the 1960s and 1970s, the theory of explosive electron emission. Unlike a low-current electron cold-field emission, an explosive electron emission results when a thermal explosion of micro-protusions takes place on the cathode surface.  The resulting dense plasma then serves as the main supplier of electrons into the diode acceleration gap \cite[p.67]{BYSTR1995B}.

As it originally started in the Soviet Union, e.g., \cite{PAVLO1974-, PAVLO1976-, PAVLO1980-}, and was later implemented with virtually no modifications in the United States at the Sandia National Laboratory \cite[p.70]{BYSTR1995B}, the development of iron-free linear induction accelerators, and most certainly their use for beam propagation experiments, must have continued in Russia.  Indeed, the LIA-10 accelerator of the 1970-1980s was upgraded to yield 50 kA for a 25 MeV pulse of 20~ns duration in 1993, and the construction of a new accelerator, LIA-30, producing a beam of 100 kA at 40 MeV, initiated \cite{BOSSA1993-}.  Similarly, the development of iron-less betatrons continued, and was even proposed as a new technology for the flash x-ray radiography of fast-going processes \cite{PAVLO1996-}.

However, there are no published results on long-range beam propagation experiments using these facilities. Moreover, while there is substantial published work on Russian beam propagation experiments through low-pressure air or even laser formed channels, e.g., \cite{KASSE1975-, WELLS1986-}, there appears to be no open publications available on experiments similar to those made with ATA and RADLAC.\footnote{The Rand Corporation report \cite{WELLS1986-} mentions that: ``A follow-on report will discuss Soviet research on the propagation of intense relativistic electron beams through higher-pressure air and gases ($P > 10^{-2}$ torr).''}

\section{PHERMEX at Los Alamos National Laboratory}
\label{phe:0}

A third technology suitable for making compact high-power accelerators is the more conventional standing-wave radio-frequency linac technology used in most high-energy linear accelerators built for fundamental nuclear and elementary-particles research.  Using this technology, it is theoretically possible to accelerate a 10 kA beam pulse from 10 to 1000 MeV in a 30-meter-long linear accelerator \cite{FAHEL1982-}.

This technology was in fact used in the first high-power flash x-ray facility, the PHERMEX accelerator completed in 1963, built at the Los Alamos National Laboratory to study the implosion of nuclear weapon's primaries  \cite{BOYD-1965-}.  As this facility was expected to be superseeded by more powerful induction linacs (such as the DARHT, which is now operating) it was suggested in the early 1980s that PHERMEX could possibly be upgraded and used to study the endo-atmospheric propagation of electron beams \cite{MOIR-1981-}. 

While the RF-linac technology has a few disadvantages compared to the induction-linac technology (e.g., a relatively short pulse length), it has an intrinsic high repetition rate and multi-pulse capability.  Besides, PHERMEX is located ``in a blast-proof building at a remote, controlled access site [where] a clear line of site of approximately 2000 meters exists'' \cite[p.2]{MOIR-1981-}.

In fact, some preliminary propagation experiments were performed using the available beam \cite{MOIR-1980-}, in particular to provide data \cite{MOIR-1981-} for validating theoretical models of two-stream instability \cite{NEWBE1982-}.  PHERMEX was subsequently upgraded to operate in the 20 to 60 MeV energy, up to 3~kA intensity range which was anticipated to be theoretically possible \cite{MOIR-1981-, STARK1983-}.   This enabled further data to be taken and to compare measurements at 7 and 21 MeV to a unified theory of the two-stream and filamentation instabilities \cite{LEE--1983B}.

Further publications include emittance measurements for typical 300 to 500 A, 26 MeV, 3.3 ns micropulses \cite{MOIR-1985A}; results on the first use of an ultra-violet-laser-ionized channel to guide multiple 30 MeV pulses over distances of 13.5 m \cite{CARLS1987-}; as well as several papers indicating that there are plans to inject the PHERMEX beam into a circular accelerator \cite{MOIR-1985B, GISLE1987-, PETIL1987-}. Besides increasing the beam energy, such an accelerator would have the advantage of providing a means to accumulate and condition the beam pulses before sending them into the atmosphere with a suitably larger energy, current, and duration.

There are, however, no subsequent publications on these developments.  In particular, as with LLNL's ATA, it is not known whether or not PHERMEX is now used as an injector to a higher-energy circular (or even possibly linear, see \cite[p.2]{FAHEL1982-}) accelerator, and there is no published information on related beam propagation experiments.

\section{Other electron-beam propagation experiments in the USA}
\label{usa:0}

Apart from the large dedicated accelerators such as ETA and ATA at LLNL, and RADLAC at SNL, which have been specially built to study electron beam propagation in the atmosphere, there are numerous other accelerators in the United States which like PHERMEX at LANL can be used to study various aspects of high-power beam generation and propagation physics.  These other accelerators comprise various proton and ion beam machines, of which some will be referred to in Sec.~\ref{hip:0}, as well as electron beam machines of which we mention a few examples here.  For each of them we will give their nominal current, voltage or energy, and pulse-length, in order to illustrate how this diversity of accelerators enables to explore the wide range or parameters that pertain to particle beam-weapons's propagation physics.

\subsection{DARHT ---  2 kA, 3.5 MV, 2 $\mu$s}

The DARHT (Dual Axis Radiographic Hydrodynamics Test) facility at LANL is an example of an advanced, high-resolution, 2~kA, 20~MeV, 70~ns flash x-ray machine characterized by a high quality ($\epsilon_{\tr}=1.2\times 10^{-3}$ m$\cdot$rad, $\Delta p/p < 0.01$)  beam produced by a linear induction accelerator \cite{YU---1996-}.  A feature of interest to the subject of this report is that the DARHT 3.5~MeV injector has a comparatively long pulse-duration of 2 $\mu$s, so that it can be used to study convective instabilities such as the `ion hose,' which is particularly important in the ion-focused regime \cite{BUDKE1956A,ROSE-2004-}, and which has properties similar to the `resistive hose' instability \cite{LEE--1978-}. It is expected that electron impact ionization of the residual background gas in the accelerator ($\approx 1.5 \times 10^{-7}$ torr average) will result in a fractional electric neutralization of the order of $10^{-4}$.  Even at this relatively low ion density, potential troublesome coherent transverse displacements (ion-hose oscillations) of the beam and channel can result due to their mutual electrostatic restoring forces.  However,  according to 3-dimensional simulations, it is expected that instability growth, which increases linearly from head to tail of the beam pulse, should be suppressed by nonlinear effects because the ion oscillation is several times larger than that of the beam, a conclusion that will have to be verified in actual measurements \cite{HUGHE2001-}.

\subsection{Hermes III --- 19 MeV, 700 kA, 25 ns}

Hermes III is a 13~TW, 19~MeV, 700~kA, 25~ns pulsed electron accelerator at the Sandia National Laboratory that produces intense bremsstrahlung doses and dose-rates over large areas for the study of nuclear effects induced by $\gamma$ rays \cite{SANFO1992A,SANFO1992B}.  This beam, with current near the Alfv\'en limit, was used to measure and model beam transport over distances up to 11.5~m in gas-cells filled with nitrogen spanning six decades in pressure range from $10^{-3}$ to $10^{3}$ torr \cite{SANFO1993,SANFO1994,SANFO1995}.

  The existence of two regimes of stable transport was confirmed: A low-pressure window (between $\sim 10^{-3}$ and $\sim 10^{-1}$ torr) that is dominated by propagation in the semi-collisionless ion-focused regime, and a high-pressure window (between $\sim 1$ and $\sim 100$ torr) that is dominated by propagation in the resistive collisional-dominated regime.  Below $\sim 10^{-3}$ torr, there is insufficient ionization to confine the beam; between the windows, the two-stream and hollowing instabilities disrupt propagation; and above $\sim 100$ torr, the resistive hose instability degrades propagation.

\subsection{IBEX --- 70 kA, 4 Mev, 20 ns}

IBEX is a 70~kA, 4~Mev, 20~ns electron accelerator at the Sandia National Laboratory that has been used to study intense electron beam hollowing instabilities, which although routinely observed in axisymmetric computer simulations of beam propagation in air, are not often seen in experiments because of competing non-axisymmetric effects \cite{EKDAH1985-, GODFR1987-}. 

   Previous experiments were made with the 10~kA, 1.5~MeV, FX-25 accelerator at the Lawrence Livermore National Laboratory \cite{LAUER1978-,JOYCE1983-}, the 40~kA, 1.5~MeV, FX-100 accelerator, and the 20~kA, 1.1~MeV, VISHNU accelerator, both at the Air Force Weapons Laboratory. Attempts to understand the results of these experiments were not very successful, especially with regards to the theoretical prediction that the hollowing instability threshold should scale as the time derivative of the beam current divided by the air density.  For this reason the IBEX experiment was undertaken, leading to good agreement between experiment and simulation, therefore validating existing hollowing instability scaling laws \cite{JOYCE1983-,EKDAH1985-, GODFR1987-}.

\subsection{MEDEA II --- 13 kA, 1.2 MV, 2 $\times$ 10 ns}

An essential requirement for endo-atmospheric systems, making possible to extend their range from a few hundred meters to a few kilometers, is to be capable of sending a burst of carefully timed pulses through a channel guiding them towards the target.  This requires accelerators capable of generating consecutive pulses at a flexible high-rate, and experiments to optimize the guiding effect of preceding pulses on subsequent ones.  In view of this, as mentioned in Sec.~\ref{ata:0}, the ATA has been designed to produce such bursts of pulses.  But there is no published information on any related or any other propagation experiment performed with the ATA.  Similarly, as mentioned in Sec.~\ref{snl:0}, a two-pulse configuration for RADLAC II has been investigated, but it is not known whether it has been implemented and used to study multi-pulse channel-tracking. 

  It is therefore interesting that the results of a two-pulse experiment, performed at McDonnell Douglas Research Laboratories (MDRL), has been published \cite{WELCH1990-}, while a similar experiment using the Pulserad 310 electron beam generator at the Naval Research Laboratory (NRL) \cite[Ref.6]{WELCH1990-} was published only two years later \cite{MURPH1992-}.

  The MDRL experiment was performed with the MEDEA II electron-beam generator, which consists of two pulse lines in series, each independently charged, which produce two independent 1.2~MV electron beam pulses from the same diode with interpulse delays as short as 0.2~ms.

  Typical reported `density-channel tracking' guiding experiments with MEDEA II consist of sending a first pulse into air at a pressure between about 250 and 550 torr, which produces a channel with a 2.2~cm radius with 26\% density reduction.  The second pulse is then injected after a delay of 1.75 ms.  The higher temperatures found in a channel of reduced density produce a greater electron-neutral momentum-transfer collision frequency $\nu$ and, hence, according to Eq.~\eqref{bcn:7}, a reduced conductivity, which result in a reduced plasma return current in that region.  This results into a magnetic guiding force which unlike the electric guiding force in `conductivity-channel tracking' has only a weak dependence on channel radius and depth.

   The magnetic guiding effect observed with MEDEA II is therefore the process of importance for guiding particle beam pulses through high-density air, which in the case of the MEDEA II beam is maximum at an ambient pressure of 400 torr, and negative at pressures below 250 torr.  At pressures above 550 torr, air scattering increases the beam radii and reduces guiding.  As the first pulse reduces the channel density by about 25\%, this implies that a third pulse would find nearly optimum guiding conditions.

\subsection{Pulserad 310 --- 5--10 kA, 1 MeV, 35 ns}

   In 1987-88 the NRL performed a series of beam `density-channel tracking' experiments that were publish in 1992 only  \cite{MURPH1992-}.  In these experiments a technique  different from that used with MEDEA II was applied because the Pulserad 310 generator is a single pulse accelerator.  Instead of studying a beam tracking a channel produced by a preceding pulse, the Pulserad 310 experiment measured the magnetic attraction between a channel pre-formed by a laser beam and a beam pulse launched along a trajectory parallel to the channel, but offset by a varying amount relative to the channel axis.  These experiments confirmed the existence of the density tracking force, but were difficult to perform and interpret since propagation along a channel offset by just one or two cm was likely to amplify instabilities that destroyed the beam, or made the interpretation of the measurements ambiguous.

  Beam conditioning prior to injection into the atmosphere was therefore essential, especially to reduce the level of perturbations that seed the resistive hose instability, and to introduce head-to-tail taper in the beam radius in order to detune that instability \cite[p.3409]{MURPH1992-}.  These experiments were therefore as much a success in demonstrating the existence of the density-channel tracking force, as in showing the paramount importance of beam conditioning.

\subsection{Febetron 706 --- 5 kA, 0.4 MeV, 3 ns}

The Febetron 706 is a relatively old pulsed electron beam generator, originally manufactured by Field Emission Corporation \cite{CHARB1967-} and later by Hewlett-Packard Corporation  \cite[Ref.11]{GREEN1985-}, both at McMinville, Oregon. The particularity of this accelerator is to produce a pulse of only $\sim$3~ns duration, approximatively ten times shorted that generally used in beam propagation experiments.  With such a short pulse-length a number of instabilities have no time to develop, while leading and falling edge effects such as beam head erosion and tail loss are exacerbated.  This enables to make a number of useful measurements, which were carried out in either a 7.6~cm diameter, 300~cm long glass drift tube, or in a 3.4~m diameter, 6~m long vacuum chamber, at McDonnell Douglas Research Laboratories (MDRL) \cite{GREEN1985-}.

   In particular, measurements confirmed the existence of a relatively high-pressure propagation window between about 2 and 8~torr; and showed that propagation over distances between 80 and 180~cm appears to be limited by erosion of the beam head at the lower pressures (so that the ion-focused regime propagation window could not be observed) and by loss of the beam due to sausage and hose instabilities at higher pressures, with the loss of tail predominant.  This led the experimenters to carefully measure the beam nose erosion rate as a function of pressure, which today remain the only published experimental data on the beam front velocity of a high intensity electron beam propagating in pinched mode \cite[Table~I\,]{GREEN1985-}.  As a matter of fact, the availability of this data seems to have escaped the attention of the authors of later publications on beam head erosion, except those of the French paper \cite{BOUCH1988-}.

\subsection{Stanford Mark III --- 10 A, 42 MeV, 4 ps}

While the accelerators so-far mentioned in this sections were all high-current but relatively low-energy machines, the last one is an example of a low-current but comparatively much high-energy accelerator typical of the many radio-frequency linear-accelerators built for research in nuclear and elementary particles physics --- the Mark~III accelerator at Stanford University \cite{FISHE1988-}.

The 10 A, 42 MeV, 4 ps beam pulses of the Mark III are three to four orders of magnitude shorter than those of the accelerators considered in the previous subsections.  The current is also $10^3$ to $10^4$ times lower, although the current density is similar, with the electrons forming a bunch 1~mm in diameter and 1.2~mm long.  Nevertheless, the beam was observed to propagate without serious degradation through 1~m of hydrogen at pressures from $10^{-3}$ to 1.25~atm, a property which can be attributed to the short time scale of the pulse.  In particular, the beam current was fully transmitted through the gas, with little or no pinching and with no evidence of a reverse current or instabilities.  The only significant effect of the gas was an emittance growth consistent with multiple scattering from the neutral gas molecules.

Therefore, since the conclusions of this experiment with the Mark~III can be extended to much higher beam energies without restriction, it  can be considered as a proof of principle experiment showing that short-pulse, GeV to TeV energy, electron or proton beams are able to propagate over very-large distances through the upper-atmosphere, and for that reason are potentially usable as outer-space beam weapons.



\section{Propagation experiments in other countries}
\label{oth:0}

In the previous sections we have mostly discussed the main beam propagation experiments that have been, or that are still being, carried out in the United States.  This is because these experiments are the most visible and possibly the most advanced in the world.  

    In fact, many technologically advanced countries could in principle carry out high-power beam propagation experiments.  This is because the required electron-beam generators and accelerators have characteristics similar to those used in flash x-ray radiography, whether for industrial or military purposes \cite{HUGHE2001-}.  The list of potential countries would therefore include all those having developed nuclear weapons, and a few others such as, in particular, Japan, South Korea, Germany, and Poland.

   Other countries which could potentially have an interest in developing particle beam weapons are those where missile defence programmes are underway.  According to the \emph{SIPRI Yearbook 2004}, these include, in 2003, the United States, Israel, and Russia; India and South Korea expressed interest in developing their own missile defences; and Japan announced an ambitious plan to develop a multi-layer missile defence system in cooperation with the USA \cite{SIPRI2004-}.  

   However, apart from the United States and Russia, the only country to have regularly published papers on research (explicitly supported by a defense-funding agency) which can unambiguously be related to particle beam weapons is France, e.g., reference \cite{BOUCH1988-} in which 10 kA beams with energies of 50 to 500 MeV are considered.

   In the case of Japan, which is well known to keep up with all new technological developments, and which has several powerful high-intensity electron and ion accelerators, world-class research is being done on all aspects of high-energy beam technology --- without being shy about referring to published work done at foreign weapons laboratories, e.g., \cite{TAKA1988-}.  This includes active participation in space-based beam propagation experiments, the development of powerful electron and ions generators for particle-beam fusion, early studies of the ion-focused regime in collaboration wuth a Dutch group \cite{YAMAG1982-}, beam propagation experiments in channels \cite{OZAKI1985-}, pioneering self-pinched proton beam propagation experiments in collaboration with a German team at the Karlsruhe Light Ion Fusion (KALIF) facility \cite{HOPPE1996-}, extensive development of the linear and circular induction accelerator technology, ambitious very-high energy accelerator programs, etc.

  Finally, while the United States and France are the only countries with an explicitly stated particle beam weapons research program, a possible exception is India.  Indeed, in the year 2000, ``the (Indian) Department of Atomic Energy intend(ed) to use part of its additional 2,270 million rupees to develop intense electron-beam machines that can potentially knock out enemy missiles'' \cite{JAYAR2000-}.  In fact, India has already a number of high-current electron beam generators, such as the 20 kA, 0.3 MV, and 200 Joules ``Kilo-Ampere Linear Injector'' (KALI-200), which was used to study beam propagation and current enhancement in the ion-focused regime in a 1~m long drift tube \cite{PAITH1988-}.\footnote{The name of the first author of this paper is mentioned in the preprint distribution list of the report \cite{UHM--1982R}, published in 1982, confirming the long term interest of Indian scientists in the subject.}  Similarly, Indian scientists have published a number of theoretical and experimental papers showing that they are closely following what is being done abroad in this context, e.g., \cite{IYYEN1989-,VENUG1992-}.


\section{High-intensity proton and ion beams}
\label{hip:0}

So far in this chapter we have only considered experiments in which high-energy, high-intensity \emph{electron} beams are injected into a gas or the atmosphere.  The main reason for this is that the technology for generating and accelerating electron beams is much more mature and readily available then the corresponding technology for heavy-ion or proton beams.  This is partly the consequence of political technological factors such as the military need for high-power electron machines for applications like flash x-ray radiography and nuclear weapons effects simulation, and partly the consequence of fundamental physical factors such as the large mass difference between electrons and ions, which makes that generation and acceleration of ion beams are generally much more difficult than of electron beams --- especially in the low-energy sections of the machines where protons and heavy ions are non-relativistic.

   Nevertheless, in view of the potentially larger range of \emph{proton} versus electron beams (see Figures~\ref{fig:exp} and~\ref{fig:ran}) and of applications such as ion-driven thermonuclear fusion, plasma heating, and laser pumping, research and development of high-current proton beam sources, and of accelerators suitable for accelerating such beams to high-energy, are under way since the early 1970s.  This effort is dominated by activities in the United States and Japan, and is followed at a much lower level by a few other countries.  According to a 1988 survey, there are at least three other countries which have developed high-current proton sources with powers of at least 1~TW : France, Germany, and Poland \cite{COOK1988-}.  This list does not include Russia, because (apart from some limited developments) Soviet pulsed power research never made the transition from high-current electron beams to high-current ions beams, as did the United States \cite[p.69]{BYSTR1995B}.

   This effort led to a number of basic concepts and developments which are described in several reviews (e.g., \cite{NATIO1979-, GOLDE1981-, KEEFE1981-, YU---1996-}) and books (e.g., \cite{OLSON1979-}, \cite[Sec.~2.4 and Chap.~7]{MILLE1982-}).  Basically, there are three general methods for generating ion beams with currents in the kiloampere range: (1) accumulation and pulse compression of lower intensity beams, (2) collective acceleration, and (3) diode-like sources.

\subsection{Accumulation and pulse compression}

In most relatively-low-intensity accelerator systems used in fundamental nuclear or elementary-particle physics research, bursts of particles generated at low-energy are stacked and accelerated to higher energies by linear or circular machines which at the same time greatly increase the peak current within the pulses.  State-of-the-art conventional accelerators (RF linacs, synchrotron, storage rings) have produced proton currents larger than 100~A at many GeV (storage rings) and about 1~A at 100~MeV (RF linacs), albeit in multi-purpose machines that are in general of very large size and weight.  In colliding beam machines, the peak currents can even reach many kA, and the energies many 100~GeV.

   Using similar techniques of particle accumulation and processing, as well as by combining separate beams \cite{YU---1996-}, it is possible to design dedicated high-intensity high-energy machines producing beams able to propagate over distances on the order of one or more Nordsieck lengths in dense gases, or over much longer distances in tenuous plasmas typical of outer-space conditions.  It is also possible to generate high-power beams of exotic particles (antiproton, muons, positrons, etc.) starting from relatively low-intensity initial sources.  Let us give two examples, where the beam particles are electrons and respectively protons:

  1. The electron-gun of the PHERMEX accelerator at Los Alamos National Laboratory produces micropulses with current of about 350~A and duration of about 3~ns, i.e., containing an average charge of about 1~$\mu$C.  One proposal for obtaining a much higher current beam is to inject 100 such pulses accelerated to 20~MeV into a modified betatron \cite{MOIR-1985B,TAGGA1984-}, so that the total charge would add to 100~$\mu$C.  This corresponds to an initial betatron current of 35~kA, which would be amplified to 140~kA at the exit of the betatron, where the energy would increase to about 60~MeV, and the pulse length shrink to about 0.7~ns.

   2. Various type of medium intensity (i.e., 10 to 100~A) ion sources have been successfully developed for tokamak neutral beam heating.  Starting from a 50~keV, 100~A, 3~$\mu$s proton beam extracted form such a source, a proposal has been put forward at the Naval Surface Weapons Center (NWSC, White Oak, Maryland) to accelerate this beam to 5~MeV with a proton induction linac, in order to inject it into a betatron-like device \cite{NAMKU1985-}.  A longitudinal pulse compression by a factor of 100 in this betatron would yield a current of 10~kA, and a pulse length of 30~ns, suitable for an endoatmospheric beam weapon.  However, a similar compression may also be achieved by the precise modulation of the voltage driving a number of induction modules in a linear accelerator, a technique that is actively developed in Japan to achieve 10~kA, 10~GeV, current and energy levels at the exit of the accelerator \cite{HORIO1998-}.

\subsection{Collective acceleration}

This concept uses in various ways an unneutralized high-intensity electron beam to accelerate ions to energies much higher than the electron energy \cite{OLSON1979-}. If the beam electron density is large enough, a sizable electrostatic potential well is formed.  Positive ions with kinetic energy less than the well depth are trapped in the well, and, if the well is accelerated, so are the ions \cite{GOLDE1981-}.\footnote{This and various related processes may also be used to accelerate clusters of ions \cite{DOGGE1969-}.}  This acceleration mechanism was discovered in 1970 in early experiments in which electron beams were propagated through a gas-filled region \cite{RANDE1970-}, and was later used as a method (mentioned in Sec.~\ref{pfa:0}) for producing plasmoid beams, e.g., \cite{DESTL1988-}.

   However, as can be anticipated by elementary considerations, the ion current in collective accelerators is generally much lower than that of the driving electron beam, and there are great practical difficulties in loading the ions in the well and keeping them trapped during their acceleration \cite{GOLDE1981-}.  This had the consequence that many experiments started in the late 1970s early 1980s, see Refs.~\cite[Chap.7]{MILLE1982-} and \cite{KEEFE1981-}, were not very successful.

   On the other hand, the basic idea of using a high-intensity low-energy electron beam to accelerate another beam to much higher energies has survived, and will probably be an essential ingredient of future ultra-high energy accelerators, such as the `two beam' and `wake field' accelerators that will be discussed in Sec.~\ref{uhe:0}.

\subsection{Diode-like sources}

The method which at present appears to be the most economical to generate kA to MA ion beam pulses, is the one which simply consists of inverting the polarity of a diode so that positive ions rather than electrons (and possibly negative ions)  are extracted from the cathode.  In the most simple configuration, a one-dimensional theoretical model predicts that the currents of nonrelativistic electrons and ions are related by the expression \cite{GOLDE1981-}, \cite[p.57]{MILLE1982-}
\begin{equation} \label{hip:1}
\frac{I_i}{I_e} = \sqrt{\frac{m_e}{m_i}},
\end{equation}
so that, for a proton source, the ratio of proton to electric current is only 2.3\%.  Consequently, various techniques (such as `reflexing,' `pinching,' or `magnetic insulation') for the suppression of the unwanted electron flow had to be developed in order to improve efficiency \cite{NATIO1979-,GOLDE1981-}.  This implies than ion sources are necessarily more complex than electron sources, what, in view of the complexity of the methods presented in the two previous subsections, corroborates the empirical fact that any electron-beam device is generally simpler than any proton- or ion-beam device of similar current and/or energy. 

Nevertheless, motivated by several applications in which beams of protons or ions are superior to beams of electrons, considerable effort has been devoted to developing relatively efficient high-intensity proton sources with about 1~MeV energy.  A typical pulsed power machine, primarily designed for generating electron beams, which has been successfully applied to the generation of ion beams is the GAMBLE II pulser at the Naval Research Laboratory.  In 1976, charge-neutralized beams of 30--50~ns duration consisting of 150--200~kA, 0.5--0.8~MeV protons were routinely produced \cite{STEPH1976-}.  Another example is the 6.0 kA, 1.1 MV proton beam obtained at Cornell University Laboratory of Plasma Studies \cite{ROTH-1985-}.  This beam, which was better than 98\% charge neutral (and could therefore have been used for plasmoid beam research), was primarily produced to study beam generation and transport to a possible high-energy linear induction proton  accelerator, a technology that is under development in the United States \cite{IVERS1981-, YU---1996-} and in Japan \cite{KAWAS1983-, HORIO1998-, HASEG2000-}.

   However, in the United States, the leading laboratory for research on high-intensity light-ion beams is the Sandia National Laboratory, where beam species such as H or Li are accelerated to more than 10~MeV at the 36~beams,  100~TW, Particle Beam Fusion Accelerator (PBFA II) facility \cite{JOHNS1989-}, as well as at the more recent SABRE and PBFA-X accelerators \cite{CUNEO1998-}.  A major challenge in this program, which goes back to the mid-1970s, has been the development of a diagnostic package that can adequately measure the parameters of such intense ion beams \cite{MCDAN1988-}. In part, this difficulty is due to the fact that the intense beams generated on PBFA are nearly 100\% space-charge neutralized and >70\% current neutralized. This precludes many electrical measurements invoking charge collection or the measurement of self-magnetic fields.  Another aspect of the difficulty is that the diagnostics must be able to operate in hard (several MeV) x-ray bremsstrahlung backgrounds of some $10^9$ to $10^{10}$~rad/s produced by electron losses in the ion diodes.  

\subsection{High-intensity proton beam propagation experiments}

  As we have said in the introduction to this section, and seen in the previous subsections, the development of high-current proton machines is considerably more difficult than that of electron machines, especially if the goal is to develop compact low-cost devices rather than big expensive facilities for fundamental research.  It is therefore not surprising that a number of basic propagation experiments, which have been made long ago with electron beams, have only recently been made with proton beams.  This is the case, for instance, of the transport of high-intensity proton beams in a Bennett pinch state, where ``the first long distance ion beam self-pinched transport experiments have been carried out'' in 1999 only \cite[p.356]{OTTIN2000-}.  In these experiments the proton beams propagated a distance of 50~cm, i.e., a factor of ten improvement over the only previous self-pinch transport experiment, carried out by a German-Japanese team at the Karlsruhe Light Ion Fusion (KALIF) facility, in which a proton beam propagated a distance of only 5~cm \cite{HOPPE1996-}.

   The essence of the difficulty with propagating low-energy proton beams in a gas is that ion-beam-induced gas ionization is substantially more rapid than electron-beam-induced gas ionization at the same energy.  This is immediately seen in Bethe's stopping power formula, Eq.~\eqref{nor:5}, which shows that ionization at low-energy scales with the inverse squared power of the velocity, that is (for a given kinetic energy) in direct proportion to the ion to electron mass ratio --- a factor of 1836 for a proton beam.  Therefore, a non-relativistic proton beam injected in a gas will rapidly charge and current neutralize, so that its effective current --- and consequently the pinch force --- will be zero.  This led to the conclusion that a non-relativistic ion beam could not propagate as a Bennett pinch, and no such transport experiments have been attempted \cite{ROSE-1999-}.

   However, by carefully studying and modeling ion-beam-induced gas ionization, including beam-ion-impact ionization,  secondary-electron-impact ionization, gas breakdown, late-time Ohmic heating, and full gas chemistry, e.g., \cite{YOUNG1994-, WELCH1996-}, it was found that ion beam self-pinched transport is possible for a limited range of gas pressure \cite{OLIVE1999-}.  Indeed, using a high-current focused proton beam produced by the GAMBLE II pulsed-power accelerator,  the self-pinched transport of a 100~kA, 1.1~MeV proton beam was demonstrated, and found to be in good agreement with simulations that predicted self-pinching in a pressure window between 35 and 80~mtorr helium \cite{OTTIN2000-}.  In this experiment, the propagation distance of the 5~cm radius proton beam was 50~cm.

   These propagation experiments should not be confused with numerous earlier experiments that demonstrated, in the United States, e.g., \cite{OLSEN1982-,NERI-1993-},  and in Japan, e.g., \cite{OZAKI1985-}, efficient propagation of 50--500~kA, 1~MeV protons beams over distances of 2--5~meter in z-discharge channels.  In these `discharge channel transport' experiments a high-conductivity channel is pre-formed in a gas by a laser (or some other means) and a z-discharge, to create a frozen magnetic field before the ion beam is injected.  Neither should they be confused with the `neutralized ballistic transport' mode in which an ion beam is neutralized to reduce its space-charge expansion prior to injection into an evacuated chamber \cite{ROY--2004-}.  The key advantage of the `self-pinched transport' mode demonstrated in Ref.~\cite{OTTIN2000-} is that it minimizes the use of ancillary equipments to transport an intense low-energy beam from a source to an accelerator (e.g., an induction linac) in which it may be accelerated to a much higher energy.

  To conclude this section, we recall that the physics of proton beam propagation is fundamentally well understood and similar to that of high-intensity electron beams, but more complex because ultra-relativistic approximations are inappropriate \cite{LAMPE1987-}.  There also significant differences in the details due to the difference in the sign of the electric charge.  Some of these have been discussed in Secs.~\ref{inv:0}, \ref{bcn:0}, \ref{mag:0}, and \ref{pla:0}, and are further discussed in Ref.~\cite{OTTIN2000-} and references therein.  However, none of these differences are such that they could prevent the propagation of high-energy, high-intensity proton beams if electron beams of similar characteristics are shown to be able to propagate successfully.  This also enables to have full confidence in computer calculations, such as simulations of beam erosion of 10~kA, 0.1 to 2~GeV proton beam pulses propagating in full pressure air \cite{ROSE-2002-}, which confirm that the possibility of using proton instead of electron beams is a serious option.

  Therefore, in the case of an equal availability of high-power electron and protons beams, the decisive factor will be the beam interaction with the target, where the lethality is primarily due to an electromagnetic cascade in the former, and a nuclear cascade in the later case, which have different properties depending on beam energy and target composition.

\section{High-intensity muon beams}
\label{him:0}

Since about twenty years considerable research is underway for producing high-intensity low-emittance \emph{muon} beams for fundamental research \cite{NEUFF1999-}.  If the techniques used for producing these beams could be extrapolated to yield sufficiently powerful pulses of muons, a single such pulse could in principle propagate (according to Figures~\ref{fig:exp} and~\ref{fig:ran}) to ranges of up to several km in open air. 

   The main problem with muons, however, is that they are short-lived particles, with a life-time $\tau_{\mu}$ of only 2.2 microseconds at rest.  Nevertheless, if the muons are accelerated, say to 1000 MeV (which since the mass of the muon is $m_{\mu} = 105$ MeV/c$^2$ corresponds to a Lorentz factor of $\gamma \approx 10$) they would be able to propagate to a distance of $\gamma c \tau_{\mu} \approx 6.6$ km before decaying.  Therefore, in theory, a series of high-energy muon beam pulses could be used either to strike a target at a distance of tens of kilometers, or else to bore a channel in the atmosphere to guide a more powerful electron or proton beam to a distant target \cite{GSPON1987B}.

   Another potential application of muon beams is that of warhead/decoy discrimination.  This is because contrary to electrons or protons, muons are not much absorbed when they pass through an object. They just loose some kinetic energy as given by Bethe's term and ohmic losses in formula \eqref{nor:4}, and they are slightly deflected by multiple scattering as given in first approximation by Rossi's formula \eqref{nor:2}.  Thus, a high-energy muon beam emerging from the rear side of an object will have a somewhat smaller kinetic energy and a somewhat larger emittance, i.e., a larger Bennett pinch radius according to equation \eqref{ben:4}.

   Therefore, if the beam-sensing system used to direct and track the muon beam is sensitive enough to measure small variations of its radius (or emittance), it is possible to measure the amount of heavy material contained in the object.\footnote{A direct measure of emittance is provided by the Cherenkov light emitted by a beam as it traverses the air.}  In this case, the muon beam could be used to probe a potential target at the same time as it could bore a channel to guide a more powerful and destructive proton or electron beam.  In fact, muon-radiography using cosmic-ray muons has recently been proposed as an alternative to x-ray radiography for detecting nuclear weapons possibly hidden in large containers. The principle is basically the same as for warhead/decoy discrimination: By measuring the amount of deflection, the object's density can be reconstructed, rather as x-rays reveal varying density through differing amounts of absorption \cite{BOROZ2003-}.

While such concepts may look very futuristic at present, they should nevertheless be seriously assessed in view of several synergistic factors which relate muon beam technology to other advanced beam technologies of importance to the subject of this report.  For example, there is strong similarity between the techniques used to produce muons (see, e.g., chapters 2 to 4 in reference \cite{NEUFF1999-}) and those used to produce antiprotons (see, e.g., \cite{GSPON1987A}) :  In both cases the simplest method consists of striking a production-target with a sufficiently high-energy proton beam to generate copious amounts of either muons or antiprotons during the collisions, to collect as many as possible of these particles with a magnetic lens, and to decrease their kinetic energy dispersion (i.e, to `cool' them) in order to form a sufficiently low-emittance beam, which can then be stored or directly used for some purpose.\footnote{Another link between muon and antimatter technology, which was already mentioned in Chap.~8, is that antimatter is the sole portable source of muons, in the sense that upon annihilation every antiproton yields on average between 2 and 3 muons, which could be fed to a collecting and cooling system without needing a high-energy beam to produce the muons in the first place \cite{GSPON1987B}.}  Moreover, if instead of producing the muons with a high-energy proton beam, a high-intensity electron beam is used instead, one is led to a system involving a multi-megawatt muon production-target driven by a multi-kiloampere electron beam, which requires the kind of technologies and experience associated with accelerators such as the ATA \cite{BARLE1994-}.\footnote{The first author of this reference led the ATA project for a number of years, see \cite{BARLE1981A,BARLE1981B}. It should be remarked that in an electron-beam based system, the same high-power electron accelerator could be used as a driver for either a free-electron-laser or a muon-factory, which would both provide a means to guide the more powerful electron-beam towards a distant target. For more examples of such synergies, see, e.g., \cite{BARLE1986-}.}

In practice, the kind of muon production and cooling systems envisaged or under construction at present, e.g., \cite{NEUFF1999-, GEER-2001-}, are not expected to be fully operational before 2006 to 2010.  Also, their muon yield will be orders of magnitudes below what would be needed to envisage concentrating a bunch of them into a pulse suitable for propagation in self-pinched mode over a significant distance in open air.  Nevertheless, this situation is quite similar to that of antiproton `factories,' which have increased (and are still increasing) their output by many orders of magnitude over the years.

\section{Ultra-high-energy particle beams}
\label{uhe:0}

Ultra-high-energy particle beams have specific merits because the large kinetic-energy per particles, i.e., $\gamma \gg 1$, is favorable in terms of stability and reduced bending in an external magnetic field.  Even though such beams have a reduced intensity for a given total power, they have a strong potential for applications in outer-space, and possibly even more for sending the very high energy beam straight up through the atmosphere.  For example, the report \cite{BLOOM1983-} considers a GeV-energy proton beam sent from ground to space (through an exponential atmosphere) with platforms/targets located at altitudes of 0, 15, 100, 400, and 32'000 km.

Until recently the main practical obstacle to considering ultra-high-energy particle beams has been the size of accelerators needed to generate them.  But with the demonstration of the feasibility of table-top laser-driven particle accelerators, see Fig.\,\ref{fig:db}, very compact accelerators became a reality.  Since theses first proof-of-principle experiments the generation of stable GeV electron beams from a centimeter-scale accelerator has been confirmed \cite{LEEMA2006-}.  The technology has now been adapted to accelerate light-ions \cite{HEGEL2006-} and protons  \cite{FUCHS2006-,ROBSO2007-}. 

The development of compact laser-based particle accelerators is a rapidly evolving field which is attracting large global interest.  This interest is motivated by their potential for civilian applications in fundamental research \cite{KATSO2004-} and medicine \cite{ROBSO2007-}, as well as in defense applications such as inertial confinement fusion \cite{HEGEL2006-}, radiography of dense objects \cite{FUCHS2006-}, and beam weapons.  It is a typical example of a dual-purpose technology in which the potential for civilian applications is highlighted, whereas the potential for military applications is minimized in open scientific publications.


\chapter{Conclusion}
\label{con:0}

In this report we have reviewed the theory and the proof-of-principle experiments which have been carried out to demonstrate that high-intensity high-energy particle beam propagation in full or reduced-density atmosphere was possible, and that the range and stability of the beam pulses were in agreement with theory.

We have, however, not analyzed the practical consequencies of these conclusions, which determine with certainty the range of physical parameters compatible with realistic high-power directed-energy particle beam systems;  neither have we investigated the R\&D related to the possible construction and deployment of such systems. These are the subjects of a companion report,\footnote{\emph{Particle beam weapons: A review and assessment of current R\&D}, report ISRI-82-05.  (This report has never been completed.)} of which only a small part, concerned with the radiological effects of directed high-intensity high-energy electron beams, has so far been published \cite{GEER-1983-}.

Nevertheless, by going through the theory and the details of several important experiments, we have followed several complete `research cycles,' that is several paths going from some initial ideas to their verification by means of a suitable experimental program.  In the present case the outcomes of these research cycles can be qualified as `scientifically successful,' meaning that the technical feasibility of some concepts have been established.  While this is important, it does not mean that the associated technologies should necessarily be developed and applied, which is precisely the motivation for writing this report.\footnote{``And let us not forget that a great breakthrough in military technology, like the invention of the H-bomb, can quickly come back to haunt us.'' Hans Bethe, Physics Today, October 1978, p.13.}


\chapter{About this report}
\label{abo:0}

 My interest in the topic of this report originates from the accidental discovery, in the Spring of 1978, of the existence of PBW research by spotting highly visible classification stamps on a unusually looking report that was inadvertently left open on a desk by physicist Burton Richter\footnote{Burton Richter (1931--) received in 1976 with Sam Ting the Nobel prize for the discovery of the $J/\Psi$ particle.  For an appreciation, see P.A.\ Moore, \emph{Honouring Burton Richter}, CERN Courier (April 2000) 23--24.} who at the time was working at CERN in the same collaboration as me, and sharing the office of my group leader, Valentine L.\ Telegdi.  This was very troublesome since I had not the slightest idea that elementary particle physics and particle accelerators could have military applications. This was something that I never heard in any lecture or seminar I had listen to, or read in any text book or scientific paper.  Moreover, this discovery was almost unthickable considering that at the age of thirty I was just coming back from the University of Chicago after completing my PhD experiment at Fermilab, and had chosen to work in fundamental research precisely because it claimed to be totally unrelated to any practical and, \emph{a fortiori}, military applications...  

On that day I went to the library and started looking for documents which could prove or disprove that there were military applications to particle physics and accelerators.  Soon I found dozens of references, including in pure science journals, showing that there were many such applications.  The reason why I had never taken notice of them is that the military connection was generally indirect, or hidden in footnotes, acknowledgments, or the affiliations of the authors.  After a few hours I became an expert in transposing pure-science papers into their real-world context, and it was clear that the particle-beam weaponry mentioned in Richter's classified report was not science fiction!

But this was not enough for me, I also wanted to know on which side Richter was sitting: Was he just interested in this subject by curiosity?  Was he a critic and an opponent to such military developments?  Or was he somebody actively collaborating with the military establishment?  I therefore tried to find a reference to a paper that would associate his name to some obvious military technology.  This is how I found his name on a `JASON report' assessing the prospect of heavy-ion driven inertial fusion, reference \cite{CALLA1977-}, i.e.,
\begin{quote}
C.G. Callan, Jr., R.F. Dashen, R.L. Garwin, R.A. Muller, B. Richter, M.N. Rosenbluth, \emph{Heavy-ion-driven inertial fusion}, JASON report JSS-77-41 (Stanford Research Institute, Arlington, Va., 31 October 1977) 14\,pp.
\end{quote}

It took me more than six months to obtain a copy of this report, which was the first confirmation that Richter was not just working on high-energy physics, i.e., `pure science' as is supposedly done at CERN or SLAC, but was also actively involved in military related research.  Moreover, what was most shocking for me in the list of authors of that report was that the names of all but one of them (M.N. Rosenbluth\footnote{Marshall N. Rosenbluth (1927--) is a well known plasma physicist who's name is on many thermonuclear and particle beam weapons physics papers. For a short biography see \emph{Hannes Alfv\'en prize to Marshall N. Rosenbluth}, Europhysics News (July/August 2002) 146. }) were well-known to me and to any particle physicists of my generation:  Callan for his work with Gross on electron-parton scattering, Dashen for calculating the neutron-proton mass difference, Garwin for his contributions to the discovery of parity violation, Muller for his theory of inclusive reactions, Richter for the discovery of the $J/\Psi$ particle.  It was as if my heroes suddenly turned into traitors of the cause of pure science...\footnote{Much later I learned, from the overwhelmingly negative reactions of my former colleagues, that I had also become a traitor in their eyes:  The normal attitude of scientists is to ignore such things as the military implications of their work, and to emphasize the luminous side of science.  Breaking the silence on the negative implications means becoming a pariah in the scientific community.  To follow his moral inclinations or the path leading to professional recognition is the dilemma of the whistle blower.}

I got the JASON report from Peter Jenni, now spokesman for the ATLAS experiment at the CERN Large Hadron Collider, who after working in the same group as me at CERN went to SLAC to collaborate with Burton Richter who had returned to the US.  He sent me the JASON report on May 9, 1979, with the remark:
\begin{quote}
``Dear Andre, I am sorry to have you had waiting so long for the JASON report.  But I was extremely busy writing the $e^+ ~ e^- \rightarrow e^+ ~ e^- ~ \eta'$ paper.  Furthermore, it was not so simple to get a copy of the report.  Finally, Burton gave me the report, from which I send you a copy. [...] Sal\"u, Peter.''
\end{quote}

   I received this document at about the time Iraqi engineers came to CERN and enquired about the technology of the rather unique and large magnet used in the experiment I was working on, a technology which could only be of interest to them if they were intending to use electromagnetic isotope separation for producing enriched uranium for a nuclear weapon \cite{ERKMA2005-}.  It meant that between Spring 1978 and Spring 1979 I had learned more than enough about the military implications of particle-accelerator technology to decide to leave CERN.  While my vocation had always been to spend my life working on pure science, these events convinced me that if the science I was working on, and the techniques I was using, were directly related to existing or new types of weapons I should make --- in the interest of peace and disarmament --- these connections known to other scientists and to the public.

  This is how, as a result of discovering the existence of particle beam weapons research and of Iraq's interest in large-scale electromagnetic isotope separation, I decided to try to create, on the model of the `Stockholm International Peace Research Institute' (SIPRI), a `Geneva International Peace Research Institute' (GIPRI).  With considerable support from a number of CERN colleagues and from Frank Barnaby the Director of SIPRI, and in collaboration with a few University of Geneva academics and Geneva personalities, GIPRI was founded at the end of 1979.

  The initial research program of GIPRI included a collaboration with SIPRI on the military applications of particle accelerators, e.g., particle beam weapons, simulation of nuclear weapons effects, and use of particle accelerator technology in the nuclear fuel cycle \cite{GSPON1983-}.  Thus, the first GIPRI paper to be submitted to a high-impact scientific journal, i.e., \emph{Nature}, was co-written with a SIPRI researcher (see pages \pageref{p-F} to \pageref{p-L}).

  The GIPRI-SIPRI collaboration also included visits to and from SIPRI.  This led me to get the definitive confirmation that 1976 Nobel laureate Burton Richter had been doing classified work on particle beam weapons in the late 1970s, and that the classified report that I saw in 1978 at CERN was indeed related to particle beam weapons.  This happened in 1981 when I was working at SIPRI as a visiting scientist on a early version of the first part of the present report.  Going through unclassified military journals that were not available in libraries such as CERN's I could read, in the July/August 1977 issue of \emph{Electronic Warfare},
\begin{quote}

[...] Both the Soviets and the US military have been investigating particle beam weapons for more than a decade.

Dr.\ John L.\ Allan, deputy for research and advanced technology at DDR\&E, recently told the executive session of the Senate Armed services Committee: ``Particle beams --- beams of electrons, for example --- are not directly affected by weather and may provide longer ranges that the HEL (high energy laser) in adverse weather'' (\emph{EW}, May/June, p.\,12). [...]

A nuclear particle beam weapon would have one decided advantage over lasers: Much more of the energy consumed by the device goes into the beam than in the extremely inefficient HEL.  However, focusing the beam is more difficult and imprecise.  Propagation through the atmosphere involves ``nasty, difficult plasma physics'' according to Burton Richter, a recent Nobel prize winner at the nonmilitary Stanford Linear Accelerator Center (SLAC) in Palo Alto, CA.  Richter would be surprised if anything of this sort could be made to travel through the atmosphere for long distances.  There would be ``very sever problems with the stability of the beam.''  It would be extremely unstable, darting about th earth's fluctuating magnetic fields ``like a lightning bolt.''  Allan echoes this by saying: ``Charged particle beams have a tendency to be unstable.  They are also deflected by magnetic fields, so pointing and tracking uncertainties exist.'' But he believes that: ``If these problems can be solved, a viable weapon system could result.''  \cite[p.\,31-32]{EW-JF1977-}.

\end{quote}

Later, in the course of the study of particle beam weapons, and of other advanced weaponry, I came across many more papers authored by well known elementary-particle physicists, of all nationalities and of both sides of the `Iron curtain.'  For a physicist particularly fond of fundamental research like me, possibly the most shocking example, however, is that of 1979 Nobel laureate Steven Weinberg (1933--) who published in 1967 a 35-page long paper on the stability of long-range high-intensity particle beams, in which he explicitly acknowledged that:
\begin{quote}
``This work was performed by the author as a member of the Jason Division of the Institute for Defense Analyses, Arlington, Virgina'' \cite[p.\,635]{WEINB1967A}.
\end{quote}
This paper, directly related to the particle beam weapons research (code-named `seesaw') at the Lawrence Livermore and Los Alamos National Laboratories, was written in 1967.  This must have been the `golden year' of Steven Weinberg:  In that year, at the age of 34, he published his most famous physics paper, \emph{A model of leptons} \cite{WEINB1967B}, as well as a 55-page long secret Jason report advising the U.S.\ Government at the highest level on the ``military consequences of a U.S. decision to use tactical nuclear weapons in South East Asia'' \cite{DYSON1967-}.

    In conclusion, the present report is an example of a `science-based technology assessment' which, in the spirit of  `peace-research,' is not written as a classified document for exclusive use by governmental executives, but as an open document available to the scientific community and to the public at large.  Indeed, it is a strong conviction of the author that scientific and technological development can only benefit a democratic society if decisions on future weapons are made on a truly informed basis.

\newpage


\noindent{\Huge {\bf Appendix}}
\label{p-F}

\vspace{2\baselineskip}

\noindent One of the first papers I wrote after leaving CERN and creating GIPRI at the end of 1979 was in collaboration with Bhupendra Jasani, a physicist working at SIPRI.  It was a commentary on a review article entitled \emph{Particle beam weapons --- a technical assessment} by MIT scientists G.\ Bekefi, B.T.\ Feld, J.\ Parmentola, and K.\ Tsipis, published in \emph{Nature} {\bf 284} (20 March 1980) 219--225.  Our commentary, \emph{Particle beam weapons: A need for re-assessment}, was submitted to \emph{Nature}.  But it was never published despite a long argument with the Editor, the MIT scientists, and their mentor Victor F.\ Weisskopf who happened to be also a member of GIPRI's honorary committee.

Since the \emph{Nature} review article and its companion paper in \emph{Scientific American} {\bf 240} (April 1979) 38--49 were quite influential in dismissing a truly science-based discussion of the subject, it is perhaps important to recall that Bekefi, Feld, Parmentola, and Tsipis's  were invited at the 1980 PBW Workshop of the U.S. Air Force Academy \cite{GUENT1980-}, and given the possibility to deliver a key-note address:

\begin{quote}

``The workshop was opened with overviews of DoD's interest in particle beam research and development by Dr.~George Gamota, and a perspective on the viability of particle beams as weapons by Dr.~John Parmentola of MIT (see, John Parmentola and Kosta Tsipis, \emph{Particle Beam Weapons}, Scientific American {\bf 240} (4) 54, April 1979).  These were followed by invited presentations in each of the five areas which defined the working panels of this meeting: Power generation and Conditioning; Sources and Injectors; Accelerators; Propagation; Beam/Material Interactions'' \cite[p.\,1]{GUENT1980-}


\end{quote}
Over the years Bekefi, Feld, Parmentola, and Tsipis's papers have been regularly cited in `military-professional' papers, usually with a touch of irony.  For example:

\begin{itemize}

\item Ref.\,1 of \cite{MURPH1987-}, `interactions of beams with channels;'

\item Ref.\,1 of \cite{WELCH1990-}, `tracking with MEDEA II;'

\item Ref.\,4 of \cite{FERNS1991A}, `tracking with Pulsread 310;'

\item Ref.\,60 of \cite{HUBBA1993-}, `beam conditioning.'

\end{itemize}

\vspace{2\baselineskip}

\begin{center}

{\Large \bf Particle beam weapons:}

{\Large \bf \emph{A need for re-assessment} }

\vspace{1\baselineskip}

{\large Andre Gsponer and Bhupendra Jasani}

{\large GIPRI-80-04 ~~ (9 August 1980) ~~ Submitted to \emph{Nature}}

\end{center}

\noindent  Recently a number of articles have been published in various scientific, technical and military journals on the possible use of high energy particle beams as weapons [1*]. One of the most recent was a review article published in \emph{Nature} early this year [2*].  Although one of the few objective articles written on the subject, it contained a number of technical errors.  Since the consequences of such weapons, should they become a reality, are so far reaching, an accurate as well as objective analysis of the state of the technology and its future prospect is very important.  In this letter, therefore, we wish to point out some errors and omissions made in the \emph{Nature} article so that any future studies on the subject do not suffer from possible, if only slight, loss in credibility.

   For instance, the beam dispersion calculations are done using formulas outside their range of validity; the conclusion on endo-atmospheric proton beam propagation can be misleading; the heating of the air in the region of the beam comes mostly from ionization losses, even in the case of electrons the bremsstrahlung's contribution is negligible; and the figure used for the atomic density of air at STP is incorrect so that errors are introduced in several calculations.

   The article also suffers from a lack of adequate discussion on the endo-atmospheric use of particle beams (i.e., for point defense against cruise-missiles and incoming re-entry vehicles) and their target damage capabilities.

   The problems of beam propagation in the atmosphere are complex and different particles (e.g., electrons, protons, and ions) of similar kinetic energies behave differently with different propagation ranges so that they have to be treated separately.  In the relative vacuum of outer space, propagation is mainly affected by the Earth's magnetic field and, therefore, discussion in this case are confined only to neutral particles (i.e., hydrogen atoms).

   As for the propagation of high energy particles in the atmosphere, contrary to the suggestion made in the \emph{Nature} article, a proton beam, in principle, has better atmospheric propagation properties than an electron beam.  We have numerically integrated the full equation of the beam expansion theory by Lee and Cooper [3*], taking beam energy loss and beam intensity loss into account.  According to our estimates, a proton beam pulse with an intensity of 10 kA and a kinetic energy of 1000 MeV would propagate in air at standard temperature and pressure up to 350 m.  After this distance the beam breaks up.  Under the same conditions, a similar pulse of an electron or an alpha ion beam would propagate to about 150 m.  For 10 kA and 10'000 MeV beam pulses, we find ranges of 200, 500, and 600 m, respectively for electron, alpha ion, and proton beams.

   In the same \emph{Nature} article, little consideration is given to the intense electromagnetic (or nuclear) cascade that is generated by a beam of electrons (or protons) as it propagates through the atmosphere or interacts with the target.  In the case of an electron beam, the electromagnetic shower produced by a 100 ns long, 10 kA, 1000 MeV beam would result in a radiation dose of about 2000 Rem per pulse within a diameter of 10 m around the beam axis and over a distance of about 400 m.  A few such pulses would thus direct towards the target a narrow radiation beam of an intensity sufficient to upset the electronics or to kill people, without the beam having to score a direct hit.

   Moreover, contrary to laser or heavy ion beams, which interact with the target only at its surface, electron, proton, or light ion beams can penetrate the target and thus inflict potentially more severe damage.  This is particularly important as it renders shielding against such beams more difficult.  Furthermore, a beam of protons in the 1000 MeV energy range on hitting a heavy target would generate a nuclear cascade with a substantial flux of spallation and evaporation neutrons [4*].  If the target contains a nuclear warhead, the neutrons could enter the fissile material causing the atoms to fission and so generate large amounts of heat.

   All these various aspects show the complex nature of such weapons and point to the fact that apart from the long range strategic applications discussed in the \emph{Nature} article, short range tactical applications of these weapons should also be considered.  This is particularly needed because in this mode particle beam weapons may be less prone to countermeasures.  Also, a proton or light ion beam has some potential as an ABM system and may compare favorably with other fast and short range ABM systems.

   The use of particle beams as weapons is an old idea, probably going back to World War Two [5*].  Even at this time it was realized that one of the major problems was the question of beam stability.  Very extensive theoretical studies were therefore conducted in the 1960's on this question by many prominent U.S.\ physicists working in collaboration with the JASON division of the Institute of Defence Analysis [6*].  It was found that the propagation of charged particle beams through ionized gases may be hindered by many possible instabilities.  It was not until the construction of powerful high energy accelerators that these theories were put to test experimentally.  It was discovered in 1967 that stable beam propagation of short but intense pulses of charged particles were possible through air, but at a reduced density corresponding to a few Torr [7*].  Similar propagation experiments were subsequently performed in the U.S.S.R.\ investigating the stability question at atmospheric pressures at well [8*].

   These results, together with the considerable development of accelerator technology led to renewed speculations on the military potential of particle beams, and triggered an extensive coverage of the subject [1*,15*].  It may be premature to draw any final conclusion on the feasibility of particle beam weapons, but the rather pessimistic conclusions drawn in the \emph{Nature} article do not seem to be supported by the large amount of effort been put at present into the research and development on such weapons and related subjects.

   Research directly related to particle beam weapons includes further testing of high intensity beam propagation in air at various pressures at the Naval Research Laboratory [9*], the study of reduced gas channel formation by pulsed laser [10*], the development of new accelerators such as the 50 MeV, 10 kA induction linac at the Lawrence Livermore Laboratory [11*], or the autoresonnant accelerator (supported by the U.S.\ Army Balistic Missile Defense Advanced Technology Center) [12*] which is expected to accelerate protons to an energy of 1000 MeV in a few tens of meters [13*].  Fot fiscal year 1981, approximately \$35 million is budgeted for particle-beam technology by the Department of Defense.  The major objective of this program is to demonstrate the feasibility of stable exo- and endo-atmospheric propagation of high power beams [14*].  Probably similar efforts are being made in the U.S.S.R.\ also [15*].

   Research indirectly related to particle beam weapons includes of course that devoted to laser beam weapons and the considerable resources being poured into the development of inertial confinement fusion devices triggered by high-energy lasers and various particle beams.  Similarly, the construction of large and powerful accelerators for both military and civilian purposes will surely help to solve many technical challenges implied in the construction of accelerators suitable for beam weapons applications.

   Finally, it is worth noting that the current proliferation of accelerator technology has serious implications on several other arms control issues.  The development of particle beam weapons, particularly their tactical use on Earth and their long range use in outer space by technologically advanced nations will certainly add a new dimension to the arms race.  It is this very important to properly re-asses the beam weapons question, especially in view of the clear need of international agreement to prevent the misuse of the modern particle accelerator technology.

\vspace{2\baselineskip}

\noindent{\Large \bf References}
\label{p-L}

\begin{enumerate}

\item[1*] C.A. Robinson, Aviat. Week \& Space Technol. (November 13, 1978) and references therein.\\ 
R.L. Garwin, Bull. Atom. Scient. {\bf 34} (1978) 24.\\
J. Parmentola and K. Tsipis, Scient. Am. {\bf 240} (1979) 54.\\
P. Laurie, New Scientist (26 April 1979) 248.\\
D. Richardson, NATO's Fifteen Nations {\bf 24} (January 1980) 66--70.

\item[2*] G. Bekefi, B.T. Feld, J. Parmentola, and K. Tsipis, Nature {\bf 284} (1980) 219.

\item[3*] E.P. Lee and R.K Cooper. Particle Accel. {\bf 7} (1975) 83.

\item[4*] P. Grand, Nature {\bf 278} (1979) 693.

\item[5*] N. Wade, Science {\bf 196} (1977) 407.

\item[6*] S. Weinberg, J. of Math. Phys. {\bf 8} (1967) 635.

\item[7*] W.T. Link, IEEE Trans. Nucl. Sci. {\bf 14} (1977) 777.

\item[8*] L.I. Rudavov et al., JETP Lett. {\bf 15} (1972) 382.

\item[9*] A.W. Ali et al;., Journal de Physique {\bf 40} (1979) C7--773

\item[10*] Several abstracts submitted to the 21th Annual Meeting of the Division of Plasma Physics, Bull. of the A. Phys. Soc. {\bf 24} (1979) 921.

\item[11*] Energy and Technology Review, report UCRL-52000-79-9 (Lawrence Livermore National Laboratory, 1970) 16.

\item[12*] B.B. Godfrey and R.J. Faehl, report LA-7567-MS (Los Alamos National Laboratory, 1978).

\item[13*] W.A. Proctor and T.C. Genoni, J. Appl. Phys. {\bf 49} (1978) 910.

\item[14*] Physics Today (June 1980) 84.

\item[15*] C.A. Robinson and P.J. Klass, Aviat. Week \& Space Technol. (July 28 and August 4, 1978)

\end{enumerate}



\chapter{References}
\label{bibl:0}

\end{document}